%% file: main.tex
\documentclass[12pt,a4]{article}
\usepackage{latexsym,amssymb,amscd}
\usepackage{lscape,fancyhdr,array,stmaryrd,euscript,wrapfig,cite}
\usepackage{ifthen,empheq}
\usepackage{sidecap}
\usepackage{rotating,framed}
\usepackage{ifpdf}
\usepackage{tikz}
\usepackage{multirow}
\usetikzlibrary{calc,fadings,decorations.pathreplacing}
\usepackage{verbatim,hyperref}
\renewcommand{\theequation}{\arabic{section}.\arabic{equation}}
\input{YoungDiagrams.tex}

\input{notation.tex}

\input{figures.tex}

\newcommand{\mysection}[1]{\section{#1}}
\newcommand{\mysectionstar}[1]{\section*{#1}}
\newcommand{\mysubsection}[1]{\subsection{#1}}

\begin{document}
\renewcommand{\thefootnote}{\fnsymbol{footnote}}
\begin{flushright}
AEI-2014-006\\
NSF-KITP-14-008
\end{flushright}

\vspace{1cm}

\begin{center}
{\bf \LARGE
Elements of Vasiliev theory
}

\vspace{2cm}

\textsc{V.E. Didenko\footnote{didenko@lpi.ru}
and E.D. Skvortsov\em${}^*$\footnote{skvortsov@lpi.ru}}

\vspace{2cm}

{\em${}^*$Lebedev Institute of Physics, Moscow, Russia }
\vspace*{.5cm}

{\em${}^\dag$  Albert Einstein Institute, Potsdam, Germany}

\vspace{1cm}
\end{center}

\vspace{0.5cm}
\begin{abstract}
We propose a self-contained description of Vasiliev higher-spin
theories with the emphasis on nonlinear equations. The main
sections are supplemented with some additional material, including
introduction to gravity as a gauge theory; the review of the
Fronsdal formulation of free higher-spin fields; Young diagrams
and tensors as well as sections with advanced topics.
The shortest route to Vasiliev equations covers 40 pages.

The general discussion is dimension independent,
while the essence of the Vasiliev formulation is discussed on the
base of the four-dimensional higher-spin theory. Three-dimensional
and $d$-dimensional higher-spin theories follow the same logic.
\end{abstract}

\newpage
\renewcommand{\thefootnote}{\arabic{footnote}}
\setcounter{footnote}{0}

\tableofcontents
\newpage


\mysection{Introduction}

One of the goals of quantum field theory is to explore the
landscape of consistent theories. Normally, but not always, one starts with some
set of free, noninteracting fields and then tries to make them
interact. Given the fact that the free fields are characterized by
spin and mass we can ask the following question: which sets of
fields specified by their spins and masses admit consistent
interaction? Massless fields of spins $s=1,\frac32,2,...$ being
gauge fields are of particular interest because their interactions
are severely constrained by gauge symmetry. The well known
examples include: Yang-Mills theory as a theory of massless
spin-one fields; gravity as a theory of a massless spin-two;
supergravities as theories of a number of spin-$\frac32$ fields,
graviton and possibly some other fields required for consistency;
string theory which spectrum contains infinitely many fields of
all spins, mostly massive being highly degenerate by spin.

There is a threshold value of spin, which is $2$. Once a theory
contains fields of spins not greater than two its spectrum can be
finite. If there is at least one field of spin greater than two, a
higher-spin field, the spectrum is necessarily infinite,
containing fields of all spins. String theory is an example of
such theory. The Vasiliev higher-spin theory is the missing link
in the evolution from the field theories of lower spins, $s\leq2$,
to string theories. The Vasiliev theory is the minimal theory
whose spectrum contains higher-spin fields. Its spectrum consists
of massless fields of all spins $s=0,1,2,3,...$, each appearing
once in the minimal theory, and in this respect it is much simpler
than string theory. The important difference of Vasiliev theory in
comparison with string theory is that while the latter has
dimensionful parameter $\ga'$, the former does not being the
theory of gauge fields based on the maximal space-time symmetry.
This higher-spin symmetry is an ultimate symmetry in a sense that it cannot result from spontaneous symmetry breaking. The Vasiliev theory
therefore has no energy scale and can be thought of as a toy model
of the fundamental theory beyond Planck scale.

In this note we would like to give a self-contained review on some
aspects of higher-spin theory. The subject dates back to the work
of Fronsdal, \cite{Fronsdal:1978rb}, who has first found the
equations of motion and the action principle for free massless
fields of arbitrary spin. His equations naturally generalize those
of Maxwell for $s=1$ and the linearized Einstein equations for
$s=2$. At the same time, as spin gets larger than two (in
bosonic case) the Fronsdal fields reveal certain new features. It
is a subject of many no-go theorems stating that interacting
higher-spin fields cannot propagate in Minkowski space. As a
yes-go result is available thanks to Fradkin and Vasiliev,
\cite{Fradkin:1986ka, Fradkin:1986qy, Fradkin:1987ks} we are not
going to discus these theorems in any detail referring to
excellent review \cite{Bekaert:2010hw}. The crucial idea that
allowed one to overcome all no-go theorems was to replace flat
Minkowski background space with the anti-de Sitter one, i.e. to
turn on the cosmological constant. That higher-spin theory seems
to be ill-defined on the Minkowski background is not surprising
from the point of view of absence of the dimensionful parameter.
In other words the interaction vertices that carry space-time
derivatives would be of different dimension in that case. The
$AdS$ background simply introduces such a dimensionful parameter,
the cosmological constant.

A canonical way of constructing interacting theories is order by
order. In the realm of perturbative interaction scheme one begins
with a sum of quadratic Lagrangians for a given spectrum of spins
and masses (these are zero since we consider gauge fields) and
tries to deform them by some cubic terms while maintaining the
gauge invariance, then by some quartic terms, etc. The gauge
transformations perturbatively get deformed as well. If a nontrivial
solution to cubic deformations is found we are said to have cubic
interaction vertices. A lot is known about cubic interactions both
in the metric-like approach of Fronsdal \cite{Berends:1984rq,
Berends:1984wp,Metsaev:1993mj,
Metsaev:1993ap,Metsaev:2005ar,Metsaev:2007rn,Metsaev:2012uy,Manvelyan:2010je,
Manvelyan:2010jr,Boulanger:2006gr,Boulanger:2008tg,
Bekaert:2010hp,Taronna:2010qq,Joung:2011ww,Joung:2012hz,Henneaux:2012wg,Zinoviev:2008ck,Buchbinder:2012iz}
and in the frame-like approach of Vasiliev and Fradkin-Vasiliev
\cite{Fradkin:1986ka, Fradkin:1986qy, Fradkin:1987ks,
Vasilev:2011xf, Alkalaev:2002rq,Zinoviev:2010cr, Alkalaev:2010af,
Boulanger:2011qt,Boulanger:2012dx,Boulanger:2011se}.

The cubic level, however, is insensitive to the spectrum of
fields, i.e. given a set of consistent cubic couplings among
various fields one can simply sum up all cubic deformations into a
single Lagrangian, which is again consistent up to the cubic
level. For the simplest case of a number of spin-one fields one
finds that the cubic consistency relies on some structure
constants $f_{abc}$ being antisymmetric. The Jacobi identity
telling us that there is a Lie algebra behind the Yang-Mills
theory arises at the quartic level only.

The technical difficulty of a theory with at least one higher-spin
field is that it is not even possible to consistently consider an
interaction of a finite amount of fields, \cite{Berends:1984rq,Metsaev:1991mt,Bekaert:2010hp,Boulanger:2013zza}. Altogether, this makes
Fronsdal program (construction of interacting theory for
higher-spin gauge fields) extremely difficult to implement. At
present, the traditional methods of metric-like approach  has led
to little progress in this direction, basically their efficiency
stops at the level of cubic interaction, see however
\cite{Metsaev:1991mt, Taronna:2011kt}. This state of affairs made
it clear that some other tools were really relevant to push the
problem forward.

A very fruitful direction initiated by
Fradkin and Vasiliev and then largely extended by Vasiliev that
eventually foster the appearance of complete nonlinear system for
higher-spin fields, \cite{Vasiliev:1990en, Vasiliev:1990vu, Vasiliev:1992av, Vasiliev:1995dn, Vasiliev:1999ba, Vasiliev:2003ev}, is the so called unfolded approach, \cite{Vasiliev:1988xc, Vasiliev:1988sa}, to dynamical
systems. It rests on the frame-like concept rather than the
metric-like and the differential form language which makes the
whole formalism explicitly diffeomorphism invariant. This is the
branch of higher-spin theory that we want to discuss in these
lectures. As we will see the concept of gauge symmetry
intrinsically resides in the unfolded approach and, therefore, it
suits perfectly for the analysis of gauge systems.

The other advantage
is that being applied to a free system it reveals all its
symmetries and the spectrum of (auxiliary) fields these symmetries
act linearly on. Particularly, the higher-spin algebra is
something that one can discover from free field theory analysis
using the unfolded machinery. This is one of the corner
stones towards the nonlinear higher-spin system. Nevertheless, the
unfolded approach is just a tool that controls gauge symmetries,
degrees of freedom and coordinate independence while containing no
extra physical input. It turned out to be extremely efficient for
higher-spin problem providing us with explicit nonlinear
equations, still it gives no clue for their physical origin.

The Vasiliev equations, \cite{Vasiliev:1990cm,Vasiliev:1990en,Vasiliev:1990bu,Vasiliev:1992av,Vasiliev:2003ev},  are background independent. The $AdS$
vacuum relevant for propagation of higher-spin gauge fields arises as some
particular exact vacuum solution, while propagation around other
vacuums have received no physical interpretation so far, neither
the geometry of space-time is known. Another issue about the
unfolded formalism is its close relation to integrability. In its
final form the space-time equations acquire zero-curvature
condition which states that space-time dependence gets
reconstructed from a given point in a pure gauge manner. Although
in principle any dynamical system can be put into the unfolded
form, in practice it is rarely possible to do so explicitly.
Surprisingly, the unfolded form of unconstrained Yang-Mills
$(s=1)$ and gravity $(s=2)$ in four dimensions is still not known unlike the complete
theory of all massless spins.

We consummate these lectures with the equations of motion for
nonlinear higher-spin (HS) bosonic fields which are known in any space-time
dimension. A substantial progress in these theories has been
achieved in lower dimensions, $d=3$ and $d=4$. These are the
dimensions where the two-component spinor formalism is available.
Not only it simplifies formulation of the equations, it as well
reduces technical difficulties in their analysis drastically. We
restrict ourselves to the simplest case of four dimensional
bosonic system in these notes. The main reason why this
simplification really takes place is due to unconstrained
spinorial realization of HS algebras accessible in lower
dimensions. To give an idea how it works, we briefly touch on HS
algebras. The effect of spinorial realization is to a large extent
similar to the one in General Relativity in four dimensions when using
Newman-Penrose formalism.

There are plenty of topics in higher-spin theory left untouched in
this review. Particularly, we do not discuss questions on the
structure of higher-spin cubic vertices, string inspired,
\cite{Francia:2006hp,Francia:2010qp,Sagnotti:2010at}, and
BRST-type formulations \cite{Buchbinder:2007vq,Fotopoulos:2010ay, Fotopoulos:2008ka,Dempster:2012vw}. We totally left
aside issues related to alternative parent type formalisms,
\cite{Barnich:2004cr,Alkalaev:2008gi,Barnich:2010sw}, developing
side by side with the unfolded approach. And even within the
Vasiliev theory there are a lot of interesting problems that we
consciously evade. Among those are problems of $AdS/CFT$
correspondence, \cite{Sezgin:2002rt,
Klebanov:2002ja,Giombi:2009wh, Giombi:2010vg}, and related
subjects of exact solutions,
\cite{Prokushkin:1998bq, Sezgin:2005pv, Sezgin:2005hf, Didenko:2009td,Iazeolla:2011cb,Didenko:2012tv}. This field
has recently received great deal of attention especially in $d=3$,
\cite{Campoleoni:2010zq, Henneaux:2010xg, Gaberdiel:2010pz, Ammon:2011nk,
Hijano:2013fja, Campoleoni:2011hg}, and develops rapidly. We
believe it deserves a separate review, the beginning of the story
is in \cite{Gomez:2013tta} with the up-to-date summary in \cite{Gaberdiel:2012uj}.

Our goal was to make the reader familiar with the apparatus of
unfolding approach which sometimes gives an impression as being
bordered on tautology with unexpected power for unconstrained
spinorial systems like those in three and four dimensions. Finally
and most importantly we wanted to introduce the Vasiliev nonlinear
equations and the technique to operate with them. In our own
perception of the field the absence of underlying physical grounds
and strictly speaking the absence of derivation of the equations
themselves have always been a great source of confusion. It made
us wonder of any smooth way of presenting the subject. In writing
this self-contained and non-technical review we tried to emphasize
the issues we had problems with ourselves while studying
higher-spin theory. The review looks quite lengthy, but it is the
price we pay for being elementary and we believe that some parts
can be dropped depending on the reader's background.

There is a number of reviews devoted to different aspects of
higher-spin theory, \cite{Vasiliev:1999ba, Sorokin:2004ie,
Bekaert:2005vh, Giombi:2012ms, Sezgin:2012ag}. These notes are
based on the lectures on Vasiliev higher-spin theories given by
the authors at the Galileo Galilei Institute, Florence. We
appreciate any comments, suggestions on the structure of the
lectures as well as pointing out missing references. Please feel
free to contact us in case you have any questions or found some
places difficult to understand.

The outline of these notes is as follows. We begin with two
sections that are not directly related to the Vasiliev higher-spin
theory --- the review of the Fronsdal formulation and introduction
to the frame-like formulation of gravity. The logic of the rest of
the sections is first to convert the Fronsdal theory into the unfolded
form, Section \ref{sec:UnfldGravity} for the gravity case and Section
\ref{sec:UnfoldingSpins} for arbitrary spin fields. Once a number
of unfolded examples is collected we turn to a more abstract
description of what the unfolded approach is, Section
\ref{sec:Unfolding}, where we emphasize its relation to Lie
algebras and representation theory. The reason why lower
dimensions are more tractable is thanks to exceptional
isomorphisms, of which we need $so(3,1)\sim sl(2,\mathbb{C})$. The
dictionary for tensors of $so(3,1)$ and spin-tensors of
$sl(2,\mathbb{C})$ is explained in Section \ref{sec:SpinVectDict}.
In Section \ref{sec:AdSLinear} using the vector-spinor dictionary
we reformulate the unfolded equations found for any $d$ in Section
\ref{sec:UnfoldingSpins}, which allows one to switch the
cosmological constant on easily. The $AdS_4$ unfolded equations
for all spins already contain certain remnants of higher-spin
algebra, which is discussed in Section \ref{sec:HSAlgebra}. With all
ingredients being available we proceed to Vasiliev equations in
Section \ref{sec:VasilievEquations}.

The shortest route that ends up at the Vasiliev equations, which should suit an experienced reader better, covers about 40 pages only and includes Sections
\ref{sec:AdSLinear}, \ref{sec:HSAlgebra} and \ref{sec:VasilievEquations}.

There are also a number of extra sections, e.g. the one devoted to
the MacDowell-Mansouri-Stelle-West formulation of gravity, which
are not necessarily needed to proceed to Vasiliev equations. Other
extra sections are devoted to more advanced topics. There is also
a number of appendices containing our index conventions and an
introduction to Young diagrams and tensors which is of some
importance since the higher-spin theory is first of all a theory
of arbitrary rank tensors.



\mysection{Metric-like formulation for free HS fields}\label{sec:Fronsdal}\setcounter{equation}{0}
In this section we collect some useful facts about metric-like
description of spin-$s$ fields, \cite{Fang:1978wz, Fronsdal:1978vb, Fronsdal:1978rb,Fang:1979hq}. Following traditional field theory the subject is pretty standard and has been reviewed many times,
see e.g. \cite{Sorokin:2004ie,Francia:2011qa,Campoleoni:2012th}. For the sake of simplicity we deal with bosonic
fields only. Description of fermions is in many respects
qualitatively similar. Although interacting massless higher spin
fields are believed not to exist in flat space-time, which is a subject of many no-go theorems, see e.g.
\cite{Bekaert:2010hw} for a review, the free fields do --
making it useful firstly to discuss the case of spin-$s$ fields in
Minkowski space and then proceed with (anti)-de Sitter.

\mysubsection{Massless fields on Minkowski background}

Standard way of thinking of {\it massless fields} is as those that
admit local gauge invariance responsible for a reduced number of
physical degrees of freedom as compared to nongauge, massive,
fields. This is generally true except for matter fields, $s=0,
1/2$ which being nongauge still can be massless.  In Minkowski
space-time a massless spin-zero field $\phi(x)$ is the one that
obeys $\square \phi(x)=0$, i.e. has zero mass-like term. Spin-one field
is a gauge field that is described by a gauge
potential\footnote{Lowercase Latin letters $a,b,c,...$ are
for the indices in the flat space, which are raised and lowered
with $\eta_{mn}=diag(-,+,...,+)$.} $A_m(x)$ obeying Maxwell equation \be \square A_m-\pl_m
\pl^n A_n=0\ee which remains invariant under local gauge
transformation $A_m\sim A_m+\pl_m\xi$. These are the well known
examples of lower spin massless fields $s=0,1$ which, as we will
see, naturally fit the general spin-$s$ free field description
developed first by Fronsdal, \cite{Fang:1978wz, Fronsdal:1978rb}.
Still, these lower spin examples are to some extent degenerate
exhibiting no special features characteristic for $s\geq 3$.

The less degenerate case of spin-two field can be described by a
symmetric tensor $\phi_{mn}=\phi_{nm}$ with gauge symmetry $\delta
\phi_{mn}=\pl_m \xi_n+\pl_m \xi_n$. This can be easily achieved
from the fully nonlinear classical theory of Einstein gravity
through its linearization. To do so, one identifies $\phi_{mn}$
with the fluctuations $g_{mn}=\eta_{mn}+\kappa\phi_{mn}$ of the
metric field $g_{mn}$ over the Minkowski background $\eta_{mn}$.
Here $\kappa$ is a formal expansion parameter. The gauge symmetry
$\delta \phi_{mn}=\pl_m \xi_n+\pl_m \xi_n$ comes about as
linearized diffeomorphism $\delta g_{mn}=\pl_m \xi^c g_{cn}+\pl_n
\xi^c g_{cm}+\xi^c\pl_c g_{mn}$. Indeed, to the lowest order we
have
\be \delta g_{mn}=\delta \phi_{mn}=\pl_m \xi^c \eta_{cn}+\pl_n \xi^c \eta_{cm}=\pl_m \xi_n+\pl_m \xi_n\,.\ee
The equations for $\phi_{mn}$ can be obtained via linearization of
the Einstein equations $R_{mn}-\frac12 g_{mn}R=0$ and coincide
with the Fronsdal equations for $s=2$, see below. Of course, the
free spin-two field can be defined without any reference to
Einstein-Hilbert action, differential geometry and
diffeomorphisms.

As it was shown by Fronsdal, \cite{Fang:1978wz, Fronsdal:1978rb},
a massless spin-$s$ field can be described by a totally symmetric
rank-$s$ tensor field\footnote{Since the number of indices that a tensor can carry is now
arbitrary we need a condensed notation. All indices in which a tensor is symmetric or
needs to be symmetrized are denoted by the same letter and a group of $s$ symmetric indices $a_1...a_s$ is abbreviated to $a(s)$. The operator of symmetrization sums over all {\it necessary} permutations only, e.g. $V^aU^a\equiv V^{a_1}U^{a_2}+V^{a_2}U^{a_1}$. More info is in Appendix \ref{app:multiindex}.}
$\phi^{a(s)}\equiv \phi^{a_1...a_s}$ which obeys an unusual trace
constraint
\begin{align}\label{tr-constr}
\phi^{a(s-4)bcde}\eta_{bc}\eta_{de}\equiv\phi\fud{a(s-4)mn}{mn}\equiv0\,.
\end{align}
Clearly, the trace constraint becomes effective starting from
$s=4$ being irrelevant for $s=0,1,2,3$. It tells us that the
Fronsdal field consists of two irreducible (symmetric and traceless) Lorentz
tensors of ranks $s$ and $s-2$.  Indeed, having an arbitrary
symmetric tensor $\phi_{m_1\dots m_k}$ one can always decompose it
into a sum
\be\phi_{m(k)}=\phi_{m(k)}'+\eta_{mm}^{\vphantom{'}}\phi_{m(k-2)}''
+\eta_{mm}^{\vphantom{'}}\eta_{mm}^{\vphantom{'}}\phi_{m(k-4)}'''+\dots\,,\ee
where all 'primed'
fields are traceless. Eq. \eqref{tr-constr} states then that
Fronsdal field $\phi_{m(s)}$ is only allowed to have
$\phi_{m(s)}'$ and $\phi_{m(s-2)}''$ to be non-zero. It is already
this odd constraint that puzzles and complicates things a lot at
interacting level as it would be more natural to work with fully
unconstrained tensors or totally traceless ones. We will see in
Section \ref{sec:UnfoldingSpins} that \eqref{tr-constr} arises naturally from the
extension of the frame-like formulation of gravity to fields of
any spin.

The dynamical input is given by Fronsdal equations
\begin{align}\label{FronOper}
F^{a(s)}&=\square \phi^{a(s)}-\pl^a \pl_m\phi^{ma(s-1)}+\pl^a \pl^a \phi\fud{a(s-2)m}{m}=0\,,
\end{align}
which are invariant under the gauge transformations (the verification of this fact is in Appendix \ref{app:multiindex})
\begin{align}
\delta \phi^{a(s)}&=\pl^a \xi^{a(s-1)}\,, && \xi\fud{a(s-3)m}{m}\equiv0\label{gaugetransFronsdal}\,,
\end{align}
where the gauge parameter is traceless (this becomes effective for $s\geq3$).

This generalizes the cases of spin-$0,1,2$ to any $s$. The
Fronsdal equations are valid for $s=2$ as well, coinciding in this
case with the linearized Einstein equations. They are also valid for
$s=1$ and $s=0$ if we notice that the third and the second terms
are absent for $s=0,1$ and $s=0$, respectively.

The tracelessness of $\xi^{a(s-1)}$  goes hand in hand with the
double-tracelessness of the Fronsdal field. Indeed, from
\eqref{gaugetransFronsdal} it follows that $\xi^{a(s-1)}$ should
be traceless once $\phi^{a(s)}$ is double traceless since the second
trace of $\phi^{a(s)}$, which vanishes identically, cannot be affected by the gauge symmetry. In
playing with trace constraints for field/gauge parameter one finds
that the Fronsdal theory is essentially unique\footnote{In
principal one can relax the tracelessness of the gauge parameter
preserving double-tracelessness of the field. However, as is seen
from \eqref{gaugetransFronsdal}, this would result in differential
constraint on a gauge parameter which is not a safe thing for it
typically affects the number of physical degrees of freedom and
having a differential constraints on fields/gauge parameters
complicates the study of interactions a lot, see, however \cite{Francia:2013sca}. If the field were
irreducible, i.e. it were symmetric and traceless, one would
derive $\pl_m \xi^{a(s-2)m}=0$ by taking trace of
\eqref{gaugetransFronsdal}, which is again a differential
constraint. Imposing such a constraint makes sense at the fully
nonlinear level in the case of $s=2$, which corresponds to
volume-preserving diffeomorphisms, see e.g. \cite{Blas:2007pp}.
The generalization for the free spin-$s$ field is also possible,
\cite{Skvortsov:2007kz,Campoleoni:2012th}. Alternatively, one
could try to project $\pl^a \xi^{a(s-1)}$ onto the traceless
component, in which case one would find no gauge invariant
equations, so this option is unacceptable. Let us also mention that the double-trace constraint
can be relaxed \cite{Francia:2007qt, Francia:2012rg} by extending the set of physical degrees of freedom, by enlarging the field content, or by allowing for higher derivatives.}. To this effect it is
instructive to look at the variation of the Fronsdal operator
under \eqref{gaugetransFronsdal} with $\xi^{a(s-1)}$ not
satisfying any trace constraints
\be\delta F^{a(s)}=3\pl^a\pl^a\pl^a\xi\fud{a(s-3)m}{m}\label{unconstvariation}\,.\ee

\paragraph{Going on-shell.} In order to see that the solutions to the Fronsdal equation do carry a spin-$s$ representation of the Poincare group and nothing else we need to solve equations and
quotient by the gauge symmetries. Usually one imposes various gauges to simplify equations
and then applies Fourier transform. It is useful to define de-Donder tensor
\besubeqs\begin{align}
D^{a(s-1)}&=\pl_m \phi^{ma(s-1)}-\frac12 \pl^a \phi\fud{a(s-2)m}{m}\,,\\
\delta D^{a(s-1)}&=\square \xi^{a(s-1)}\,,\label{de-Donder gauge}\\
F^{a(s)}&=\square \phi^{a(s)}-\pl^a D^{a(s-1)}\,,
\end{align}\esubeqs
which transforms in a simple way under the gauge transformation
and constitutes the non-$\square \phi$ part of the Fronsdal
operator. Since the de-Donder tensor carries as many components as
the gauge parameter \eqref{de-Donder gauge} it is possible to
gauge it away, i.e. to impose $D^{a(s-1)}=0$ as a gauge condition. Then, one is left with the gauge
transformations $\xi^{a(s-1)}$ obeying $\square \xi^{a(s-1)}=0$.
Since $\xi^{a(s-1)}$ is still nontrivial one can further impose one
more condition, $\phi\fud{a(s-2)m}{m}=0$. Indeed, $\phi^{a(s)}$ is
now on-shell, i.e. $\square \phi^{a(s)}=0$, and $\delta
\phi\fud{a(s-2)m}{m}=2\pl_m\xi^{ma(s-2)}$. The gauge-fixed
equations and constraints read
\begin{subequations}
\label{onshell}
\begin{align}
\square\phi^{a(s)}&=0\,, &\square\xi^{a(s-1)}&=0\,,\label{onshellA}\\
\pl_m \phi^{ma(s-1)}&=0\,, & \pl_m \xi^{ma(s-2)}&=0\,,\label{onshellB}\\
\phi\fud{a(s-2)m}{m}&=0\,, & \xi\fud{a(s-3)m}{m}&=0\label{onshellC}\,,\\
\delta \phi^{a(s)}&=\pl^a \xi^{a(s-1)}\,.\label{onshellD}
\end{align}
\end{subequations}
These are all the Lorentz covariant conditions one can impose
without trivializing the solution space. There is still a leftover
on-shell gauge symmetry, which manifests indecomposable structure
of the representation carried by $\phi^{a(s)}$.

\paragraph{Counting degrees of freedom.} The
representation corresponding to massless particle can be shown to
be induced from a finite-dimensional representation of the
Wigner's little algebra $so(d-2)$. It is the representations of
$so(d-2)$ that specify the spin. See, e.g. \cite{Bekaert:2006py} or
the second chapter of the Weinberg's QFT textbook, for the review
of the Wigner's construction. To count degrees of freedom, or
better to say to identify the spin of the field, one solves \eqref{onshellA} by
performing the Fourier transform
\be\phi^{a(s)}(x)=\int d^dp\, \delta(p^2)\, \varphi^{a(s)}(p) e^{ipx}\ee
and analogously for $\xi^{a(s-1)}$.

Since the equations are Lorentz covariant all points in momentum
space are equivalent and we can look at any $p_m$, $p_m p^m=0$ to
count the number of independent functions. A convenient choice is
given by light-cone coordinates
\be
x_{\pm}=\ff{1}{\sqrt{2}}(x_1\pm x_0)\,,\qquad
x_{i}=\{x_2,...,x_{d-1}\}\,.
\ee
Indices range $a=\{+,-,i\}$, $i=1...(d-2)$ in these coordinates
with the metric $\eta^{+-}=\eta^{-+}=1$, $\eta^{ij}=\delta^{ij}$
and all other components being zero. Let us take
$p_m=e\delta^+_m$, where $e$ is some constant and $\delta^+_m$ is
the Kronecker delta. The components of $\varphi^{a(s)}$ can be
split into
\be\varphi^{\overbrace{+...+}^N\overbrace{-...-}^M i(s-M-N)}\,.\ee
Eq. \eqref{onshellB} tells that all components with at least one
$+$ direction vanish, $\varphi^{+a(s-1)}=0$ (analogously for
$\xi^{a(s-1)}$). Now one can use gauge symmetry \eqref{onshellD}
$\delta \varphi^{a(s)}=e\eta^{+a}\xi^{a(s-1)}$, i.e.
$\delta\varphi^{-a(s-1)}=e\eta^{+-}\xi^{a(s-1)}$  to set
$\varphi^{-...-i...i}=0$. Finally, we are left with
$\varphi^{i(s)}(p)$ that is symmetric and $so(d-2)$-traceless,
i.e. traceless with respect to $\delta^{ij}$. Indeed, $\varphi^{a(s)}$
is $so(d-1,1)$-traceless, which can be rewritten as
\be\varphi\fud{a(s-2)m}{m}\equiv\varphi^{a(s-2)ij}\delta_{ij}+2\varphi^{a(s-2)+-}\eta_{+-}\equiv0\,.\ee
Then, we note that the last term carries at least one index along '$+$'-direction, so it is zero by \eqref{onshellB}.
To conclude, the degrees of freedom are those of an irreducible rank-$s$ $so(d-2)$-tensor times the dependence on $p_m$ that lives on $(d-1)$-dimensional cone. That $\varphi^{i(s)}(p)$ is an irreducible rank-$s$ tensor at each $p$, namely it is symmetric and traceless, implies that it describes a single spin-$s$ particle.

\paragraph{Lagrangian.} The Fronsdal Lagrangian reads
\begin{align}S=-\frac{1}{2}\int_{M^d}&\left(\pl_m\phi^{a(s)}\pl^m\phi_{a(s)}-
\frac{s(s-1)}{2}\pl_m\phi^{n\phantom{n}a(s-2)}_{\phantom{n}n}\pl^m\phi_{k\phantom{k}a(s-2)}^{\phantom{k}k}\right.+\notag\\
&\qquad+s(s-1)\pl^m\phi^{n\phantom{n}a(s-2)}_{\phantom{n}n}\pl^k\phi_{kma(s-2)}-s\pl_m\phi^{ma(s-1)}\pl^n\phi_{na(s-1)}+\label{FlatFronsdalLagrangian}\\
&\qquad\qquad\qquad\left.-\frac{s(s-1)(s-2)}4\pl_m\phi^{n\phantom{n}ma(s-3)}_{\phantom{n}n}\pl^r\phi_{r\phantom{r}ka(s-3)}^{\phantom{k}k}\right)\notag\,.\end{align}
It is fixed up to an overall factor and total derivatives by the gauge symmetry, \eqref{gaugetransFronsdal}, \cite{Curtright:1979uz}. It can be put into a more compact form by integrating by parts
\begin{align}\label{FronLEq}
S&=\frac12\int_{M^d}\, \phi_{a(s)}\, G^{a(s)}\,, &G^{a(s)}&= F^{a(s)}-\frac12 \eta^{aa}F\fud{a(s-2)m}{m}\,,
\end{align}
where the trace of the Fronsdal operator is
\begin{align}
F[\phi]\fud{a(s-2)m}{m}&=2\square \phi\fud{a(s-2)m}{m}-2\pl_n \pl_m\phi^{mna(s-2)}+\pl^a \pl_m \phi\fud{a(s-3)mn}{n}\,.
\end{align}
The gauge invariance of the action implies certain Bianchi identities
\begin{align*}
0=\delta S=s\int_{M^d}\, \pl_b\xi_{a(s-1)}\, G^{a(s-1)b}=-s\int_{M^d}\, \xi_{a(s-1)}\, \pl_m G^{a(s-1)m}=-s \int_{M^d}\, \xi_{a(s-1)} B^{a(s-1)}\,,
\end{align*}
i.e. the following linear operator annihilates the \lhs of the equations of motion\footnote{Let us note that $\pl_m G^{a(s-1)m}$ has one more term, $-\frac12 \eta^{aa}\pl_m F\fud{a(s-3)mn}{n}$, which is projected out thanks to traceless $\xi^{a(s-1)}$, which again shows the importance of $\xi\fud{a(s-3)m}{m}\equiv0$.}
\begin{align}
B[F]^{a(s-1)}&=\pl_m F^{a(s-1)m}-\frac12 \pl^a F\fud{a(s-2)m}{m}\,, & B[F[\phi]]\equiv&0\,.
\end{align}
Again, it is instructive to see how the Bianchi identities get violated if $\phi^{a(s)}$ does not obey the double-trace constraint,
\be B[F[\phi]]^{a(s-1)}=-\frac32 \pl^a\pl^a\pl^a\phi\fud{a(s-4)mn}{mn}\,.\label{BianchiViolation}\ee

Let us note that the equation that comes from Lagrangian, \eqref{FronLEq} or \eqref{FlatFronsdalLagrangian}, is $G^{a(s)}=0$ and is a little bit different from \eqref{FronOper}. They are equivalent in fact. Indeed, taking the trace
\begin{align}
G\fud{a(s-2)m}{m}=-\frac{d+2s-6}{2}F\fud{a(s-2)m}{m}\,,
\end{align}
we see that $G^{a(s)}=0$ implies $G\fud{a(s-2)m}{m}=0$ unless $s=2$ and the dimension is too low, $d=2$, for spinning fields to propagate. Therefore, $F\fud{a(s-2)m}{m}=0$ follows form the action. On substituting this back to $G^{a(s)}=0$ one finds $F^{a(s)}=0$. It was important that both $G^{a(s)}$ and $F^{a(s)}$ are double traceless as a consequence of $\phi\fud{a(s-4)mn}{mn}\equiv 0$.

\paragraph{The long and winding road from representations to Lagrangians.}
It is worth stressing that a systematic approach to Lagrangians
can be quite difficult requiring to answer a priori four different
questions.

(i) to classify all unitary irreducible representations of the
space-time symmetry group, Poincare in our case. Postulates of
Quantum Mechanics combined with the Special Relativity (i.e. the
idea that the physical laws are covariant under the Poincare
algebra) result in the statement that all systems, e.g.
particles, must carry a unitary representation of the Poincare
algebra, \cite{Wigner:1939cj}. This is where the notion of spin
and mass comes out as parameters specifying a representation. The
representation theory of Poincare algebra has a little to do with
the space-time directly.

(ii) to realize these representations on the solutions of certain
P.D.E's imposed on certain tensor fields over the Minkowski space.
We have seen that a spin-$s$ representation is realized on
$\varphi^{i(s)}(p)$ where $p^2=0$ and has again little to do with
the space-time. An on-shell description is given by
\eqref{onshell} in terms of traceless $\phi^{a(s)}(x)$
that is defined up to a gauge transformation. At this stage we see
that $\varphi^{i(s)}$ comes as projection/factor of
$\varphi^{a(s)}$ and in principle one can imagine embedding
$\varphi^{i(s)}$ into an $so(d-1,1)$-tensor with more indices such
that the equations/gauge symmetries project out redundant
components. The number of indices that a field may carry is not
directly related to the spin as a parameter of an irreducible
representation, one has to take equations/gauge symmetries into
account. There are generally infinitely many descriptions of one
and the same representation by different combinations of
field/P.D.E./gauge-symmetry. The simplest example is a spin-one
particle, photon, which can be equally well described by gauge
potential $A_m$, $\square A_m-\pl_m \pl^n A_n=0$, $\delta
A_m=\pl_m \xi$ or by nongauge field strength $F_{mn}=-F_{nm}$,
$\pl^nF_{mn}=0$, $\pl_a F_{bc}+\pl_b F_{ca}+\pl_c F_{ab}=0$. There
is a generalization of this example to fields of all spins higher
than one, which is in Section \ref{subsec:Weylmodule}.

All possible descriptions of the same representation are
known as dual descriptions. While the gauge potential is capable of realizing all
possible types of local interactions a particle can have, this is not so for the rest of
the dual descriptions. As an example, the interactions with E/M field are introduced
by means of $\pl\rightarrow \pl_m+A_m$ and not in terms of $F_{mn}$.

(iii) to find an off-shell description, i.e. to extend
fields/gauge parameters in such a way that no differential
constraints like \eqref{onshellA}-\eqref{onshellC} remain. An
off-shell description as we have seen requires adding a traceless
rank-$(s-2)$ tensor to the traceless $\phi^{a(s)}$ to be combined
together into a double traceless Fronsdal field.

(iv) to get these P.D.E.'s (or equivalent to them) as variational
equations for certain Lagrangian.   Coincidentally, in the case of
massless spin-$s$ fields in Minkowski or anti-de Sitter space the same
field content as we used for an off-shell description is
sufficient to write down a Lagrangian, which is not true for the
massive spin-$s$ field, \cite{Singh:1974rc, Singh:1974qz}.

\mysubsection{Massless fields on (anti)-de Sitter background}
The Fronsdal theory can be easily extended to the constant
curvature backgrounds, \cite{Fronsdal:1978vb, Fang:1979hq}, which
are the maximally symmetric solutions of Einstein equations with
cosmological constant $\Lambda$. These are known as de Sitter,
$\Lambda>0$, and anti-de Sitter, $\Lambda<0$ spaces. The algebraic
double trace constraint \eqref{tr-constr} remains unchanged but
$\eta_{mn}$ gets replaced with\footnote{From now on we reserve
lowercase Latin letters $a,b,c,...$ for the indices in the flat
space, which are raised and lowered with $\eta_{mn}$. Underlined
lowercase Latin letters $\ua,\mm,\nn,\rr,\kk,...$ are the 'world'
indices being raised and lowered with some generally nonconstant
metric $g_{\mm\nn}$. } the (anti)-de Sitter
metric\footnote{Equivalently we can transfer all indices to the
fiber with the help of the tetrad/frame/vielbein field $h^a_\mm$,
which will be introduced systematically in the next Section. In
fact, it has to be introduced once we would like to include
fermions. Then the algebraic constraints do not change at all
\begin{align}
\phi^{a(s-4)bbcc}\eta_{bb}\eta_{cc}&\equiv0\,, & \phi^{a(s)}&=\phi_{\mm(s)}h^{\mm a}...h^{\mm a}\,.
\end{align}
In taking derivatives we can use
$[D_\mm,D_\nn]V^{a(s)}=\Lambda h^a_\mm h^{\vphantom{a}}_{\nn b}V^{ba(s-1)}-\Lambda h^a_\nn h^{\vphantom{a}}_{\mm b}V^{ba(s-1)}$.
} $g_{\mm\nn}(x)$
\begin{align}
\phi_{\mm(s-4)\nn\nn\rr\rr}\,g^{\nn\nn}g^{\rr\rr}&\equiv0\,.
\end{align}
All derivatives should be covariantized and we use the following
normalization
\begin{align}\label{covderadsrule}
[D_\mm,D_\nn]V^{\ua}&=\Lambda \delta{}^\ua_\mm
g{}^{\phantom{a}}_{\nn \ub}V^{\ub }-\Lambda \delta{}^\ua_\nn
g{}^{\phantom{a}}_{\mm \ub}V^{\ub }\,,
\end{align}
where $\Lambda$ is the cosmological constant. The gauge
transformation law now reads
\begin{align}
\delta \phi_{\ua(s)}=\nabla_\ua \xi_{\ua(s-1)}\,.
\end{align}
The Fronsdal operator is promoted to
\begin{align}\label{adsfron}\notag
F^{\ua(s)}[\phi]&=\square \phi^{\ua(s)}-\nabla^\ua \nabla_\mm\phi^{\mm \ua(s-1)}+\frac12\nabla^\ua \nabla^\ua \phi\fud{\ua(s-2)\mm}{\mm}-m^2_{\phi}\phi^{\ua(s)}+2\Lambda g^{\ua\ua}\phi\fud{\ua(s-2)\mm}{\mm}\,,\\
m^2_{\phi}&=-\Lambda ((s-2)(d+s-3)-s)\,,
\end{align}
where we note the appearance of the mass-like terms. Mind that
$\pl^a\pl^a$ in the third term has changed to $\frac12\nabla^\ua
\nabla^\ua$ since covariant derivatives do not commute and
$s(s-1)$ terms are now needed to symmetrize over $\ua(s)$ as contrast
to $s(s-1)/2$ terms in flat space. In checking the gauge
invariance of the Fronsdal equations we find that the leading terms vanish
thanks to the gauge invariance of the Fronsdal operator in flat
space. However, in order to cancel the gauge variation we have to
commute some of the derivatives. The commutators produce certain
new terms that can fortunately be canceled by adding mass-like
terms. The strange value of $m^2_{\phi}$ can be derived using the
representation theory of $so(d-1,2)$ or $so(d,1)$,
\cite{Metsaev:1995re, Metsaev:1997nj}, which are the symmetry
algebras of anti-de Sitter and de Sitter spaces, respectively. It
is related to the Casimir operator in the corresponding
representation. We will have more to say about (anti)-de Sitter
later on. The lesson is that masslessness that implies gauge
invariance does not necessarily imply the absence of mass-like
terms. Protection of gauge invariance is more important than the
presense/absence of mass-like terms as it guarantees that the
number of physical degrees of freedom does not change when
switching on the cosmological constant. The constant curvature of
(anti)-de Sitter space acts effectively as a harmonic potential
that renormalizes the value of the mass term. Let us also note,
that the precise value of the mass-like term depends also on the
way the covariant derivatives in the second term of
\eqref{adsfron} are organized.

The action has formally the same form
\begin{align}\label{FronLEqAdS}
S&=\frac12\int\, \phi_{\ua(s)}\, G^{\ua(s)}\,, &G^{\ua(s)}&= F^{\ua(s)}-\frac12 g^{\ua \ua}F\fud{\ua(s-2)\mm}{\mm}\,,
\end{align}
where the trace of the Fronsdal operator reads
\begin{align}\notag
F[\phi]\fud{\ua(s-2)\mm}{\mm}&=2\square \phi\fud{\ua(s-2)\mm}{\mm}-2\nabla_\nn \nabla_\mm\phi^{\mm \nn \ua(s-2)}+\nabla^\ua \nabla_\mm \phi\fud{\ua(s-3)\mm \nn}{\nn}-m_1^2\phi\fud{\ua(s-2)\mm}{\mm}\,,\\
m_1^2&=-2\Lambda (s-1)(d+s-3)\,.\label{FronsdalEqAdS}
\end{align}
Finally let us note that \eqref{unconstvariation} and \eqref{BianchiViolation} are still valid upon replacing $\pl_a\rightarrow \nabla_\ua$.

The process of imposing gauges is analogous to the case of the
Minkowski background. Given this, without going into details, we
can conclude that equation \eqref{FronsdalEqAdS} describes the same number of
physical degrees of freedom since it preserves the same amount of
gauge symmetry and the order of equations and Bianchi identities
remains unchanged.

It is worth stressing that by going from Minkowski to more general
backgrounds one can lose certain amount of physical
interpretation. For example, in anti-de Sitter space the
space-time translations, $P_a$, do not commute so one cannot diagonalize
all $P_a$ simultaneously. In particular, $P_aP^a$ is no longer a
Casimir operator as it is the case in the Minkowski space.
Nevertheless, in some sense anti-de Sitter algebra $so(d-1,2)$ is
better than Poincare one in being one of the classical Lie
algebras, while Poincare algebra is not semi-simple, which leads
to certain peculiarities in constructing representations of the
latter. The Poincare algebra can be viewed as a contraction of
$so(d-1,2)$. The particles in the case of anti-de Sitter algebra
$so(d-1,2)$ should be defined as Verma modules with spin and mass
being related to the weights of $so(d-1,2)$, which is somewhat technical and we
refer to \cite{Metsaev:1995re, Brink:2000ag,Skvortsov:2009zu}.

As for the field description the best one can do on a general
background is to ensure that the number and order of gauge
symmetries/equations/Bianchi identities remains unchanged (or get
changed in a coherent way) so as to preserve the number of degrees of
freedom, \cite{Barnich:1993vg,Kaparulin:2012px}.

Once the gravity is dynamical or the background is different from (anti)-de Sitter or Minkowski, the Fronsdal operator is no longer gauge invariant. Indeed, in verifying the gauge invariance we have to commute covariant derivatives $\nabla$'s. For the case of the Minkowski space $\nabla$'s just commute. For the case of (anti)-de Sitter (constant curvature) space the commutator is proportional to the background metric, so the commutators produce mass-like terms. In generic background we are left with \begin{align}
\delta F=R_{...} \nabla \xi^{...}+\nabla R_{...} \xi^{...}\neq0
\end{align}
where $R_{...}$ is the Riemann tensor. Therefore the Fronsdal operator becomes inconsistent on more general configurations of metric in the sense that the lack of gauge invariance brings in extra degrees of freedom (usually these come as negative norm states).

In particular when the metric $g_{\mm\nn}$ becomes a dynamical field we face the problem of how to make higher-spin fields interact with gravity. This was the starting point for the no-go \cite{Aragone:1979hx} by Deser and Aragone and then yes-go results by Fradkin and Vasiliev \cite{Fradkin:1986qy,Fradkin:1987ks} (see review \cite{Bekaert:2010hw} on various no-go theorems related to higher-spins). Some comments on the Fronsdal theory on general Riemannian manifolds can be found in extra Section \ref{extra:FronsdalonRiemann}.

We see that there is something special about higher-spin fields, the threshold being $s=2$, since all lower-spin fields, $s=0,\frac12,1$ can propagate on any background, $g_{\mm\nn}$, and the graviton is self-consistent on any background of its own.

\paragraph{Summary.} There is a well-defined theory of
free fields of any spin-$s$ on the specific backgrounds, which are
Minkowski and (anti)-de Sitter --- maximally symmetric solutions
of Einstein equations with/without cosmological constant. The fields and
gauge parameters have to obey certain trace constraints,
\eqref{tr-constr}, (\ref{gaugetransFronsdal}.b).



\mysection{Gravity as gauge theory}\label{sec:Gravity}\setcounter{equation}{0}
Among theories of
fundamental interactions there are Yang-Mills gauge theories
based on (non)abelian Lie algebras and General Relativity (GR) that
stands far aside and is typically viewed as essentially different
from gauge theories. Particularly, the way it was formulated by
Einstein, GR does not rest on any gauge group. On the other hand,
gravity clearly has a gauge symmetry represented by arbitrary
coordinate transformations and diffeomorphisms. From that
perspective it seems quite natural to address a question of a
gauge form of GR.

This section is aimed to demonstrate that gravity
can in many respects be thought of as a gauge theory. The relevant
variables to see this are the so called vielbein $e^a_\mm$ and
spin-connection $\omega^{a,b}_\mm$, which can to some extent be
treated as components of a Yang-Mills connection of Poincare,
$iso(d-1,1)$, de Sitter, $so(d,1)$, or anti-de Sitter, $so(d-1,2)$, algebras. The
reader familiar with Cartan formulation of gravity can skip the
entire section. We begin with a very short and elementary
introduction to the Cartan geometry, then proceed to various ways
of thinking of gravity as a gauge theory. The MacDowell-Mansouri-Stelle-West
formulation of gravity is left to the extra Section \ref{extra:MMSW}. The relevant references include
\cite{MacDowell:1977jt, Stelle:1979aj, Vasiliev:2001wa,
Ortin:2004ms} and \cite{Hehl:1976kj} for the references on the
original papers by Cartan, Weyl, Sciama, Kibble.

\mysubsection{Tetrad, Vielbein, Frame, Vierbein,....}
In differential geometry one deals with manifolds -- something
that can be built up from several copies of the Euclidian space.
The point is that not every hyper-surface we can imagine is
homeomorphic to a Euclidian space and hence can be covered by some
global coordinates. Therefore we have to cut a generic manifold
into smaller overlapping pieces each of which can be thought of as
a copy of Euclidian space. We need transition functions that allow
us to identify the regions of two copies of Euclidian space whose
images overlap on the manifold. A manifold itself then comes as a
number of copies of Euclidian space (patches) together with the
transition functions that are defined for certain pairs of copies
and obey certain consistency relations.

In differential geometry framework the objects, tensors, transform
properly under the change of the coordinates so that the scalar
(physical) quantities we compute do not depend on the choice of
coordinates. Despite the fact that differential geometry is
designed in a democratic way with respect to different
coordinates, this is not fully so for tensors. Indeed, given a
tensor $T$ its components $T\fud{\mm...}{\nn...}$ are given with
respect to the basis in the tangent space that is induced from the
coordinates in the current chart. The basis vectors at a given
point are vectors that are tangent to the coordinate lines, see
the figure below. \putfigureifpdf{\figureTangetBase} We will refer to
such 'bare' tensors as to world tensors and to the indices they
carry, $\mm,\nn,...$, as to world indices. To disentangle the
basis in the chart and in the tangent space we may introduce an
auxiliary nondegenerate matrix $e^a_\mm(x)$ that transfers tensor
indices from the basis induced by the particular coordinates to
some other basis in the tangent space we may prefer more. With
the help of $e^a_\mm(x)$ each world tensor acquires an avatar
\begin{align}
T\fud{\mm...}{\nn...}\longrightarrow T\fud{a...}{b...}=e^a_\mm...\,T\fud{\mm...}{\nn...}\,(e^{-1})_b^\nn...
\end{align}
and we refer to the tensor in the new basis as to the fiber
(tangent) tensor and to the indices it carries, $a,b,...$ as to
fiber (tangent) indices. In principle, tensors of mixed type, i.e.
those that carry both world and fiber indices simultaneously are
possible and such tensors do appear in our study. But the rule of
course is that only indices of the same type can be contracted
with either $\delta^a_b$ or $\delta^\mm_\nn$.

If no derivatives are around it is obvious that one can use either
of the bases for algebraic computations, e.g. taking tensors
products or contracting indices, i.e. the following diagram
commutes
\begin{align*}
\begin{CD}
\mbox{set of world tensors } T\fud{\mm...}{\nn...},... @>e, e^{-1}>> \mbox{set of fiber tensors } T\fud{a...}{b...},...\\
@VV \mbox{operations} V @VV\mbox{operations} V\\
\mbox{derived set of world tensors } T\fud{\nn...}{\nn...},... @>e, e^{-1}>> \mbox{derived set of fiber tensors } T\fud{a...}{a...}
\end{CD}
\end{align*}
In other words, having a set of world tensors first, we can either
do some algebraic computations like taking tensor products or
contracting indices and then transfer all indices left free into
the fiber ones with the help of $e^a_\mm$, $(e^{-1})^\mm_a$ or we
can first transfer all indices to the fiber ones and then perform
identical computations but in the fiber.

There is a strong motivation from physics to introduce $e^a_\mm$
--- the equivalence principle. In the famous Einstein's thought
experiment an experimentalist, when put into a freely falling
elevator without windows, cannot tell whether she is falling
freely in gravitational field or is left abandoned in the open
space far away from any sources of gravitational
field. Equivalently, gravitational field is locally
indistinguishable from the accelerating frame. This has led
Einstein to the equivalence principle (EP). EP implies that
locally one can always eliminate the gravitational field by taking
a freely falling elevator. This statement lies at the core of all
problems in defining stress-tensor of gravity. As we are going to
consider gravity, there is a preferred set of bases given by
Einstein's elevators --- elevators that are freely falling in a
local gravitational field --- the physics in this elevator is
locally as in the Special Relativity (SR). The latter is true for
physically small elevators, i.e. up to tidal forces, etc.

The EP tells us that the metric in the new basis, which is associated with the elevator, is constant, for example, $\eta^{ab}=\mbox{diag}(-++...+)$, i.e.
we have
\begin{align}
\eta^{ab}&=e^a_\mm(x)\,g^{\mm\nn}(x)\,e^b_\nn(x)\,, & \eta&=e^T\,g\, e\,, \label{etaeveywhere}
\end{align}
or, equivalently, one can always recover the original metric
$g_{\mm\nn}(x)$
\begin{align}
g_{\mm\nn}(x)&=e^a_\mm(x)\, \eta_{ab}\,e^b_\nn(x)\, .
\end{align}
The object $e^a_\mm(x)$, i.e. the Einstein's elevator, is called
{\it tetrad} or {\it vierbein} in the case of four-dimensional
space-time; {\it vielbein, soldering form} or {\it frame} in
arbitrary $d$; {\it zweibein, dreibein, etc.} in case of two,
three, {\it etc.} dimensions.

It is worth noting that the metric as a function of the vielbein
is  defined in such a way that different ways of raising and
lowering indices lead to the same result. For example, the inverse
vielbein $e^\mm_a$ is just the matrix inverse of $e^a_\mm$, but it
can also be  viewed as $e^a_\mm$ whose indices were raised/lowered
with $g^{\mm\nn}$ and $\eta_{ab}$,
\be e^\mm_a=(e^{-1})^\mm_a=g^{\mm\nn}\,e_\nn^b\, \eta_{ba}\,.\ee

Obviously, $e^a_\mm(x)$, being a $d\times d$ matrix that depends
on $x$, has enough components to guarantee \eqref{etaeveywhere}. If we forget about the $x$ dependence,
$e^a_\mm$ is a matrix that diagonalizes the given quadratic form $g_{\mm\nn}$.  A
change of coordinates $x^\mm\rightarrow y^\mm$ amounts to defining
$d$ functions $y^\mm=f^\mm(x^\nn)$, i.e. it has less 'degrees of
freedom' as compared to the vielbein. Indeed, in order for
$e^a_\mm(x)$ to be equivalent to a change of coordinates it must
be $e^a_\mm=\pl_\mm f^a(x)$. The integrability of this condition,
i.e. $0\equiv(\pl_\nn\pl_\mm-\pl_\mm\pl_\nn) f^a(x)$, implies
$\pl_\mm e^a_\nn-\pl_\nn e^a_\mm=0$, which is generically not
true. It is obvious that one cannot remove gravitational field
everywhere just by a coordinate transformation since there are
tensor quantities like Riemann tensor $R_{\mm\nn,\rr\kk}$ and
$R_{\mm\nn,\rr\kk}=0$ is a coordinate independent statement.

The equivalence principle leads to an idea of General Relativity
(GR) as being a localization (gauging) of Special Relativity (SR)
and this is the idea we would like to follow and to generalize to
fields of all spins. SR can be thought of as the theory of the
global Poincare invariance, i.e. a theory of $ISO(d-1,1)$ as a
rigid symmetry. Which amount of this symmetry gets localized in
GR? Apparently the elevators form an equivalence class since given
an elevator $e^a_\mm(x)$ one can rotate it and boost it at any
velocity $v$. These transformations belong to the Lorentz group
$SO(d-1,1)$. At each point (or physically speaking at small
neighborhood of each point) we have a different elevator and hence
the Lorentz transformations can depend on $x$. To put it formally,
$e^a_\mm(x)$ and \be e'^a_\mm(x)=A\fud{a}{b}(x)
e^b_\mm(x)\label{framelargetr}\,,\ee where $A(x)\in SO(d-1,1)$,
i.e. $A^T\eta A=\eta$, produce the same $g_{\mm\nn}$ and do not
change $\eta^{ab}$. If the transformation is small, i.e. $A\fud{a}{b}$ is
close to the unit matrix, we can write
$A\fud{a}{b}=\delta\fud{a}{b}-\epsilon\fud{a,}{b}$ and
$\epsilon^{a,b}\equiv \epsilon\fud{a}{c}\eta^{cb}$ is
antisymmetric, $\epsilon^{a,b}=-\epsilon^{b,a}$. Then a small
change in $e^a_\mm$ results in
\be\delta e^a_\mm(x)=-\epsilon\fud{a,}{b}(x)e^b_\mm(x)\label{lorentzrottetrad}\,,\ee
which is a localized version of \eqref{framelargetr}.

However, we lost translations of $ISO(d-1,1)$ as the local
symmetry. Translation brings an elevator to another point where
the gravitational field may differ. As we will see local
translations are not genuine symmetries.

Now the metric $g_{\mm\nn}$ can be viewed as a derived object and
not as a fundamental. Every statement in the language of
$g_{\mm\nn}$ can be always rewritten in the language of $e^a_\mm$
and not vice verse because $e^a_\mm$ is defined up to an
$x$-dependent Lorentz rotation in accordance with the fact the
Einstein's elevator is not unique. We can also see this by
counting independent components, $g_{\mm\nn}$ has $d(d+1)/2$
components, while $e^a_\mm$ has $d^2$ components. The Lorentz
transformations form a $d(d-1)/2$-dimensional group, so
\be \#\mbox{vielbein}-\#\mbox{Lorentz}=\#\mbox{metric}\,,\ee
which means that we did not lose or gain any new 'degrees of
freedom'.

There is one more fundamental reason to introduce the vielbein ---
matter fields, e.g. electrons, protons, neutrons, which are
fermions and thus are represented by spinors. They
do experience gravitational interaction and we have to deal
with this experimental fact. Let us emphasize that the very
definition of spinors relies on the representation theory of the
Lorentz algebra $so(d-1,1)$, which in the Minkowski space of
Special Relativity is a subalgebra of the full Poincare symmetry
algebra $iso(d-1,1)$. The notion of spin and mass rests on the
representation theory of $iso(d-1,1)$ too. These are the
parameters that define unitary irreducible representations of
$iso(d-1,1)$. The existence of spinors, which is due to the first
homotopy group of $SO(n)$ being nontrivial, makes it possible
to consider the action of the group up to a phase which
distinguishes between contractible and non-contractible paths on
the group. This leads to a bizarre consequences, e.g. the electron
wave function changes its sign upon $2\pi$-rotation.

A theory formulated in terms of some tensor representations of the
Lorentz algebra, which are then used to define tensor fields over
the Minkowski space, can be straightforwardly extended to a theory
that has $gl(d)$ as a symmetry algebra and then to a
diffeomorphism invariant theory. Clearly, having an
$so(d-1,1)$-tensor $T^{abc...}$ in some theory we can replace it
with a tensor of $gl(d)$ of the same type. Then, having a tensor
of $gl(d)$ we can turn it into a field $T^{abc...}(x)$ and make it
transform under diffeomorphisms, see the table below for some
examples. However, there is no straightforward lift of spin-tensor
representations of $so(d-1,1)$ to $gl(d)$.
Apparently\footnote{Formally the fundamental group of $SL(d)$,
$GL(d)=SL(d)\times GL(1)$, is the same as for $SO(d)$, because it
is determined by the maximal compact subgroup. However, the
double-valued representations of $SL(d)$ are infinite-dimensional.
A possible way out is to take infinite-dimensional spinorial
representations of $SL(d)$ seriously, \cite{Ne'eman:1978gj}. Such
representations, when restricted to $SO(d)$, decompose into an
infinite sum of spin-tensor representations of all spins and hence
contain higher-spin fields. As we will learn the
consistency of higher-spin theory requires infinite number of
higher-spin fields, so at the end of the day $SL(d)$-spinors may
not be so far away, \cite{Vasiliev:2007yc}.} we do not know of
what replaces spin-tensors in the case of $gl(d)$. The vielbein
solves this problem as we can put ourselves into the reference
frame where the symmetry algebra is $so(d-1,1)$, the difference is
that it is a local statement. In order to construct Lagrangians
and field equations we need to extend the covariant derivative to
tensors with fiber indices.

\begin{table}
\begin{tabular}{|c|c|c|}
  \hline
  & SR & GR \tabularnewline
  general Lagrangian & \rule{0pt}{20pt}$\int d^dx\, \mathcal{L}(\phi,\pl_\mm\phi,\eta^{ab})$ & $\int \sqrt{|g|}\, d^dx\, \mathcal{L}(\phi,\nabla_\mm\phi,g^{\mm\nn})$ \tabularnewline
  spin-zero& \rule{0pt}{20pt}$\frac12 \int d^dx\, \pl_a \phi\, \pl_b \phi\, \eta^{ab}$ &
  $\frac12 \int \sqrt{|g|}\, d^dx\, \pl_\mm \phi\, \pl_\nn \phi\, g^{\mm\nn}$\tabularnewline
  spin-one & \rule{0pt}{20pt}$\frac14\int d^dx\, F_{ab}\, F^{ab}$ &
  $\frac14\int \sqrt{|g|}\,  d^dx\, F_{\mm\nn}\, F_{\rr\kk}\, g^{\mm\rr} g^{\nn\kk}$\tabularnewline
  spin-half & \rule{0pt}{20pt}$\int d^dx\, \bar{\psi} \gamma^a \pl_a \psi$ & ???, wait for \eqref{ActionDirac}\tabularnewline
  \hline
\end{tabular}
\end{table}

\paragraph{Covariant derivative.} It is necessary to define the covariant derivative
in the fiber, then we can make it act in any representation of the
Lorentz algebra, $so(d-1,1)$, in particular on spin-tensors and
hence be able to write down the Dirac Lagrangian in the
gravitational field. The covariant derivative needs to be defined
in a way that the following diagram commutes, otherwise there will
be too many problems in comparing the results of differentiation
in the two bases (we still think that the simple recipe to replace
$\pl$ with $\DL=\pl+\Gamma$ works well for world tensors so we do
not want to abandon this knowledge),
\begin{align}
\begin{CD}
V_\mm @>e^\mm_a>> V_a\\
@VVD_\nn V @VVD_\nn V\\
D_\nn V_\mm @>e^\mm_a>> D_\nn V_a
\end{CD}\label{CovDerDiagram}
\end{align}
The diagram implies that we can first differentiate a tensor, then
transfer it to another basis, or first transfer it to another
basis and then differentiate. The results must coincide. Since in
the world basis a vector in two coordinate frames can be related
by any $GL(d)$ matrix the Christoffel symbol $\Gamma^\mm_{\nn\rr}$
is a generic matrix in $\mm,\rr$. In the fiber basis any change of
coordinates must be a Lorentz transformation. For the same reason
that we used to introduce $\Gamma$ we introduce the
spin-connection $\omega_\nn{}\fud{a,}{b}$. It has two types of
indices, the world index is due to $D_\nn$ and the two fiber
indices makes it a matrix in the fiber. Inside the covariant
derivative each fiber index is acted by the spin connection and
each world index by the Christoffel symbol, e.g.
\begin{align*}
D_\nn V_\mm&=\pl_\nn V_\mm +\Gamma^{\rr}_{\nn\mm} V_\rr^{\vphantom{a}}\,, \\
D_\nn V^\mm&=\pl_\nn V^\mm -\Gamma^{\mm}_{\nn\rr} V^\rr\,, \\
D_\nn V^a&=\pl_\nn V^a+\omega_\nn{}\fud{a,}{b} V^b\,,
\end{align*}
and for the most general case
\be D^{\vphantom{a}}_\nn T^{abc...}_{\mm...}=\pl^{\vphantom{a}}_\nn T^{abc...}_{\mm...}+\omega^{\vphantom{a}}_\nn{}\fud{a,}{u}T^{ubc...}_{\mm...}+
\omega^{\vphantom{a}}_\nn{}\fud{b,}{u}T^{auc...}_{\mm...}+\Gamma^{\rr}_{\nn\mm}T^{abc}_{\rr}+...\,.\ee
Since only Lorentz rotations are allowed in the fiber we must have
$\omega_\mm{}^{a,b}=-\omega_\mm{}^{b,a}$, where we have used the
right to raise and lower fiber indices with the help of
$\eta^{ab}$. Equivalently we can impose $D_\mm \eta^{ab}=0$ to
find $\omega_\mm{}^{a,b}$ antisymmetric. The consistency
condition, the condition for diagram \eqref{CovDerDiagram} to commute, leads to
\begin{align}
e^{a}_{\mm} D_\nn^{\vphantom{a}} V_a &=D^{\vphantom{a}}_\nn V^{\vphantom{a}}_\mm\,, & V_\mm=&\,e^{a}_{\mm} V_a\,.
\end{align}
Since this must hold for any $V_\mm$ we get
\begin{align}\label{De}
D^{\vphantom{a}}_\nn e^a_\mm&=\pl^{\vphantom{a}}_\nn e^a_\mm+\Gamma^{\rr}_{\nn\mm}
e^a_\rr+\omega^{\vphantom{a}}_\nn{}\fud{a,}{b} e^b_\mm=0\,.
\end{align}
This is called the vielbein postulate. Several comments can be made about the postulate
\begin{itemize}
  \item
The vielbein postulate is analogous to $D_\mm g_{\nn\rr}=0$
postulate in that it is designed to disentangle algebraic
manipulations with the help of $e$ (or $g$) and covariant
derivatives, i.e. it ensures that contractions of indices commute
with covariant derivatives. Note that \eqref{De} implies $D_\mm
g_{\nn\rr}=0$. \item \eqref{De} can be solved both for $\Gamma$
and $\omega$ as functions of $e$ and its first derivatives, see
Appendix \ref{app:spinconnection}. This is supported by comparing
the number of equations $d^3$ with the total number of components
of $\#\Gamma=d\times d(d+1)/2$ and $\#\omega=d\times d(d-1)/2$,
$\#\Gamma+\#\omega=\# eqs$.
  \item In the solution $\Gamma(e)$ the vielbein comes all the way in combinations that can be recognized as $g$ and $\pl g$. One recovers the usual Christoffel symbols.
  \item On the contrary, the solution $\omega(e)$ cannot be rewritten in terms of the metric $g_{\mm\nn}$, which supports the vielbein being a fundamental field.
\end{itemize}

If we anti-symmetrize in \eqref{De} over $\mm\nn$ and use that $\Gamma^{\rr}_{\mm\nn}$ is symmetric, we find
\begin{align}
T^a_{\nn\mm}=\pl^{\vphantom{a}}_\nn e^a_\mm-\pl^{\vphantom{a}}_\mm e^a_\nn+\omega^{\vphantom{a}}_\nn{}\fud{a,}{b} e^b_\mm- \omega^{\vphantom{a}}_\mm{}\fud{a,}{b} e^b_\nn=0\,, \label{vpostulateA}
\end{align}
i.e. $\Gamma$ disappears and the system of equations turns out to
have a triangular form. We can first solve for $\omega$ and then
for $\Gamma$. Explicit solution for $\omega$ is given in Appendix \ref{app:spinconnection}. In case there is no need for $\Gamma$ we can use
\eqref{vpostulateA}. It can be more compactly rewritten if we hide
the world indices by saying that $e^a_\mm$ and
$\omega_\mm{}\fud{a,}{b}$ are differential forms. A short introduction to
the language of differential forms can be found in Appendix \ref{app:diffforms}.

Thinking of $e^a_\nn$ and  $\omega_\nn{}\fud{a,}{b}$ as degree-one
differential forms, $e^a=dx^\nn{}\, e^a_\nn $,
$\omega\fud{a,}{b}=dx^\nn{}\,\omega_\nn{}\fud{a,}{b} $ one can
rewrite \eqref{vpostulateA} as
\begin{align} \label{ZeroTorsion}
T^a&=d e^a+\omega\fud{a,}{b}\wedge e^b=De^a=0\,.
\end{align}
Two-form $T^a\equiv \frac12 T^a_{\mm\nn}dx^\mm\wedge dx^\nn$ is
called the torsion. We can check the integrability of
\eqref{ZeroTorsion} applying $d$ to \eqref{ZeroTorsion} and using
that $d^2\equiv0$ and then using \eqref{ZeroTorsion} again to
express $de^a$. We have nothing to say on how $d\omega^{a,b}$
looks like so we keep it as it is. The
result\footnote{\label{ftn:Integrability}In its simplest form this
is just the Frobenius integrability condition. Given a set of
PDE's $\pl_\mu \phi(x)=f_{\mu}(x)$ the commutativity of partial
derivatives imply $0\equiv(\pl_\mu
\pl_\nu-\pl_\nu\pl_\mu)\phi(x)=\pl_\mu f_\nu-\pl_\mu f_\nu=0$. The
last equality does not hold for a generic vector-function $f_\mu$,
which means that the system can be inconsistent.} is
\begin{align}\label{RiemannBianchiA}
F\fud{a,}{b}\wedge e^b&=0\,, & F\fud{a,}{b}&=d\omega\fud{a,}{b}+\omega\fud{a,}{c}\wedge \omega\fud{c,}{b}\,.
\end{align}
The two-form $F\fud{a,}{b}$ has four indices in total and is in fact related to the Riemann tensor
\begin{align}
R_{\mm\nn,\rr\uu}=F_{\mm\nn}{}\fud{a,}{b}\, e_{a\rr}\, e^b_{\uu}\,.
\end{align}
It is a painful computation to solve $T^a=0$ for $\omega^{a,b}$ as a
function of $e^a$ and then compute $F^{a,b}$ to see that $e^a$ appears in
combinations that can be rewritten in terms of the metric.
Fortunately, there is a back-door. Let us compute the  commutator
of two covariant derivatives on some vector $V^\mm$ and the same
for $V^a=V^\mm e^a_\mm$, i.e. $[D_\mm,D_\nn]V^\rr$ and
$[D_\mm,D_\nn]V^a$. The two results must match after transferring
all the indices to fiber ones or to world ones. We already know
that $[D_\mm,D_\nn]V^\rr$ is expressed in terms of the Riemann
tensor. Analogously, $[D_\mm,D_\nn]V^a$ can be expressed in terms
of $F_{\mm\nn}{}\fud{a,}{b}$, which gives
\begin{align}
R_{\mm\nn}\fud{\rr}{\uu} V^\uu=\left(F_{\mm\nn}{}\fud{a}{b} V^b\right) e_a^\rr=\left(F_{\mm\nn}{}\fud{a}{b} e^b_\uu V^\uu\right) e_a^\rr=\left(F_{\mm\nn}{}\fud{a}{b} e^b_\uu e_a^\rr \right)V^\uu\,.
\end{align}

The identity \eqref{RiemannBianchiA} can be then recognized as the
first Bianchi identity for the Riemann tensor, being a three-form
it anti-symmetrizes over the three indices in square brackets,
\be\label{RiemannFirstBianchi} R_{[\mm\nn,\rr]\uu}\equiv
R_{\mm\nn,\rr\uu}+R_{\nn\rr,\mm\uu}+R_{\rr\mm,\nn\uu}\equiv0\,.\ee
Since everything in the metric-like formulation can be derived
from the frame-like one, it is not surprising that the
Einstein-Hilbert action\footnote{We omit the gravitational
constant everywhere from our formulae. Our excuse is that we are
not going to compute the precession of the perihelion of Mercury
or anything like that in these notes.}
\begin{align}
S_{EH}&=\int \sqrt{\det{g}}\, R
\end{align}
can be rewritten in Cartan-Weyl form\footnote{How to integrate
differential forms is explained at the end of Appendix
\ref{app:diffforms}.}
\begin{align}
S_{CW}&=\int F^{a,b}(\omega(e))\wedge e^c\wedge...\wedge e^u\, \epsilon_{abc...u}\label{FrameLikeActionA}\,.
\end{align}
The integrand is a top-form, i.e. the form of maximal degree,
which is the space-time dimension, and can be integrated. Let us
note that $\omega$ used in the action is assumed to be expressed
in terms of $e$ via the vielbein postulate, \eqref{De}, or the
torsion constraint, \eqref{ZeroTorsion}, which obscures the
interpretation of $\omega$ as a gauge field of the Lorentz
algebra. This is what we would like to improve on.

\mysubsection{Gravity as a gauge theory}

\paragraph{Short summary on Yang-Mills.} The deeper we go into the gravity
the more similarities with the Yang-Mills theory we find with some
important differences though. From this perspective let us collect
basic formulas of Yang-Mills theory. The main object in Yang-Mills
theory is the gauge potential $A_\mm$ that takes values in some
Lie algebra, say $\mathfrak{g}$. We treat it as a degree-one form
$A=A_\mm dx^\mm$ with values in the adjoint representation of
$\mathfrak{g}$. The index of the Lie algebra is implicit but we
can always recover it $A=A^\aI t_\aI$ with $t_\aI$ being the
generators of $\mathfrak{g}$, i.e. there is a Lie bracket
$[t_\aI,t_\aJ]=f\fdu{\aI\aJ}{\aK} t_\aK$.

There can also be matter fields, i.e. fields taking values in arbitrary representation of $\mathfrak{g}$. For example, let $\phi(x)=\phi^\Aa(x)$ be a vector in some vector space $V$ that carries a representation $\rho$ of $\mathfrak{g}$, i.e. $\rho: \mathfrak{g}\rightarrow End(V)$, which means that we have matrices $\rho(t_\aI)\fud{\Aa}{\Ab}$ associated with each of the generators $t_\aI$ such that $[\rho(t_\aI),\rho(t_\aJ)]=\rho([t_\aI,t_\aJ])=f\fdu{\aI\aJ}{\aK} \rho(t_\aK)$, i.e. the matrix commutator is expressed via the Lie bracket and hence in terms of the structure constants.

In the table below we collect some formulae that we will use many times in what follows\\

\noindent\begin{tabular}{|w{8cm}|x{6.4cm}|}
  \hline
 \centering description & formula \tabularnewline\hline
  \rule{0pt}{20pt} gauge transformation \\($\epsilon$ is a zero-form with values in
   $\mathfrak{g}$, $\epsilon=\epsilon^\aI t_\aI$) &  $\delta A=D\epsilon\equiv d\epsilon+[A,\epsilon]$ \\ $\delta \phi=-\rho(\epsilon)\phi$\tabularnewline
  \rule{0pt}{20pt} curvature or field strength & $F(A)=dA+\frac12[A,A]$\tabularnewline
  \rule{0pt}{20pt} covariant derivative & $D\phi=d\phi+\rho(A)\phi$\tabularnewline
  \rule{0pt}{20pt} generic variation $\delta A$ of $F$ &  $\delta F=D\delta A\equiv d\delta A+[A,\delta A]$ \tabularnewline
  \rule{0pt}{20pt} gauge variation of $F$& $\delta F=[F,\epsilon]$ \tabularnewline
  \rule{0pt}{20pt} gauge variation of $D\phi$  & $\delta D\phi=-\rho(\epsilon)D\phi$ \tabularnewline
  \rule{0pt}{20pt} Bianchi identity & $DF\equiv dF+[A,F]\equiv0$\tabularnewline
  \rule{0pt}{20pt} Jacobi identity & $[A,[A,A]]\equiv0$ \tabularnewline
  \rule{0pt}{20pt} the commutator of two $D's$ \\($\bullet$ is a placeholder)  & $D^2\bullet=F\bullet$, $D^2 \phi=\rho(F)\phi$\tabularnewline

  \hline
\end{tabular}
\vspace{0.3cm}

For example, the Jacobi identity acquires  a simpler form
$[A,[A,A]]\equiv0$ because $A$ is a one-form and hence,
$[A_\mm,[A_\nn,A_\kk]]\, dx^\mm\wedge dx^\nn\wedge dx^\kk$
implicitly imposes anti-symmetrization over the three slots, which
is the Jacobi identity. Analogously, $D^2$ computes the commutator
of two $D$'s, $DD=D_\mm D_\nn \, dx^\mm\wedge dx^\nn\equiv \frac12
[D_\mm, D_\nn] \, dx^\mm\wedge dx^\nn$, which is the
field-strength.

\paragraph{Back to gravity.}
The theory of gravity in terms of vielbein/spin-connection variables must be invariant under the local Lorentz transformations. Now we can simply say that $\omega^{a,b}$ is a gauge field (Yang-Mills connection) of the Lorentz algebra, $so(d-1,1)$. Denoting the generators as $L_{ab}=-L_{ba}$ we have the following commutation relations
\begin{align}
[L_{ab},L_{cd}]=L_{ad}\eta_{bc}-L_{bd}\eta_{ac}-L_{ac}\eta_{bd}+L_{bc}\eta_{ad}\,.
\end{align}
The Yang-Mills connection is then $\omega=\frac12 \omega^{a,b}L_{ab}$, which already looks like spin-connection. For a moment we will treat $e^a$ as a vector matter, i.e. with $\rho$ given by $\rho(L_{ab})\fud{c}{d}=-\eta_{ad}\delta^c_b+\eta_{bd}\delta^c_a$. As a connection, $\omega$ possesses its own gauge parameter $\epsilon=\frac12 \epsilon^{a,b}L_{ab}$.

Specializing the formulas from the table above we find the gauge transformations
\besubeqs\begin{align}
\delta \omega^{a,b}&=d\epsilon^{a,b}+\omega\fud{a,}{c} \epsilon^{c,b}+\omega\fud{b,}{c} \epsilon^{a,c}\equiv D\epsilon^{a,b}\,,\\
\delta e^a&=-\epsilon\fud{a,}{b}e^b\label{lorentzrottetradB}\,,
\end{align}\esubeqs
which correspond to infinitesimal Lorentz rotations. The last line
is exactly \eqref{lorentzrottetrad}. The transformation law for
the spin-connection can be derived without making any reference to
the Yang-Mills rules --- one can apply the same reasonings as for
the Christoffel symbols, i.e. use \eqref{lorentzrottetradB} and
the fact that $D_\mm V^a$ must be a tensor quantity (Lorentz
vector in the index $a$).

The Yang-Mills field-strength $F(\omega)$ is exactly $F^{a,b}(\omega)$ found above, \eqref{RiemannBianchiA}. The torsion constraint $T^a=0$, \eqref{ZeroTorsion}, is just the condition for the covariant derivative $De^a$ of $e^a$ to vanish. We also find that $DF^{a,b}\equiv0$ as a Bianchi identity. Taking into account the relation between $F^{a,b}$ and the Riemann tensor we recover the second Bianchi identity $D_{[\mm}R_{\nn\rr],\kk\uu}\equiv0$.

In particular we can now solve the problem of extending Dirac Lagrangian to curved manifolds since the covariant derivative can act in any representation of the Lorentz algebra. To be precise, we define fiber spinor field $\psi^\Aa(x)$, the fiber $\gamma$-matrices $\gamma_a=\gamma_a{}\fud{\Aa}{\Ab}$, $\{\gamma_a,\gamma_b\}=2\eta_{ab}$. Then the generators of the Lorentz algebra in the spinor representation are given by $\rho(L_{ab})=\frac14 [\gamma_a,\gamma_b]$ and the covariant derivative acts as \be D_\mm\psi^\Aa=\pl_\mm \psi^\Aa+\frac12 \omega^{a,b}_\mm\rho(L_{ab}){}\fud{\Aa}{\Ab}\psi^\Ab\,.\ee
Finally, the Dirac action on a curved background reads,
\be S_D[\psi,e,\omega]=\int \det{e}\, (i\bar{\psi} \gamma^a e^\nn_a \overrightarrow{D}_\nn \psi-i\bar{\psi} \gamma^a e^\nn_a \overleftarrow{D}_\nn \psi-m \bar{\psi}\psi)\,.\label{ActionDirac}\ee

What are the symmetries of the frame-like action \eqref{FrameLikeActionA}? All the fiber indices are contracted with the invariant tensor $\epsilon_{ab...u}$ of the Lorentz algebra and $e^a$ as well as $F^{a,b}(\omega)$ transform homogeneously under local Lorentz rotations, i.e. like a vector and a rank-two antisymmetric tensor. This implies that the action has local $so(d-1,1)$-symmetry. It is also diffeomorphism invariant since it is an integral of a top-form.

There are still some subtleties that prevent one from simply stating
that gravity is the Yang-Mills theory. Namely, $\omega^{a,b}$ is a
function of vectorial matter $e^a$ via the torsion constraint, \eqref{ZeroTorsion};
$e^a$ is a one-form rather than pure vector matter; $e^a_\mm$ must
be invertible since $\det{g}\neq0$; the action does not have the
Yang-Mills form. Nevertheless, by going to the first-order
formulation of gravity one can further improve the interpretation
of gravity as a gauge theory.

\paragraph{Note on first-order actions.}
We are not aiming at rigorous definitions here. The field
equations are usually second-order P.D.E.s for bosonic fields. We
call the actions that immediately lead to second-order equations
the second-order actions. For example, classical action for a free
particle $\int\frac12 \dot{q}_i \dot{q}^i$, the Fronsdal action, \eqref{FronLEq}, or
the Einstein-Hilbert action are second-order actions because the
variational equations are of the second order.

Let us begin with the free particle. The Hamiltonian is $H=\frac12 p_i p^i$, $p_i=\dot{q}_i$. We can express the Lagrangian back using $L=p\dot{q}-H$, where we would like to treat $p^i$ as an independent variable for a moment, so we have
\begin{align}
S(q,p)=\int (\dot{q}^i-\frac12 p^i)p_i\,.
\end{align}
Now there are two variational equations
\begin{align}
\frac{\delta S}{\delta p^i}&=\dot{q}^i- p^i=0\,, & \frac{\delta S}{\delta q^i}&=\dot{p}^i=0\,.
\end{align}
The first equation is algebraic with respect to momenta $p^i$ and is solved as $p^i=\dot{q}^i$. Then the second equation reduces to $\dot{p}^i=\ddot{q}^i=0$ as desired.

To make notation coherent we can use $dq^i$, where
$d=dt\frac{\pl}{\pl t}$ instead of $\dot{q}^i dt$, so that we
treat $q^i$ as a vector valued zero-form over one-dimensional
manifold, which is the world-line of the particle parameterized by
$t$. We can also introduce a one-dimensional einbein $e=dt$ to
write
\begin{align}
S(q,p)=\int (dq^i-\frac12 e p^i)p_i\,.
\end{align}
This is how a typical first-order action looks like. The ideology
is that one introduces additional fields, the analogs of momenta
$p$, such that the new, first-order, action depends on the
original fields and momenta. The action now contains first-order
derivatives only. The equations for momenta are algebraic and
express momenta as first-order derivatives of the original fields.
On substituting the solutions for the momenta into the action one
gets back to the original action. In many cases the advantage of
the first-order approach is that the action is simpler, less
nonlinear and the new fields, momenta, as independent fields may
have certain interpretation (this is what happens to
$\omega^{a,b}$).

As an example, it is well-known that in the case of gravity one can treat $\Gamma^\kk_{\mm\nn}$ as an independent variable in the action (Palatini formulation), writing
\begin{align}
S_{P}(g,\Gamma)=\int \sqrt{\det{g}}\, g^{\mm\nn}R_{\mm\nn}(\Gamma)\,.
\end{align}
The equations of motion for $\Gamma$ are equivalent to $\nabla_\mm
g^{\nn\kk}=0$ and imply that $\Gamma^\kk_{\mm\nn}$ is the
Levi-Civita connection. On substituting this to the action one
gets back to the pure Einstein-Hilbert.

One can do more by replacing $g_{\mm\nn}$ as independent
variable with
$\mathsf{g}^{\mm\nn}=\sqrt{\det{g}}\, g^{\mm\nn}$, which is a purely algebraic change of
variable that is invertible in $d>2$. Then the action
for gravity is schematically $\mathsf{g}(\pl\Gamma+\Gamma^2)$,
i.e. at most cubic. All non-polynomial nonlinearities of gravity
are removed by using the first-order approach and new appropriate
variable $\mathsf{g}$. There are two sources of nonlinearities.
The first one, is in $\sqrt{\det{g}}\, g^{\mm\nn}$. The second one
arises when solving for $\Gamma$ as one has to invert the metric.

\paragraph{Back to gravity again.}
Let us take the route of first-order actions and see if we can
treat $\omega^{a,b}$ as an independent variable (the analog of
Palatini formulation in terms of vielbein and spin-connection).
Generally, one cannot just isolate a bunch of derivatives and
fields (the expression for $\omega$ in terms of $e^a$) and call it
a new field to get the first-order formulation. Fortunately, this
is not the case with $\omega$. The variation of the action
\begin{align}
S_{f}(e,\omega)&=\int F^{a,b}(\omega)\wedge e^c\wedge...\wedge e^u\, \epsilon_{abc...u}\label{FrameLikeActionB}\,,
\end{align}
where $\omega$ is now an independent field, reads\footnote{There is a version
of the Stokes theorem for covariant derivatives, so we may not split
$D$ into two pieces $D=d+\omega$ when integrating by parts. In the usual
Stokes theorem $0=\int d(A\fm{p}\wedge B\fm{q})=\int
dA\fm{p}\wedge B\fm{q}+(-)^p A\fm{p}\wedge dB\fm{q}$, where
$p+q=d$, one can replace $d$ with covariant derivative $D$
provided the integrand is a scalar, i.e. in a trivial representation of the Lie algebra
we are considering. First, one checks that if $I$
is a scalar then $dI\equiv DI$. Next, if $I$ is a
composite, e.g., $I=A\fm{p}\wedge B\fm{q}$, then $D$ satisfies the
chain rule, which gets modified by a sign factor in front of the
second term since all the objects are differential forms, i.e.
$d(A\fm{p}\wedge B\fm{q})=D(A\fm{p}\wedge B\fm{q})=DA\fm{p}\wedge
B\fm{q}+(-)^p A\fm{p}\wedge DB\fm{q}$. }
\begin{align}
\delta S_{f}(e,\omega)&=(d-2)\int\left( \delta\omega^{a,b}\wedge T^c\wedge...\wedge e^u\,  + F^{a,b}(\omega)\wedge \delta e^c\wedge...\wedge e^u\, \right)\epsilon_{abc...u}\,,
\end{align}
where we used that $\delta F^{a,b}=D\delta\omega^{a,b}$ and integrated by parts to find $T^a=De^a$. Assuming the frame field be invertible we find the following equations
\besubeqs\begin{align}
\delta \omega^{a,b}_\mm&: & T^a=De^a=d e^a+\omega\fud{a,}{b}\wedge e^b&=0\,,\\
\delta e^a_\mm&: & F^{a,b}_{\mm\nn}\, e_b^{\nn}&=0\,.
\end{align}\esubeqs
The first equation allows us to solve for the spin-connection.
Under the condition that $T^a=0$ the second equation, when
rewritten in the metric-like language, gives $R_{\mm\nn}=0$, which
is the vacuum Einstein equation. Up to the moment when we have to
solve the torsion constraint, \eqref{ZeroTorsion}, $\omega^{a,b}$ can be treated as a
Yang-Mills connection of the Lorentz algebra.

Let us note that the first-order action is polynomial as compared
to the second order action where the nonlinearities come from
$\omega(e)$ that involves inverse of the vielbein.

It is worth stressing that second-order and first-order approaches
may lead to different results under certain conditions. For
example, if we add matter, such as spin-$\frac12$ fields,
\eqref{ActionDirac}, to the gravity action, i.e. $S=S_f+S_D$ then
the torsion constraint gets modified inasmuch as $\omega^{a,b}$
contributes to the matter action, $S_D=S_D[\psi,e,\omega]$. In the
second order approach $\omega=\omega(e)$. In the first order
approach instead of $T^a=0$ we find $T\sim\bar{\psi}\psi$, i.e.
the torsion is fixed in terms of the matter fields. One can still
solve for $\omega=\omega(e,\psi)$. Restoring the gravitational
constant one finds the difference between the action for fermions coupled to the first-order and
second order gravity to be quadratic in the gravitational constant,
which has never been tested experimentally. Within the
second-order approach one can always reproduce the corrections due
to $\omega(e)-\omega(e,\psi)\neq0$ by adding them to the action by
hand. In supergravity, thinking of $\omega$ as an independent
field leads to more compact expressions. Namely one can start from
the second-order approach and then find that certain terms have to
be added to the action to make it invariant under
super-transformations. These terms can be reproduced automatically
by considering $\omega$ as an independent field, i.e. within the
first-order approach.

The cosmological term $\Lambda \sqrt{\det{g}}$ can be represented in the frame-like form as
\be S_\Lambda= \Lambda \int e^a\wedge e^b\wedge...\wedge e^u\, \epsilon_{abc...u}\,.\label{cosmoterm}\ee

\paragraph{More on gravity as a gauge theory.}
That spin-connection $\omega^{a,b}$ and vielbein $e^a$ are both
one-forms makes one expect they should have a similar
interpretation. Later we find that it is possible to consider
all one-forms as taking values in some Lie algebra. Now we look for a
unifying connection $A=\frac12\omega^{a,b}L_{ab}+e^aP_a$, where
$P_a$ are generators associated with gauge field $e^a$. We already
know the commutator $[L,L]$ and also know that $e^a$ behaves as a
vector of $so(d-1,1)$, which fixes the commutator $[L,P]$. There
are not so many things one can write for $[P,P]$ and we are left
with a one-parameter family
\begin{align}\begin{split}
[L_{ab},L_{cd}]&=L_{ad}\eta_{bc}-L_{bd}\eta_{ac}-L_{ac}\eta_{bd}+L_{bc}\eta_{ad}\,,\\
[L_{ab},P_c]&=P_a\eta_{bc}-P_b\eta_{ac}\,,\\
[P_a,P_b]&=-\Lambda L_{ab}\,,
\end{split}\label{dsadspoincare}
\end{align}
where $\Lambda$ is some constant and the Jacobi identities are
satisfied for any $\Lambda$. Freedom in rescaling the generators
leaves us with three distinct cases: $\Lambda>0$, $\Lambda<0$ and
$\Lambda=0$. These three cases can be easily identified with
de-Sitter algebra $so(d,1)$, anti-de Sitter algebra $so(d-1,2)$
and Poincare algebra $iso(d-1,1)$, respectively.

That
$\Lambda=0$ corresponds to $iso(d-1,1)$ is obvious. Let $T_{\aAs\aBs}=-T_{\aBs\aAs}$,
where $\aAs,\aBs,...$ range over $a$ and one additional direction,
denoted by $5$, i.e. $\aAs=\{a,5\}$, be the generators of $so(d,1)$ or $so(d-1,2)$.
They obey
\be[T_{\aAs \aBs},T_{\aCs\aDs}]=T_{\aAs\aDs}\eta_{\aBs\aCs}-T_{\aBs\aDs}\eta_{\aAs\aCs}-
T_{\aAs\aCs}\eta_{\aBs\aDs}+T_{\aBs\aCs}\eta_{\aAs\aDs}\label{AdSAlgebra}\,.\ee
Defining $L_{ab}=T_{ab}$, $\sqrt{|\Lambda|}P_a=T_{a5}$ we find
\eqref{dsadspoincare} with the last relation being
$[P_a,P_b]=-\eta_{55}|\Lambda|L_{ab}$, which explains the minus.

The Yang-Mills curvature\footnote{We reserve $F^{a,b}$ for the
Riemann two-form, while $R^{a,b}$ contains the cosmological term.}
$F=\frac12R^{a,b}L_{ab}+T^aP_a$ and gauge transformations $\delta
A=D\epsilon$, where
$\epsilon=\frac12\epsilon^{a,b}L_{ab}+\epsilon^aP_a$ read
\besubeqs\begin{align}
R^{a,b}&=d\omega^{a,b}+\omega\fud{a,}{c}\wedge\omega^{c,b}-\Lambda e^a\wedge e^b\,, & T^a&=De^a\label{adsLorentzCurvatures}\,,\\
\delta \omega^{a,b}&=D\epsilon^{a,b}-\Lambda e^a \epsilon^b+\Lambda e^b \epsilon^a\,, & \delta e^a&=D\epsilon^a -\epsilon\fud{a,}{b}e^b\label{adsLorentzGauge}\,,
\end{align}\esubeqs
where $D$ is the Lorentz covariant derivative $d+\omega$. The torsion is now one of the components of the Yang-Mills field-strength! According to the general Yang-Mills formulae, the curvatures transform as
\besubeqs\begin{align}
\delta R^{a,b}&=-\epsilon\fud{a,}{c} R^{cb}-\epsilon\fud{b,}{c} R^{a,c}-\Lambda  T^a \epsilon^b+\Lambda T^b\epsilon^a\,,\\
\delta T^a&=-\epsilon\fud{a,}{b}T^b+ R\fud{a,}{b}\epsilon^b\,.
\end{align}\esubeqs
The Bianchi identity for the Yang-Mills field-strength, $DF=0$, when written in components, is
\besubeqs\begin{align}
DT^a-e_m\wedge R^{a,m}\equiv0\,,\label{BianchiAdSA}\\
DR^{a,b}+\Lambda T^a\wedge e^b -\Lambda T^b\wedge e^a\equiv0\label{BianchiAdSB}
\end{align}\esubeqs
If the torsion constraint \eqref{ZeroTorsion} is imposed the first
identity simplifies to $e_m\wedge F^{a,m}\equiv0$, i.e. the first Bianchi identity \eqref{RiemannBianchiA} (Mind that $e_b\wedge R^{a,b}$ can be replaced by $e_b\wedge F^{a,b}$ since $e_a\wedge e^a\equiv0$.). The second one simplifies to $DF^{a,b}\equiv0$, which is the second Bianchi
identity for the Riemann tensor, $\nabla_{[\uu}
R_{\mm\nn],\kk\rr}\equiv0$. Consequently, all useful relations
automatically arise when both $e$ and $\omega$ are combined into a
single connection.

Let us consider the following action, where instead of $F^{a,b}$ we use the field-strength of the (anti)-de Sitter algebra, $R^{a,b}$, and we also add a cosmological term with an arbitrary coefficient $\alpha$:
\begin{align}
S&=\int R^{a,b}(\omega)\wedge e^c\wedge...\wedge e^u\, \epsilon_{abc...u}+\alpha S_\Lambda\label{FrameLikeActionC}\,.
\end{align}
Note that a specific cosmological term, which is $-S_\Lambda$, \eqref{cosmoterm}, is already included into the action above through $R^{a,b}$.

When $e^a$ is joined with $\omega^{a,b}$ into a single Yang-Mills connection there appears a new gauge symmetry with a parameter $\epsilon^a$, the local translations, which we did not observe in gravity before. However,
the action \eqref{FrameLikeActionC} is invariant under
$so(d-1,1)$-part of gauge transformations, i.e. $\epsilon^{a,b}$,
and it is not invariant under local translations with $\epsilon^a$ as is seen after taking the variation
\begin{align}\label{FrameLikeVariationTr}\begin{aligned}
\delta S_f= -(d-2)(d-3)&\int R^{a,b}(\omega)\wedge T^c\wedge \epsilon^c\wedge e^f\wedge...\wedge e^u\,\epsilon_{abcf...u}+\\
+\Lambda&(d-1)(-2+\alpha d)\int  T^a \wedge \epsilon^b\wedge e^c\wedge...\wedge e^u\, \epsilon_{abc...u}\,.\end{aligned}
\end{align}
It is not a new symmetry. Local translations become a symmetry of the action when
torsion is zero\footnote{Let us note that there is something
special about $d=3$. For example, for $\Lambda=0$ we have $\delta \int F^{a,b}
\wedge e^c\,\epsilon_{abc}=\int F^{a,b}\wedge D \epsilon^c \,
\epsilon_{abc}$ which vanishes upon integrating by parts and using
$DF^{a,b}\equiv0$. When $\Lambda\neq0$ we can choose $\alpha=\tfrac23$ so that \eqref{FrameLikeVariationTr} vanishes, i.e. we observe that $S=\int (F^{a,b} e^c-\tfrac{\Lambda}3e^ae^b)e^c\epsilon_{abc}$ is invariant under local translations. This is the action that can be obtained as a difference of two Chern-Simons actions for $so(2,1)$.}, $T^a=0$. We stress that $T^a=0$ is not a
dynamical equation, it is a constraint that allows one to solve
for $\omega^{a,b}$ as a function of $e^a$.

When torsion is zero the local translations can be identified with
diffeomorphisms, so they do not make a new symmetry. Indeed, there is a general
identity\footnote{Remember that $\mathcal{L}_\xi=di_\xi+i_\xi d$, where $\mathcal{L}_\xi$ and $i_\xi$
are Lie and inner derivatives, respectively, see Appendix \ref{app:diffforms}.
Then, one completes $i_\xi d A$ to $F(A)$ which completes $d(i_\xi A)$ to a
gauge transformation.}
\be \mathcal{L}_\xi A=D (i_\xi A) +i_\xi F(A)\,, \ee
i.e. the Lie derivative of any Yang-Mills connection $A=A^\aI t_\aI$ can be represented as a sum of the gauge transformation with $\epsilon=i_\xi A$, i.e. $\epsilon^\aI=\xi^\mm A_\mm^\aI$, and
a curvature term. Specializing to our case we derive
\begin{align}
\mathcal{L}_\xi e^a&=\delta_\xi e^a+i_\xi T^a\,, &
\mathcal{L}_\xi \omega^{a,b}&=\delta_\xi \omega^{a,b}+i_\xi R^{a,b}\,,
\end{align}
where $\delta_\xi$ means the gauge variation with
$\epsilon^a=\xi^\mm e^a_\mm$ and
$\epsilon^{a,b}=\xi^\mm\omega^{a,b}_\mm$. When torsion is zero we
have $\mathcal{L}_\xi e^a=\delta_\xi e^a$, i.e. diffeomorphisms
acting on $e^a$ can be represented as a particular gauge
transformations. This is in accordance with the invariance of the
action under local translations for vanishing torsion.
Diffeomorphisms acting on $\omega^{a,b}$ are not equivalent to
gauge transformations because $R^{a,b}$, which is related to the
Riemann tensor, is generally non-zero. This does not cause a
problem since the dynamical variable is the vielbein. A
diffeomorphism performed on $e$ induces a diffeomorphism for
$g_{\mm\nn}=e^a_\mm \eta_{ab} e^b_\nn$.

That there are three ways, \eqref{dsadspoincare}, to unify
$\omega^{a,b}$ and $e^a$ within one Lie algebra is directly
related to the fact that there are three most symmetric solutions to
Einstein equations with cosmological constant $\Lambda$. These are
de Sitter space, $\Lambda>0$, anti-de Sitter space, $\Lambda<0$,
and Minkowski space, $\Lambda=0$.

Despite the unification of $\omega^{a,b}$ and $e^a$ into a single
Yang-Mills connection there is an important difference between the
two. We use $\omega^{a,b}$ to construct Lorentz-covariant
derivative $D$ and couple matter to gravity, e.g. as in
\eqref{ActionDirac}, but we do not use $e^a$ inside $D$. The frame
field is always outside and is used to built a volume form and
contract indices. Let us mention that within the higher-spin
theory the difference between $\omega^{a,b}$ and $e^a$ to some
extent vanishes as we will see that $e^a$ does contribute to the
covariant derivative!

\paragraph{Most symmetric background is equivalent to \texorpdfstring{$\boldsymbol{dA+\frac12[A,A]=0}$}{flat connection}. }
The important observation is that de Sitter, anti-de Sitter and
Minkowski space-times are solutions of $F(A)=0$ with
$A=\frac12\omega^{a,b}L_{ab}+e^aP_a$ being the gauge field of the corresponding symmetry
algebra where the commutation
relations are given in \eqref{dsadspoincare} and $\Lambda$
distinguishes between the three options. In terms of the
Riemann tensor these space-times are defined by the following constraint\be
R_{\mm\nn,\rr\kk}=\Lambda
(g_{\mm\rr}g_{\nn\kk}-g_{\nn\rr}g_{\mm\kk})\label{adsMetricLike}\,.\ee
Within the frame-like approach this corresponds to
\begin{align} F(A)&=0 && \Longleftrightarrow&& &&
\left\{
  \begin{array}{ll}
     T^a=0\\
    F^{a,b}(\omega(e))=\Lambda e^a\wedge e^b
  \end{array}
\right.\label{adsFrameLike}\,, \end{align}
where $\omega$ is expressed in terms of $e$ via the torsion
constraint and the second equation imposes \eqref{adsMetricLike} once
we remember the relation between $F^{a,b}$ and $R_{\mm\nn,\rr\kk}$. This is
equivalent to $F(A)=0$. As always it is implied that $\det e^a_\mm
\neq 0$.

It is not hard to write down some explicit solutions. If $\Lambda=0$, i.e. the space-time is Minkowski, a useful choice is given by Cartesian coordinates
\begin{align}
e^a_\mm\, dx^\mm&= \delta^a_\mm\, dx^\mm=dx^a\,, &&\omega^{a,b}_\mm \, dx^\mm=0\,,
\end{align}
i.e., the vielbein is just a unit matrix and there is no difference between world and fiber. The spin-connection is identically zero. This choice leads to $g_{\mm\nn}=\eta_{\mm\nn}$ and $\Gamma^\kk_{\mm\nn}=0$.

If $\Lambda\neq0$ a useful choice is given by the Poincare
coordinates $x^\mm=(z,x^i)$ where
\be g_{\mm\nn}=\frac1{|\Lambda|}\frac1{z^2} (dz^2+dx^i dx^j \eta_{ij})=\frac1{|\Lambda|}\frac1{z^2}(dx^\mm dx^\nn \eta_{\mm\nn})\,,\ee
where $z$ is an analog of radial direction and $x^i$ are the coordinates on the leaves of constant $z$. Then we can use, for example,
\begin{align}\label{PoincareCoordVector}
e^a_\mm\, dx^\mm&= \frac1{\sqrt{|\Lambda|}}\frac1z\delta^a_\mm\, dx^\mm &&\omega^{a,b}_\mm \, dx^\mm=-\frac1{z} \left(\delta^a_\mm\eta^{bz}-\delta^b_\mm\eta^{az}\right)dx^\mm\,.
\end{align}
Note that $g_{\mm\nn}=e^a_\mm e^b_\nn \eta_{ab}$ applies, of course.

\paragraph{Summary.} We have shown that the tetrad $e^a$ and spin-connection $\omega^{a,b}$ can be unified as gauge fields of Poincare or (anti)-de Sitter algebra, $A=\frac12 \omega^{a,b}L_{ab}+e^aP_a$. The Yang-Mills curvature then delivers constituents of various actions and contains Riemann two-form and torsion.

There are at least three ways to treat Yang-Mills connections:
\begin{itemize}
  \item As in the genuine Yang-Mills theory, i.e. $\int tr (F_{\mm\nn} F^{\mm\nn})+\mbox{matter}$.
  \item As in Chern-Simons theory. We are in $3d$ with the action $\int tr (FA-\frac13 A^3)$.
  \item As in $d>3$ gravity. Here
we found several options how to treat vielbein and
spin-connection. The least we can do is to say that $\omega^{a,b}$
is an $so(d-1,1)$-connection and $e^a$ is a vector-valued
one-form. Another option is to unify $\omega^{a,b}$ and $e^a$ as
gauge fields of one of the most symmetric Einstein's vacua, i.e.
(anti)-de Sitter or Minkowski. The case of (anti)-de Sitter is
less degenerate since the symmetry algebras are semi-simple. Any
theory can be re-expanded over one of its vacuum and the expansion
is covariant with respect to the symmetries of the vacuum. In the
case of Einstein theory, depending on the cosmological constant,
there are three maximally symmetric vacua.
\end{itemize}

We found several reasons to replace metric with the vielbein or frame and spin-connection.
\begin{enumerate}
  \item To have freedom in introducing general basis in the tangent space.
  \item
To make transition from a generic curved coordinates to the ones
of the Einstein's elevators, where the local physics is as in SR.
This leads us to the idea that GR is a localized (gauged) version
of SR. At any rate we expect that we should gauge the Lorentz
algebra $so(d-1,1)$. This makes us feel that GR should be close to
the Yang-Mills theory. There are also important differences with
the Yang-Mills theory, which we discuss below.
  \item To make half-integer spin fields, in particular matter fields, interact with gravity. This is the strongest motivation, of course. Since it is the frame-like approach that allows matter fields to interact with gravity it is promising to stick to this approach and look for its generalization to fields of all spins.
\end{enumerate}
The similarities and distinctions between gravity and Yang-Mills theory include
\begin{itemize}
  \item[$\boldsymbol{+}$] Spin-connection is a gauge field of the Lorentz algebra.
  \item[$\boldsymbol{-}$] On-shell it is not an independent propagating field, rather it is expressed in terms of the vielbein field via the torsion constraint $T^a=0$.
  \item[$\boldsymbol{+}$] The matter fields interact with $\omega$ through the covariant derivative, i.e. minimally, e.g. $D\psi$, like in Yang-Mills theory.
  \item[$\boldsymbol{+}$] The action is cooked up from the Yang-Mills curvatures.
  \item[$\boldsymbol{-}$] In any case the action does not have the Yang-Mills form.
  \item[$\boldsymbol{-}$] There is a condition $\det{e^a_\mm}\neq0$, i.e. $\det{g_{\mm\nn}}\neq0$, that is hard to interpret within the Yang-Mills theory.
  \item[$\boldsymbol{+}$] One can unify both vielbein and spin-connection as gauge fields of some Lie algebra.
  \item[$\boldsymbol{-}$] There are several options to achieve that (Poincare, de Sitter or anti-de Sitter).
\item[$\boldsymbol{-}$]  While $\omega^{a,b}$ appears in covariant
derivative $D$ only, vielbein $e^a$ is always around when building
a volume form or contracting indices, so the unification of $e$
and $\omega$ within one gauge field is not perfect --- their
appearance is different, both in the gravity and in the matter
Lagrangians.
   \item[$\boldsymbol{-}$] The local translation symmetry associated with frame field $e^a$ becomes a symmetry of the action only when the torsion constraint is imposed, then it can be identified with diffeomorphisms.
   \item[$\boldsymbol{-}$] The full group structure for a diffeomorphism invariant theory with some internal local symmetry (Yang-Mills or gravity in terms of $e^a$ and $\omega^{a,b}$) is that of a semidirect product of diffeomorphisms and local symmetry group, see extra Section \ref{extra:semidirect}.

  \end{itemize}

Some other gravity-like actions are mentioned in extra Section \ref{extra:Lovelock}.
There are further improvements possible when cosmological constant is non-zero, the MacDowell-Mansouri-Stelle-West approach, which is reviewed in extra Section \ref{extra:MMSW}.



\mysection{Unfolding gravity}\setcounter{equation}{0}\label{sec:UnfldGravity}
Let us abandon the action principle, i.e. the Einstein-Hilbert action, and concentrate on the
equations of motion. The appropriate variables we need are
vielbein $e^a$ and $so(d-1,1)$ gauge field $\omega^{a,b}$,
spin-connection. As we have already seen, they can be viewed as
the gauge fields associated with either Poincare or (anti)-de
Sitter algebra, the gravity then shares many features of
Yang-Mills theory. We aim to write the Einstein equations by
making use of the language of differential forms. Particularly, it
means that field equations are necessarily of first order. It may
seem to be too restrictive since many known dynamical equations of
interest contain higher derivatives. However, pretty much as any
system of differential equations can be reduced to the first order
form by means of extra variables, so is any classical field theory
can be rewritten in differential form language by virtue of
auxiliary fields. In practice one typically needs infinitely many
of those. Such equations are called {\it unfolded},
\cite{Vasiliev:1988xc, Vasiliev:1988sa}, and it is in this form
that the Vasiliev theory is given. The formulation of gravity
obtained in this section admits a natural extension to all
higher-spin fields.

Our starting point is the torsion constraint, \eqref{ZeroTorsion}, and the definition of
the $so(d-1,1)$-curvature, \eqref{RiemannBianchiA},
\besubeqs\begin{align}
T^a&=De^a=de^a+\omega\fud{a,}{b}\wedge e^b=0\,,\label{TorsionForm}\\
F^{ab}&=d\omega^{a,b}+\omega\fud{a,}{c}\wedge \omega^{c,b}\,.\label{CurvatureForm}
\end{align}\esubeqs
The Einstein equations without matter and cosmological
constant
\begin{align}
{R}_{\mm\nn}-\frac12 g_{\mm\nn} {R}&=0
\end{align}
are equivalent in $d>2$ to $R_{\mm\nn}=0$ and hence
\begin{align} \label{EinsteinUgly}R_{\mm\nn}&=0 && \Longleftrightarrow&& &F^{a,b}_{\mm\nn} (e^{-1})^\nn_b&=0\,.\end{align}
There are two clumsy properties of the latter expression: (i) we
had to undress the differential form indices of the curvature
two-form; (ii) we needed the inverse of $e^a_\mm$ to contract
indices, i.e. from the Yang-Mills point of view we had to take the
'inverse of the Yang-Mills field'. One may ask a naive question:
whether is it possible to formulate the gravity entirely in the
language of differential forms and connections? Indeed, this is
possible and it is the starting point for the higher-spin
generalizations. In Section \ref{sec:Unfolding} we explain that
the equations formulated solely in the language of differential
forms, {\it unfolded} equations, have a deep algebraic meaning. For a moment let us just
explore this path blindly.

First of all, the Riemann tensor $R_{\mm\nn,\kk\rr}$ is traceful and has the following decomposition
\begin{align}\notag
R_{\mm\nn,\kk\rr}&=W_{\mm\nn,\kk\rr}+\alpha
(g_{\mm\kk}R_{\nn\rr}-g_{\nn\kk}R_{\mm\rr}-g_{\mm\rr}R_{\nn\kk}+g_{\nn\rr}R_{\mm\kk})+
\beta( g_{\mm\kk}g_{\nn\rr}-g_{\nn\kk}g_{\mm\rr})R\,,\\\label{WorldRiemannDec}
\alpha&=\frac{1}{d-2}\,,\qquad\qquad\quad \beta=-\frac{1}{(d-2)(d-1)}\,,\\
R_{\mm\nn,\kk\rr} g^{\nn\rr}&=R_{\mm\kk}\,, \qquad\qquad R_{\mm\kk}g^{\mm\kk}=R\,,\notag
\end{align}
where $R_{\mm\kk}$ is the Ricci tensor, $R$ is the scalar
curvature and $W_{\mm\nn,\kk\rr}$ is the traceless part of the
Riemann tensor, called Weyl tensor. The coefficients are fixed by
the normalization in the last line. Weyl tensor has the same
symmetry properties as the Riemann one, i.e.
\be\label{Ws}
W_{\mm\nn,\kk\rr} =-W_{\nn\mm,\kk\rr}=-W_{\mm\nn,\rr\kk}\,, \qquad
\qquad W_{[\mm\nn,\kk]\rr}\equiv0\,,
\ee
where the second property in \eqref{Ws} is the algebraic Bianchi
identity. The Weyl tensor is by definition traceless
\be W_{\mm\nn,\kk\rr} g^{\nn\rr} \equiv0\,.\ee
In the Young diagram language\footnote{An introductory course on
the Young diagrams language is in the Appendix \ref{app:Young}. It
is really necessary to have some understanding of what the
possible symmetry types of tensors are in order to proceed to the
higher-spin case.} the Weyl tensor is depicted as the
'window'-like diagram
\begin{align}
W_{\mm\nn,\kk\rr}&:\YoungpBB  && R_{\nn\rr}:\YoungpB\oplus\bullet&& R:\bullet
\end{align}
The vacuum Einstein equations imply $R_{\mm\nn}=0$, but
the whole Riemann tensor, of course, may not be zero. Vanishing
Riemann tensor, $R_{\mm\nn,\kk\rr}=0$, describes empty Minkowski
space. While $R_{\mm\nn}=0$ has a rich set of solutions
corresponding to various configurations of the gravitational
field, e.g. gravitational waves, black holes etc. The difference
between very strong $R_{\mm\nn,\kk\rr}=0$ and $R_{\mm\nn}=0$ is
exactly the Weyl tensor. One can say that it is the Weyl tensor
that is responsible for the richness of gravity. The trivial but crucial step
is the equivalence of the two equations
\begin{align}\label{EinsteinTwoForms}
R_{\mm\nn}&=0 &&\Longleftrightarrow &&
R_{\mm\nn,\kk\rr}=\mathcal{W}_{\mm\nn,\kk\rr}\,,
\end{align}
where $\mathcal{W}_{\mm\nn,\kk\rr}$ has the algebraic properties of
the Weyl tensor otherwise left unspecified at a point. While in
the first form the equation directly imposes $R_{\mm\nn}=0$, in
the second form it tells us that the only non-zero components of
the Riemann tensor are allowed to be along the Weyl tensor direction, i.e.
the Ricci, $R_{\mm\nn}$ must vanish. Formally, taking trace of the
second equation one finds $R_{\mm\nn,\kk\rr}
g^{\nn\rr}=R_{\mm\kk}=\mathcal{W}_{\mm\nn,\kk\rr} g^{\nn\rr}=0$.
It is also important that the second Bianchi identity,
$\nabla_{[\uu} R_{\mm\nn],\kk\rr}\equiv0$,  implies
\be\label{BianchiforWeyl}\nabla_{[\uu}\mathcal{W}_{\mm\nn],\kk\rr}\equiv0\,,\ee
i.e. $\mathcal{W}_{\mm\nn,\kk\rr}$ is arbitrary at a point but its
first-derivatives are constrained. Formally,
\eqref{BianchiforWeyl} can be viewed as one more consequence of
\eqref{EinsteinTwoForms}.

The general idea behind the 'unfolding' of gravity is that {\it
instead of specifying which components of the Riemann tensor and
its derivatives have to vanish we can parameterize those that do
not vanish by new fields}. This is applicable to any set of fields
subject to some differential equations. Instead of imposing
equations directly we can specify which derivatives of the fields
may not vanish on-shell.

Let us transfer the above consideration to the frame-like
approach. Instead of $R_{\mm\nn,\kk\rr}$ we have two-form
$F^{ab}(\omega)$. Converting the differential form indices of the
$F^{ab}\equiv F^{ab}_{\mm\nn}\, dx^\mm\wedge dx^\nn$ to the fiber
we get a four-index object
\be
F^{ab|cd}=-F^{ba|cd}=-F^{ab|dc}=F^{ab}_{\mm\nn} e^{\mm c} e^{\nn d}\,.
\ee
When no torsion constraint is imposed, $F^{ab|cd}$ has more
components than the Riemann tensor. It is antisymmetric in each
pair and there is no algebraic Bianchi identity implied. We find
the following decomposition into irreducible components
\begin{align}
F^{ab|cd}\sim \YoungpAA\otimes\YoungpAA&=\YoungpAAAA\oplus\left(\YoungpBAA\oplus\YoungpAA\right)\oplus
\left(\YoungpBB\oplus\YoungpB\oplus\bullet\right)
\end{align}
When torsion constraint is imposed one derives the following
consequence, the algebraic Bianchi identity, \eqref{RiemannBianchiA}, \eqref{RiemannFirstBianchi}, which we will refer to as the integrability constraint
\begin{align}\label{AlgBianchi}
0\equiv d de^a&=-d(\omega\fud{a}{b}\wedge e^b) & &\Longrightarrow
& &F^{a,b}\wedge e_b\equiv0\,.
\end{align}
The components with the symmetry of the first three diagrams do
not pass the integrability test as they are too antisymmetric. For
example, if $F^{ab}=e_m\wedge e_n C^{a,b,m,n}$, where $C^{a,b,m,n}$ is antisymmetric, then
$F^{ab}\wedge e_b=e_m\wedge e_n\wedge e_b\, C^{a,b,m,n}$, which implies $C^{a,b,m,n}=0$ since
$e^a_\mm$ is invertible. Analogously\footnote{It is obvious that $e_{\mm}^a V_a=0$ implies $V_a=0$ since the vielbein is invertible, e.g. we can always choose $e^a_\mm=\delta^a_\mm$ at a point. Similarly, operators of the form $e_m \wedge...\wedge e_n C^{...m...n}$ just antisymmetrize over the indices $m,...,n$ that are contracted with vielbein one-forms. Decomposing $C^{...m...n}$ into irreducible symmetry types one checks using the Young properties which of the components are annihilated and which are not. Those that are not should be removed since they are
not solutions to $e_m \wedge...\wedge e_n C^{...m...n}=0$. Say, for $C^{a|b}$ equation $e_m\wedge e_n C^{m|n}=0$ implies that the antisymmetric component of $C^{a|b}$ must vanish while the symmetric one is arbitrary.}, if $F^{ab}=e_m\wedge e_n C^{am,n,b}$, where $C^{aa,b,c}$
has the symmetry of the second component and contains the trace,
which is the third component, we find $F^{ab}\wedge e_b=e_m\wedge
e_n\wedge e_b\, C^{am,n,b}$, which does not vanish identically.
Since $F^{ab}$ is related to the Riemann tensor when the torsion
constraint is imposed, it comes as no surprise that the last three
components form the decomposition of the Riemann tensor into
irreducible tensors. From now on the torsion constraint is implied. Analogously to
\eqref{WorldRiemannDec}, the (fiber) Riemann tensor $F^{ab|cd}$
can be decomposed as
\begin{align}\notag
F^{ab|cd}&=C^{ab,cd}+\alpha
(\eta^{ac}\mathcal{R}^{bd}-\eta^{bc}\mathcal{R}^{ad}-\eta^{ad}\mathcal{R}^{bc}+\eta^{bd}\mathcal{R}^{ac})
+\beta(\eta^{ac}\eta^{bd}-\eta^{bc}\eta^{ad})\mathcal{R}\,,\\ \label{FiberRiemannDec}
F\fud{am|b}{m}&=\mathcal{R}^{ab}\,, \qquad\qquad \mathcal{R}\fud{n}{n}=\mathcal{R}\,,\\
C^{ab,cd}&:\YoungpBB\,, \qquad\mathcal{R}^{ab}:\YoungpB\oplus\bullet\,,\qquad \mathcal{R}:\bullet\notag
\end{align} We treat $C^{ab,cd}$, $\mathcal{R}^{ab}$, $\mathcal{R}$ as zero-forms.
$C^{ab,cd}$, which is the fiber Weyl tensor, has the symmetry of the Riemann tensor, i.e. of the window Young diagram, but it is traceless, $C\fud{am,b}{m}\equiv0$. The trace of the fiber Riemann tensor is the fiber Ricci tensor $\mathcal{R}^{ab}$, whose trace is the scalar curvature $\mathcal{R}$, the latter coincides with $R$ because it is a scalar.

It is easy to see that \eqref{FiberRiemannDec} can be rewritten solely in terms of differential forms as
\begin{align}\label{FDFDec}
F^{ab}&=e_m\wedge e_n C^{ab,mn}+
\alpha( e^{a}\wedge e_n \mathcal{R}^{b,n}-e^{b}\wedge e_n \mathcal{R}^{a,n})+\beta e^a\wedge e^b \mathcal{R}
\end{align}
and it does not restrict $F^{ab}$, just telling which components
of $F^{ab}$ may in principle be non-zero when the torsion
constraint is imposed.

The Einstein equations are equivalent to $\mathcal{R}^{ab}=0$. The
same reasoning as in \eqref{EinsteinTwoForms}, compared to
\eqref{EinsteinUgly} forces us to require that the only non-zero
components of $F^{ab}$ should be given by the Weyl tensor. Namely, we
just remove $\mathcal{R}^{ab}$, $\mathcal{R}$ from the \rhs of
\eqref{FDFDec}
\begin{align}\label{Onmassshellsa}
F^{ab}_{\mm\nn}&=e_{\mm m} e_{\nn n} C^{ab,mn} && \Longleftrightarrow &&  &&F^{ab}=e_m\wedge e_n C^{ab,mn}\,,
\end{align}
which is equivalent to \eqref{EinsteinUgly} in the same way as it
was the case for \eqref{EinsteinTwoForms}. Now we would like to change the basis for the tensors,
to make them carry more symmetric indices.

The tensors that have symmetry of Young diagrams different from one-row/one-column, i.e.
the tensors that are not totally symmetric/antisymmetric are called mixed-symmetry tensors. For example, the Weyl tensor has mixed-symmetry. There is a special feature of mixed-symmetry tensors that they admit several presentations, i.e. several seemingly unrelated realizations of the same symmetry type. For example, in one presentation the tensor has several groups of symmetric indices while in the other one the indices are split into groups of antisymmetric ones. More detail on mixed-symmetry tensors can be found in Appendix \ref{app:Young}.

Using the ambiguity in presentation of mixed-symmetry tensors,
instead of $C^{ab,cd}$, which is manifestly antisymmetric in pairs of indices, we can switch to $C'^{ac,bd}$ that is manifestly symmetric in pairs,
\begin{align}
&C^{ab,cd}=-C^{ba,cd}=-C^{ab,dc} &&
\bep(20,20)\put(0,0){\YoungBB}\put(2,12){$a$}\put(2,2){$b$}\put(12,2){$d$}\put(12,12){$c$}\eep&&
C'^{ac,bd}=C'^{ca,bd}=C'^{ac,db}\,.
\end{align}
In what follows we will use the symmetric basis for tensors, which is more convenient for higher-spin fields, and we write $C^{aa,bb}$ instead of $C'^{aa,bb}$. In the symmetric basis\footnote{A typical exercise on Young symmetry is to show that the \rhs is antisymmetric in $ab$, which is not manifest. The expression is antisymmetric if its symmetrization vanishes. Symmetrizing over $ab$ we get $C^{am,bn}+C^{bm,an}$ which is almost the Young condition $C^{am,bn}+C^{bm,an}+C^{ab,mn}\equiv0$. Hence we get $0\equiv F^{ab}+F^{ba}=-e_m\wedge e_n C^{ab,mn}$. Remembering that $e_m\wedge e_n$ is antisymmetric in $mn$ while $C^{ab,mn}$ is symmetric we get the desired $e_m\wedge e_n C^{ab,mn}\equiv0$. } \eqref{Onmassshellsa} reads
\begin{align}\label{Onmassshellsb}
F^{ab}&=e_m\wedge e_n C^{am,bn}\,.
\end{align}
The \rhs of \eqref{Onmassshellsb} manifestly satisfies the algebraic Bianchi identity \eqref{AlgBianchi}, but it is not so for the differential Bianchi identity, an analog of \eqref{BianchiforWeyl}. The second Bianchi identity is a consequence of $DF^{ab}\equiv0$, which implies (recall that $De^a=0$)
\be
0\equiv DF^{ab}=D(e_m\wedge e_n C^{am,bn})=e_m\wedge e_n\wedge
DC^{am,bn}\label{spinbcon}\,.
\ee

We can either stop here and supplement \eqref{Onmassshellsb} with
\eqref{spinbcon} or try to analyze this integrability condition. Eq
\eqref{spinbcon} is a restriction on the derivatives of the Weyl
tensor. Were $C^{ab,cd}$ arbitrary we would find several tensor
types to appear in $D_\mm C^{ab,cd}$. This is equivalent to
analyzing the second term of the Taylor expansion of $C^{aa,bb}(x)$
and decomposing Taylor coefficients into irreducible $so(d-1,1)$
tensors. In doing so it is convenient to transfer the world index
of $D_\mm$ to the fiber and define
\be B^{aa,bb|c}=e^{\mm c}D_\mm C^{aa,bb}\,.
\ee
We use the slash notation within a group of tensor indices (e.g.,
$B^{aa,bb|c}$) to split the indices (tensor product) into the
groups of indices in which the tensor is irreducible. If we
ignore the second Bianchi identity for a moment there are no
relations between $aa,bb$ and $m$, i.e. as the tensor of the
Lorentz algebra it has the following decomposition into
irreducibles
\be B^{aa,bb|m}\sim \YoungpBB\otimes\YoungpA=\YoungpCB\oplus\left(\YoungpBBA\oplus\YoungpBA\right)\label{SpinTwoBianchiA}\ee
To parameterize all three components we can introduce three
zero-forms $C^{aaa,bb}$, $C^{aa,bb,c}$ and $C^{aa,b}$ that are irreducible as fiber tensors, i.e. with the Young symmetry as indicated above and traceless. Then, we can write
\begin{align}\begin{aligned}
DC^{aa,bb}&=e_m \left(C^{aam,bb}+\frac12 C^{aab,bm}\right)+e_m C^{aa,bb,m}+e^bC^{aa,b}+e^a
C^{bb,a}\\&\qquad-\frac{2}{d-2}e_m\left(\eta^{aa}C^{bb,m}+\eta^{bb}C^{aa,m}-\frac12\eta^{ab}C^{ab,m}\right)\,.\end{aligned}
\label{SpinTwoBianchiAB}
\end{align}
The first two terms project\footnote{One more feature of tensors is that the operation of adding one index (taking tensor product) or removing it (contracting the index with an external
object) does not preserve the symmetry properties.
Indeed, the tensor product contains in general several irreducible
components and we are free to keep them all or to project onto one
of them. For example, given $T^{aa,b}$ of \YoungmBA{} symmetry
type, we can contract it with vielbein to get $e_m T^{am,b}$. The
resulting tensor is neither symmetric nor antisymmetric in $ab$
and contains both \YoungmB{} and \YoungmAA{} symmetry types. It
can be projected onto \YoungmB{} simply as $e_m T^{am,b}+e_m
T^{bm,a}=-e_m T^{ab,m}$ and onto \YoungmAA{} as $e_m T^{am,b}-e_m
T^{bm,a}$. The operation of adding one index and projecting onto
irreducible component is more complicated. For example, $e^b
T^{aa,b}$ neither has any definite symmetry type nor is it
traceless. Typically a number of terms is needed to project a
contraction/tensor product of two irreducible tensor objects onto
an irreducible component. If we need an irreducible tensor of
orthogonal algebra like we do above, rather than $gl(d)$, we have in addition to
project out its traces. } onto \YoungmBB, so do the first two
terms in the second line as well. The resulting tensor is not
traceless, the trace projector is imposed with the help of the
last group of terms. The projector for $e_m C^{aa,bb,m}$ is
trivial. We see that it is easy to say what are the Young diagrams
of $so(d-1,1)$-irreducible tensors that $C^{aa,bb|c}$ contain,
this amounts to computing the tensor product, \eqref{SpinTwoBianchiA}. It is much more
complicated to handle this decomposition in the language of
tensors due to Young- and trace-projectors that are generally
there, \eqref{SpinTwoBianchiAB}. Luckily, many statements can be proven in terms of Young
diagrams without appealing to the tensor language.

Back to \eqref{SpinTwoBianchiA}, it is easy to see that the
presence of the last two components is not consistent with
\eqref{spinbcon}, which is equivalent to $e_m\wedge e_n \wedge e_c
B^{am,bn|c}\equiv0$. So we have to keep the first term only in order to write the
solution to the differential Bianchi identity
\begin{align}\label{SpinTwoBianchiB}
DC^{aa,bb}&=e_m \left(C^{aam,bb}+\frac12 C^{aab,bm}\right)\,,
&&{aaa,bb}:\YoungpCB
\end{align}
$C^{aaa,bb}$ is the first 'descendant' of the Weyl tensor that
allows us to solve the differential constraint in a constructive
way. The second term together with $\frac12$ in front of it
ensures that the r.h.s. has the right Young symmetry. Again, we
can either stop here or check if there are differential
constraints for $C^{aaa,bb}$. So far all the equations were exact
in the sense that we had not neglected any terms. In continuing
the process we find a technical complication. Indeed, $D^2=F\sim
eeC$, \eqref{Onmassshellsb}, so checking the integrability of the
equation last obtained we get stuck with
\begin{align*}
D D C^{aa,bb}&= e_m\wedge e_n C\fud{am,n}{c}C^{ac,bb}+e_m\wedge e_n C\fud{bm,n}{c}C^{aa,cb}=-e_m \left(D C^{aam,bb}+\frac12 D C^{aab,bm}\right)
\end{align*}
That $D^2\sim eeC$ makes the \lhs nonlinear, so we have to solve
$ee CC=eDC^1$ where $C^1$ denotes $C^{aaa,bb}$. Hence we have to
introduce quadratic terms on the \rhs of $DC^{aaa,bb}$. This is
what should be expected since gravity is  a nonlinear theory.

We will use the example of gravity in the frame-like formalism as
a starting point for higher-spin generalization. The extension to
higher-spin fields is to be done at the free level first and then
we review the Vasiliev solution to the nonlinear problem. At the
nonlinear level fields of all spins interact with each other and
no truncations of the full system to a finite subset of fields is
possible. Given that we will continue the gravity part of the
story at the linearized level only, i.e. we will neglect $D^2$, assuming that $D^2\sim0$.

Repeating the derivation of the constraint on the first derivatives of $C^{aaa,bb}$ from \eqref{SpinTwoBianchiB} in the approximation $D^2=0$ we now find
\begin{align}
D D C^{aa,bb}&=0=-e_m \left(D C^{aam,bb}+\frac12 D
C^{aab,bm}\right)\,.
\end{align}
It can be solved analogously to the way we did before by
considering all possible \rhs of $DC^{aaa,bb}$ resulting in
\begin{align} DC^{aaa,bb}&=e_m \left(C^{aaam,bb}+\frac13 C^{aaab,bm}\right)+O(eC^2) & aaaa,bb&=\YoungpDB\end{align}
Continuing this process we can derive one by one the following set of equations/constraints that describes Einstein's gravity without the cosmological constant and matter
with higher-order corrections due to $D^2\neq0$ neglected
\begingroup\allowdisplaybreaks\besubeqs\begin{align}
&\left\{
  \begin{array}{ll}
    T^a=De^a=0 \\
    \delta e^a=D\epsilon^a-\epsilon\fud{a,}{b}e^b
  \end{array}
\right.\label{LinEinstAA}\\
&\left\{
  \begin{array}{ll}
    F^{a,b}=d\omega^{a,b}+\omega\fud{a,}{c}\wedge \omega^{c,b}=e_m\wedge e_n C^{am,bn} \label{LinEinstAB}\\
    \delta \omega^{a,b}=D\epsilon^{a,b}+(\epsilon_m e_n-\epsilon_n e_m )C^{am,bn}
  \end{array}
\right. \\
&DC^{aa,bb}=e_m \left(C^{aam,bb}+\frac12 C^{aab,bm}\right) & aa,bb&=\YoungpBB\notag\\
&DC^{a(k+2),bb}=e_m \left(C^{a(k+2)m,bb}+\frac1{k+2} C^{a(k+2)b,bm}\right)+O(eC^2) &a(k+2),bb&=\bep(50,20)\put(0,-7){\YoungBB}\put(20,3){\RectARow{3}{k}}\notag\eep\\
&\delta C^{a(k+2),bb}=-\epsilon\fud{a,}{m}C^{a(k+1)m,bb}-\epsilon\fud{b,}{m}C^{a(k+2),bm}\notag\\
&\phantom{\delta C^{a(k+2),bb}=}+\epsilon_m \left(C^{a(k+2)m,bb}+\frac1{k+2} C^{a(k+2)b,bm}\right)+O(\epsilon C^2) \label{LinEinstAD}
\end{align}\esubeqs\endgroup
where we marked those equations that are not
exact with $+O(C^2)$ label. Notice that all $C^{a(k+2),bb}$ transform under local Lorentz transformations.

It is worth stressing that zero-forms, like the Weyl tensor or matter fields of Yang-Mills, do not have their own gauge parameters. Nevertheless, they take advantage of gauge fields' parameters, but the gauge transformations
do not contain derivatives.

Notice the local translation symmetry $D\epsilon^a$ in
\eqref{LinEinstAA}, \eqref{LinEinstAB}, \eqref{LinEinstAD}. It gets restored since the torsion constraint
is imposed and we can interpret local translations as
diffeomorphisms. Hence, we automatically gauge the Poincare
algebra when considering equations of motion, while at the level
of the action principle we had certain problems in interpreting it
as resulting from gauging of the Poincare algebra in Section \ref{sec:Gravity}.

Going to nonlinear level we find $O(eC^2)$ and
higher order corrections on the \rhs of equations due to $D^2\sim
eeC$. Similar corrections we will find in the equations
describing higher-spin fields.

Equivalently, we can consider the first order expansion of gravity
over the Minkowski background. The background Minkowski space can
be defined by vielbein $h^a_\mm$ and by spin-connection
$\varpi^{a,b}_\mm=-\varpi^{b,a}_\mm$ obeying the torsion
constraint and zero-curvature
\begin{align} \label{ZeroTorsionMink}
T^a&=d h^a+\varpi\fud{a,}{b}\wedge h^b=0\,, & d\varpi\fud{a,}{b}+\varpi\fud{a,}{c}\wedge \varpi\fud{c,}{b}&=0\,,
\end{align}
the latter implies that the whole Riemann tensor vanishes. A convenient choice in the case of Minkowski is given by Cartesian coordinates, where
\begin{align}
g_{\mm\nn}&=\eta_{\mm\nn}\,, & \Gamma^\rr_{\mm\nn}&=0\,, & h^a_\mm&=\delta^a_\mm\,, & \varpi^{a,b}_\mm&=0\,.
\end{align}

It is useful to define the background Lorentz-covariant
derivative $D=d+\varpi$, which we denote by the same letter as the
full Lorentz-covariant derivative above. That
\eqref{ZeroTorsionMink} implies $D^2=0$ supports this notation as
we were going to neglect $D^2$ anyway.

The linearization of \eqref{LinEinstAA}-\eqref{LinEinstAD} over Minkowski background reads
\begingroup\allowdisplaybreaks\besubeqs\begin{align}
&\left\{
  \begin{array}{ll}
    T^a=De^a-h_m \wedge \omega^{a,m}=0 \\
    \delta e^a=D\epsilon^a-\epsilon\fud{a,}{b}e^b
  \end{array}
\right.\label{LinEinstA}\\
&\left\{
  \begin{array}{ll}
    F^{a,b}=D\omega^{a,b}=h_m\wedge h_n C^{am,bn} \label{LinEinstB}\\
    \delta \omega^{a,b}=D\epsilon^{a,b}
  \end{array}
\right. & aa,bb&=\YoungpBB\\
&DC^{aa,bb}=h_m \left(C^{aam,bb}+\frac12 C^{aab,bm}\right) & aaa,bb&=\YoungpCB\notag\\
&DC^{a(k+2),bb}=h_m \left(C^{a(k+2)m,bb}+\frac1{k+2} C^{a(k+2)b,bm}\right) &a(k+2),bb&=\bep(50,20)\put(0,-7){\YoungBB}\put(20,3){\RectARow{3}{k}}\notag\eep\\
&\delta C^{a(k+2),bb}=0\label{LinEinstC}
\end{align}\esubeqs\endgroup
Let us make few comments on this system. $D$ is defined with
respect to the background, $d+\varpi$. The linearized torsion
constraint has a slightly different form because $\omega\fud{a,}{m}\wedge e^{m}$ yields, when
linearized, two types of terms, $\varpi e$ and $\omega h$, the first being hidden inside the background
Lorentz derivative, $D$. Analogously, when linearized, $F^{ab}$
has lost its $\omega\omega$ piece and is simply $D\omega^{a,b}$.

The Minkowski background implies $D^2=0$, so the linearized curvature $D\omega^{a,b}$ is gauge invariant.
Therefore, $C^{aa,bb}$ on the \rhs of \eqref{LinEinstB} should not transform under $\epsilon^{a,b}$ anymore,
the same being true for all $C^{a(k+2),bb}$. This is due to the fact that $D$ has lost its dynamical spin-connection $\omega^{a,b}$, of which $\epsilon^{a,b}$ is a gauge parameter. The absence of mixing of the form $\omega C$ implies there is no need to rotate $C^{a(k+2),bb}$ anymore and similarly for the $\epsilon^a$-symmetry. This all follows from linearization, of course.

The linearized Einstein equations, i.e. the Fronsdal equations for
$s=2$, are imposed by \eqref{LinEinstA}-\eqref{LinEinstB},
which we will show for the spin-$s$ generalization later. The rest
of equations (and the vanishing of the torsion) are constraints in a
sense that they do not impose any differential equations merely
expressing one field in terms of derivatives of the other.



\mysection{Unfolding, spin by spin}\setcounter{equation}{0}\label{sec:UnfoldingSpins}
In this section we begin to move towards the linearized Vasiliev
equations that describe an infinite multiplet of free HS fields in
$AdS$. Some preliminary comments are below.

When fields of all spins are combined together into the multiplet
of a higher-spin algebra the equations they satisfy turn out to be
much transparent and revealing than the equations for individual
fields. We will not follow this idea in this section rather
consider those fields individually, spin by spin. Moreover the
technique that happened to be extremely efficient for HS fields
may look superfluous for some simple cases like a free scalar
field, which we consider at the end.

The idea we blindly follow in this section is to look for the
frame-like (tetrad-like, vielbein-like, ...) formulation for
fields of all spins. The simple guiding principle is to express
everything in the language of differential forms. All fields in
question are differential forms that may take values in some
linear spaces that are viewed as the fiber over the space-time
manifold. The equations of motion are required to have the
following schematic form
\be d(\mbox{field})=\mbox{exterior products of the fields themselves}\ee
This is what we have already achieved for the case of gravity.
Equations of this form are called the {\it unfolded equations}, \cite{Vasiliev:1988xc, Vasiliev:1988sa}. It
is this simple idea that could have been used to discover the
frame-like formulation of gravity and yet it also leads to the
nonlinear frame-like formulation of higher-spin fields. The
detailed and abstract discussion is left to Section
\ref{sec:Unfolding}, where we show that such equations are intimately connected with the theory of Lie
algebras.

The structure on the base manifold that we will need is quite poor
--- only differential forms are allowed to be used along with
the operations preserving the class of differential forms, i.e.
the exterior derivative $d$ and the exterior product, $\wedge$.

The spin-two case corresponds to the gravity itself. At the
nonlinear level we find a perfect democracy among fields of all
spins. This is not quite so at the free level because all fields
propagate over Minkowski (this section) or (anti)-de Sitter space
(Section \ref{sec:AdSLinear}) which is the vacuum value of the
spin-two. We have the background non-propagating gravitational
fields defined by $h^a$, $\varpi^{a,b}$, which obey
\eqref{adsFrameLike} with $\Lambda=0$, i.e. the Yang-Mills field
strength of the Poincare algebra is zero. These fields are
considered to be the zero order vacuum ones within the
perturbation theory, while propagating fields of all spins,
including the spin-two itself, are of order one, so that the
equations are linear in perturbations. If we had a master field,
say $W$, whose components correspond to fields of all spins and
the full theory in terms of $W$, then we could say that we
expanded it as $W_0+g W_1+\mbox{higher orders}$, where $g$ is a
small coupling constant and
\be W_0=\{\underbrace{0}_{s=0},\underbrace{0}_{s=1}, \underbrace{h^a, \varpi^{a,b}, C^{aa,bb}=0,...,C^{a(k+2),bb}=0,...}_{s=2},\underbrace{0}_{s=3},\underbrace{0}_{s=4},...\}\,,\ee
where the sector of spin-two is non-degenerate and contains $h^a$, $\varpi^{a,b}$ that obey
$Dh^a=0$ and $D^2=0$, $D=d+\varpi$, which is equivalent to having Minkowski space.

We would like to read off the part of the theory that is linear in
$W_1$ and determine $W_1$ itself. Recall that in the case of
linearized gravity $W_1$ was found to contain one-forms $e^a$,
$\omega^{a,b}$ and infinitely many zero-forms $C^{a(k+2),bb}$ that
start from the Weyl tensor $C^{aa,bb}$ for $k=0$. For the
Minkowski vacuum all $C^{a(k+2),bb}$ vanish. If we wish to expand
the theory over the space with say a black hole inside, then the
Weyl tensor would be nonvanishing.

Each of the cases considered below consists of two parts: a quasi-derivation explaining why the solution
has its particular form and a part with the results, where the
system of equations is written down. The relevant original references include \cite{Vasiliev:1980as, Lopatin:1987hz, Vasiliev:2001wa,Vasiliev:2003ev}.

\mysubsection{\texorpdfstring{$\boldsymbol{s=2}$}{spin-two} retrospectively}\label{sec:SpinTwoRetro} By
the example of gravity we would like to show the main steps of how
one could have discovered the frame-like gravity from the Fronsdal
theory for $s=2$ using the idea that the theory should be
formulated in terms of differential forms and bearing in mind the
Yang-Mills theory. The starting point is a symmetric traceful field $\phi^{aa}$, the Fronsdal field at $s=2$, that
has a gauge symmetry, \eqref{gaugetransFronsdal},\footnote{In Minkowski space there is no clear distinction
between world and fiber indices, so the next few sentences are rather heuristic. }
\begin{align}
\delta\phi^{aa}&=\pl^a \epsilon^{a} && \Longleftrightarrow && \delta\phi_{\mm\mm}=\pl_\mm \epsilon_{\mm}
\end{align}
and the Fronsdal equations, \eqref{FronOper}, specialized to $s=2$. We would like to
replace $\phi^{aa}$ with a yet unknown  differential form $e^{*}$ of
a certain degree $q$ taking values in some tensor representation
of the Lorentz algebra, denoted by placeholder $*$. The gauge
transformations are then $\delta e^*=d \epsilon^*$, where
$\epsilon^*$ is a differential form of degree $q-1$ that takes
values in the same Lorentz representation $*$. Comparing
\begin{align}
\delta\phi^{aa}&=\pl^a \epsilon^{a} && \Longleftrightarrow && &\delta e^*=d \epsilon^*=dx^\mm \pl_\mm \epsilon^*
\end{align}
we see that one index $a$ carried by $\pl^a$ should turn into $\pl_\mm$, then
the leftover index should belong to $\epsilon^*$, i.e. $*=a=\YoungpA$. Since $e^*$
must carry the same indices as its gauge parameter we have $e^a_\mm$ and hence $\delta e^a_\mm = \pl_\mm \epsilon^a$. That the gauge parameter is a
Lorentz vector immediately tells us that the frame field has a
vector index too.
Since the world indices of differential forms are distinct from
from fiber indices, to write $\pl_\mm \epsilon^a+\pl^a
\epsilon_\mm$ is meaningless. We also see that the frame field must be a one-form
because the gauge parameter is naturally a zero-form.
Consequently, we found
\be\delta e^a=d\epsilon^a\,.\ee
That world and fiber indices in $e^a_\mm$ are distinct types of indices implies
that there are no symmetry/trace conditions between $a$ and $\mm$
in $e^a_\mm$. In particular $e^a_\mm$ contains more components as
compared to original $\phi^{aa}$, which is symmetric. In Minkowski space in Cartesian
coordinates there is no difference between world and fiber
indices, but formally we can always use the background $h^{\mm a}$ to
convert the world indices. With the help of $h^{\mm a}$ we see that in
addition to the totally symmetric component to be identified with
$\phi^{aa}$ the frame field, $e^{a|b}=e^a_\mm h^{\mm b}$, contains
an antisymmetric component too, i.e.
\be e^{a|b}\sim \YoungpA\otimes\YoungpA=\YoungpAA\oplus\left(\YoungpB\oplus\bullet\right)\,.\ee
The symmetric part of the frame field transforms as the Fronsdal
field. Indeed,
\begin{align} &\delta e^{a|b} =\pl^b\epsilon^a && \Longrightarrow&&  \delta (e^{a|b}+e^{b|a}) =\pl^b\epsilon^a+\pl^a\epsilon^b\,.\end{align}
The antisymmetric component can be a propagating field unless we
manage to get rid of it. The simplest solution is to
introduce a new gauge symmetry, which acts algebraically and whose
purpose is to gauge away the antisymmetric component completely,
\be \delta e^a_\mm= \pl_\mm \epsilon^a -\epsilon\fud{a,}{\mm}\,,\ee
where $\epsilon^{a,b}=-\epsilon^{b,a}$ and we do not care about
the difference between world and fiber indices since we can always
set $h^a_\mm=\delta^a_\mm$ at a point. To make the last expression
meaningful we can cure it as
\begin{align} dx^\mm\, \delta  e^a_\mm&= dx^\mm\, \pl_\mm \epsilon^a -dx^\mm\, h^b_{\mm}\epsilon\fud{a,}{b} && \Longleftrightarrow&& \delta e^a=d\epsilon^a-h_b \wedge\epsilon^{a,b}\,.\end{align}
Now we can gauge away the antisymmetric part of the frame field
\begin{align} \delta (e^{a|b}-e^{b|a})=\pl^b\epsilon^a-\pl^a\epsilon^b-2\epsilon^{a,b}\,.\end{align}
Indeed, the gauge symmetry with $\epsilon^{a,b}$ is algebraic and obviously $\epsilon^{a,b}$ has the same number of components. Therefore, we can always impose $(e^{a|b}-e^{b|a})=0$ and the condition for the left-over gauge symmetry
\begin{align} 0=\delta (e^{a|b}-e^{b|a})=\pl^b\epsilon^a-\pl^a\epsilon^b-2\epsilon^{a,b}\end{align}
expresses $2\epsilon^{a,b}=\pl^b\epsilon^a-\pl^a\epsilon^b$ and
does not restrict $\epsilon^a$. So, when $\YoungmAA=0$ gauge is
imposed the whole content of the frame field is given by the
Fronsdal field with its correct gauge transformation law.

In practice we do not need to impose $\YoungmAA=0$ gauge or to go to
the component form and convert indices with the inverse background
frame field $h^{\mm b}$. What we need is a guarantee that the
theory can be effectively reduced to the Fronsdal one (at least at
the linearized level) and that there are no extra propagating
degrees of freedom.

We used the Fronsdal theory as a starting point for the frame-like
generalization, but all the statements, e.g. that the equations to
be derived below do describe a spin-$s$ representation of the
Poincare algebra, can be made without any reference to the
Fronsdal theory. Since the frame field is needed anyway, e.g., to
make fermions interact with gravity, there is no reason to go back
to the Fronsdal theory once the frame-like higher-spin
generalization is worked out.

Since the background may not be given in Cartesian coordinates the
form of the gauge transformations valid in any coordinate system
reads
\begin{align} \delta e^a=D\epsilon^a-h_b \wedge\epsilon^{a,b}\label{RetrogaugeA}\,,\end{align}
where $D=d+\varpi$ is the background Lorentz derivative and we recall that $D^2=0$.

Now we can recognize \eqref{RetrogaugeA} as the combination of
local translations and local Lorentz transformations. The local
translation symmetry has its roots in the Fronsdal's symmetry,
while the purpose for local Lorentz transformations is to
compensate the redundant components resided inside the frame
field.

We found a reason for the gauge parameter $\epsilon^{a,b}$. In the
realm of the Yang-Mills theory, there are no homeless gauge
parameters. Hence, there must be a gauge field $\omega^{a,b}$,
which is a one-form and its gauge transformation law starts as
$\delta \omega^{a,b}=d\epsilon^{a,b}+...$ or equivalently
\be\delta \omega^{a,b}=D\epsilon^{a,b}+...\quad \,.\label{RetrogaugeB}\ee
The existence of the spin-connection goes hand in hand with the
existence of $\epsilon^{a,b}$.

Given gauge transformations \eqref{RetrogaugeA}-\eqref{RetrogaugeB} it is easy to guess the gauge invariant field-strengths to be
\begin{align} T^a&=De^a-h_b\wedge \omega^{a,b}\,, && F^{a,b}=D\omega^{a,b}+...\quad\,,\end{align}
which is the torsion and the linearized Riemann two-form. The
question of whether we should impose $T^a=0$ and what we should
write instead of $...$ in $F^{a,b}$ is dynamical. One can see that
setting $T^a$ to zero expresses $\omega$ in terms of $e$ and imposes no
differential equations on the latter. At this point one can repeat
the analysis of the previous section to find that
$F^{a,b}=h_m\wedge h_n C^{am,bn}$ imposes the Fronsdal equations,
etc.

The case of $s=2$ is deceptive since $h^a_\mm$ is already a
vielbein. The illustration with spin-two is not self contained
because the background space must have been already defined in
terms of $h^a$ and $\varpi^{a,b}$. We can imagine that we know how
to define the background geometry but we are unaware of how to put
propagating fields on top of this geometry. Eventually we find out
that propagating field $e^a$ can be naturally combined with the
background $h^a$ into the full vielbein.

\mysubsection{\texorpdfstring{$\boldsymbol{s\geq2}$}{higher-spin}} \label{subsec:UnfldAnySpin}
We would like to systematically look for an
analog of the frame-like formalism for the fields of any spin
$s\geq2$. The starting point is the gauge transformation law and
the algebraic constraints on the Fronsdal field and its gauge parameter
\begin{align}
\delta\phi^{a(s)}&=\pl^a \epsilon^{a(s-1)}\,, &&\phi\fud{a(s-4)mn}{nm}\equiv0 \,, &&\epsilon\fud{a(s-3)n}{n}\equiv0\,.
\end{align}
The Fronsdal field has to be embedded into a certain generalized
frame field $e\fm{q}^{a...}$ with the form degree and the range of
fiber indices to be yet determined. We  understand already that
writing $de\fm{q}^{a...}=...$ implies a gauge transformation of the
form $\delta e\fm{q}^{a...}=d\xi\fm{q-1}^{a...}$ with
$\xi\fm{q-1}^{a...}$ being the form of degree $q-1$ valued in the
same representation of the fiber Lorentz algebra. This gauge
symmetry is in general reducible
$\delta\xi\fm{q-1}^{a...}=d\chi\fm{q-2}^{a...}$ unless $q=1$. We
know that there are no reducible gauge symmetries in the case of
totally-symmetric fields. Therefore, $q=1$ and gauge parameter is a zero-form, $\xi^{a...}$,
i.e. it has no differential form indices. In order to match
\be dx^\mm\, \pl_\mm\xi^{a...}\qquad \mbox{with}\qquad\pl^a\epsilon^{a(s-1)}\ee
the gauge parameter must be $\xi^{a(s-1)}$, i.e. symmetric and traceless in the
fiber indices. So, the frame field must be one-form $e^{a(s-1)}\fm{1}$
while the gauge transformation law now reads
\begin{align}\label{UnfldFrondsGTA}
\delta e^{a(s-1)}\fm{1}&=d\xi^{a(s-1)}+... \,, &&e\fud{a(s-3)m}{m}=\xi\fud{a(s-3)m}{m}\equiv0\,.
\end{align}
Let us convert the world indices in the last formula to the fiber
\begin{align}\label{UnfldFrondsGT}
&\delta e^{a(s-1)|b}=\pl^b \xi^{a(s-1)}\,, && e^{a(s-1)|b}=e^{a(s-1)}_\mm h^{\mm b}\,.
\end{align}
As in the spin-two case the frame field contains more components than the original metric-like, Fronsdal, field. The irreducible content of $e^{a(s-1)|b}$ is given by\footnote{We refer to the Appendix \ref{app:YoungTensorProd}, where the tensor product rules are discussed.}
\begin{align}
\AYoungp{4}{s-1}\otimes \YoungpA=\left(\AYoungp{5}{s}\oplus\AYoungp{4}{s-2}\right)\oplus\BYoungp{4}{1}{s-1}{}\,,
\end{align}
where the first two components, when put together, are exactly the
content of the Fronsdal tensor since a double-traceless rank-$s$
tensor $\phi^{a(s)}$ is equivalent to two traceless tensors
$\psi^{a(s)}$ and $\psi^{a(s-2)}$ of ranks $s$ and $s-2$
\begin{align*}
\phi^{a(s)}=\psi^{a(s)}+\frac{1}{d+2s-4}\eta^{aa}\psi^{a(s-2)}\,, && \phi\fud{a(s-2)m}{m}=\psi^{a(s-2)}\,, &&\phi\fud{a(s-4)mn}{nm}\equiv0\,,
\end{align*}
where the coefficient is fixed by the relation in the middle. In terms of field components the decomposition of $e^{a(s-1)|b}$ into a Fronsdal-like field and the leftover traceless tensor $\psi^{a(s-1),b}$ with the symmetry of \parbox{28pt}{\BYoung{4}{1}{s-1}{}} reads
\begin{align} e^{a(s-1)|b}=\frac1s\phi^{a(s-1)b}&+{\psi}^{a(s-1),b}+\notag\\
&+\frac{1}{s(d+s-4)}\left(\frac{(s-2)}2\eta^{ab}\phi\fud{a(s-2)m}{m}-\eta^{aa}
\phi\fud{ba(s-3)m}{m}\right)\,.\label{FrameFieldDecomp}\end{align} The overall normalization is fixed in such a way that
\begin{align} e^{a(s-1)|a}&=\phi^{a(s)} \label{UnfldsFronsdalEmb}\,.\end{align}
(remember that $\psi^{a(s-1),a}\equiv0$ and it drops out). Note
the group of terms in the second line of \eqref{FrameFieldDecomp}
which are absent in the spin-two case. The coefficients are fixed
from \eqref{UnfldsFronsdalEmb} and from the tracelessness of
$e^{a(s-1)|b}$ in $a$ indices. In the language of differential
forms with the Fronsdal field $\phi^{a(s)}$ and $\psi^{a(s-1),b}$
treated as zero-forms the embedding of the Fronsdal field reads
\begin{align}\notag
e^{a(s-1)}\fm{1}=\frac1s h_b\,\phi^{a(s-1)b}&+h_b\psi^{a(s-1),b}+\\&+\frac{1}{s(d+s-4)}\left(\frac{(s-2)}{2} h^a\phi\fud{a(s-2)c}{c}-\eta^{aa}h_m \phi\fud{a(s-3)mc}{c}\right)\,.
\end{align}

Such an unnatural within the metric-like approach condition as vanishing of the second trace
\be g^{\mm\mm}g^{\nn\nn}\phi_{\mm\mm \nn\nn \rr(s-4)}\equiv0\label{doubletracelessnessmetric}\ee
comes from a very natural condition within the frame-like approach
--- the frame field is an irreducible fiber tensor. The condition
of the type \eqref{doubletracelessnessmetric} is a source of
problems in the interacting theory as it has to be either
preserved or deformed when $g_{\mm\nn}$ is not Minkowski or (anti)-de Sitter but a dynamical field. In contrast, $\eta_{bb}e^{bba(s-3)}\equiv0$, \eqref{UnfldFrondsGTA}, causes no problem since $\eta_{aa}$ is a non-dynamical object, the
dynamical gravity is described by the frame field $e^a\fm{1}$. The
identification of the Fronsdal tensor as the totally symmetric
component of the frame field we discussed is essentially linear.

Combining \eqref{UnfldFrondsGT} with \eqref{UnfldsFronsdalEmb} we
recover the Fronsdal gauge transformation law for the totally
symmetric component of the frame field.

As in the case of spin-two, the frame field contains an additional
component $\psi^{a(s-1),b}$. To prevent it from becoming a
propagating field we can introduce an algebraic gauge symmetry
with a zero-form parameter $\xi^{a(s-1),b}$, so as to make it
possible to gauge away $\psi^{a(s-1),b}$. As a tensor $\xi^{a(s-1),b}$ is irreducible, i.e. Young and traceless, since the unwanted component is like that. This correction can be
written as follows
\begin{align}
\delta e^{a(s-1)}\fm{1}=-h_m \xi^{a(s-1),m}\label{UnfldsSecondGA}
\end{align}
and coincides for $s=2$ with the action of local Lorentz
rotations. Once we had to introduce $\xi^{a(s-1),b}$ there must be
an associated gauge field $\omega^{a(s-1),b}\fm{1}$, the
generalization of the spin-connection. This allows us to write the
first equation immediately as
\begin{align}
&de^{a(s-1)}\fm{1}=h_m\wedge \omega^{a(s-1),m}\fm{1} &&  \label{UnfldsFirstA}\end{align}
together with the gauge transformations
\begin{align}
\delta e^{a(s-1)}\fm{1}&=d\xi^{a(s-1)}-h_m\xi^{a(s-1),b}\,, && \delta \omega^{a(s-1),b}\fm{1}=d\xi^{a(s-1),b}\,.
\end{align}
Already these equations as
well as all others below, point towards some algebra, whose
connection $A$ contains the gravity sector in terms of
vielbein/spin-connection and higher-spin connections, of which
$e^{a(s-1)}$ is a particular component. What we see is the
linearization of $F=dA+\frac12[A,A]$ under $A_0+g A_1$, which
gives $F=dA+[A_0,A]$, $A_0=\{h,\varpi\}$

Once we have \eqref{UnfldsFirstA} we can check its integrability to read off the restrictions on $d\omega^{a(s-1),b}\fm{1}$
\begin{align}
&0\equiv d de^{a(s-1)}\fm{1}=d\left( h_m\wedge \omega^{a(s-1),m}\fm{1}\right)= -h_m\wedge d\omega^{a(s-1),m}\fm{1}\,.
\end{align}
In principle one should write down all the tensors that can appear on the \rhs of $d\omega^{a(s-1),m}\fm{1}$ to see which of them passes trough the integrability condition. We are not going to present this analysis and just claim that the solution has a nice form of
\begin{align}\label{UnfldsSecondB}
d\omega^{a(s-1),b}\fm{1}&=h_m\wedge \omega^{a(s-1),bm}\fm{1}\,, && a(s-1),bb=\parbox{50pt}{{\RectBYoung{5}{$ s-1$}{\YoungB}}}\,,
\end{align}
where the new field is a one-form $\omega^{a(s-1),bb}\fm{1}$ with
values in the irreducible representation (Young and traceless) of
the Lorentz algebra specified above. The integrability holds
thanks\footnote{The following trivial identities are used here and
below
\be\label{Unfldhidentities} h^a\wedge h^b+h^b\wedge h^a\equiv0\,, \qquad\qquad h_m \wedge h^m\equiv0\,.\ee} to $h_m\wedge h_n \wedge\omega^{a(s-1),mn}\fm{1}\equiv0$.
A new one-form field comes with the associated gauge parameter, so the last equation ensures the invariance under
\begin{align}
\delta \omega^{a(s-1),b}\fm{1}=d\xi^{a(s-1),b}-h_m\xi^{a(s-1),bm}\,, && \delta \omega^{a(s-1),bb}\fm{1}=d\xi^{a(s-1),bb}\,.
\end{align}

Let us emphasize that we found a new field that was absent in the
case of gravity! To have a vielbein and a spin-connection was
enough for a spin-two field. The field $\omega^{a(s-1),bb}\fm{1}$ is the first of
the {\it extra fields} that appear in the frame-like formulation
of higher-spin fields, \cite{Fradkin:1986qy,Lopatin:1987hz}. The necessity for the {\it extra fields}
can be seen already in \eqref{UnfldsFirstA}. From the pure gravity
point of view this is a torsion constraint, which allows one to
express spin-connection as a derivative of the vielbein. This is
not so for $s>2$. One can either find that spin-connection
$\omega\fm{1}^{a(s-1),b}$ cannot be fully solved from
\eqref{UnfldsFirstA} or observe that \eqref{UnfldsFirstA} is
invariant under $\delta
\omega\fm{1}^{a(s-1),b}=h_m\xi^{a(s-1),bm}$. These are equivalent statements
as the component of the
spin-connection affected by the extra gauge symmetry cannot be
solved for. The appearance of an additional symmetry tells us
there is an associated gauge field, the first extra field. The
extra gauge symmetry can also be seen in the quadratic action
built from $e^{a(s-1)}\fm{1}$ and $\omega^{a(s-1),b}\fm{1}$, \cite{Vasiliev:1980as}.

At this stage it is necessary to see if the Fronsdal equations
appear inside \eqref{UnfldsSecondB}. It is a right time for them to emerge since $\omega^{a(s-1),b}\fm{1}$
is expressed as $\pl e^{a(s-1)}\sim \pl \phi^{a(s)}$ via \eqref{UnfldsFirstA} and \eqref{UnfldsSecondB} contains $\pl\pl \phi^{a(s)}$. The Fronsdal equations are imposed by
the same trick as in gravity --- the \rhs of \eqref{UnfldsSecondB} parameterizes those derivatives
that can be non-zero on-shell. We postpone
this check till the summary section. Once we have found the second
equation we can check its integrability. This process continues
smoothly giving
\begin{align}
&d \omega^{a(s-1),b(k)}\fm{1}=h_m\wedge \omega\fm{1}^{a(s-1),b(k)m}\,, && a(s-1),b(k)=\BYoungp{4}{3}{s-1}{k}\,,
\end{align}
until the second row in the Young diagram is saturated
\begin{align}
&d \omega^{a(s-1),b(s-2)}\fm{1}=h_m\wedge \omega\fm{1}^{a(s-1),b(s-2)m}\,, && a(s-1),b(s-1)=\BYoungp{4}{4}{s-1}{}\,,
\end{align}
the integrability of which implies $h_k\wedge d\omega\fm{1}^{a(s-1),b(s-2)k}\equiv0$, the unique solution being
\begin{align}
d\omega\fm{1}^{a(s-1),b(s-1)}&=h_m\wedge h_n C^{a(s-1)m,b(s-1)n}\,, && a(s),b(s)=\BYoungp{5}{5}{s}{}\,,
\end{align}
where $C^{a(s),b(s)}$ is a zero-form. It is easy to see that it is
a solution. Indeed, $h_k\wedge h_m\wedge h_n\wedge
C^{a(s-1)m,b(s-2)kn}\equiv0$ since two anticommuting vielbein
one-forms are contracted with two symmetric indices, \eqref{Unfldhidentities}.

On-shell $C^{a(s),b(s)}$ is expressed as order-$s$ derivative of
the Fronsdal field. It is called the (generalized) Weyl tensor for
a spin-$s$ field and it coincides with the (linearized) Weyl
tensor if we set $s=2$. Apart from the Fronsdal operator the Weyl
tensor is also gauge invariant. One can prove that there are two
basic gauge invariants, the Fronsdal operator and the Weyl tensor.
The rest of invariants are derivatives of these two.

Again we have the integrability condition for the Weyl tensor,
\be h_m\wedge h_n \wedge d  C^{a(s-1)m,b(s-1)n}\equiv 0\,.\label{dCthreelines} \ee
This we can easily solve. First, we decompose $\pl^c C^{a(s),b(s)}$ into
irreducible tensors of the Lorentz algebra
\begin{align}
\BYoungp{5}{5}{s}{}\otimes \YoungpA=\parbox{60pt}{\bep(50,20)\put(50,10){\YoungA}\put(0,0){\RectBRow{5}{5}{$s$}{}}\eep}
\oplus\BYoungp{5}{4}{s}{s-1}\oplus\parbox{50pt}{\RectCRow{5}{5}{1}{$s$}{}{}}
\end{align} All of these can in principle appear on the \rhs of $dC^{a(s),b(s)}=h...$.
The third component, call it $C^{a(s),b(s),c}$ does not pass the
integrability test, giving $h_m\wedge h_n \wedge h_k
C^{a(s-1)m,b(s-1)n,k}\neq0$. The first component
$C^{a(s+1),b(s)}$ passes the test $h_m\wedge h_n \wedge h_k
C^{a(s-1)km,b(s-1)n}\equiv0$. The reason is simple, three
vielbeins form a rank-three antisymmetric tensor and there is not
enough room in the tensor having the symmetry of a two-row diagram
to be contracted with it\footnote{It is the property of tensors
having the symmetry of one- and two-row Young diagram that
anti-symmetrization over any three indices vanishes.}. It may seem
that the second component, $C^{a(s),b(s-1)}$ passes the test too,
this is not true, however. The reason is that it appears as the
trace. Had we been not interested in traces, the decomposition
would have contained two components only, given by the $gl_d$
tensor product rule,
\begin{align}
\BYoungp{5}{5}{s}{}\otimes_{gl_d} \YoungpA=\parbox{60pt}{\bep(50,20)\put(50,10){\YoungA}\put(0,0){\RectBRow{5}{5}{$s$}{}}\eep}
\oplus\parbox{50pt}{\RectCRow{5}{5}{1}{$s$}{}{}}
\end{align}
The point is that both the components in the decomposition above
have their traces of the type $C^{a(s),b(s-1)}$ and these are the
same $\pl_m C^{a(s-1),b(s-2)m}$. From the $so(d-1,1)$
decomposition, we know that there is only one trace. The traces
are irrelevant for the integrability condition, so if the second
component above does not go through the integrability test, so do
all its traces, i.e. $C^{a(s),b(s-1)}$. More precisely, on the r.h.s. of \eqref{dCthreelines} one finds
a group of terms involving $C^{a(s),b(s-1)}$
\begin{align}
\eqref{dCthreelines}\ni h_a\wedge h_b \wedge h_c \left(2C^{a(s),b(s-1)}\eta^{bc}-\frac{1}{s-1}C^{a(s),b(s-2)c}\eta^{bb}+...\right)\neq0
\end{align}
that does no vanish unless $C^{a(s),b(s-1)}\equiv0$ because of the second term. Eventually, one is left with $C^{a(s+1),b(s)}$ and the equation
now reads
\be
\label{UfldHSWeylFirst}dC^{a(s),b(s)}=
h_m\left(C^{a(s)m,b(s)}+{\frac{1}2}C^{a(s)b,b(s-1)m}\right)\,.
\ee
Again, $C^{a(s)m,b(s)}$ alone does not have the symmetry of the
{\it l.h.s}, hence we have to project it appropriately by hand,
which is done with the help of the second term in the brackets.
Proceeding this way we arrive at
\begingroup\allowdisplaybreaks\begin{align}\notag
dC^{a(s+k),b(s)}&=
h_m\left(C^{a(s+k)m,b(s)}+\frac{1}{k+2}C^{a(s+k)b,b(s-1)m}\right)\,, && a(s+k),b(s)=\parbox{70pt}{\bep(60,20)\put(40,10){\RectARow{3}{$k$}}
\put(0,0){\RectBRow{4}{4}{$s$}{}}\eep}\,.
\end{align}\endgroup
The fields $C^{a(s+k),b(s)}$ are expressed as derivatives $\pl^a...\pl^a C^{a(s),b(s)}$ of the Weyl tensor.

\paragraph{Unfolded equations for any \texorpdfstring{$\boldsymbol{s}$}{spin}.} Summarizing, we get the following diverse equations
\begin{align}
&\begin{cases}
    \DL \omega^{a(s-1),b(k)}\fm{1}=h_c\wedge\omega^{a(s-1),b(k)c}\fm{1}\,, \\
    \delta \omega^{a(s-1),b(k)}\fm{1}=\DL\xi^{a(s-1),b(k)}-h_c\,\xi^{a(s-1),b(k)c}\,, \end{cases} && 0\leq k<s-1\,,\label{UnfldOneFormsA}\\
&\begin{cases}
    \DL \omega^{a(s-1),b(s-1)}\fm{1}=h_c\wedge h_d\,C^{a(s-1)c,b(s-1)d}\,, \\
    \delta\omega^{a(s-1),b(s-1)}\fm{1}=\DL\xi^{a(s-1),b(s-1)}\,,
\end{cases} && k=s-1\,,\\
&\begin{cases}
\DL C^{a(s+i),b(s)}=h_c\left(C^{a(s+i)c,b(s)}+\frac{1}{i+2}C^{a(s+i)b,b(s-1)c}\right)\,, \\
\delta C^{a(s+i),b(s)}=0\,,
\end{cases}&& i=0,1,2,3,...\,,\label{UnfldZeroFormsEq}
\end{align}
where we have replaced $d$ with the background Lorentz derivative
$\DL=d+\varpi$, which makes the system valid in any coordinate
system in Minkowski space. Note that $\DL^2=0$. Let us  enlist
once again the spectrum of fields (differential form degree and
the Young shape of the fiber indices)
\begin{align*}
    \begin{tabular}{l}
      grade\\
    \end{tabular}&:&& 0 && 1 &&...&& s-1 & &s & &s+1 & & ...\\
    \begin{tabular}{l}
      Young\\
      shape\\
    \end{tabular}
    &: & & \AYoungp{4}{s-1} && \BYoungp{4}{1}{s-1}{} && ...&&   \BYoungp{4}{4}{s-1}{} & & \BYoungp{5}{5}{s}{} & & \parbox{60pt}{\bep(50,20)\put(50,10){\YoungA}\put(0,0){\RectBRow{5}{5}{$s$}{}}\eep} & & ... \\
    \begin{tabular}{l}
      degree\\
    \end{tabular}&:&& 1 && 1 &&...&&  1 & & 0 & & 0 & & ...,
\end{align*}

We recover below the Fronsdal equations from the unfolded ones.
What is left aside is that the most of the fields are not
dynamical. Except for those components of the frame field which are
in one-to-one correspondence with the Fronsdal field all the other
components are either auxiliary or Stueckelberg. The auxiliary fields are
those that can be expressed as derivatives of the Fronsdal field
by virtue of the equations of motion, while the Stueckelberg fields can
be gauged away with the help of the algebraic gauge symmetry
(similarly to how the extra components of the frame field can be
gauged away by local Lorentz transformations). A rigorous proof of these facts
requires an advanced technology, called \Sigm-cohomology, \cite{Lopatin:1987hz}.

\paragraph{Fronsdal equations from unfolded ones.} We already know
the way the Fronsdal field is embedded into the frame-like field.
Let us now show where the Fronsdal equations reside. We need the
first two unfolded equations with the form indices revealed
\begin{align*}
\pl_{\mm}^{\vphantom{a}}e^{a(s-1)}_\nn-\pl_{\nn}^{\vphantom{a}}e^{a(s-1)}_\mm&=
h_{c\mm}^{\vphantom{a}}\omega^{a(s-1),c}_\nn-h_{c\nn}^{\vphantom{a}}\omega^{a(s-1),c}_\mm,\\
\pl_\mm^{\vphantom{a}}\omega^{a(s-1),b}_\nn-\pl_\nn^{\vphantom{a}}\omega^{a(s-1),b}_\mm&=
h_{c\mm}^{\vphantom{a}}\omega^{a(s-1),bc}_\nn-h_{c\nn}^{\vphantom{a}}\omega^{a(s-1),bc}_\mm\,.
\end{align*}
Converting all indices to the fiber
\begin{align*}
&e^{a(s-1)|b}\equiv e^{a(s-1)}_\mm h^{b\mm}, && \omega^{a(s-1),b|c}\equiv\omega^{a(s-1),b}_\mm h^{c\mm}, &&
\omega^{a(s-1),bb|c}\equiv\omega^{a(s-1),bb}_\mm h^{c\mm}
\end{align*}
we  get
\besubeqs\begin{align}
\label{UnfldHSEqA}\pl^ce^{a(s-1)|d}-\pl^de^{a(s-1)|c}&=\omega^{a(s-1),c|d}-\omega^{a(s-1),d|c},\\
\label{UnfldHSEqB}\pl^c\omega^{a(s-1),b|d}-\pl^d\omega^{a(s-1),b|c}&=
\omega^{a(s-1),bc|d}-\omega^{a(s-1),bd|c}\,.
\end{align}\esubeqs
By virtue of the first equation the spin-connection can be expressed in terms of the first derivatives of the frame field which contains the Fronsdal tensor together with a pure gauge components. Then, the second equation imposes the Fronsdal equation and expresses the second spin-connection. In order to project onto the Fronsdal equations we contract (\ref{UnfldHSEqB}) with $\eta_{bd}$ and symmetrize over $c$ and $a(s-1)$, which gives
\be\label{UnfldHSEqC}F^{a(s)}=\pl_c\omega^{a(s-1),c|a}-\pl^a\omega^{a(s-1),c|}_{\phantom{a(s-1),c|}c}=0.\ee Note that the extra field disappeared because of the specific projection made. Now we symmetrize $a(s-1)$ and $d$ in (\ref{UnfldHSEqA}) \be
\omega^{a(s-1),c|a}=\pl^ce^{a(s-1)|a}-\pl^ae^{a(s-1)|c}\,,\ee so that we can express the first term in \eqref{UnfldHSEqC}. To express the second term of \eqref{UnfldHSEqC} we contract (\ref{UnfldHSEqA}) with $\eta_{da}$,
\be\omega\fud{a(s-2)m,c|}{m}=\pl^ce\fud{a(s-2)m|}{m}-\pl_m e^{a(s-2)m|c}\,.\ee
then symmetrize $a$'s with $c$ and use Young symmetry $\omega^{a(s-2)m,a}=-\omega^{a(s-1),m}$,
\be\omega^{a(s-1),m|}_{\phantom{a(s-1),m|}m}=
\left(\pl_me^{a(s-2)m|a}-\pl^ae^{a(s-2)m|}_{\phantom{a(s-2)m|}m}\right)\,.\ee
Plugging the last two equations into (\ref{UnfldHSEqC}) we find
\be\square
e^{a(s-1)|a}-\pl^a\pl_c\left(e^{a(s-2)c|a}+e^{a(s-1)|c}\right)+2\pl^a\pl^ae^{a(s-2)c|}_{\phantom{a(s-2)c|}c}=0\,.\ee
Now we have to remember how the Fronsdal field is embedded into
the frame field
\begin{align*}
e^{a(s-1)|a}&=\phi^{a(s)}\,, && e^{a(s-2)c|}_{\phantom{a(s-2)c|}c}=\frac12\phi^{a(s-2)c}_{\phantom{a(s-2)c}c}\,, && e^{a(s-2)c|a}+e^{a(s-1)|c}=\phi^{a(s-1)c}\,.
\end{align*}
Magically, all the terms come exactly in the combinations above,
which leads to the Fronsdal equations, (\ref{FronOper}),
\be\label{UnfldFronsdal}\square\phi^{a(s)}-\pl^a\pl_c\phi^{a(s-1)c}+\pl^a\pl^a\phi^{a(s-2)c}_{\phantom{a(s-2)c}c}=0.\ee
We  have seen already that the gauge transformations for the
totally symmetric component of the frame field $e^{a(s-1)|a}$ are
those of the Fronsdal field while the extra component can be
gauged away.

\paragraph{Rough structure of HS connections and zero-forms.} As it was argued, all of the fields in
the unfolded equations above except for certain components of higher-spin vielbein are expressed as derivatives of the Fronsdal field. Let us illustrate in the language of Young diagrams what are the
derivatives of $\phi^{a(s)}$ that $\omega^{a(s-1),b(k)}$ and $C^{a(s+k),b(s)}$ turn out to contain. For simplicity we shall ignore the traces, i.e. we will work with tensors as if they were tensors of $gl(d)$. Firstly, the connections contain derivatives up to order-$(s-1)$\\
\rule{0pt}{12pt}\\
\begin{tabular}{w{4.2cm}|ccccc}
  \multirow{2}{*}{\parbox{4.0cm}{derivative of Fronsdal and its Young shape}} & $\phi$ & $\pl\phi$ & $\pl\pl\phi$ & ... & $\pl^{s-1}\phi$ \tabularnewline
  & \rule{0pt}{25pt}$\bep(50,20)\put(0,10){\RectARow{5}{$s$}}\eep$ & $\RectBRow{5}{1}{$s$}{}$ & $\bep(50,20)\put(0,0){\YoungB}\put(0,10){\RectARow{5}{$s$}}\eep$ &  & $\RectBRow{5}{4}{$s$}{$s-1$}$ \tabularnewline
  \multirow{2}{*}{\parbox{4.0cm}{HS connection and its Young shape}} & \rule{0pt}{25pt}$e^{a(s-1)}_\mm$ & $\omega^{a(s-1),b}_\mm$ & $\omega^{a(s-1),bb}_\mm$ &... & $\omega^{a(s-1),b(s-1)}_\mm$  \tabularnewline
   & \rule{0pt}{30pt}$\bep(50,20)\put(0,10){\RectARow{4}{$s-1$}}\eep$ & $\RectBRow{4}{1}{$s-1$}{}$ & $\bep(50,20)\put(0,0){\YoungB}\put(0,10){\RectARow{4}{$s-1$}}\eep$ & & $\RectBRow{4}{4}{$s-1$}{$s-1$}$
\end{tabular}\\
Each cell in the second row is produced by the index on partial derivative. Let us note that the order-$k$ derivative of the Fronsdal field has the following decomposition into irreducible Young shapes
\be \pl^b...\pl^b \phi^{a(s)}\sim \RectARow{4}{$k$}\otimes \RectARow{5}{$s$}=\parbox{50pt}{\RectBRow{5}{4}{$s$}{$k$}}\oplus ...\oplus \RectARow{6}{$s+k$}\ee
Therefore, only a subset of derivatives is embedded into higher-spin connections. It turns out that these are enough. The same picture for zero-forms looks as follows\\
\rule{0pt}{12pt}\\
\begin{tabular}{w{4.2cm}|cccc}
  \multirow{2}{*}{\parbox{4.0cm}{derivative of Fronsdal and its Young shape}} & $\pl^s\phi$ & $\pl^{s+1}\phi$ &  $\pl^{s+1}\phi$ & ...  \tabularnewline
  & $\RectBRow{5}{5}{}{$s$}$ & $\bep(60,20)\put(0,0){\RectBRow{5}{5}{}{$s$}}\put(50,10){\YoungA}\eep$ & $\bep(70,20)\put(0,0){\RectBRow{5}{5}{}{$s$}}\put(50,10){\YoungB}\eep$ \tabularnewline
  \parbox{4.0cm}{zero-form} & \rule{0pt}{25pt}$C^{a(s),b(s)}$ & $C^{a(s+1),b(s)}$ & $C^{a(s+2),b(s)}$ & ...
\end{tabular}\\
\rule{0pt}{5pt}\\
In particular it is obvious that the zero-forms contain gauge-invariant derivatives of the Fronsdal field, which can be seen from the position of indices inside the Young diagram. For example, if we compute the gauge variation of the Weyl tensor we need to replace $\phi^{a(s)}$ with $\pl^a\xi^{a(s-1)}$ add $\pl^{b(s)}$ and impose the Young projection.
\be \delta C^{a(s),b(s)}=\delta\,\parbox{50pt}{\bep(50,20)\put(0,0){\RectBRow{5}{5}{}{}}%
\put(23,12){$\phi$}\put(2,2){$\pl$}\put(18,2){$\cdots$}\put(42,2){$\pl$}\eep}=
\parbox{50pt}{\bep(50,20)\put(0,0){\RectBRow{5}{5}{}{}}%
\put(25,12){$\xi$}\put(2,12){$\pl$}\put(2,2){$\pl$}\put(18,2){$\cdots$}\put(42,2){$\pl$}\eep}\equiv0\ee
Then two derivatives will inevitably show up in the same column, which vanishes identically if we change the symmetric basis to the anti-symmetric one.

We see that some of the non-gauge invariant derivatives of the Fronsdal tensor are embedded into the higher-spin connections, while gauge-invariant ones are encoded by zero-forms. This is all true on-shell, i.e. when all of the unfolded equations above are imposed.

\paragraph{Lower-spins.} It is obvious that $s=2$ is a particular case of the above construction.
One simply sets $s=2$ to get the unfolded equations describing free graviton of Section \ref{sec:UnfldGravity} and \ref{sec:SpinTwoRetro}. While one always needs infinitely-many zero-forms, the number of one-forms equals the spin of the field, $s$. We will see below that lower-spin cases, $s=1$ and $s=0$ are just a degenerate and a very degenerate cases of the spin-$s$ unfolded equations. In particular, there are no one-forms in the unfolded formulation of $s=0$ in accordance with the fact that scalar field is not a gauge field. It is worth elaborating on the $s=0,1$ cases from scratch.

\mysubsection{\texorpdfstring{$\boldsymbol{s=0}$}{scalar}}\label{sec:UnfldsScalar}

A free massless scalar field in Minkowski space seems to be the
case where the unfolded approach gives little advantage. At some
point the formulation looks tautological. The only reason to
consider the scalar case is because it is in this form the scalar
field turns out to be embedded into the full higher-spin theory.

Given a scalar field $C(x)$, it is easy to put the Klein-Gordon equation $\square C(x)=0$ in the first order form
\begin{align}
\pl_\mm C(x)&=C_\mm(x)\,, & \eta^{\nn\mm}\pl_\nn C_\mm(x)&=0=\eta^{\nn\mm}\,\pl_\nn\pl_\mm C(x)\,,
\end{align}
which is not yet what we need. We can replace an auxiliary field $C_\mm(x)$ with its fiber representative $C^a(x)$,  $C^a=h^{\mm a}\,C_\mm$ and rewrite the first equation as
\begin{align}\label{UnfldScalarAA}
d C&=h_a C^a && \Longleftrightarrow&& dx^\mm \left(\pl_\mm C=h_{\mm a}C^a\right)\,.
\end{align}
The difficulty is with the second equation, it must be of the form
\begin{align}
d C^a=...
\end{align}
where $...$ means some one-form. We can enumerate all the fields that can appear on the {\it r.h.s.} Indeed, \lhs reads $dx^\mm\, \pl_\mm C^a$. Raising all indices with the inverse vielbein we get $\pl^a C^b$, i.e. just a rank-two tensor. As such it can be decomposed into three irreducible components: traceless symmetric -- \YoungmB, trace -- $\bullet$ and antisymmetric -- \YoungmAA
\begin{align}
\YoungpB&=\pl^a C^a-\frac1d\eta^{aa}\pl_m C^m\,, & & \bullet=\pl_m C^m\,, &&
\YoungpAA=\pl^aC^b-\pl^b C^b\,.\label{ScalarDecompA}
\end{align}
Therefore, introducing $C^{aa}$, $C'$ and $C^{a,b}$ in accordance with the pattern above, the most general \rhs reads
\begin{align}
d C^a= h_m C^{am}+h_m C^{a,m}+\frac1d h^a C'\label{UnfldScalarA}\,.
\end{align}
However, $C^a$ is not an arbitrary fiber vector, we see that it came as a
derivative of the scalar and hence $\pl^a C^b-\pl^b C^a\equiv 0$,
i.e. $C^{a,b}\equiv0$. By virtue of \eqref{UnfldScalarA} we see that $\square
C=C'$. Once $C'$ is present on the \rhs there is no Klein-Gordon
equation imposed. Therefore, we have to set $C'$ to zero too\footnote{Another option is to express $C'$
as a function of $C$. For example, $C'=m^2C$, gives Klein-Gordon equation with a mass term.
One can also introduce an arbitrary potential $C'=V(C)$.}.
Finally, $C^{aa}$ parameterizes those second order derivatives
that are generally non-zero on-shell and the second equation reads
\begin{align}
d C^a=h_m C^{am}\label{UnfldScalarC}\,.
\end{align}
Actually, the absence of $C^{a,b}$ can be seen in checking the
integrability of \eqref{UnfldScalarAA}. Nilpotency of the exterior
derivative $d^2=0$ implies
\be 0\equiv d d C= d( h_a C^a)=-h_a\wedge (h_m C^{am}+h_m C^{a,m}+\frac1d h^a C')=-h_a\wedge h_m C^{a,m}\,, \ee
where we used \eqref{Unfldhidentities}. Since the vielbein is
invertible we get $C^{a,m}\equiv0$.

Can we stop at $C^{aa}$? By applying $d$ to \eqref{UnfldScalarC} we find that $C^{aa}$ is not unconstrained
\begin{align}\label{UnfldScalarBianchiA}
0\equiv d d C^a=d( h_m C^{am})=-h_m\wedge dC^{am}\,.
\end{align}
So we are led to consider possible \rhs in $dx^\mm \pl_\mm
C^{aa}$. Decomposing the tensor product $\YoungmB \otimes
\YoungmA$ with \YoungmB{} and \YoungmA{} representing $C^{aa}$ and
$\pl^m$, respectively, one finds
\begin{align}
\YoungpB \otimes \YoungpA=\YoungpBA\oplus \YoungpC\oplus \YoungA\,,
\end{align}
i.e. the most general equation for $C^{aa}$ reads
\begin{align*}
d C^{aa}=h_m C^{aam}+h_m C^{aa,m}+\left(h^a \tilde{C}^a-\frac2d\eta^{aa}h_m\tilde{C}^m\right)\,.
\end{align*} Note that $h^a \tilde{C}^a$ is not traceless and has to be supplemented with the second term to agree with vanishing trace of $C^{aa}$.
Both the hook \YoungmBA{} component $C^{aa,b}$ and the trace $\tilde{C}^a$ are inconsistent with \eqref{UnfldScalarBianchiA} since
\be h_n\wedge dC^{an}=h_n\wedge h_m C^{an,m}-\frac{d-2}{2}h^a\wedge h_m \tilde{C}^m\neq0\,.\ee
Therefore we have to exclude $C^{aa,b}$ and $\tilde{C}^a$ from the {\it r.h.s.}. The only term left parameterizes third-order derivatives of the scalar field on-shell
\begin{align*}
d C^{aa}=h_m C^{aam}\,.
\end{align*}
Continuing this way one one arrives at the final answer. A typical
situation in obtaining the unfolded equations is that one makes
use of dynamical equations at first few levels which defines the
pattern of first auxiliary fields and then at each next level
there are only Bianchi identities that further constraint the
final form of equations.

\paragraph{Unfolded equations for \texorpdfstring{$\boldsymbol{s=0}$}{spin-zero}.} The full system of equations that describe a free massless scalar field reads
\begin{align}{\label{UnfldScalar}
\DL C^{a(k)}=h_m C^{a(k)m}\,, \qquad\qquad\qquad C^{a(k-2)mn}\eta_{mn}\equiv0}\,.
\end{align}
where the set of fields consists of totally-symmetric traceless zero-forms $C^{a(k)}$.

By virtue of \eqref{UnfldScalar} the field $C^{a(k)}$ with $k>0$ is expressed as the $k$-th order derivative
$\pl^a...\pl^a C$ of the lowest field $C(x)$ which can be
identified with the original scalar field we started with. All
derivatives $\pl^a...\pl^a C$ are traceless since
$\eta_{mn}\pl^{mna(k-2)}C=\pl^{a(k-2)}\square C=0$, which explains
why $C^{a(k)}$ has to be traceless. The Klein-Gordon equation
appears thanks to the vanishing trace of $C^{aa}$. Simply, the
second equation, with $C^a$ already solved as $\pl^a C$, implies
$\pl^a \pl^a C=C^{aa}$. Since the \rhs has vanishing trace, by
contracting with $\eta_{aa}$ we derive $\square C=0$.

In the table below we list the spectrum of fields. The form degree is always zero.
\begin{align*}
    \mbox{grade}&:&& 0 & &1 & &2 & &3 & & ...\\
    \mbox{Young shape}&: & & \bullet & & \YoungA & & \YoungB & & \YoungC & & ... \\
    \mbox{degree}&:&& 0 & & 0 & & 0 & & 0 & & ...
\end{align*}
This is what one gets upon setting $s=0$ in \eqref{UnfldOneFormsA}-\eqref{UnfldZeroFormsEq} and dropping
tensors with $(s-1)=-1$ indices.

\paragraph{The scalar is almost tautological.} In this simple case of a scalar field there is a way to make the unfolded system tautological. Let us contract all the fiber indices with an auxiliary variable $y_a$ to build $C(y|x)$
\begin{align}
C(y|x)=\sum_k\frac1{k!} C^{a(k)}(x)\, y_a...y_a
\end{align}
then \eqref{UnfldScalar} is rewritten as
\begin{align}
&\left(\DL-h^m \frac{\pl}{\pl y^m}\right) C(y|x)=0\,, && \frac{\pl}{\pl y^m}\frac{\pl}{\pl y_m}C(y|x)\equiv0\,,\label{UnfldScalarTau}
\end{align}
where the last condition makes $C^{a(k)}$ traceless, see Appendix \ref{app:YoungGenfunc}. Going to
Cartesian coordinates where there is no difference between the
world and fiber indices we find
\begin{align}
&\left(\frac{\pl}{\pl x^m} - \frac{\pl}{\pl y^m}\right) C(y|x)=0\,, && \square^y C(y|x)\equiv0\,.
\end{align}
The first equation tells us that the dependence on $x$ is exactly the dependence on $y$, the solution being $C(y|x)=C(x-y)$, while the second condition imposes Klein-Gordon equation in the fiber. Therefore, the fiber Klein-Gordon equation is mapped to the $x$-space Klein-Gordon equation, which looks like a tautology. If we switch to a massive scalar or put a scalar on $AdS$
instead of Minkowski or proceed to fields with spin the
correspondence gets far from being tautological!

It is certainly true that one can write a tautological system that
imposes the original equations in the fiber and then translates
the dynamics to the base in a trivial way, like above
$\pl_x=\pl_y$. The point is that such system will not have an
unfolded form apart from a simple example above. The reason as we
will see is that the unfolded system contains more fields, especially connections, which
brings in the geometry.

\mysubsection{\texorpdfstring{$\boldsymbol{s=1}$}{spin-one}} \label{subsec:spinone}

The starting point is the gauge transformation law
\begin{align}
\delta A_\mm=\pl_\mm\xi\,,
\end{align}
which suggests to treat $A_\mm$ as a one-form $A\fm{1}=A_\mm
dx^\mm$, which is what usually done. The gauge parameter is then a
zero-form and we can simply write $\delta A\fm{1}=d\xi$. The
equations have to start with $dA=...$ and there are three options
for what $...$ might be:
\begin{align}
dA\fm{1}&=R\fm{2}\,, & dA&=h_m\wedge \omega^m\fm{1}\,, & dA&=h_m\wedge h_n C^{m,n}\,,
\end{align}
where $R\fm{2}$ is a two-form; $\omega^m\fm{1}$ is some
vector-valued one-form and $C^{m,n}$ is a zero-form that is
antisymmetric in its fiber indices, thus belonging to \YoungmAA.
Note that these three options partially overlap as all of the
fields on the \rhs contain \YoungmAA, whenever all the world
indices are converted to the fiber with the inverse vielbein. No
matter how natural the first option is, it fails. Indeed, the
first equation is invariant under $\delta A\fm{1}=d \xi+
\xi\fm{1}$, $\delta R\fm{2}=d \xi\fm{1}$ where $\xi\fm{1}$ is some
scalar one-form, the gauge parameter for $R\fm{2}$. The latter means that one can gauge away
$A\fm{1}$ completely with the help of $\xi\fm{1}$, which is not
what we need. The second option fails in a similar way once we
observe that there is an additional gauge symmetry $\delta
A\fm{1}=-h_m \xi^m$, $\delta \omega^a\fm{1}=d \xi^a$ with a
vector-valued zero-form $\xi^a$. Therefore, we have to use the
third option
\begin{align}\label{SpinOneFirstA}
dA\fm{1}&=h_m\wedge h_n C^{m,n}\,,
\end{align}
which is just a way to parameterize the Maxwell tensor, $\pl_\mm A_\nn-\pl_\nn A_\mm=C^{a,b}h_{a\mm}h_{b\nn}$. The integrability of this equation implies
\begin{align}
0\equiv d dA&=d( h_m\wedge h_n C^{m,n})=h_m\wedge h_n\wedge dC^{m,n}
\end{align}
Again, as in the scalar case, $dC^{m,n}$ is restricted. The most general \rhs of $dC^{m,n}$ is given by evaluating the tensor product $\YoungmA\otimes\YoungmAA$, where $\YoungmA$ and $\YoungmAA$ correspond to $\pl^a$ and $C^{a,b}$
\begin{align}
\YoungpA\otimes\YoungpAA=\YoungpBA\oplus\YoungpAAA\oplus\YoungA\,,\label{SpinOneRHS}
\end{align}
\begin{align*}
\YoungpBA&=\pl^{c}C^{a,b}-\pl^{[c}C^{a,b]}+\frac2{d-1}\eta^{c[a}\pl_dC^{b,]d}\,,\\
\YoungpAAA&=\pl^{[c}C^{a,b]}\,,\\
\YoungpA&=\pl_m C^{a,m}=\square A^a-\pl^a\pl_m A^m\,,
\end{align*}
so introducing $C^{aa,b}$, $C^{a,b,c}$ and $C^a$ in accordance with the pattern above $dC^{a,b}$ reads
\begin{align}
dC^{a,b}=h_m \left(C^{am,b}+\frac12 C^{ab,m}\right)+h_m C^{a,b,m}+\left(h^a C^b-h^b C^a\right)\,.
\end{align}
The terms in brackets make up the rank-two antisymmetric fiber tensor. We see that the presence of the antisymmetric component on the \rhs violates integrability as $h_m\wedge h_n\wedge h_k C^{m,n,k}\neq0$. The same time we know that $\pl_{[\mm} C_{\nn,\rr]}=\pl_{[\mm}\pl_\nn A_{\rr]}\equiv0$, which is the Bianchi identity, and hence there is no such component in the jet of $A_\mm$.

We would like to impose the Maxwell equations, but we see that
keeping $C^a$ on the \rhs we rather get $\square A^a-\pl^a\pl_m
A^m=C^a$. Therefore, $C^a$ has to be set to zero. The only field left is
$C^{aa,b}$ with the symmetry of $\YoungmBA$. It parameterizes the
second order derivatives of $A_\mm$ that do not vanish on-shell.
Now the second equation in the hierarchy reads
\begin{align}\label{SpinOneFirst}
dC^{a,b}=h_m \left(C^{am,b}+\frac12 C^{ab,m}\right)\,.
\end{align}
The integrability condition for this equation implies that $C^{aa,b}$ cannot be arbitrary. A straightforward analysis shows that we need to introduce fields $C^{a(k),b}$ that have the symmetry of \smallpic{\RectAYoung{5}{$\scriptsize k$}{\YoungAA}} and are traceless.

\paragraph{Unfolded equations for \texorpdfstring{$\boldsymbol{s=1}$}{spin-one}.}
The full system reads \besubeqs\begin{align}\label{UnfldMaxwellEquationsA}
\DL A\fm{1}&=h_m\wedge h_n C^{m,n}\,, && \delta A\fm{1}=\DL \xi\,,\\
\DL C^{a(k),b}&=h_m\left(C^{a(k)m,b}+\frac1{k+1} C^{a(k)b,m}\right)\,, \label{UnfldMaxwellEquationsB}
\end{align}\esubeqs
where let us stress again that all fields are $so(d-1,1)$-irreducible as fiber tensors.
It is useful to list the fields in order of their
appearance. As compared to the scalar
case we have more diverse structure of fiber tensors and there is
one field which is not a zero-form
\begin{align*}
    \mbox{grade}&:&& 0 & &1 & &2 & &3 & & ...\\
    \mbox{Young shape}&: & & \bullet & & \YoungAA & & \YoungBA & & \YoungCA & & ... \\
    \mbox{degree}&:&& 1 & & 0 & & 0 & & 0 & & ...
\end{align*}
Again, this is the specialization of \eqref{UnfldOneFormsA}-\eqref{UnfldZeroFormsEq} to $s=1$.

\mysubsection{Zero-forms}\label{subsec:Weylmodule}
Let us have a closer look at the sector of zero-forms of unfolded equations \eqref{UnfldOneFormsA}-\eqref{UnfldZeroFormsEq}. The first observation is that the equations for zero-forms
$C^{a(s+k),b(s)}$, \eqref{UnfldZeroFormsEq}, are self-consistent
and do not require the presence of any one-forms apart from the background fields $h^a$
and $\varpi^{a,b}$. What do they
describe? The answer can be read off from the spin-one case. The
zero-forms begin with the Maxwell tensor $C^{a,b}$. Since we do
not supplement it with gauge potential $A\fm{1}$ and the
corresponding equation \eqref{SpinOneFirstA}, it is the
fundamental field now. Then \eqref{SpinOneFirst} implies the
following two equations
\begin{align}\label{SpinOneFirstB}
\pl^aC^{b,c}+\pl^b C^{c,a}+\pl^c C^{a,b}&=0\,, & \pl_mC^{a,m}&=0\,.
\end{align}
Indeed, the most general \rhs of \eqref{SpinOneFirst} could involve three components according to \eqref{SpinOneRHS}, but it contains only one, \eqref{SpinOneFirst}. Therefore, the two combinations of first derivatives of $C^{a,b}$ are set to zero by \eqref{SpinOneFirst}, giving rise to the equations above. These are the standard equations that impose $dF=0$ and $\pl_m F^{n,m}=0$ for two-form $F_{m,n}dx^m\wedge dx^n$, $F_{m,n}=C_{m,n}$. The first equation is the Bianchi identity which implies $F=dA$ for some $A$, so the gauge potential is implicitly there. The second one is the Maxwell equations without any sources.

The same story for the spin-two case gives the Weyl tensor $C^{aa,bb}$ as the fundamental variable and imposes
\begin{align}\label{SpinTwoFirst}
\pl^{[a}C^{bc],de}&=0\,, & \pl_mC^{ab,cm}&=0\,,
\end{align}
which can be recognized as the two components of the differential Bianchi identities for the Weyl tensor. In the case of a spin-$s$ field we find the following equations
\begin{align}\label{SpinSFirst}
\pl^{[u}C^{a(s-1)u,b(s-1)u]}&=0\,, & \pl_mC^{a(s),b(s-1)m}&=0\,,
\end{align}
where the anti-symmetrization over three $u$'s is implied. The
rest of equations \eqref{UnfldZeroFormsEq} merely expresses the fields as derivatives of the
generalized Weyl tensor. The simplest way to solve equations is to
perform Fourier transform to find
\begin{align}
C^{a(s),b(s)}(p)=\varphi^{a(s)}(p)\, p^b...p^b+\mbox{permutations}\,,
\end{align}
where $\varphi^{a(s)}(p)$ is a symmetric tensor that depends on momentum $p^a$ and the following additional conditions are true
\begin{align}
p_mp^m&=0\,, & \varphi\fud{a(s-2)m}{m}&=0\,, & p_m\varphi^{a(s-1)m}&=0\,.
\end{align}
We almost reproduced the on-shell conditions
\eqref{onshellA}-\eqref{onshellC} for the Fronsdal field. Where is
the gauge symmetry? The Young properties imply that
$\varphi^{a(s)}$ of special form $p^a\xi^{a(s-1)}$ yields a
vanishing Weyl tensor. This is the on-shell Fronsdal gauge
transformations in momentum space. It is not possible to build an
antisymmetric tensor with $p^a p^b$, which is automatically
symmetric. Likewise, it is not possible to build $C^{a(s),b(s)}$
out of $p^b...p^b p^a \xi^{a(s-1)}$. The problem is that two $p$'s
show up in the same column of the Young diagram, $p^a p^b$. By
going into antisymmetric presentation, where $C^{a(s),b(s)}$ has
$s$ pairs of antisymmetric indices, $C^{ab,cd,...}$, we see that
the gauge transformation of $\phi^{a(s)}$ does not affect the Weyl
tensor. Therefore, $\phi^{a(s)}$ is defined up to a gauge
transformation and we recover the on-shell Fronsdal theory.

The upshot is that the equations for zero-forms also describe a spin-$s$ massless field, but in
a non-gauge way. The examples of $s=1,2$ tells us that it is the gauge potentials $A_\mm$ and $g_{\mm\nn}$ (which needs to be replaced by $e^a$, $\omega^{a,b}$) that mediate interactions. Therefore, while the non-gauge description of massless fields is perfectly fine at the free level it is crucial to supplement it with gauge potentials in order to proceed to interactions. The full unfolded equations tells us that the gauge potentials are determined up to a gauge transformation by the gauge invariant Weyl tensors (field-strengths).

\mysubsection{All spins together}\label{subsec:AllSpinstogether}
One can write a relatively simple system that describes fields of
all spins $s=0,1,2,3,...$ together. This is a Minkowski prototype
of the unfolded equations in $AdS_4$ of Sections
\ref{sec:AdSLinear} and \ref{sec:HSAlgebra}. First we sum up all
the fields that appeared above into generating functions of two
auxiliary variables, $y^a,p^a$,
\besubeqs\begin{align}
\omega\fm{1}(y,p|x)&=\sum_{s>0}\sum_{k=0}^{k=s-1} \frac1{(s-1)!k!}\,\omega\fm{1}^{a(s-1),b(k)} y_a...y_a\, p_b...p_b\,,\label{GenFuncMinkA}\\
C(y,p|x)&=\sum_{s=0}\sum_{k=0} \frac1{(s+k)!s!}\,C^{a(s+k),b(s)} y_a...y_a\, p_b...p_b\,.\label{GenFuncMinkB}
\end{align}\esubeqs
An analogous decomposition is used for the collective gauge parameter
$\xi(y,p|x)$. One of the crucial observations is that one-forms
$\omega^{a(s-1),b(k)}\fm{1}$ when summed over all spins, cover all the
irreducible Lorentz representations with the symmetry of
two-row Young diagrams, each appearing in one copy
\be \{\omega^{a(s-1),b(k)}\fm{1}\} \qquad\qquad a(s-1),b(k): \BYoungp{5}{3}{s-1}{k}\ee
The zero-forms cover the same variety of Young diagrams (all two-row)
\be \{C^{a(s+k),b(s)}\} \qquad\qquad a(s+k),b(s): \BYoungp{6}{3}{s+k}{s}\ee
Therefore, the fiber spaces of one-forms and zero-forms are isomorphic!

We need the Taylor coefficients of \eqref{GenFuncMinkA}-\eqref{GenFuncMinkB} to obey Young symmetry and trace constraint, which implies\footnote{See Appendix \ref{app:YoungGenfunc} for the implementation of Young-symmetry/trace constraints in terms of generating functions. It is these constraints that will be shown to be automatically solved with the help of spinorial auxiliary variables later on, which makes $d=3,4$ HS theories simpler.}
\begin{align}
\left\{y^n\frac{\pl}{\pl p^n},\frac{\pl^2}{\pl y_m \pl y^m},\frac{\pl^2}{\pl y_m \pl p^m},\frac{\pl^2}{\pl p_m \pl p^m}\right\}\left(\omega\fm{1}(y,p|x), C(y,p|x), \xi(y,p|x)\right)&\equiv0\label{AllspinsSpace}
\end{align}
Then the unfolded system of equations together with the gauge
transformations can be rewritten as
\besubeqs
\begin{align}
\DD \omega\fm{1}(y,p|x)&=\delta_{N_y,N_p}h^m\wedge h^n \frac{\pl^2}{\pl y^m \pl p^n} C(y,p|x)\,,\label{AllspinsMinkA}\\
\delta \omega\fm{1}(y,p|x)&=\DD \xi(y,p|x)\label{AllspinsMinkB}\,,\\
\widetilde{\DD} C(y,p|x)&=0\label{AllspinsMinkC}\,,\\
\DD&=\DL-h^m\frac{\pl}{\pl p^m}\label{AllspinsMinkD}\,,\\
\widetilde{\DD}&=\DL-h^m\left(\frac{\pl}{\pl y^m}+\frac{1}{N_y-N_p+2}p^n\frac{\pl}{\pl y^n}\frac{\pl}{\pl p^m}\right)\label{AllspinsMinkE}\,,\\
\DL&=d+\varpi^{a,b}\left(y_a\frac{\pl}{\pl y^b}+p_a\frac{\pl}{\pl
p^b}\right)\,,\label{AllspinsMinkF}
\end{align}
\esubeqs
where $N_y=y^m\pl/\pl y^m$, $N_p=p^m\pl/\pl p^m$ are Euler
operators that just count the number of tensor indices, i.e. $[N_y,y_a]=+y_a$, $[N_y,\pl/\pl y^a]=-\pl/\pl y^a$, etc. The Lorentz-covariant
derivative, $\DL$, must hit every index contracted with $y_a$ or $p_b$. This is achieved
with the help of the differential operator in the last line. In fact, it makes use of the standard representation of the orthogonal algebra on functions
\begin{align}&\varpi^{a,b}\left(y_a\frac{\pl}{\pl y^b}+p_a\frac{\pl}{\pl
p^b}\right)=\frac12 \varpi^{a,b} L_{ab}  &&L_{ab}=L^y_{ab}+L^p_{ab}\,,\label{LorentzCovMink}
\end{align}
where the generators of the Lorentz algebra split into $y$ and $p$ parts
\begin{align}&&L^y_{ab}=\left(y_a\frac{\pl}{\pl y^b}-y_b\frac{\pl}{\pl y^a}\right) && &L^p_{ab}=\left(p_a\frac{\pl}{\pl p^b}-p_b\frac{\pl}{\pl p^a}\right)\,.
\end{align}
As is to be expected, the spin-connection comes contracted with the Lorentz generators in a given representation. Analogously we can see that \eqref{AllspinsMinkD} and \eqref{AllspinsMinkE} deliver two different representations of the translation generators
\besubeqs\begin{align}
P_m&=\frac{\pl}{\pl p^m}\label{AllspinsMinkAD}\,,\\
\widetilde{P}_m&=\left(\frac{\pl}{\pl y^m}+\frac{1}{N_y-N_p+2}p^n\frac{\pl}{\pl y^n}\frac{\pl}{\pl p^m}\right)\label{AllspinsMinkAE}\,.
\end{align}\esubeqs
Altogether we have two realizations of the Poincare algebra $L_{ab}$, $P_c$ and $L_{ab}$, $\widetilde{P}_c$ on space \eqref{AllspinsSpace}. Therefore, the system  \eqref{AllspinsMinkA}-\eqref{AllspinsMinkF} has a plain representation theory meaning.

Taylor expanding system  \eqref{AllspinsMinkA}-\eqref{AllspinsMinkF} we find all the unfolded equations \eqref{UnfldOneFormsA}-\eqref{UnfldZeroFormsEq} that we derived above. The Kronecker symbol $\delta_{N_y,N_p}$ ensures
that only the Weyl tensors of $C$ contribute to the
\rhs of the equations for $\omega$. One of the effective realization of $\delta_{N_y,N_p}$
could be
\be \left.C(yt,pt^{-1})\right|_{t=0}\ee

The system is gauge invariant and consistent thanks to
$\DD^2=0$, $\widetilde{\DD}^2=0$. It is important that the term that sources one-forms $\omega$ by
zero-forms $C$ does not spoil the consistency. The system can be projected onto a particular spin-$s$ subsector with the help of
\begin{align}
&N_y\omega\fm{1}(y,p|x)=(s-1)\omega\fm{1}(y,p|x)\,, &&N_p C(y,p|x)=sC(y,p|x)\,.
\end{align}
The system describes free fields of all spins
$s=0,1,2,3,...$ over the Minkowski space.
It is a very nontrivial problem to find a nonlinear theory that leads to this equations in
the linearization.  There is even a simpler set of equations, \cite{Vasiliev:2005zu}, that does the same with the help of additional auxiliary fields. For the system above it was
proved, \cite{Vasiliev:2005zu}, under certain assumptions, that it is not possible to find
a nonlinear completion. But it becomes possible in anti-de Sitter space.

\paragraph{Summary.} We have found generalizations of the frame/vielbein
and spin-connection variables known for the spin-two case of gravity to the spin-$s$ case.
We have outlined the derivation of the unfolded equations, \eqref{UnfldOneFormsA}-\eqref{UnfldZeroFormsEq},
that are equivalent to the Fronsdal equations in the sense that the Fronsdal field
can be identified with certain component of the spin-$s$ frame field and the unfolded equations
impose Fronsdal equations and express other fields as derivatives of the Fronsdal field, up to
pure gauge terms, of course. The equations for the fields of all spins together have a simpler form \eqref{AllspinsMinkA}-\eqref{AllspinsMinkF} where various terms are related to the representation theory of the Poincare algebra.



\mysection{Unfolding}\label{sec:Unfolding}\setcounter{equation}{0}
In this section we would like to give a systematic overview of the
unfolded approach, \cite{Vasiliev:1988xc, Vasiliev:1988sa}, which underlies the Vasiliev HS theory. The unfolded approach is a universal method of formulating
differential equations on manifolds. Any set of differential
equations can be put into this form in principle. Once it is done
one can learn a lot about the symmetries of the original system.
The symmetries that were not at all obvious in the original
formulation become manifest when the unfolded form of equations is
available. This turned out to be very useful for the HS problem
analysis since the HS theory is so much constrained by the
symmetry.

The unfolded approach is one of the general methods and as such it
is not a cure-all. It can provide some substantial progress once
there is a rich hidden symmetry behind the system, but it may not
give any advantage as compared to other approaches when dealing
with systems that have a poor symmetry. Besides, the unfolding of
some nonlinear dynamical equations can be of a great technical
challenge let alone there are only few examples of nonlinear
equations in the literature that acquire explicit unfolded form.

A few of the nice features of the unfolded approach include: its
manifest diffeomorphism invariance; it is specifically suited to
account for gauge symmetries, conserved currents and charges.
Generally when looking for the interactions starting from a linear
gauge theory, one is faced with several problems at once: find a
deformation of the Lagrangian or the equations of motion and a deformation
of the gauge symmetry in such a way as to leave the Lagrangian or the
equations of motion invariant. These deformations are governed by
the same set of structure constants within the unfolded approach.

The variety of unfolded equations is at least as rich as all Lie
algebras together with all representations thereof and
cohomologies. As we will see all of the structure constants that
are at the bottom of any unfolded equations have certain
interpretation within the Lie algebra theory.

Though at present it is far from being clear, the unfolded
approach have a potential to be the generalization of the notion
of integrability in higher dimensions. Certain aspects of the
unfolded approach showed up in the context of supergravity
\cite{D'Auria:1982nx, D'Auria:1982pm, Nieuwenhuizen:1982zf} and in
topology \cite{Sullivan77} under the name of Free Differential
Algebras.

\mysubsection{Basics}
Let $W^\aAb$ be a set of differential forms over some
manifold $\mathcal{M}_d$. The index $\aAb$ is a formal one that we
let run over some set of fields assuming the Einstein summation
convention. In practice it runs over a set of linear spaces\footnote{For example,
$\aAb$ runs over Lorentz tensors with the symmetry of \parbox{30pt}{\smallpic{\RectBRow{5}{4}{$\scriptstyle s-1$}{$\scriptstyle k$}}}, $k=0,...,s-1$ with $q_\aAb=1$ and \parbox{40pt}{\smallpic{\RectBRow{6}{4}{$\scriptstyle  s+n$}{$\scriptstyle s$}}}, $n=0,1,...$ with $q_\aAb=0$ in the case of a spin-$s$ field as we have seen in Section \ref{sec:UnfoldingSpins}.}. The differential form degree of $W^\aAb$ is denoted by
$|\aAb|$. Equations of the following form are referred to as the
unfolded equations
\be\label{UnfldEquationsA} d W^\aAb = F^\aAb(W),\ee
where $d$ is the exterior (de-Rham) differential and the function
$F^\aAb(W)$ has degree $|\aAb|+1$ and is assumed to be
expandable in terms of the exterior products of the fields
themselves
\be
F^\aAb(W)=\sum_{n}\sum_{|\aBb_1|+...+|\aBb_n|=|\aAb|+1}
f^\aAb_{\phantom{\aAb} \aBb_1... \aBb_n}W^{\aBb_1}\wedge...\wedge
W^{\aBb_n},\ee The structure constants
$f^\aAb_{\phantom{\aAb} \aBb_1... \aBb_n}$ are certain elements
of $\mbox{Hom}(\aBb_1\otimes...\otimes\aBb_n,\aAb)$, i.e. maps from
tensor product $\aBb_1\otimes...\otimes\aBb_n$ to $\aAb$. They satisfy the
symmetry condition induced by the form degree of the fields
\be f^\aAb_{\phantom{\aAb} \aBb_1...\aBb_i\aBb_{i+1}... \aBb_n}=(-)^{|\aBb_i||\aBb_{i+1}|}f^\aAb_{\phantom{\aAb} \aBb_1...\aBb_{i+1}\aBb_i... \aBb_n}\,.\ee
The crucial point is to impose the integrability condition
\be \label{UnfldJacobi} F^\aBb\wedge\frac{\delta F^\aAb}{\delta
W^\aBb}\equiv0,\ee that arises when applying $d$ to both sides of
\eqref{UnfldEquations}, using $d^2=0$ and the equations of motion
again on the \rhs (we will use $\pl_\aBb$ instead of
$\frac{\delta}{\delta W^\aBb}$)\footnote{The derivatives carry the same grading as the fields, i.e. $\pl_\aAb\pl_\aBb=(-)^{|\aAb|\,|\aBb|}\pl_\aBb\pl_\aAb$.}
\be 0\equiv d^2W^\aAb=d F^\aAb(W)=dW^\aBb\wedge \pl_\aBb F^\aAb(W)=F^\aBb\wedge \pl_\aBb F^\aAb(W)\,. \ee
The integrability condition implies that the equations are
formally consistent and contain all their differential
consequences, i.e. in checking that $[\pl_\mm,\pl_\nn]\equiv0$ with the help
of equations of motion one finds no additional restrictions on the fields. In deriving
\eqref{UnfldJacobi} we also assume the absence of manifest
space-time dependence in $F^\aBb$.

When written in terms of the structure constants the integrability condition reads
\be \sum_{} f^\aCb_{\phantom{\aAb} \aDb_1... \aDb_m}\,f^\aAb_{\phantom{\aAb} \aCb \aBb_2... \aBb_n} \, W^{\aDb_1}\wedge....\wedge W^{\aBb_n}=0\,,\ee
and reminds the Jacobi identity. Let us note that some of these
equations may hold automatically if the total differential form
degree exceeds the dimension of the manifold $\mathcal{M}_d$. The
useful concept introduced by Vasiliev is that of the {\it
universal} unfolded system, \cite{Bekaert:2005vh,
Vasiliev:2005zu}. The unfolded equations are called universal iff
the system is integrable irrespectively of the dimension, i.e.
formally for any $d$. In other words, the integrability constraint
in some fixed dimension $d=d_0$ may admit nontrivial solution  due
to the fact that any differential form of the degree $d_0+1$ is
identically zero. The {\it universal} unfolded systems are
supposed to be consistent for any $d$.

The bonus of the universal system, which has been already shown to
be useful in applications, is that it remains meaningful on a
manifold of any dimension. In particular one can consider the same
unfolded equations over different space-times (say, $AdS_{d+1}$
and its conformal boundary $M_d$, which could be a way to make
AdS/CFT tautological, \cite{Vasiliev:2012vf}).

As the unfolded equations are formulated entirely in the language
of differential forms, there is no need for extra data, like a
metric or Christoffel symbols, that $\mathcal{M}_d$ is typically
equipped with.

It is convenient to rewrite the unfolded equations as the zero-curvature condition
\be\label{UnfldEquations} R^\aAb=0\,, \qquad\qquad R^\aAb\equiv d W^\aAb - F^\aAb(W)\,,\ee
where $R^\aAb$ will be referred to as the curvature.

The unfolded equations turn out to implement the concept of gauge symmetry automatically. Indeed, let us define the following gauge variations
\besubeqs
\begin{align}\label{UnfldGauge}\delta_{\epsilon} W^\aAb&=d\epsilon^\aAb+\epsilon^\aBb\pl_\aBb F^\aAb,&& \mbox{if }|\aAb|>0,\\
\label{UnfldGaugeZeroForm}\delta_\epsilon
W^\aAb&=+\epsilon^{\aBb'}\pl_{\aBb'} F^\aAb,&& \aBb':
|\aBb'|=1,\qquad \mbox{if }|\aAb|=0,\end{align}
\esubeqs
where we have to
distinguish between forms of degree greater than zero and
zero-forms. Each $W^\aAb$ with $|\aAb|>0$ has an associated gauge
symmetry with parameter $\epsilon^\aAb$ that is a degree $|\aAb|-1$
form taking values in the same space as $W^\aAb$ does. There are
no genuine gauge symmetries associated with zero-forms. They
transform without the $d\epsilon^\aAb$-piece similar to matter
fields and with the parameters of the degree-one fields among $W^\aAb$
they are coupled to.

One can check that the variation of the curvature $R^\aAb$
vanishes on-shell, i.e. when $R^\aAb=0$, since
\be \delta R^\aAb=(-)^{\aCb+1}\epsilon^{\aCb} R^{\aBb} \pl_\aBb \pl_\aCb F^\aAb\,.\ee
This is a generalization of $\delta F =[F,\epsilon]$ in the Yang-Mills theory with $\pl_\aBb \pl_\aCb F^\aAb$ playing the role of the structure constants of the Lie algebra, which in general are not constants and do depend on the fields.

The integrability condition can be rewritten as a Bianchi identity for the curvature
\be
d R^\aAb+R^\aCb \pl_{\aCb} F^\aAb\equiv0\,,
\ee
which is a generalization of $DF\equiv0$ in the Yang-Mills.

There is also a generalization of the fact that within a gauge
theory a diffeomorphism can be represented as a gauge
transformation plus a correction due to nonvanishing curvature
\be \mathcal{L}_\xi A=D (i_\xi A)+i_\xi F(A)\,, \qquad\qquad F(A)=dA+\frac12[A,A]\,.\ee
The analog within the unfolded approach reads
\be \mathcal{L}_\xi W^\aAb=\delta_{i_\xi\cdot W} W^\aAb +i_\xi\cdot R^\aAb\,.\ee
The difference is that within the unfolded approach any
diffeomorphism is always a particular gauge transformation
on-shell, since $R^\aAb=0$.

The notion of reducible gauge symmetries is also easily implemented. Indeed, the condition for $W^\aAb$ to be invariant under the gauge transformation
\be \delta_{\epsilon} W^\aAb=0\,, \qquad\qquad d\epsilon^\aAb+                 \epsilon^\aBb\pl_\aBb F^\aAb=0\,,\ee
can be interpreted as the unfolded-like system itself with the field content extended by the gauge parameters $\epsilon^\aAb$. As such it is gauge invariant under the second-level gauge transformations
\be\delta \epsilon^\aAb=d \xi^\aAb+\xi^\aBb \pl_\aBb F^\aAb\,,\ee
where we neglect terms bilinear in $\epsilon^\aAb$ by the definition of what gauge transformations are. Again, $\delta \epsilon^\aAb=0$ is an unfolded equation and so on until the zero-forms are reached. The level-two gauge parameter here $\xi^\aAb$ is a degree-$(|\aAb|-2)$ form. Therefore, a degree-$q$ field has $q$ levels of gauge (and gauge for gauge) transformations in total.

\mysubsection{Structure constants} In this
section we unveil the algebraic meaning of the structure
constants. As we will see soon, the sector of one-forms is at the
core of any unfolded-system as they necessarily belong to some Lie
algebra, say $\mathfrak{g}$. The rest of the fields can be
interpreted as taking values in various $\mathfrak{g}$-modules.
These modules can be in general glued (coupled) together via
Chevalley-Eilenberg cocycles of $\mathfrak{g}$ with coefficients in these modules.

\paragraph{Lie algebras and flatness/zero-curvature.}
Suppose there are one-forms only or a closed subsector thereof.
Let $\Omega^I\equiv \Omega^I_\mu dx^\mu$ denote the components of
the one-forms in some basis. Then the only unfolded-type equations
one can write read
\be
d\Omega^I+\frac12 f^I_{JK}\Omega^J\wedge\Omega^K=0\label{UnfldFlatness},\ee
where $f^I_{JK}$ are the structure constants that are antisymmetric in $J,K$ since one-forms anti-commute. The integrability condition (\ref{UnfldJacobi}) implies the Jacobi identity
\be
f^I_{JK}f^J_{LM}\Omega^K\wedge\Omega^L\wedge\Omega^M\equiv0\qquad\longleftrightarrow\qquad
f^I_{J[K}f^J_{LM]}\equiv0\,.\ee Therefore, $f^I_{JK}$ define
certain Lie algebra, call it $\mathfrak{g}$. Then
\eqref{UnfldFlatness} is the flatness or zero-curvature condition
for a connection of $\mathfrak{g}$. Equivalently,
\eqref{UnfldFlatness} means that the Yang-Mills field-strength
$F=d\Omega+\frac12[\Omega,\Omega]$ vanishes. The gauge
transformations associated with the zero-curvature equations are
the standard Yang-Mills transformations, $\delta \Omega^I
=D_\Omega \epsilon^I$. Such equations can describe background
geometry, e.g. Minkowski or AdS, see \eqref{adsFrameLike}.

\paragraph{Modules and covariant constancy.}
Consider now the extension of the unfolded system with one-forms
by a sector $W^\aA\fm{q}$ of $q$-forms, $q\neq1$. The most general equations
that are linear in $W^\aA\fm{q}$ read
\be \label{UnfldqFormsEquations}d W^\aA\fm{q}+{f_I}^\aA_{\phantom{\aA}\aB}\,\Omega^I\wedge W^\aB\fm{q}=0\,,\ee
with $\Omega^I$ obeying \eqref{UnfldFlatness} and
${f_I}^\aA_{\phantom{\aA}\aB}$ being some new structure constants.
The integrability \eqref{UnfldJacobi} then implies
\be\Omega^J\wedge\Omega^K\left(-\frac12 f^I_{JK}{f_I}^\aA_{\phantom{\aA}\aB}+
{f_J}^\aA_{\phantom{\aA}\aC}{f_K}^\aC_{\phantom{\aC}\aB}\right)W^\aB\fm{q}=0\,,\ee
where there is an implicit anti-symmetrization over $J,K$ in the
last term due to anti-commuting nature of $\Omega^I$. Thinking of
${f_I}^\aA_{\phantom{\aA}\aB}$ as endomorphisms
$f_I\in\mbox{Hom}(V,V)$, just matrices on the space $V$ where the
$q$-forms take value, we recognize the definition of a
$\mathfrak{g}$-module
\be f_J \circ f_K-f_K\circ f_J=[f_J,f_K]=f^I_{JK}{f_I}\,\ee
where $\circ$ denotes the usual matrix product over implicit indices $\aA$.

Therefore, writing down \eqref{UnfldqFormsEquations}-type equations is equivalent to specifying certain representation $V$ of $\mathfrak{g}$ and the corresponding equation is the covariant constancy condition for a field with values in $V$,
\begin{align} \label{UnfldCovariantConstancy}D_\Omega W^\aA\fm{q} \equiv d
    W^\aA\fm{q}+{f_I}^\aA_{\phantom{\aA}\aB}\,\Omega^I \wedge W^\aB\fm{q}&=0\,, & \delta W^\aA\fm{q}&=D_\Omega \xi^\aA\fm{q-1}\,,\end{align}
where we added gauge transformations --- the covariant derivative in the module defined by ${f_I}^\aA_{\phantom{\aA}\aB}$. The gauge invariance is thanks to $\DO\DO=0$, \eqref{UnfldFlatness}.

\paragraph{Contractible cycles and empty equations.} Consider now the linear
equations of the form
\begin{align}
dW\fm{q}^\aA&=f^\aA_{\phantom{\aA}\aBs}W^{\aBs}\fm{q+1}\,, && dW^{\aBs}\fm{q+1}=0\label{UfldContractible}.
\end{align}
By linear transformations they can always be split into two types
of subsystems $f^\aA_{\phantom{\aA}\aBs}$
\begin{align} &I:&&dW\fm{q}^\aAs+W^{\aAs}\fm{q+1}=0,\qquad dW^{\aBs}\fm{q+1}=0,\label{UfldContractibleA}\\
&II:&&dW\fm{q}^\aAt=0,\label{UfldContractibleB}\end{align}
where the last equation of \eqref{UfldContractibleA} is the integrability condition for the first one. \eqref{UfldContractibleA} is called a contractible cycle, \cite{Sullivan77}. With the help of gauge transformations  $\delta W\fm{q}^\aAs=d\xi^\aAs\fm{q-1}-\chi^\aAs\fm{q}$ we can always gauge away $W\fm{q}^\aA$ since $\chi^\aAs\fm{q}$ has the same number of components and enters algebraically. Then we are left with $W^{\aAs}\fm{q+1}=0$. Consequently we see that the equations of type \eqref{UfldContractibleA} are dynamically empty. The type-II equations can be solved in a pure gauge form $W\fm{q}^\aAt=d\xi^\aAt$ by virtue of the Poincare lemma unless $q=0$, see extra Section \ref{sec:HomotopyIntegrals}. If $q=0$ the solution is just a constant, $c^\aAt$, which cannot be gauged away. Therefore, the equations of type-II are dynamically empty unless $q=0$. In the latter case the solution space is that of $W\fm{0}^\aAt$ at a point, i.e. $c^\aAt$.

In what follows we will never meet contractible cycles as parts of the unfolded equations. Clearly their presence is redundant or at least unnecessary. Being found in some unfolded system, such contractible subsystems can always be removed. By contrast, type-II for $q=0$ and with $d$ extended to the covariant derivative in some module will be shown to be very important since they carry all degrees of freedom.

\paragraph{Cocyles and couplings.} Here we consider the most general types
of unfolded equations that can appear, while the somewhat technical discussion is left to extra Section \ref{sec:ChevalleyEilenberg}.
At the free level all the equations are linear in the dynamical fields described by some $W^\aA\fm{p}$,  $W^\aAs\fm{q}$, etc, but they can be nonlinear in the background fields, i.e the connection $\Omega$,
\besubeqs\begin{align} \label{UnfldCocycleA}
D_\Omega W^\aA\fm{p} &\equiv d W^\aA\fm{p}+{f_I}^\aA_{\phantom{\aA}\aB}\,\Omega^I\wedge W^\aB\fm{p}=
{f_{I_1...I_k}}^\aA_{\phantom{\aA}\aBs}\, \Omega^{I_1}\wedge...\wedge\Omega^{I_k}\wedge W^\aBs\fm{q}\,,\\
D_\Omega W^\aAs\fm{q} &=...
\end{align}\esubeqs
As it is explained in extra Section \ref{sec:ChevalleyEilenberg}, ${f_{I_1...I_k}}^\aA_{\phantom{\aA}\aBs}$ is a Chevalley-Eilenberg cocycle of the underlying Lie algebra, $\mathfrak{g}$, which $\Omega$ takes values in. The cocycle takes values in $\module_1\otimes \module^*_2$, where the $\mathfrak{g}$-modules $\module_1$ and $\module_2$ are associated with $W^\aA\fm{p}$ and $W^\aAs\fm{q}$, respectively.

Unfolded equations that have something to do
with interactions have to be nonlinear. Assuming that there is a well-defined
linearization, i.e. nonlinearities can be neglected in some limit, we should
get the structures of the form \eqref{UnfldCocycleA} at the free level. This equips us
with a bunch of $\mathfrak{g}$-modules, $\module_1$, $\module_2$, ... and Chevalley-Eilenberg cocyles. Nonlinear deformations of \eqref{UnfldCocycleA} can have the following form
\begin{align} \label{UnfldCocycleNonlinA}
\begin{aligned}D_\Omega W^\aA\fm{p} &\equiv d
W^\aA\fm{p}+{f_I}^\aA_{\phantom{\aA}\aB}\,\Omega^I\wedge
W^\aB\fm{p}= {f_{I_1...I_k}}^\aA_{\phantom{\aA}\aBs_1...\aBs_m}\,
\Omega^{I_1}\wedge...\Omega^{I_k}\wedge W^{\aBs_1}\fm{q_1}\wedge
...\wedge W^{\aBs_m}\fm{q_m}\,.\end{aligned}
\end{align}
Again, ${f_{I_1...I_k}}^\aA_{\phantom{\aA}\aBs_1...\aBs_m}$ is a Chevalley-Eilenberg cocycle. The gauge
transformations can also be written down.

\paragraph{Zero-forms and degrees of freedom.} The sector of zero-forms is a distinguished one. First of all, the most general linear equations for zero-forms, say $C^\aAs$, have the form of the covariant constancy condition
\begin{align} \label{UnfldGenAA}
dC^\aAs+f^\aAs_{I\aBs}\,\Omega^I\wedge C^{\aBs}&=0
\end{align}
and no sources on the \rhs are allowed by the form-degree argument. Indeed, the sources must be at most and at least linear in $\Omega$, which means that the sources are zero-forms, so we can move them on the \lhs Therefore, zero-forms are initial objects in a sense that they can source forms of higher degrees, but nothing can source zero-forms.

The most important property of zero-forms is that they are moduli of local solutions. Locally, by Poincare Lemma one can always solve $d\omega=0$ as $\omega =d \xi$ unless $\omega$ is a zero-form. For zero-forms $dC^\aAs=0$ has the solution space, $C^\aAs(x)=c^\aAs=const$, that is isomorphic to the space where $C^\aAs(x)$ takes values in, say $\module$. The analogous statement on the size of the solution space can be made about the covariant analog $\DO C^\aAs=0$ of $dC^\aAs=0$. The covariant generalization of Poincare Lemma implies that all fields that are forms of non-zero degree can be represented locally as pure gauge unless they are sourced by zero-forms, in the later case they consist of two pieces, pure gauge and another one determined by zero-forms. Zero-forms are not pure gauge. There are no gauge parameters at all associated with zero-forms. Therefore, locally a solution of any unfolded system is determined up to a gauge transformation by the values of all zero-forms at a point, i.e. by $\module$.

That is why we find an infinitely many zero-forms when unfolding
fields of different spins. Fields carry infinitely many 'degrees
of freedom' in a sense that the solutions of field equations are
parameterized by functions on Cauchy surface (usually it is the
number of functions referred to as the number of physical degrees
of freedom) and these functions are equivalent to specifying infinitely many
constants parameterized by the zero-forms at a point, i.e. by $c^\aAs$.

Combining together the fact that locally solution is reconstructed from values of zero-forms at a point and the general property of unfolded equations that zero-forms take values in certain representation, say $\module$, of the underlying Lie algebra, we can conclude that $\module$ is related to the space of one-particle states, \cite{Vasiliev:2001zy, Shaynkman:2001ip}. Remember that the solution space of field equations (one particle states) must carry a unitary irreducible representation (or a number thereof if the multiplet contains many particles), say $\module'$ of the space-time symmetry algebra. Since the solution is parameterized by zero-forms, it is actually parameterized by $\module$. So $\module$ and $\module'$ must be tightly related. In fact, they are equivalent when complexified.

Let us finish with the following illustration of the unfolded approach.
Up to the issues related to gauge symmetries and
the fact that unfolded equations are diffeomorphism invariant and could be nonlinear the idea is to identify those
derivatives of original dynamical fields that remain non-zero on-shell. Imposing equations of motions, $Eq(\phi)=0$ sets some of the derivatives $\pl^k\phi$ to zero, the rest is then parameterized by zero-forms.
\putfigureifpdf{\UnfoldedIdea}

\paragraph{Unfolded systems of HS type.} In Section \ref{subsec:AllSpinstogether} we found that the linearized unfolded equations that describe fields of all spins read schematically as
\besubeqs
\begin{align} \label{UnfldGenA}
d\Omega^I+\frac12 f^I_{JK}\,\Omega^J\wedge\Omega^K&=0\,,\\
d\omega^\aA+f^\aA_{J\aB}\,\Omega^J\wedge\omega^\aB&=f^\aA_{JK|\aB}\,\Omega^J\wedge\Omega^K\wedge C^\aB\label{UnfldGenAB}\,,\\
dC^\aA+\bar{f}^\aA_{I\aB}\,\Omega^I\wedge C^{\aB}&=0\,,\label{UnfldGenAC}
\end{align}
\esubeqs
where $\Omega^I=\{h^a,\varpi^{a,b}\}$ is the Poincare connection. Recall, see Section \ref{subsec:AllSpinstogether}, that the space of one-forms $\omega^\aA$, when summed over all spins, is isomorphic to that of zero-forms. They cover irreducible Lorentz representations with the symmetry of all two-row Young diagrams, each appearing once. That is why we used the same index for $\omega^\aA$ and $C^\aA$. The cocycle on the \rhs of \eqref{UnfldGenAB} was found to have the form
\be h_m\wedge h_n C^{a(s-1)m,b(s-1)n}\,.\ee

The problem of interactions can be reduced to seeking for the nonlinear terms such that the integrability is preserved. The form degree argument shows that nonlinearities can be of two types:  $f^\aA_{\aB\aC}\,\omega^\aB\wedge\omega^\aC$ on the \lhs of \eqref{UnfldGenAB} and zero-forms can source \rhs of all equations by the terms of the form $\omega\wedge \omega CC...C$, \eqref{UnfldGenAB},
and $\omega CC...C$, \eqref{UnfldGenAC}.

Already at this point we can see that there must be an
infinite-dimensional Lie algebra, $\mathfrak{g}$, of which $\omega^\aA$ are gauge
fields. Recall that $\Omega^I$ is a subset of $\omega^\aA$
associated with the gravity sector. At the linearized level we
cannot fully see this algebra. We found a part of structure
constants. Namely, $f^I_{JK}$, which define the symmetry algebra, $\mathfrak{h}=iso(d-1,1)$, of the background space and
$f^\aA_{J\aC}$, which show $\mathfrak{g}$ as a module under its subalgebra $\mathfrak{h}$,
but not all $f^\aA_{\aB\aC}$. Therefore,
the first question is the existence of such an algebra and
secondly whether nonlinear deformations of equations are possible
or not. To solve the problem within the perturbation theory one
has to switch on the cosmological constant, which replaces $\mathfrak{h}=iso(d-1,1)$ with
$so(d-1,2)$ or $so(d,1)$. Then all questions
above can be answered in affirmative and the solution is given by the
Vasiliev equations. It is important that the spectrum of the fields we found does not change when the cosmological
constant is turned on. Like in the Fronsdal theory, there are some corrections proportional to $\Lambda$. Schematically the solution given by the Vasiliev equations reads
\besubeqs\begin{align} \label{UnfldGenBA}
d\omega^\aA+\frac12 f^\aA_{\aB\aC}\,\omega^\aB\wedge\omega^\aC&=
f^\aA_{\aB\aC|\aD}\,\omega^\aB\wedge\omega^\aC\wedge C^\aD+f^\aA_{\aB\aC|\aD\aE}\,\omega^\aB\wedge\omega^\aC\wedge C^\aD\wedge C^\aE+...\,,\\ \label{UnfldGenBB}
dC^\aA+\bar{f}^\aA_{\aB\aC}\,\omega^\aB\wedge C^{\aC}&=\bar{f}^\aA_{\aB|\aC \aD}\,\omega^\aB\wedge C^{\aC}\wedge C^\aD+...\,,
\end{align}\esubeqs
where we stress that zero-forms and one-forms take values in the
same linear space, whose basis is given by irreducible Lorentz
tensors with the symmetries of all two-row Young diagrams, each in one copy.

As was noticed in \cite{Vasiliev:1988sa}, the unfolded equations for higher-spin fields have the following structure
\besubeqs\begin{align} \label{UnfldGenCA}
d\omega^\aA&=F^\aA(\omega,C)\,,\\
dC^\aA&=C^\aB \frac{\pl}{\pl
\omega^\aB}F^\aA(\omega,C)\label{UnfldGenCA1}\,,
\end{align}\esubeqs
i.e. the equations for zero-forms are built from the same
structure function $F^\aA(\omega,C)$ as the equations for
one-forms. Formally, we applied $C^\aB \frac{\pl}{\pl \omega^\aB}$
to the first equation in order to get the second one. For example,
this operation transforms $d\omega+\frac 12 [\omega,\omega]$
into $dC+[\omega,C]$. There is a technical detail related to fact that $C$
takes values in the twisted-adjoint representation, which will
become clear around \eqref{UnfldsSpinorialDCB} and it does not
violate the statement that the equations for $C$ are determined by
those for $\omega$. Therefore, there is no doubling of
structure constants, \eqref{UnfldGenBB} is fully determined by
\eqref{UnfldGenBA}. Let us note that \eqref{UnfldGenCA1} is
nevertheless not an integrability consequence of
\eqref{UnfldGenCA}.

\paragraph{Summary.}
We found that the structure constants of any unfolded system of
equations have certain interpretation in terms of Lie theory.
There is some underlying Lie algebra, which is related to the
subsector of one-forms. The rest of the fields take values in
some modules of this Lie algebra. Couplings between different
sectors of forms are given by some representatives of the Chevalley-Eilenberg
cohomology groups with coefficients in the coupled modules. Locally, solution to any unfolded system is reconstructed up to a gauge transformation from knowing the values of all zero-forms at some point.



\mysectionstar{Higher-spin theory in four dimensions}\setcounter{equation}{0}
In the remaining we will consider the application of unfolded
approach to four dimensional HS theory. Remarkable progress has
been achieved in lower dimensions $d=3$ and $d=4$ due to effective
twistor-like language available in these dimensions. It turns out
to be perfectly suited for dealing with HS algebras and eventually
brings the one to nonlinear HS equations. We restrict ourselves to the
case of four dimensions, for $d=4$ unlike $d=3$ case admits
propagation of HS fields. To proceed in this direction we need to
reformulate our settings in terms of spinor language.

\mysection{Vector-spinor dictionary}\label{sec:SpinVectDict} It turns
out that higher-spin fields can be described by a set of one-form
connections and zero-forms that are irreducible Lorentz tensors as the fiber tensors.
In dealing with irreducible tensors we have to preserve Young
symmetry properties and tracelessness. The advantage of the
spinorial language as compared to the tensor one is that the trace
and Young symmetry constraints get automatically resolved.
Unfortunately, it is available in specific dimensions only. It is
the condition for traces to vanish that leads to complicated trace
projectors, the Young symmetry is not for free though too. We have
not faced any trace projectors yet since we developed a theory in
Minkowski space.  The unfolded equations for higher-spin fields in
$AdS$ get modified by more complicated terms which are not seen when $\Lambda=0$ (we will see them in Section \ref{sec:AdSLinear}), e.g.
\begin{align}\label{sigAp}
\DL C^{a(k),b(l)}&=...+\Lambda\left(h^{a}_{}C^{a(k),b(l)}-
\frac{k}{(d+2k-2)}h_{m}C^{a(k-1)m,b(l)}\eta^{aa}-\right.\notag\\
&-\frac{l}{d+k+l-3}h_{m}C^{a(k),mb(l-1)}\eta^{ab}+\\&+
\left.\frac{k}{l(d+2k-2)(d+k+l-3)}h_{m}C^{a(k-1)b,b(l-1)m}\eta^{aa}\right),\notag
\end{align}
getting more and more involved at each level of algebraic
manipulations. The unfolded equations for the complete multiplet
of higher-spin fields have a simpler form, but still the
$d$-dimensional higher-spin theory, \cite{Vasiliev:2003ev}, is way
more complicated than the $4d$ one.

The only condition that a spin-tensor has to obey is the symmetry
condition --- it must be symmetric in (each sort of) indices it
has. To have all indices symmetrized is much simpler than to
preserve a more general Young symmetry types. But the major benefit of using
spin-tensors rather than tensors is that the corresponding
tensors, which can be recovered by contracting all spinor indices
with $\gamma$-matrices, are automatically traceless. The latter
property is of great use for practical calculations. Being
available for free in spinorial language it is not granted and
causes a real headache in the language of tensors which is on top
of having the definite Young symmetry type. There are deeper
reasons of course as to why spinors are so effective, we just
mention those from practical standpoint.

In the next section we establish the dictionary, which is a little bit boring, so the reader
who trusts $so(3,1)\sim sl(2,\mathbb{C})$ can directly proceed to the
summary and the table in Section \ref{subsec:spinvectdict}.

\mysubsection{\texorpdfstring{${so(3,1)\sim sl(2,\mathbb{C})}$}{so(3,1)-sl(2,C) dictionary}}
The idea behind two-component representation for $so(3,1)$ Lorentz
fields is the identity $4=2\times 2$. Particularly, any Lorentz
vector $v^a$, $a=0,...,3$ can be encoded by two-by-two matrix
\be
v^{a}\leftrightarrow \begin{pmatrix} m & n \\ p & q
\end{pmatrix},
\ee
The convenient basis for $2\times2$-matrices is given by
$\gs^{a}_{\ga\gbd}=(I, \gs^{i})$, where $I$ - is the unit-matrix
and $\gs^i_{\ga\gbd}$ --- Pauli matrices
\be
\gs^{1}=\begin{pmatrix} 0 & 1 \\ 1 & 0 \end{pmatrix}\,,\quad
\gs^{2}=\begin{pmatrix} 0 & -i \\ i & 0 \end{pmatrix}\,,\quad
\gs^{3}=\begin{pmatrix} 1 & 0 \\ 0 & -1 \end{pmatrix}\,,
\ee
of these we know, that
\be
\{\gs_i, \gs_j\}=2\gd_{ij}I\,,\qquad \rm{Tr}\,\gs_i=0\,.
\ee
The corresponding $v^a$ is then a hermitian matrix, $v=v^\dag$,
\be v^{a}\leftrightarrow v^{\ga\gad}=\begin{pmatrix} v^0+v^3 & v^1-iv^2 \\ v^1+iv^2 & v^0-v^3
\end{pmatrix}\label{twobytwoherm}\,.\ee

Let us choose Minkowski metric in the form $\eta_{ab}=(-+++)$ and
introduce dual set to $\gs$-matrices $\bar{\gs}^{a\gad\gb}=(I,
-\gs^{i})$, then for any $v^a$ we can define
\be
v_{\ga\gbd}=v^{m}\gs_{m\ga\gbd}\,,
\ee
such that
\be
v^{m}=-\ff12v_{\ga\gbd}\bar{\gs}^{m\gbd\ga}\,.
\ee
Indeed,
$-\ff12v_{\ga\gbd}\bar{\gs}^{a\gbd\ga}=-\ff12v^{b}\gs_{b\ga\gbd}\bar{\gs}^{a\gbd\ga}=
-\ff12v^b{\rm{Tr}}\,(\gs_b\bar{\gs}^a)=-\ff14v^b{\rm{Tr}}\,\{\gs_b,
\bar{\gs}^a\}=v^{a}$. It is convenient then for each type of
indices dotted and undotted to introduce antisymmetric matrices
\be
\gep^{\ga\gb}=i\gs^2=\begin{pmatrix} 0 & 1\\ -1 &
0\end{pmatrix}=\gep^{\gad\gbd}\,,
\ee
with their inverses $-\gep_{\ga\gb}=(i\gs^2)^{-1}=-i\gs^2$,
$\gep_{\ga\gb}=-\gep_{\gb\ga}$,
$\gep_{\ga\gc}\gep^{\gb\gc}=\gd_{\ga}{}^{\gb}$, {\it idem.} for $\epsilon^{\gad\gbd}$, and define
symplectic $sp(2)$-form for raising and lowering
indices\footnote{The appearance of the symplectic structure is not
accidental since $sl(2)\sim sp(2)$ thanks to the following
identity valid for $2\times2$ matrices, $A^T\epsilon A=\det{A}\,
\epsilon$, which implies that an $SL(2)$ matrix $A$, $\det A=1$,
is at the same time a symplectic matrix $A^T\epsilon A=\epsilon$.}
\be
A_{\ga}=A^{\gc}\gep_{\gc\ga}\,,\qquad
A^{\ga}=\gep^{\ga\gb}A_{\gb}\,.
\ee
note the position of indices, which, unlike the case of a
symmetric metric, can be changed at the price of an additional
sign factor\footnote{There are different conventions on how to
work with antisymmetric metric tensor. Our convention may seem
unusual at first sight, but in the end it is the simplest one. For
example, raising and then lowering an index one gets back to the
original expression. A convention different from ours produces a
sign factor, which to us is unnatural and requires much care in
computations.  We are free to change the position of uncontracted
indices using our convention. More detail on symplectic calculus
can be found in Appendix \ref{app:diffcalc}.}.

Note, that
$A_{\ga}B^{\ga}=-A^{\ga}B_{\ga}$. Using the above definitions one
finds, that
\be
\gs^{m}_{\ga\gad}=\gep_{\gb\ga}\gep_{\gbd\gad}\bar{\gs}^{m\gbd\gb}\,,\qquad
\bar{\gs}^{\gad\ga}_m=\gep^{\ga\gb}\gep^{\gad\gbd}\gs_{m\gb\gbd}\,,
\ee
allowing us to identify
\be
\gs_{m}^{\ga\gad}\equiv\bar{\gs}_{m}^{\gad\ga}\,.
\ee
Matrices $\gs_{m\ga\gad}$ with indices being raised and lowered by
$\gep_{\ga\gb}$ and $\gep_{\gad\gbd}$ are called Van der Waerden
symbols. Two-component spinors $A_{\ga}$ and $A^{\gad}$ are called
chiral and anti-chiral, correspondingly. Let us note that one
should be careful with numerical factors in identifying scalar
combinations, e.g.
\be
v_{\ga\dot{\alpha}}v^{\ga\dot{\alpha}}=-2v_{a}v^{a}
\ee
To proceed we need a simple lemma. Any bi-spinor  $A_{\ga\gb}$ can
be decomposed as
\be\label{l1}
A_{\ga\gb}=A_{(\ga\gb)}+\ff12\gep_{\ga\gb}A_{\gc}{}^{\gc}\,.
\ee
The proof is straightforward, decomposing
$A_{\ga\gb}=A_{(\ga\gb)}+A_{[\ga\gb]}$ and noting, that any
antisymmetric $2\times2$-matrix $A_{[\ga\gb]}\sim\gep_{\ga\gb}$ one gets
\eqref{l1}. Using it we can further prove the following relation
for $\gs$-matrices
\be
\gs^{m}_{\ga}{}^{\gad}\gs^{n}_{\gb\gad}=\eta^{mn}\epsilon_{\ga\gb}+(\gs^{mn})_{\ga\gb}\,,\qquad
(\gs^{mn})_{\ga\gb}=-(\gs^{nm})_{\ga\gb}=(\gs^{mn})_{\gb\ga}=\ff12[\gs^{m},
\gs^n]_{\ga\gb}\,.
\ee
Indeed, consider first symmetric part $(mn)$ of \lhs
\be\gs^{m}_{\ga}{}^{\gad}\gs^{m}_{\gb\gad}=-\gs^{m}_{\gb}{}^{\gad}\gs^{m}_{\ga\gad}=
\ff12\gep_{\ga\gb}\gs^{m}_{\gc}{}^{\gad}\gs^{m\gc}{}_{\gad}=\eta^{mm}\gep_{\ga\gb}\,.\ee
For antisymmetric part $[ab]$ one obtains the  definition of the
r.h.s. The great advantage of the $4d$ spinor formalism is that it
allows one to handle easily traceless Young diagrams.
Particularly, let us state an important fact which will be very
useful in what follows. Given a traceless Lorentz tensor $C_{a_1b_1, a_2b_2, ...
,a_sb_s }$ corresponding to a rectangular two-row Young diagram, i.e. it is antisymmetric with respect to each pair of indices
$C_{..,ab,..}=-C_{..,ba,..}$, it is
traceless with respect to any indices and it satisfies Young
condition --- antisymmetrization of any three gives zero
$C_{..,[ab,c],..}=0$, then its spinor counterpart reads
\be
C_{a_1b_1, a_2b_2, ... ,a_sb_s }\quad\to\quad C_{\ga(2s)},
\bar{C}_{\gad(2s)}\,,
\ee
where the repeated indices $\ga(2s)=\ga_1\dots\ga_{2s}$ mean total
symmetrization as usual.

We will demonstrate this statement with the simplest examples. Let
us start with $s=1$, i.e. $C_{ab}=-C_{ba}$. Its spinor image by
definition is $C_{\ga\gad,
\gb\gbd}=C^{ab}\gs_{a\ga\gad}\gs_{b\gb\gbd}$. Using \eqref{l1},
one can decompose
\be
C_{\ga\gad, \gb\gbd}=C^{ab}\left(\ff12\gs_{a\ga\gad}\gs_{b\gb\gbd}
+\ff12\gs_{a\gb\gad}\gs_{b\ga\gbd}+
\ff12\gep_{\ga\gb}\gs_{a\gc\gad}\gs_{b}{}^{\gc}{}_{\gbd}\right)\,.
\ee
Using now $ab$-antisymmetry we finally obtain
\be
C_{\ga\gad,
\gb\gbd}=\gep_{\ga\gb}\bar{C}_{\gad\gbd}+\gep_{\gad\gbd}C_{\ga\gb}\,,
\ee
where
\be
C_{\ga\gb}=C_{\gb\ga}=\ff12\gs_{a\ga\gdd}\gs_{b\gb}{}^{\gdd}C^{ab}\,,\qquad
\bar{C}_{\gad\gbd}=\bar{C}_{\gbd\gad}=\ff12\gs_{a\gc\gad}\gs_{b}{}^{\gc}{}_{\gbd}C^{ab}\,.
\ee
It is crucial that $C_{\ga\gb}$ is a complex conjugate of
$\bar{C}_{\gad\gbd}$. A symmetric rank-two spin-tensor has $3$
complex components, i.e. $6$ real, which is exactly the number of
components of the Maxwell tensor $F_{mn}=-F_{nm}$. If $C_{\ga\gb}$
and $\bar{C}_{\gad\gbd}$ were not conjugated to each other, the
corresponding $F_{mn}$ would be complex. For the same reason, a
tensor is often equivalent to a pair of spin-tensors, which are
complex conjugated with the only exception for
$C_{\ga(s),\gad(s)}$ which is self-conjugated. In particular, the
spinorial version of $x^a$, which is $x^{\ga\gad}$, is a hermitian
matrix, \eqref{twobytwoherm}.

The next less trivial example is $C_{ab,cd}$ whose Young
symmetries are those of gravity Weyl tensor, taken again in the antisymmetric basis. Its spinor
counterpart can be shown to be
\be
C_{\ga\gad\gb\gbd, \gc\gdd\gd\gdl}=C^{ab,cd}
\gs_{a\ga\gad}\gs_{b\gb\gbd}\gs_{c\gc\gdd}\gs_{d\gd\gdl}=
C_{\ga\gb\gc\gd}\gep_{\gad\gbd}\gep_{\gdd\gdl}+
\bar{C}_{\gad\gbd\gdd\gdl}\gep_{\ga\gb}\gep_{\gc\gd}\,,
\ee
where
\be
C_{\ga(4)}=\ff14C^{ab,cd}\gs_{a\ga}{}^{\gdd}\gs_{b\ga\gdd}
\gs_{c\ga}{}^{\gdl}\gs_{d\ga\gdl}
\ee
is a totally symmetric spin-tensor. Note, that in deriving this
result not only the tracelessness and antisymmetry of $ab$ and
$cd$ have been used, but also the Young condition $[ab,c]=0$. On
top of that one repeatedly exploits \eqref{l1}. Analogously, one
shows that
\begin{align*}
&C_{\ga_1\gad_1\gb_1\gbd_1,...,\ga_{s}\gad_{s}\gb_{s}\gbd_{s}}=
C_{\ga(s)\gb(s)}\gep_{\gad_1\gbd_1}...\gep_{\gad_s\gbd_s}+c.c.\,,\quad\\
&C_{\ga(2s)}=\ff{1}{2^s}C^{a_1b_1, a_2b_2, ... ,a_sb_s
}\gs_{a_1\ga}{}^{\gdd_1}\gs_{b_1\ga\gdd_1}...
\gs_{a_s\ga}{}^{\gdd_s}\gs_{b_s\ga\gdd_s}\,.
\end{align*}
Finally, let us conclude this section with spinor representation
for Minkowski metric $\eta_{ab}$
\be\notag
\eta_{\ga\gad,
\gb\gbd}=\eta_{ab}\gs^{a}_{\ga\gad}\gs^{b}_{\gb\gbd}=2
\gep_{\ga\gb}\gep_{\gad\gbd}
\ee

\mysubsection{Dictionary} \label{subsec:spinvectdict}
The dictionary between tensors of the $4d$ Lorentz algebra $so(3,1)$
and spin-tensors of $sl(2,\mathbb{C})$ is summarized in the table nearby.
\begin{table}[h!]
\begin{tabular}{|x{0.1cm}x{4.5cm}|x{6cm}|x{3.5cm}|}
\hline &$so(3,1)$ tensor & $sl(2,\mathbb{C})$ tensor &
dimension\tabularnewline\hline &$\bullet$ & $\bullet$ & 1
\tabularnewline &Dirac spinor & $T^\ga\oplus T^\gad$ & 4
\tabularnewline \rule{0pt}{26pt}& $T^{a}\sim\YoungpA$ &
$T^{\ga\gad}$ & 4\tabularnewline \rule{0pt}{26pt}&
$T^{a,b}\sim\YoungpAA$ & $T^{\ga\ga}\oplus \bar{T}^{\gad\gad}$ &
6\tabularnewline \rule{0pt}{26pt}& $T^{aa}\sim\YoungB$&
$T^{\ga\ga,\gad\gad}$ & 9\tabularnewline \rule{0pt}{26pt}&
$T^{a(k)}\sim\AYoungp{4}{k}$ & $T^{\ga(k),\gad(k)}$ &
$(k+1)^2$\tabularnewline \rule{0pt}{26pt}&
$T^{a(k),b(k)}\sim\BYoungp{4}{4}{k}{}$ & $T^{\ga(2k)}\oplus
\bar{T}^{\gad(2k)}$ &  $2(2k+1)$ \tabularnewline \rule{0pt}{26pt}&
$T^{a(k),b(m)}\sim\BYoungp{6}{4}{k}{m}$ &
$T^{\ga(k+m),\gad(k-m)}\oplus \bar{T}^{\ga(k-m),\gad(k+m)}$ &
$2(k+m+1)(k-m+1)$\tabularnewline \hline
\end{tabular}
\end{table}

If we sum over all spins $s=0,1,2,3,...$, as it was done in \ref{subsec:AllSpinstogether}, then we see that
the set of one-forms is isomorphic to the set of zero-forms in a sense that they cover
the same space of tensors in the fiber space --- all two-row Young diagrams, each appearing once, see Section \ref{subsec:AllSpinstogether}. The main statement of this section is that the same space is isomorphic to the space of all spin-tensors of even ranks, each appearing once, i.e.
\begin{align}\notag
\sum_{s>k}\omega^{a(s-1),b(k)}\sim \sum_{s,i} C^{a(s+i),b(s)} \sim
\sum_{k+m=\mbox{even}} \omega^{\ga(k),\gad(m)}\sim \sum_{k+m=\mbox{even}}
C^{\ga(k),\gad(m)}\,.
\end{align}
Allowing for spin-tensors of odd ranks brings fermions. The same space can be described as the space of all functions of two auxiliary two-component variables, $y_\ga$, $\bry_\gad$, since the Taylor coefficients cover the same range of spin-tensors,
\begin{align}\label{sltwopacking}
f(y,\bry)=\sum_{k,m} \frac{1}{k!m!}f^{\ga(k),\gad(m)}
y_\ga...y_\ga \, \bry_\gad...\bry_\gad\,.
\end{align}
The main advantage of using spin-tensor generating function is
that $f(y,\bry)$ is an unconstrained function (with fermions) or obeys a simple constraint $f(-y,
-\bar{y})=f(y, \bar{y})$ (bosons only), while to describe
the same space in terms of vector-like variables we need a number
of differential constraints, see Section \ref{subsec:AllSpinstogether} and Appendix \ref{app:YoungGenfunc}, which are difficult to deal with, especially in interactions.

There is an alternative way to describe the same space. One can
take just one auxiliary variable $Y_A$ with the index $A$ running
over $4$ values. We can think of it as of composite index
$A=\{\ga,\gad\}$. Then
\begin{align}\label{spfourpacking}
f(Y)=\sum_s\frac{1}{s!} f^{A(s)}\,
Y_A...Y_A=\sum_{k,m}\frac{1}{k!m!} f^{\ga(k),\gad(m)}
y_\ga...y_\ga \, \bry_\gad...\bry_\gad
\end{align}
This fact has again a group-theoretical meaning, $so(3,2)\sim
sp(4,\mathbb{R})$, which is useful to remember of. It is reviewed
in Appendix \ref{app:moreonso} together with the branching rules
for $so(3,2)\downarrow so(3,1)$. The bosonic projection $f(-y,
-\bar{y})=f(y, \bar{y})$ means $f(Y)=f(-Y)$.



\mysection{Free HS fields in \texorpdfstring{$\boldsymbol{AdS_4}$}{AdS-four}}\label{sec:AdSLinear}\setcounter{equation}{0}

There is no conceptual problem to extend the unfolded equations
for a field of any spin to $AdS_d$. However, the unfolded
equations for the individual fields are more complicated than for
all spins gathered into an infinite multiplet of the higher-spin
algebra. Thanks to the effective spinorial language the unfolded
equations in $AdS_d$ with $d=3,4,5$ are simpler. We will focus on
the higher-spin theory in $AdS_4$. There are no propagating
higher-spin fields in $d<4$, so the case of $AdS_4$ is the first
nontrivial one and historically the first example of higher-spin
theory.

A good warm up starting point on the way is to consider
Klein-Gordon equation in Minkowski space-time, i.e. the spinorial
version of \eqref{UnfldScalar}, and then generalize the elaborated
machinery to $AdS_4$ and to any spin. A content of this section
should be compared with Section \ref{sec:UnfoldingSpins} as we
simply convert our $d$-dimensional findings into spinorial form
for $d=4$ and then extend the equations to $AdS_4$, which is a
minor modification in the spinorial notation but a challenge in
the vector one. Fortunately, one does not need to apply
$\sigma$-matrices to every equation in Section \ref{sec:UnfoldingSpins},
reaching an equivalent result entirely in the language of
spin-tensors.

\mysubsection{Massless scalar and HS zero-forms on Minkowski}\label{subsec:SpinorScalarMink}
We are interested in obtaining first order differential equations
of motion for a free massless scalar that will be equivalent to
\be\label{KG}
\Box C=0\,,\qquad \Box=\p_{m}\p^{m}
\ee
The equations that we are looking for should be formulated in
terms of differential forms. The spinorial avatar of coordinate $x^m$ is a bispinor $x^{\ga\gad}$.
So, let us introduce differential
$d=dx^{m}\ff{\p}{\p x^{m}}= dx^{\ga\gad}\ff{\p}{\p
x^{\ga\gad}}$. The most general equation that one can write down
for $dC$ has the form
\be\label{C1}
dC=C_{\ga\gad}h^{\ga\gad}\qquad\Longleftrightarrow\qquad
\p_{\ga\gad}C=C_{\ga\gad}\,,
\ee
where $h_{\ga\gad}=dx_{\ga\gad}$ is the vielbein in Cartesian coordinates in Minkowski space-time.
Formally, $h^{\ga\gad}_m=\sigma^{\ga\gad}_m$ or $h^{\ga\gad}_{\gb\gbd}=\epsilon\fdu{\ga}{\gb}\epsilon\fdu{\gad}{\gbd}$. The field
$C_{\ga\gad}$ is some arbitrary field, whose only purpose is to parameterize the first derivative of $C$. Proceeding further, one
writes down the most general equation for $dC_{\ga\gad}$, c.f. \eqref{UnfldScalarA},
\be\label{C2}
dC_{\ga\gad}=C_{\ga\gad|\gb\gbd}h^{\gb\gbd}\qquad\Longleftrightarrow\qquad
\p_{\gb\gbd}C_{\ga\gad}=C_{\ga\gad|\gb\gbd}\,,
\ee
where we once again have introduced a new field
$C_{\ga\gad|\gb\gbd}$, which can be decomposed into traceless and
traceful parts, c.f. \eqref{ScalarDecompA},
\be\label{C22}
C_{\ga\gad|\gb\gbd}=C_{\ga\gb,\gad\gbd}+T_{\ga\gb}\gep_{\gad\gbd}+
\bar{T}_{\gad\gbd}\gep_{\ga\gb}+T\gep_{\ga\gb}\gep_{\gad\gbd}\,.
\ee
On the other hand, from \eqref{C1}, \eqref{C22} one has
$C_{\ga\gad|\gb\gbd}=\p_{\gb\gbd}\p_{\ga\gad}C$. Using then
\eqref{KG} one finds that $T_{\ga\gb}=T_{\gad\gbd}=T=0$ and so,
c.f. \eqref{UnfldScalarC},
\be
dC_{\ga\gad}=C_{\ga\gb, \gad\gbd}h^{\gb\gbd}
\ee
Keep going in similar fashion we have the decomposition
\begin{align*}
dC_{\ga(2), \gad(2)}=C_{\ga(2), \gad(2)|\gb\gbd}h^{\gb\gbd}= \big(
C_{\ga(2), \gad(2)\gbd}+\gep_{\ga\gb}T_{\ga,
\gad(2)\gbd}+\gep_{\gad\gbd}T_{\ga(2)\gb,
\gbd}+\gep_{\ga\gb}\gep_{\gad\gbd}T_{\ga, \gad}\big)h^{\gb\gbd}\,.
\end{align*}
Observing that $\p_{\gb\gbd}C_{\ga(2),
\gad(2)}=\p_{\gb\gbd}\p_{\ga\gad}C_{\ga\gad}$ and using
$[\p_{\ga\gad}, \p_{\gb\gbd}]=0$ we immediately find that $T_{\ga,
\gad(3)}=T_{\ga(3), \gad}=T_{\ga, \gad}=0$ and so,
\be
dC_{\ga(2), \gad(2)}=C_{\ga(2)\gb, \gad(2)\gbd}h^{\gb\gbd}\,.
\ee
The chain of the obtained equations is infinite and at each level
$n$ reads
\be\label{Cn}
dC_{\ga(n), \gad(n)}=C_{\ga(n)\gb,
\gad(n)\gbd}h^{\gb\gbd}\,,\qquad n\geq 0\,.
\ee
Its dynamical content is equivalent to that of Klein-Gordon
equation for the lowest component $n=0$, $C(x)$, expressing the
higher level fields via derivatives of the physical field $C$, c.f. \eqref{UnfldScalar},
\be
\Box C(x)=0\,,\qquad C_{\ga(n),
\gad(n)}=\underbrace{\p_{\ga\gad}\dots\p_{\ga\gad}}_{n}C(x)
\ee
It is said, that the fields $C_{\ga(n), \gad(n)}$ for $n\geq 0$
constitute a scalar module. It can be shown that the obtained set
of fields that describe a scalar field is the same for any
space-time background. The dynamical equations for general
background of course will be different, yet their form is known
only for $(A)dS_d$ and Minkowski, \cite{Shaynkman:2000ts}.

It is convenient to pack the obtained scalar module into a
generating function
\be\label{gen}
C(y, \bar{y}|x)=\sum_{n=0}^{\infty}\ff{1}{(n!)^2}C_{\ga(n),
\gad(n)}(y^{\ga})^n(\bar{y}^{\gad})^n\,,
\ee
this leads to the following equation\footnote{Peculiarities of differential calculus when the metric is symplectic, i.e. antisymmetric, which is our case of $\epsilon_{\ga\gb}$, are discussed in Appendix \ref{app:diffcalc}.}, c.f. \eqref{UnfldScalarTau},
\be\label{u0}
dC(y, \bar{y}|x)-h^{\ga\gad}\ff{\p^2}{\p y^{\ga}\p\bar{y}^{\gad}}
C(y, \bar{y}|x)=0\,,
\ee
which by construction is consistent with integrability condition
$d^2=0$. Indeed, we need to check that
\begin{align}
0\equiv dd C(y, \bar{y}|x)=h^{\ga\gad}\ff{\p^2}{\p y^{\ga}\p\bar{y}^{\gad}} \wedge h^{\gb\gbd}\ff{\p^2}{\p y^{\gb}\p\bar{y}^{\gbd}}C(y, \bar{y}|x)
\end{align}
The latter is obvious with the help of the following identity
\begin{align}
h^{\ga\gad}\wedge h^{\gb\gbd} \equiv \frac12H^{\ga\gb} \epsilon^{\gad\gbd}+\frac12H^{\gad\gbd} \epsilon^{\ga\gb}
\end{align}
where $H^{\ga\gb}=H^{\gb\ga}=h\fud{\ga}{\gdd}\wedge h^{\gb\gdd}$, idem. for $H^{\gad\gbd}$, and the fact that $\xi_\ga \xi_\gb \epsilon^{\ga\gb}\equiv 0$ for any two-component commuting object $\xi_\ga$, even ${\pl}/{\pl y^\ga}$.

The consistency does not require specific for a scalar
module grading of generating function, i.e. it does not rely on the number of dotted indices being equal
to that of undotted as expressed by
\be
\Big( y^{\ga}\ff{\p}{\p y^{\ga}}-\bar{y}^{\gad}\ff{\p}{\p
y^{\gad}}\Big)C(y, \bar y|x)=0\,,
\ee
This allows us to consider a generating
function of the form
\be
C(y, \bar{y}|x)=\sum_{n,m=0}^{\infty}\frac1{m!n!}C_{\ga(m),
\gad(n)}(y^{\ga})^m(\bar{y}^{\gad})^n
\ee
and impose the condition \eqref{u0} to see if it reproduces
anything reasonable. It is convenient to present $C(y, \bar y)$ as
$C=\sum_{2s=0}^{\infty}C_{s}(y, \bar y|x)$, where
\be
C_{s}=\sum_{n=0}^{\infty}\frac1{(2s+n)!n!}(C_{\ga(n+2s),
\gad(n)}(y^{\ga})^{n+2s}(\bar{y}^{\gad})^{n}+\bar{C}_{\ga(n),
\gad(n+2s)}(y^{\ga})^{n}(\bar{y}^{\gad})^{n+2s})\,,
\ee
such that
\be
\Big( y^{\ga}\ff{\p}{\p y^{\ga}}-\bar{y}^{\gad}\ff{\p}{\p
\bar{y}^{\gad}}\Big)C_{s}(y, \bar y|x)=\pm 2s\, C_{s}\,,
\ee
Note that the grading operator above commutes with the equations
of motion, so \eqref{u0} decomposes into independent subsystems
for each $s$. Using the spinor-tensor dictionary derived in
Section \ref{sec:SpinVectDict} and the field content used in
Section \ref{subsec:UnfldAnySpin} it is easy to see that
\eqref{u0} is the spinor version of the unfolded equations for the
zero-forms $C^{a(s+k),b(s)}$, \eqref{AllspinsMinkC}. The equations
also make sense for half-integer $s$, which gives unfolded equations for fermions for free.

The module $C_{1/2}$ describes left and right massless
$s=1/2$ fields $C_{\ga}, \bar{C}_{\gad}$ according to Dirac
equations
\be
\p_{\gb\gad}C^{\gb}=0\,,\qquad \p_{\ga\gbd}\bar{C}^{\gbd}=0\,,
\ee
with all of the rest fields in the module $C_{1/2}$ expressed
via derivatives of the fermion matter fields $C_{\ga}$ and
$\bar{C}_{\gad}$. The $s=1$ module is equivalent to Maxwell equations
and Bianchi identities for (anti)self-dual parts of the Maxwell tensor
$C_{\ga\gb}\oplus\bar{C}_{\gad\gbd}$
\be
\p_{\gc\gbd}C^{\gc}{}_{\ga}=0\,,\qquad
\p_{\gb\gdd}\bar{C}_{\gad}{}^{\gdd}=0\,,
\ee
which are spinorial versions of \eqref{SpinOneFirstB} and look more symmetric than \eqref{SpinOneFirstB}.
For $s=2$ we arrive at Bianchi identities for gravity Weyl tensor
\be
\p_{\gc\gbd}C^{\gc}{}_{\ga(3)}=0\,,\qquad
\p_{\gb\gdd}\bar{C}_{\gad(3)}{}^{\gdd}=0\,,
\ee
which are spinorial versions of \eqref{SpinTwoFirst}. One
reproduces equations for generalized HS curvatures $C_{\ga(2s)},
\bar{C}_{\gad(2s)}$ in this fashion, c.f. \eqref{SpinSFirst},
\be\label{bM}
\p_{\gc\gbd}C^{\gc}{}_{\ga(2s-1)}=0\,,\qquad
\p_{\gb\gdd}\bar{C}_{\gad(2s-1)}{}^{\gdd}=0\,.
\ee
Similar analysis can be implemented for $AdS_4$ space-time. We
will see that it does not lead to much complication.

\mysubsection{Spinor version of \texorpdfstring{${AdS_4}$}{AdS-four} background}

To write down spinor form of equation \eqref{adsFrameLike} we need to find
spinor counterparts for $so(3,2)$ generators and connection
fields. Recall that Lorentz vector $x^a$ is represented by
bi-spinor $x_{\ga\gad}$, while antisymmetric Lorentz tensor
$T_{ab}=-T_{ba}$ by a pair of symmetric and conjugate bispinors, $T_{\ga\gb}$ and
$\bar{T}_{\gad\gbd}$. Therefore, the spinorial version of $AdS_4$
generators $P_a$, $L_{ab}$,  \eqref{dsadspoincare}, are the translations $P_{\ga\gbd}$ and Lorentz
generators $L_{\ga\gb}$, $\bar{L}_{\gad\gbd}$. The representation
in terms of spinor generators gives rise explicitly to the well
known isomorphism\footnote{Understanding of
$sp(4,\mathbb{R})\sim so(3,2)$ is not required in this section as
all formulae will be written down using the Lorentz covariant
basis of $sl(2,\mathbb{C})\sim so(3,1)$. The discussion of
$sp(4,\mathbb{R})\sim so(3,2)$ can be found in Appendix
\ref{app:sofivespfour}.} $so(3,2)\sim sp(4, \mathbb{R})$. Indeed,
gathering $L_{\ga\gb}, \bar{L}_{\gad\gbd}, P_{\ga\gbd}$ together
into symmetric matrix, $A,B,...=1,...,4$,
\be\label{spfourmatrix}
T_{AB}=T_{BA}=\begin{pmatrix} L_{\ga\gb} & P_{\ga\gbd}\\
P_{\gb\gad} & \bar{L}_{\gad\gbd}\end{pmatrix}
\ee
and defining the invariant $sp(4,\mathbb{R})$-form as
\be
\gep_{AB}=-\gep_{BA}=\begin{pmatrix} \gep_{\ga\gb} & 0\\ 0 &
\gep_{\gad\gbd}\end{pmatrix}\,.
\ee
the $sp(4)$ commutation relations
\be[T_{AB},T_{CD}]=\epsilon_{BC}T_{AD}+\epsilon_{AC}T_{BD}+\epsilon_{BD}T_{AC}+
\epsilon_{AD}T_{BC}\label{spfourrelations}\ee when rewritten in
terms of $sl(2,\mathbb{C})$ with $A=\{\ga,\gad\}$
give\footnote{Four terms, $2\times2$, in the first line are needed
to symmetrize over the indices denoted by the same letter. Two
terms are in the second line with no additional terms in the third
line. Note that $\ga$ is totally different from $\gad$, so that no
symmetrization over $\ga$, $\gad$ is possible without breaking
the Lorentz symmetry.}
\begin{subequations}
\label{AdSSpinorial}
\begin{align}
&[L_{\ga\ga}, L_{\gb\gb}]=\gep_{\ga\gb}L_{\ga\gb}\,,&&
[\bar{L}_{\gad\gad},
\bar{L}_{\gbd\gbd}]=\gep_{\gad\gbd}L_{\gad\gbd}\,,\label{AdSSpinorialA}\\
&[L_{\ga\ga}, P_{\gb\gbd}]=\gep_{\ga\gb}P_{\ga\gbd}\,,&&
[\bar{L}_{\gad\gad}, P_{\gb\gbd}]=\gep_{\gad\gbd}P_{\gb\gad}\,,\label{AdSSpinorialB}\\
&[P_{\ga\gad},
P_{\gb\gbd}]=\gl^2(\gep_{\ga\gb}\bar{L}_{\gad\gbd}+\gep_{\gad\gbd}L_{\ga\gb})\,,\label{AdSSpinorialC}
\end{align}
\end{subequations}
where for historical reason the cosmological constant $\Lambda$
was replaced by $\lambda^2$. Its appearance requires rescaling of
the translation generators $T_{\ga\gad}\rightarrow \lambda
P_{\ga\gad}$. For connection fields we have $\varpi^{a,b}\to
\varpi^{\ga\gb}, \bar{\varpi}^{\gad\gbd}$, $h^a\to h^{\ga\gad}$.
The zero-curvature condition \eqref{adsFrameLike}, specialized to
$so(3,2)$ in terms of spinor connections
\be\label{AllFlatB}\Omega=\frac12\omega^{\ga\gb}L_{\ga\gb}+h^{\ga\gad}P_{\ga\gad}+\frac12\bar{\omega}^{\gad\gbd}L_{\gad\gbd}\ee
acquires the following form
\begin{subequations}
\label{Cartan}
\begin{align}&dh^{\ga\gbd}+\varpi^{\ga}{}_{\gc}\wedge
h^{\gc\gbd}+\bar{\varpi}^{\gad}{}_{\gdd}\wedge h^{\ga\gdd}=0\,,\label{CartanA}\\
&d\varpi^{\ga\gb}+\varpi^{\ga}{}_{\gc}\wedge\varpi^{\gc\gb}=-\gl^2
h^{\ga}{}_{\gdd}\wedge h^{\gb\gdd}\,,\label{CartanB}\\
&d\bar{\varpi}^{\gad\gbd}+\bar{\varpi}^{\gad}{}_{\gdd}\wedge\bar{\varpi}^{\gdd\gbd}=
-\gl^2h_{\gc}{}^{\gad}\wedge h^{\gc\gbd}\,.\label{CartanC}
\end{align}
\end{subequations}
Let us define Lorentz-covariant derivative $D=d+\varpi$ as follows
\be\label{LorentzDerSpinorial}
DA^{\ga\gad}=dA^{\ga\gad}+\varpi^{\ga}{}_{\gb}\wedge A^{\gb\gad}+\bar{\varpi}^{\gad}{}_{\gbd}\wedge A^{\ga\gbd}\,,
\ee
where $A^{\ga\gad}$ is any $q$-form. The upshot is that each index of a spin-tensor is
acted by the appropriate half of the spin-connection, undotted indices are rotated with $\varpi^{\ga\gb}$,
while dotted indices with $\bar{\varpi}^{\gad\gbd}$.
From \eqref{Cartan} then we have
\be\label{R}
D^2A^{\ga\gad}=-\gl^2 H^{\ga}{}_{\gb}\wedge A^{\gb\gad}-\gl^2 \bar{H}^{\gad}{}_{\gbd}\wedge A^{\ga\gbd}\,,
\ee
where we recall that $H^{\ga\gb}=h\fud{\ga}{\gdd}\wedge h^{\gb\gdd}$.

Altogether we have four equivalent ways of presenting the background $4d$ anti-de Sitter geometry
in terms of flat connection $\Omega$, which has $10$ components, in accordance with various choices
of the base
\begin{subequations}
\label{AllFlat}
\begin{align}
so(3,1):& & \Omega=&\frac12\omega^{a,b} L_{ab}+h^aP_a &&6+4\label{AllFlatA}\\
sl(2,\mathbb{C}):& & \Omega=&\frac12\omega^{\ga\gb}L_{\ga\gb}+h^{\ga\gad}P_{\ga\gad}+\frac12\bar{\omega}^{\gad\gbd}L_{\gad\gbd} && 3+4+\bar{3}\\
so(3,2):& & \Omega=&\frac12 \Omega^{\aAs,\aBs} T_{\aAs\aBs} && 5\times 4/2\\
sp(4,\mathbb{R}):& & \Omega=&\frac12 \Omega^{AB} T_{AB} && 4\times 5/2 \label{AllFlatD}
\end{align}
\end{subequations}
with the commutation relations given in \eqref{dsadspoincare}, \eqref{AdSSpinorial}, \eqref{AdSAlgebra} and
\eqref{spfourrelations} in terms of  $so(3,1)$, $sl(2,\mathbb{C})$, $so(3,2)$ and $sp(4,\mathbb{R})$ bases,
respectively.

An explicit solution to \eqref{Cartan} is provided by Poincare coordinates, which are useful in applications. For example, the flat connection of the form \eqref{AllFlatB}, i.e. the spinorial
counterpart of \eqref{PoincareCoordVector}, is
\begin{align}
\label{flatP}
\varpi^{\ga\ga}&=\frac{i}{2z}d\mathrm{x}^{\ga\ga}\,, &
\varpi^{\gad\gad}&=-\frac{i}{2z}d\mathrm{x}^{\gad\gad}\,,  &
h^{\ga\gad}&=\frac{1}{2z\lambda}(-d\mathrm{x}^{\ga\gad}+i\epsilon^{\ga\gad}dz)\,,
&
\end{align}
where the coordinates split into the radial coordinate $z$ and
three boundary coordinates $x^i$ packed into a symmetric real
bispinor $\mathrm{x}^{\ga\gb}=\mathrm{x}^{\gb\ga}$. All together
$z$ and $\mathrm{x}^{\ga\gb}$ combine into hermitian
$x^{\ga\gad}=\mathrm{x}^{\ga\gb}\delta_{\gb}{}^{\gad}+i\epsilon^{\ga\gad}z$,
where $i\epsilon^{\ga\gad}$ is one of the Pauli matrices,
$-\sigma_2$. The choice of coordinates breaks manifest symmetry
down to the boundary Lorentz symmetry $so(2,1)$ that rotates
$\mathrm{x}^{\ga\gb}$ and dilatations, $\mathrm{x},z\rightarrow
\gamma\mathrm{x}, \gamma z$. The vierbein $h^{\ga\gad}$ gives the
expected metric tensor
\be
ds^2=\ff{1}{z^2\lambda^2}(dz^2+dx_i dx^i)\,.
\ee

\mysubsection{HS zero-forms on \texorpdfstring{$AdS_4$}{AdS-four}}
First of all let us extend equations \eqref{u0} written in
Cartesian coordinates to any coordinate system, where the
spin-connection may not be trivial. This is done similarly to the
way we did in Sections \ref{sec:UnfoldingSpins}.
One can replace $d$ with the Lorentz-covariant derivative
$\DL=d+\varpi$ to obtain the following equation in Minkowski
space-time
\be\label{Unfldzeroformsflat}
\DL C(y, \bar{y}|x)-h^{\ga\gad}\ff{\p^2}{\p y^{\ga}\p\bar{y}^{\gad}}
C(y, \bar{y}|x)=0\,,
\ee
where the fact that $\DL$ acts on every index, \eqref{LorentzDerSpinorial}, can be simulated by, c.f. \eqref{LorentzCovMink},
\be\DL=d+\varpi^{\ga\gb}y_\ga \pl_\gb+\varpi^{\gad\gbd}y_\gad \pl_\gbd\,,\label{LorentzDerY}\ee
where $\varpi,\bar{\varpi},h$ obey \eqref{Cartan}
with $\lambda=0$, which entails $\DL^2=0$. Our next goal is to
lift \eqref{Unfldzeroformsflat} to $AdS_4$ where $\lambda\neq0$
and the system has to be modified.

The technical difference in deriving equations analogous
to \eqref{Cn} for a scalar in $AdS_4$ is that the covariant
derivatives no longer commute in this case $D^2\neq 0$, \eqref{R}.
Consider the case of the massive scalar on $AdS_4$ in some detail,
i.e. we aim to rewrite in the unfolded form the Klein-Gordon equation\footnote{The sign in front of $m^2$ is in accordance with the signature $(+---)$.}
\be\label{KGm}
D_{\ga\gad}D^{\ga\gad}C=-m^2 C\,.
\ee
The analysis is analogous to Minkowski case, we have
\be
DC=C_{\ga\gad} h^{\ga\gad}\,,
\ee
where $C_{\ga\gad}$ is yet unconstrained. We can write then
\begin{align*}
DC_{\ga\gad}=C_{\ga\gad|\gb\gbd}h^{\gb\gbd}=\big(C_{\ga\gb,\gad\gbd}+T_{\ga\gb}\gep_{\gad\gbd}+
\bar{T}_{\gad\gbd}\gep_{\ga\gb}+T\gep_{\ga\gb}\gep_{\gad\gbd}\big)h^{\gb\gbd}\,.
\end{align*}
From $C_{\ga\gad}=D_{\ga\gad}C$ and \eqref{KGm} one finds that
$T_{\ga\gb}=\bar{T}_{\gad\gbd}=0$, while $T=\ff14 m^2 C$.
Proceeding further
\begin{align*}
DC_{\ga(2), \gad(2)}=C_{\ga(2), \gad(2)|\gb\gbd}h^{\gb\gbd}= \big(
C_{\ga(2)\gb, \gad(2)\gbd}+\gep_{\ga\gb}T_{\ga,
\gad(2)\gbd}+\gep_{\gad\gbd}T_{\ga(2)\gb,
\gbd}+\gep_{\ga\gb}\gep_{\gad\gbd}T_{\ga, \gad}\big)h^{\gb\gbd}\,.
\end{align*}
Using Bianchi identities \eqref{R} one finds, that $T_{\ga(3),
\gad}=T_{\ga, \gad(3)}=0$, and $T_{\ga\gad}\sim C_{\ga\gad}$. This
pattern persists at higher levels and allows one to write general
expression
\be\label{ScalarAnsatz}
DC_{\ga(n), \gad(n)}=C_{\ga(n)\gb,
\gad(n)\gbd}h^{\gb\gbd}+f_nC_{\ga(n-1), \gad(n-1)}h_{\ga\gad}\,,
\ee
where coefficients $f_n$ are to be determined from Bianchi identities
\eqref{R}. Note that the field content does not change when we switch on $\lambda$, the $f_n$ term
compensates for $\DL\DL\neq0$. On one hand, we have
\be
D^2C_{\ga(n), \gad(n)}=-\gl^2H\fdu{\ga}{\gb}C_{\gb\ga(n-1), \gad(n)}-\gl^2
H\fdu{\gad}{\gbd} C_{\ga(n),\gbd\gad(n-1)}
\ee
On the other, using
\begin{align*}
DC_{\ga(n)\gb, \gad(n)\gbd}=\,&C_{\ga(n)\gb\gc,
\gad(n)\gbd\gdd}h^{\gc\gdd}+f_{n+1}\Big( C_{\ga(n-1)\gb,
\gad(n-1)\gbd}h_{\ga\gad}+C_{\ga(n), \gad(n)}h_{\gb\gbd}+\\
&+ C_{\gb\ga(n-1), \gad(n)}h_{\ga\gbd}+C_{\ga(n),
\gbd\gad(n-1)}h_{\gb\gad}\Big)
\end{align*}
and
\be
DC_{\ga(n-1), \gad(n-1)}=C_{\ga(n-1)\gb,
\gad(n-1)\gbd}h^{\gb\gbd}+f_{n-1}C_{\ga(n-2),
\gad(n-2)}h_{\ga\gad}
\ee
one arrives at the following recurrent condition
\be
-\gl^2=-\frac12f_{n+1}(n+2)+\frac12n f_n\,,
\ee
which can be easily solved by
\be
f_n=+\gl^2 + \ff{A}{(n+1)n}\,,
\ee
where $A$ is related to the mass term in \eqref{KGm}
$A=\ff12(m^2-4\gl^2)$. The pure massless case, when a
scalar in question is conformal, corresponds to $A=0$ and yields the simplest $f_n$
\be\label{m}
m^2=-4\gl^2\,.
\ee
In this case we have the following chain of equations
\be
DC^{\ga(n), \gad(n)}=h_{\gb\gbd} C^{\ga(n)\gb,
\gad(n)\gbd}+\gl^2h^{\ga\gad}C^{\ga(n-1),
\gad(n-1)}\,,
\ee
which in terms of the generating function \eqref{gen} reduces to
\begin{align}\label{UnfldzeroformsAdS}
&\widetilde{\DD} C(y,\bar y|x)=0\,,\\
&\widetilde{\DD}=\DL-h^{\ga\gad}\ff{\p^2}{\p
y^{\ga}\p\bar{y}^{\gad}}-\gl^2h^{\ga\gad}y_{\ga}\bar{y}_{\gad}\,.
\label{TwsitedAdjointDer}\end{align} Just as in Minkowski
space-time, \eqref{UnfldzeroformsAdS} is consistent, i.e. $d^2=0$,
for any $C(y, \bar y|x)$ no matter if it has or has not the form
of a scalar module \eqref{gen}. In the generic case eq.
\eqref{UnfldzeroformsAdS} describes all fields with $s\geq 0$
propagating in $AdS_4$ along the lines of Section
\ref{subsec:SpinorScalarMink}. For $s\geq 1$ these read, c.f.
\eqref{bM},
\be\label{bA}
\DL_{\gb\gad}C^{\gb}{}_{\ga(2s-1)}=0\,,\qquad
\DL_{\ga\gbd}\bar{C}^{\gbd}{}_{\gad(2s-1)}=0\,.
\ee
The obtained equation has a clear algebraic meaning which we
reveal in what follows. Obviously, there is a well-defined flat
limit, $\lambda\rightarrow0$ that results in
\eqref{Unfldzeroformsflat}.

\mysubsection{HS gauge potentials on \texorpdfstring{$AdS_4$}{AdS-four}}
So far massless higher-spin fields have been described in terms of
generalized curvatures analogous to the Maxwell $s=1$ tensor and the $s=2$
Weyl tensor, which is parallel to Section \ref{subsec:Weylmodule}.
These along with $s=0$ and $s=1/2$ matter fields
reside in zero-form Weyl module $C(y,\bar y|x)$. In this section we
add gauge potentials aiming at the spinorial, extended to $AdS_4$, version of \eqref{AllspinsMinkA}-\eqref{AllspinsMinkF}.

The
simplest case one can start with is a massless $s=1$ field. Its
gauge potential $A=A_{\mu}dx^{\mu}$ defines the Maxwell tensor which
is the $C_{\ga\gb}, \bar{C}_{\gad\gbd}$ components of the Weyl module
according to
\be
dA=h^{\ga}{}_{\gdd}\wedge
h^{\gb\gdd}C_{\ga\gb}+h_{\gc}{}^{\gad}\wedge
h^{\gc\gbd}\bar{C}_{\gad\gbd}\,.
\ee
The gauge potential $A$ possesses a standard gauge invariance
\be
\gd A=d\xi\,.
\ee
Consider now the massless $s=2$ case. To do so one needs to impose the
Einstein equations and linearize them around $AdS_4$ background.
Imposing the Einstein equations is equivalent to stating that the Riemann
tensor differs from that of $AdS_4$ by the Weyl tensor. In other
words, we can write down the following equations for the $s=2$ Lorentz
connection $\omega^{\ga\gb}$, $\bar{\omega}^{\gad\gbd}$ and the
vierbein field $e^{\ga\gad}$, which are
\eqref{LinEinstAA}-\eqref{LinEinstAB} in the language of spinors,
\begin{subequations}
\begin{align}
\DL e^{\ga\gad}=\,&de^{\ga\gad}+\omega^{\ga}{}_{\gc}\wedge e^{\gc\gad}+
\bar{\omega}^{\gad}{}_{\gdd}\wedge e^{\ga\gdd}=0\,,\\
\label{s2}
&d\omega^{\ga\gb}+\omega^{\ga}{}_{\gc}\wedge\omega^{\gc\gb}+\gl^2e^{\ga}{}_{\gdd}\wedge
e^{\gb\gdd}=e_{\gc\gdl}\wedge e_{\gd}{}^{\gdl}C^{\ga\gb\gc\gd}\,,\\
&d\bar{\omega}^{\gad\gbd}+\bar{\omega}^{\gad}{}_{\gdd}\wedge
\bar{\omega}^{\gbd\gdd}+\gl^2e_{\gc}{}^{\gad}\wedge e^{\gc\gbd}=\label{s3}
e_{\gd\gdd}\wedge e^{\gd}{}_{\gdl}\bar{C}^{\gad\gbd\gdd\gdl}\,.
\end{align}
\end{subequations}
This system is exact, like \eqref{LinEinstAA}-\eqref{LinEinstAB}
is, in the sense that it is valid in the full gravity provided
that Weyl tensor obeys the Bianchi identities. Then we linearize
it around $AdS_4$ by replacing
$\omega^{\ga\gb}\rightarrow\varpi^{\ga\gb}+\omega^{\ga\gb}$,
$e^{\ga\gad}\rightarrow h^{\ga\gad}+e^{\ga\gad}$, where frame
fields $\varpi$ and $h$ are of vacuum $AdS_4$. The Weyl tensor for
the empty $AdS_4$ vanishes, so $C^{\ga(4)}\oplus \bar C^{\gad(4)}$
should be taken of the first order. The zeroth order yields the
$so(3,2)$ zero-curvature equations,
\eqref{Cartan}. The resulting linearized
equations (linear in $\omega$ and $e$) reduce to
\begin{subequations}
\label{s2lin}
\begin{align}&\DL e^{\ga\gad}+\go^{\ga}{}_{\gc}\wedge h^{\gc\gad}+
\bar{\go}^{\gad}{}_{\gdd}\wedge h^{\ga\gdd}=0\,,\label{s2linA}\\
&\DL \go^{\ga\gb}+\gl^2(h^{\ga}{}_{\gdd}\wedge
e^{\gb\gdd}+e^{\ga}{}_{\gdd}\wedge
h^{\gb\gdd})=h_{\gc\gdl}\wedge h_{\gd}{}^{\gdl}C^{\ga\gb\gc\gd}\,,\\
&\DL\bar{\go}^{\gad\gbd}+\gl^2(h_{\gc}{}^{\gad}\wedge e^{\gc\gbd}+e_{\gc}{}^{\gad}\wedge h^{\gc\gbd})=h_{\gd\gdd}\wedge h^{\gd}{}_{\gdl}\bar{C}^{\gad\gbd\gdd\gdl}\,.\label{s2linC}
\end{align}
\end{subequations}
Eq. \eqref{s2}-\eqref{s3}, which are analogous to \eqref{LinEinstA}-\eqref{LinEinstB}, are consistent provided the Bianchi identities
for the Weyl tensor hold, therefore, \eqref{s2lin} are consistent under the linearized Bianchi identities for Weyl tensor \eqref{bA}. To rewrite
\eqref{s2lin} in the generating form, we pack fields
$\go_{\ga\gb}$, $\bar{\go}_{\gad\gbd}$ and $e_{\ga\gbd}$ into
quadratic in $y,\bar{y}$ polynomial
\be\label{s2pot}
\omega_{s=2}=\frac12\go^{\ga\gb}y_{\ga}y_{\gb}+\frac12\bar{\go}^{\gad\gbd}\bar{y}_{\gad}\bar{y}_{\gbd}+
e^{\ga\gad}y_{\ga}\bar{y}_{\gad}\,,
\ee
then system \eqref{s2lin} arranges into
\begin{align}\label{s2oms}
&\DD \omega_{s=2}=h^{\gc\gad}\wedge
h_{\gc}{}^{\gbd}\ff{\p^2}{\p\bar{y}^{\gad}\p\bar{y}^{\gbd}}C_{s=2}(0,
\bar{y}|x)+h^{\ga\gdd}\wedge h^{\gb}{}_{\gdd}\ff{\p^2}{\p
y^{\ga}\p y^{\gb}}C_{s=2}(y, 0|x)\,,
\end{align}
where (since we are in $AdS_4$ till the very end, the cosmological constant is set to $1$, $\lambda=1$)
\begin{align}\label{AdjointDer}
&\DD=\DL+h^{\ga\gad}(y_\ga \pl_\gad+\bar{y}_\gad \pl_\ga)\,.
\end{align}
\eqref{s2oms} needs to be supplemented with \eqref{UnfldzeroformsAdS} in the sector of spin-two. Setting $y$ or $\bar{y}$ to zero on the \rhs of \eqref{s2oms} is a way to project onto the Weyl tensor.

Again, equation \eqref{s2oms} is
consistent without any reference to the structure of $s=2$ module
\eqref{s2pot} and so can be generalized for one-form generating
function \be \omega(y, \bar
y|x)=\sum_{n,m=0}^{\infty}\ff{1}{m!n!}\go_{\ga(m),
\gad(n)}(y^{\ga})^m(\bar{y}^{\gad})^n\,,\ee
\be\label{oms}
\DD \omega=h^{\gc\gad}\wedge
h_{\gc}{}^{\gbd}\ff{\p^2}{\p\bar{y}^{\gad}\p\bar{y}^{\gbd}}C(0,
\bar{y}|x)+h^{\ga\gdd}\wedge h^{\gb}{}_{\gdd}\ff{\p^2}{\p
y^{\ga}\p y^{\gb}}C(y, 0|x)\,.
\ee
Eq. \eqref{oms} is called the central on-mass-shell theorem,
\cite{Vasiliev:2001wa,Vasiliev:1988sa}. It is consistent since
$\DD\DD=0$. Since the system is linear it decomposes into an
infinite set of subsystems for fields of spin $s=1,2,3,...$. The
following number operator commutes with $\DD$ and allows one to
single out a subsystem for a particular spin
\be
\Big( y^{\ga}\ff{\p}{\p y^{\ga}}+\bar{y}^{\gad}\ff{\p}{\p
\bar{y}^{\gad}}\Big)\omega_{s}(y, \bar y|x)=\pm 2(s-1)\,
\omega_{s}\,,
\ee
In particular, $\omega^{\ga(s-1),\gad(s-1)}$ component is the spinorial avatar of the spin-$s$ vielbein $e^{a(s-1)}$. The set of fields $\go^{\ga(s-1\pm k), \gad(s-1\mp k)}$ are avatars of $\omega^{a(s-1),b(k)}$. The \lhs of \eqref{oms} reads
\be
\DL \go^{\ga(n), \gad(m)}+ h^{\ga}{}_{\gdd}\wedge\go^{\ga(n-1),
\gad(n)\gdd}+ h_{\gc}{}^{\gad}\wedge\go^{\ga(n)\gc, \gad(m-1)}\,.
\ee
It is set to zero everywhere except for purely holomorphic and
antiholomorphic components driven by $C(y, 0)$ and $C(0, \bar y)$. At this point
we can collect all equations together.

\paragraph{Summary.} Free higher-spin fields including the scalar can be described uniformly by
the following simple system of equations,
\begin{subequations}
\label{UnfldsSpinorial}
\begin{align}\label{UnfldsSpinorialA}
&\DD \omega(y,\bar y|x)=h^{\gc\gad}\wedge
h_{\gc}{}^{\gbd}\ff{\p^2}{\p\bar{y}^{\gad}\p\bar{y}^{\gbd}}C(0,
\bar{y}|x)+h^{\ga\gdd}\wedge h^{\gb}{}_{\gdd}\ff{\p^2}{\p
y^{\ga}\p y^{\gb}}C(y, 0|x)\\
&\delta \omega(y,\bar y|x)=\DD \epsilon(y,\bar y|x)\label{UnfldsSpinorialB}\\
&\widetilde{\DD} C(y,\bar y|x)=0\,,\label{UnfldsSpinorialC}
\end{align}
\end{subequations}
where we added the gauge transformation law for $\omega$ and
\besubeqs
\begin{align}
&\DD=\DL+h^{\ga\gad}(y_\ga \pl_\gad+\bar{y}_\gad \pl_\ga)\label{UnfldsSpinorialDA}\\
&\widetilde{\DD}=\DL-h^{\ga\gad}\pl_\ga\pl_\gad-h^{\ga\gad}y_{\ga}\bar{y}_{\gad}
\label{UnfldsSpinorialDB}\\
&\DL=d+\varpi^{\ga\gb}y_\ga \pl_\gb+\varpi^{\gad\gbd}\bar{y}_\gad
\pl_\gbd \label{UnfldsSpinorialDC}\end{align}
\esubeqs
These equations are
the spinorial version of
\eqref{AllspinsMinkA}-\eqref{AllspinsMinkF} extended to $AdS_4$.

It is a beneficial exercise to check that the system above is
consistent. First, one observes that $\DD^2=0$ and
$\widetilde{\DD}^2=0$\footnote{This formula is a bit of abuse of
notation as formally the operator in \eqref{UnfldsSpinorialDB}
does not even satisfy the chain rule.}, so the system is gauge invariant and is
consistent up to the terms on the \rhs of
\eqref{UnfldsSpinorialA}. The last thing to show is that these
terms do not spoil the consistency.

The system describes fields of spins $s=0,\frac12,1,\frac32,2,\frac52, 3,...$, each in one copy. Truncation to the bosonic sector can be done by requiring $\omega$ and $C$ be even functions of $y,\bar{y}$,
\begin{align}
\omega(y,\bry)&=\omega(-y,-\bry)\,, & C(y,\bry)&=C(-y,-\bry)\label{TruncBosons}\,.
\end{align}
After having defined the higher-spin algebra in the next section we will show that these peculiar operators, $\DD$, $\widetilde{\DD}$ are automatically generated by representation theory.

Let us conclude with a picture below that shows the links between the fields of unfolded system \eqref{UnfldsSpinorial}.
\putfigureifpdf{\UnfoldedEqFieldMap}



\mysection{Higher-spin algebras}\label{sec:HSAlgebra}\setcounter{equation}{0}
So far we have obtained linear in fluctuations unfolded equations
that describe free fields of all spins,
\eqref{UnfldsSpinorial}. At the
linearized level fields of different spins are independent and the
system simply decomposes into a set of decoupled equations each
describing a free field of certain spin. Likewise, the $su(N)$ Yang-Mills theory at zero coupling is
just a sum of $N^2-1$ Maxwell actions for noninteracting photons. In fact, there is a Lie algebra
that unifies fields of all spins,
\cite{Fradkin:1986ka,Vasiliev:1986qx,Konshtein:1988yg,Konstein:1989ij}.

As we already know from the general discussion of Section \ref{sec:Unfolding} on the unfolded
approach one-forms should take values in some Lie algebra. Let us begin with the background connection $\Omega$, obeying \eqref{AllFlat}, which contains
$h$, $\varpi$. Let $\omega$ collectively denote free fields of
all spins. So, the
$\Omega\omega$-type terms, which one sees in
\eqref{AllspinsMinkA}, \eqref{UnfldsSpinorialA}, should arise
from the linearization of $\frac12[W, W]$, where $W$ takes values
in some possibly infinite-dimensional Lie algebra, which we call
the higher-spin algebra, $\mathfrak{g}$. To recover $\Omega\omega$ one replaces $W$ with
$\Omega+g\omega$, where we introduced a formal expansion parameter
$g$. The HS algebra $\mathfrak{g}$ contains the anti-de Sitter
algebra $so(3,2)\sim sp(4,\mathbb{R})$ as a subalgebra since the graviton must
belong to the spectrum and it is described by an $so(3,2)$-connection. Flat
$sp(4,\mathbb{R})$-connection $\Omega$ describes the vacuum over
which the perturbation theory is defined. Let us expand the
Yang-Mills field strength for $W$ in terms of $\Omega+g\omega$
\begin{align}
R=dW+\frac12[W,W]=\underbrace{d\Omega
+\frac12[\Omega,\Omega]}_{\substack{\displaystyle\mbox{vanishes}\\\rule{0pt}{12pt} \displaystyle\eqref{Cartan}}}+\underbrace{g(d\omega+[\Omega,\omega])}_{\substack{\displaystyle\mbox{first
order correction}\\ \rule{0pt}{12pt}\displaystyle\eqref{UnfldsSpinorialA}}}+\underbrace{g^2\frac12[\omega,\omega]}_{\mbox{second
order}}
\end{align}
Since $sp(4,\mathbb{R})\sim so(3,2)$ is a subalgebra of
$\mathfrak{g}$, we can decompose $\mathfrak{g}$ into irreducible
$sp(4,\mathbb{R})$-modules\footnote{It is a general property of
any Lie algebra. Any Lie algebra is first of all a representation
of the algebra itself, the adjoint one. For a simple algebra, the
adjoint is irreducible. For any given subalgebra, the algebra as a
linear space is a representation again, which is reducible in
general. It is this decomposition of the algebra taken as a linear
space into a direct sum of irreducible representations of its
subalgebra that we consider.}. Since $sl(2,\mathbb{C})\sim
so(3,1)$ is a subalgebra of $so(3,2)$, we can also consider a more
fine-grained decomposition of $\mathfrak{g}$ into
$sl(2,\mathbb{C})$ irreducible representation. Depending on the
assumptions about the spectrum of fields contained
in $W$ one can find the following components of $\omega$ (we remind of the dictionary \eqref{spfourpacking}. $A,B,...=1,...,4$ are $sp(4)$ indices, which  can be
split into a pair of $sl(2,\mathbb{C})$ indices, $A=\{\ga,\gad\}$)\\

\begin{tabular}{|x{1cm}|x{8cm}|x{4cm}|}
  \hline
  spin & $sl(2,\mathbb{C})$-content & $sp(4,\mathbb{R})$-content \tabularnewline\hline
  1 & $\omega$ & $\omega\sim\bullet$ \tabularnewline
  2 & $\omega^{\ga\ga}, \omega^{\ga\gad},\omega^{\gad\gad}$ & $\omega^{AA}\sim\YoungpB$ \tabularnewline
  3 & $\omega^{\ga(4)}, \omega^{\ga(3),\gad},\omega^{\ga(2),\gad(2)},\omega^{\ga(1)\gad(3)},\omega^{\gad(4)}$ & $\omega^{A(4)}\sim\YoungpD$ \tabularnewline
  s & $\omega^{\ga(k),\gad(m)}$, $k+m=2(s-1)$ & $\omega^{A(2s-2)}\sim\AYoungp{4}{2s-2}$ \tabularnewline
  \hline
\end{tabular}
\vspace{0.2cm}

Given some putative spectrum of fields it is a valid question to address if there is a Lie algebra,
$\mathfrak{g}$, that contains $sp(4,\mathbb{R})$ as a subalgebra
and has this spectrum of generators under $sp(4,\mathbb{R})$ or
$sl(2,\mathbb{C})$. The HS algebra, is the
fundamental object in the HS theory. There are many different ways
to define it. We present the most useful for practical
computations below. That the linearized equations
\eqref{UnfldsSpinorial} take the
simplest form when a field of every spin appears once naturally
suggests this to be the property of the algebra we are looking
for\footnote{This assumption turns out to be valid for bosonic
algebra only. If one wishes to add fermions a single copy of every
spin would be not enough.}.

The existence of the algebra is important in the Yang-Mills or SUGRA type
theories. Had it not existed we would have proven a no-go
result that some particular set of fields does not admit any
interactions at all. This could have happened for a different
field multiplet, e.g., for a finite number of fields with at least
one of them having spin greater than two.

The information that the free theory tells us is not enough to
recover the algebra immediately. What we know is the decomposition
of $\mathfrak{g}$ as a linear space under its subalgebra
$\mathfrak{h}=sp(4,\mathbb{R})\sim so(3,2)$, equivalently we know how $W$
decomposes,
\begin{align} \mathfrak{g}|_{\mathfrak{h}}&=\mathfrak{h}\oplus\bigoplus_{s\neq2} V_s\,, && V_s=\AYoungp{4}{2s-2}\,,\\
W&=\frac12 (\Omega^{AB}+\omega^{AB})T_{AB}+\sum_{s\neq2} \omega^{A(2s-2)}T_{A(2s-2)}\,,\end{align}
where we singled out the adjoint representation $V_2$ of $\mathfrak{h}$ itself. In terms of fields we
see that the gravitational sector is represented by two fields, background $\Omega^{A,B}$ and dynamical $\omega^{A,B}$. By definition, we know the Lie bracket on $\mathfrak{h}$, i.e. $[\mathfrak{h},\mathfrak{h}]=\mathfrak{h}$, \eqref{spfourrelations} or \eqref{AdSSpinorial}, and we also know $[\mathfrak{h},V_s]=V_s$ which manifests $V_i$ being an irreducible representation of $\mathfrak{h}$.
The missing piece of information is $[V_i,V_j]=?$. To summarize our knowledge we give the known commutation relations for the generators
\besubeqs\begin{align}
[T_{AB},T_{CD}]&=\epsilon_{BC}T_{AD}+\epsilon_{AC}T_{BD}+
\epsilon_{BD}T_{AC}+\epsilon_{AD}T_{BC}\label{knownrelA}\,,\\
[T_{AB}, T_{C(k)}]&= \epsilon_{BC} T_{AC(k-1)}+\epsilon_{AC} T_{BC(k-1)}\,, &&k=2s-2\,.\label{knownrelB}
\end{align}\esubeqs
Splitting $A=\{\ga,\gad\}$ we can get the same relations in $sl(2,\mathbb{C})$ basis.

The most general ansatz for the missing commutators that preserves $sp(4)$ decomposition would be
\begin{align}
[T_{A(k)}, T_{B(m)}]&= \sum_i \alpha_{k,m}^i\underbrace{\epsilon_{AB}...\epsilon_{AB}}_i T_{A(k-i)B(m-i)} \label{knownrelC}
\end{align}
and we need to look for solutions to the Jacobi identity, which should give options for otherwise free coefficients $\alpha_{k,m}^i$. Here we assumed that generator of every spin can appear at most once or do not appear at all, so there are no additional 'color' indices carried by $T_{A(k)}$. One can solve the Jacobi identity and
find a unique higher-spin algebra, \cite{Fradkin:1986ka}, but we proceed to the effective realization of this algebra in terms of oscillators.

The last comment is that any element $f$ of the higher-spin
algebra $\mathfrak{g}$ can be expanded as $f^\aI e_\aI$ with the
basis vectors being $T_{A(k)}=e_\aI$
\begin{align}
f\in\mathfrak{g}&\,, &&f=\sum_k f^{A(k)}\, T_{A(k)}\label{genelement}\,,
\end{align}
where $f^{A(k)}$ are totally symmetric tensors, which are the 'coordinates' in the linear space spanned by $e_\aI$. The commutation relations $[e_\aI,e_\aJ]=f_{\aI\aJ}^\aK e_\aK$, \eqref{knownrelA}-\eqref{knownrelC}, can be rewritten in terms of the coordinates $f^\aI=f^{A(k)}$ as well,
\besubeqs\begin{align}
\eqref{knownrelA}:& &&[f^{AB}T_{AB},g^{CD}T_{CD}]=(4f\fud{A}{C}g^{CD})T_{AD}\label{knownrelBA}\,,\\
\eqref{knownrelB}:& &&[f^{AB}T_{AB},g^{C(k)}T_{C(k)}]=(2kf\fud{A}{B}g^{BC(k-1)})T_{AC(k-1)}\label{knownrelBB}\,,
\end{align}\esubeqs
which is a shorter way to encode the commutation relations.

\paragraph{\texorpdfstring{$\star$}{star}-product realization.}
Let us consider functions of an auxiliary variable $Y^A$ that is
an $sp(2n)$ vector. There is nothing special about $sp(4)$ in this
section and we can extend it to $sp(2n)$. The indices $A,B,...$
range over $2n$ values. The reason we need $sp(4)$ in the end is
because $sp(4)$ is the symmetry algebra of $AdS_4$, which becomes
relevant in the next section only. The Taylor coefficients are
symmetric $sp(2n)$-tensors now
\begin{align}
f(Y)=\sum_k\frac{1}{k!} f_{A(k)}Y^A...Y^A,
\end{align}
and, as we know, upon splitting $A=\{\ga,\gad\}$ we get the
required spin-tensor content in the case of $sp(4)$. Truncation to
bosons is equivalent to $f(Y)=f(-Y)$, c.f. \eqref{TruncBosons}.
So, at least there is a simple way to pack all the coordinates
\eqref{genelement} into a generating function, which was already mentioned
in \eqref{spfourpacking}.

We want to find some operation on the space of functions $f(Y)$ in
$Y$ that induces an operation with required properties
\eqref{knownrelA}-\eqref{knownrelC} or \eqref{knownrelBA}-\eqref{knownrelBB} and solves Jacobi. Standard
product, i.e. $f(Y)g(Y)$ is a commutative operation,
$[f(Y),g(Y)]\equiv0$ and does not even lead to
$\eqref{knownrelBA}$. We need something less trivial and less
local, involving the derivatives
in $Y$.

One more idea, that was realized, \cite{Vasiliev:1986qx}, after
the solution was found in \cite{Fradkin:1986ka}, is that the
Jacobi is solved automatically if the Lie bracket $[f,g]$ comes as
a commutator $[f,g]=f\star g-g\star f$ from some associative
product
\be
f(Y)\star(g(Y)\star h(Y))=(f(Y)\star g(Y))\star h(Y)\,,
\ee
and this is the case for higher-spin algebras. We have seen that
the standard product is too simple. There are two ways to
represent $f\star g$, which is an operation that is linear in its
two arguments, $f$ and $g$, and sends them to some other function
of $Y$,
\besubeqs\begin{align}
f(Y)\star g(Y)&=\int f(U)g(V) K(U,V;Y) d^{2n}U d^{2n}V\,,\\
f(Y)\star g(Y)&=\left.K\left(\frac{{\pl}}{\pl U},Y,\frac{{\pl}}{\pl V}\right) f(U) g(V)\right|_{U=V=Y}\,.
\end{align}\esubeqs
The differential form is perfect when the arguments are polynomials. Note that the differential form can be rewritten as
\begin{align}
f(Y)\star g(Y)&=f(Y)K\left(\frac{\overleftarrow{\pl}}{\pl Y},Y,\frac{\overrightarrow{\pl}}{\pl Y}\right) g(Y)\,.
\end{align}
The integral form is suitable when arguments are more general
integrable functions. The associativity of the product imposes
severe restrictions on the form of the kernels $K$. Still, there
are many solutions, which are equivalent up to a change of the
basis $e_\aJ\rightarrow A\fdu{\aI}{\aJ} e_\aI$. We present the
most convenient one. The solution we choose, which is called the
Moyal $\star$-product, reads
\besubeqs\begin{align}\label{StarProductA}
f(Y)\star g(Y)&=\int f(U)g(V) \exp\left(i(U_A-Y_A)(V^A-Y^A)\right) d^{2n}U d^{2n}V\,,\\
f(Y)\star g(Y)&=f(Y) \exp i
\left(\overleftarrow{\pl}_A\epsilon^{AB}
\overrightarrow{\pl}_B\right) g(Y)\,, \label{StarProductB}
\end{align}\esubeqs
and there is an equivalent form of the first formula
\begin{align}\label{StarProductAA}
f(Y)\star g(Y)&=\int f(Y+U)g(Y+V) e^{(iU_AV^A)} d^{2n}U d^{2n}V\,.
\end{align}
It is assumed that $\int$ sign includes the numerical factor in
order to ensure,
\begin{align} &1\star f(Y)=f(Y)\star 1=f(Y)\,, && 1\star1=1\,,\end{align}
this is achieved by assuming
\begin{align}
1\star 1=\int \exp\left(iU_AV^A\right) d^{2n}U d^{2n}V=\int
\delta^{2n}(U) d^{2n}U=1\,.\end{align} The second formula is
understood as
\begin{align}
f(Y)\star g(Y)&=f(Y) \left(1+i\overleftarrow{\pl}_A\epsilon^{AB}
\overrightarrow{\pl}_B-\frac12 (\overleftarrow{\pl}_A\epsilon^{AB}
\overrightarrow{\pl}_B)^2+...\right)g(Y)\,.
\end{align}

Let us prove that $\star$-product yields an example of a higher-spin
algebra with all required conditions satisfied. Firstly, with the
help of the differential form one sees that
\be Y_A\star Y_B=Y_AY_B+i \epsilon_{AB}\ee
and hence the algebra does not give a trivial Lie bracket
\be\label{osc}
[Y_A,Y_B]_\star=Y_{A}*Y_{B}-Y_{B}*Y_{A}=2i\gep_{AB}\,.
\ee
Moreover, one finds that the commutation relations for $T_{AB}$, \eqref{knownrelA}, \eqref{knownrelBA}, holds with
\begin{align}
T_{AB}=-\frac{i}2 Y_A Y_B
\end{align}
Indeed, expanding the $\exp$-formula up to the third term one finds
\begin{align}\notag
f^{AA} T_{AA}\star g^{BB}T_{BB}=-\frac14 f^{AA}g^{BB}Y_AY_AY_BY_B+2f\fud{A}{C}g^{CB}\left(-\frac{i}{2}Y_AY_B\right)+\frac12f^{AB}g_{AB}
\end{align}
and then arrives at
\begin{align}
[f^{AA} T_{AA}, g^{BB}T_{BB}]_\star=4f\fud{A}{C}g^{CB}T_{AB}\,,
\end{align}
which is \eqref{knownrelBA} and is equivalent to \eqref{knownrelA}. Therefore, $sp(2n)$ is a subalgebra\footnote{We have an associative algebra with some embedding of $sp(2n)$ into it. By definition,
such an algebra has to be related to the universal enveloping algebra of $sp(2n)$, see extra Section \ref{sec:UEA}.} and corresponds to quadratic polynomials in $Y$. The rest of generators are given by $e_\aI=T_{A(k)}=Y_A...Y_A$ up to an unessential numerical prefactor.

Using $\exp$-formula one also proves \eqref{knownrelBB},
\eqref{knownrelB}, where the following formulae are useful
\begin{align}\label{Yleftright}
Y_A\star g(Y)&=\left(Y_A +i\overrightarrow{\pl}_A\right) g(Y)\,, & f(Y)\star Y_A&=\left(Y_A -i\overrightarrow{\pl}_A\right)f(Y)\,.
\end{align}
Finally, we have a Lie algebra, constructed from an associative
one, which obeys \eqref{knownrelA}, \eqref{knownrelB} and we can
look at \eqref{knownrelC},
\begin{align}\notag
f_{A(k)}&Y^A...Y^A\star g_{B(m)} Y^B...Y^B=\\ &\sum_n
\frac{i^nk!m!}{n!(k-n)!(m-n)!}\left(f_{A(k-n)}{}^{C(n)}g_{C(n)B(m-n)}\right)
Y^A...Y^AY^B...Y^B\,.\notag
\end{align}
Note, that  in the commutator terms with even $n$ survive only.
With the same $\exp$ formula one can check that the quadratic
Casimir operator of $sp(2n)$ is a fixed number $-\frac12T_{AB}\star
T^{AB}=-\frac14n(2n+1)$.

Let us show how the integral realization of star-product works,
since it is this form that will be very handy in less trivial
calculations. Let us also stress, that the differential
realization of the star-product strictly speaking is well defined
only for polynomials, whereas its integral form acts on much
broader space of functions thus being relevant for HS analysis
with infinite set of fields.

Let us calculate $Y_{A}\star f(Y)$ ($f(Y)\star Y_{A}$ is
analogous)
\begin{align}\notag
&Y_A\star f(Y)=\left.\ff{\p}{\p p^A}e^{p^BY_B}\right|_{p=0}\star f(Y)\\
&\int\notag
e^{p^B(Y_B+U_B)}f(Y+V)e^{iU_BV^B}=\int
e^{iU_B(V^B-ip^B)+p^BY_B}f(Y+V)= \\
&=\int \gd(V-ip)e^{p^BY_B}f(Y+V)=f(Y+ip)e^{p^BY_B}\,,\notag
\end{align}
from which one gets \eqref{Yleftright}. With the help of either
realization one finds the following useful formulae
\be\label{triv}
\begin{aligned}
[Y_A, f(Y)]_\star &=2i\pl_Af(Y)\,,\\ \{Y_A, f(Y)\}_\star&=2Y_A f(Y)\,, \\
[T_{AB},f(Y)]_\star&=-\frac{i}2[Y_{A}Y_{B}, f(Y)]_\star =(Y_A\pl_B+Y_B\pl_A)f(Y)\,,\\
\{T_{AB},f(Y)\}_\star&=-\frac{i}2\{Y_{A}Y_{B}, f(Y)\}_\star =-i(Y_AY_B-\pl_B\pl_A)f(Y)\,.
\end{aligned}
\ee

\paragraph{Oscillator realization.}
Let us mention another realization of the same algebra, which is
useful for Fock-type analysis in practice. One starts with good
old quantum mechanical pairs of operators of coordinate/momentum
or creation/annihilation (we ignore $\hbar$),
\begin{align}
[\hat{p}_k,\hat{q}^j]=-i \delta^j_k\,.
\end{align}
The pair can be packed into one operator
$\hat{Y}_A=(\hat{q}^j,\hat{p}_k)$. We label $\hat{Y}_A$ with a
hat since it is an operator. Then, the commutation relations
\begin{align}
[\hat{Y}_A,\hat{Y}_B]&=2i\epsilon_{AB}\,, & \epsilon_{AB}&=\frac12\begin{pmatrix}
                 0 &  -\delta^j_k \\
                  \delta_k^j & 0 \\
               \end{pmatrix} &
\end{align}
are identical to \eqref{osc} up to a similarity transformation for $\epsilon_{AB}$. The difference is that $\hat{Y}_A$ are operators and the product is
not expected to be commutative from the very beginning, while $Y_A$ are usual commuting variables endowed in addition to the dot-product with the noncommutative $\star$-product. The two constructions are isomorphic.

The higher-spin algebra is just the algebra of all
quantum-mechanical operators one can construct with $\hat{q}$ and
$\hat{p}$, i.e. it is the algebra of all functions
$f(\hat{Y})=f(\hat{q},\hat{p})$, while $\star$-product realization
is just an effective way to compute the product of operators
$f(\hat{q},\hat{p})g(\hat{q},\hat{p})$. Keeping in mind that
$\hat{p}$ can be represented in the space of functions of $q$ as
$p_k=i\frac{\pl}{\pl q^k}$ we come to the conclusion that the
higher-spin algebra in four dimensions is the algebra of all
differential operators $f(\hat{Y})=f(q,\pl_q)$, which is called
the Weyl algebra.

\paragraph{Lorentz covariant base.} Consider the $sp(4,\mathbb{R})$ case, i.e. $A,B,...=1,...,4$, and
$A=\{\ga,\gad\}$, then we find $T_{AB}$ split as, c.f.
\eqref{spfourmatrix}, \eqref{AdSSpinorial},
\begin{align}
T_{AB} \longleftrightarrow L_{\ga\gb}=T_{\ga\gb},\quad
P_{\ga\gad}=T_{\ga\gad},\quad \bar L_{\gad\gbd}=\bar T_{\gad\gbd}
\end{align}
with the help of \eqref{triv} one finds the realization of Lorentz
and translation generators by differential operators in $y_\ga,
\bar y_\gad$
\besubeqs
\begin{align}
[L_{\ga\gb},f(Y)]_\star&=-\frac{i}2[y_\ga y_\gb, f(Y)]_\star=(y_\ga \pl_\gb+y_\gb \pl_\ga)f(Y)\,,\\
[P_{\ga\gad},f(Y)]_\star&=-\frac{i}2[y_\ga \bar y_\gad,
f(Y)]_\star=(y_\ga \pl_\gad+\bar y_\gad \pl_\ga)f(Y)\,.
\end{align}
\esubeqs
We will need in what follows one more peculiar operator, given by
an anticommutator,
\begin{align}
\{P_{\ga\gad},f(Y)\}_\star=-\frac{i}2\{y_\ga \bar y_\gad,
f(Y)\}_\star=-i(y_\ga \bar y_\gad-\pl_\ga \pl_\gad)f(Y)\,.
\end{align}

\paragraph{Back to the free equations.}
Now we can see some manifestation of the algebra in the free
equations \eqref{UnfldsSpinorial}, where
the operators can be identified as follows
\besubeqs
\begin{align}
&\DD\omega=\DL\omega+h^{\ga\gad}(y_\ga \pl_\gad+\bar y_\gad
\pl_\ga)\omega\label{UnfldsSpinorialDBA}=
\DL\omega+h^{\ga\gad}[P_{\ga\gad},\omega]_\star\,,\\
&\widetilde{\DD}C=\DL C+i h^{\ga\gad}\ff{\p^2}{\p y^{\ga}\p\bar{y}^{\gad}}C-ih^{\ga\gad}y_{\ga}\bar{y}_{\gad}C
=\DL C+h^{\ga\gad}\{ P_{\ga\gad},C\}_\star\label{UnfldsSpinorialDBB}\,,\\
&\DL=d+\varpi^{\ga\gb}y_\ga \pl_\gb+\bar\varpi^{\gad\gbd}\bar
y_\gad \pl_\gbd=
d+\frac12\omega^{\ga\gb}[L_{\ga\gb},\bullet]_\star+\frac12\bar\omega^{\gad\gbd}[\bar
L_{\gad\gbd},\bullet]_\star
\label{UnfldsSpinorialDBC}\,,\end{align}\esubeqs where $\bullet$ is a
placeholder to be replaced with the actual expression the operator
acts on. The appearance of the anti-commutator in
$\widetilde{\DD}$ is crucial and will soon be explained.
Some $i$-factors in $\widetilde{\DD}$ as compared to
\eqref{UnfldsSpinorialDB} can be compensated by rescaling
$C^{\ga(k),\gad(m)}\rightarrow g_{k,m}C^{\ga(k),\gad(m)}$, which
we did not use in Section \ref{sec:AdSLinear}. This  freedom in
rescaling was fixed somewhat arbitrary by normalizing the first
term on the \rhs of \eqref{ScalarAnsatz} not to have any
prefactors. The HS algebra tells us that the $i$-factors are more
natural.

Star-product allows us to define $AdS$ frame fields as one-form
components to various bilinears of $y$ and $\bar{y}$ supplemented
with star-product zero-curvature condition. Indeed, using
\eqref{triv} one can easily convince oneself, that the following $sp(4)$-connection
\begin{align}\Omega=\frac12\Omega^{AB}T_{AB}&=\frac12 \omega^{\ga\gb}L_{\ga\gb}+
h^{\ga\gad}P_{\ga\gad}+\frac12 \bar\omega^{\gad\gbd}\bar L_{\gad\gbd}\\
&=\label{con}
-\ff{i}{4}(\go_{\ga\gb}y^{\ga}y^{\gb}+\bar{\go}_{\gad\gbd}
\bar{y}^{\gad}\bar{y}^{\gbd}+2
h_{\ga\gad}y^{\ga}\bar{y}^{\gad})
\end{align}
substituted into
\be\label{flat}
d\Omega+\Omega\star\Omega=0
\ee
results in \eqref{Cartan}. Then, \eqref{UnfldsSpinorialDBA}-\eqref{UnfldsSpinorialDBB} read now\footnote{Note that $\{\Omega,\omega\}$ is
a mock anti-commutator, it is a commutator since one-forms anti-commute. Indeed,
\be\{\Omega,\omega\}_\star=\{\Omega_\mm dx^\mm,\omega_\nn dx^\nn\}_\star=[\Omega_\mm,\omega_\nn]_\star\, dx^\mm\wedge dx^\nn\,.\ee
This is a difference between the genuine Lie algebra bracket, where, like in Yang-Mills, we have $[\Omega,\omega]$, and the bracket constructed as a commutator in associative algebra, where $[a,b]=a\star b-b\star a$.
In the latter case we have to take into account every additional grading, like the differential form degree, which sometimes turns commutator into anti-commutator. It is easy to translate the Yang-Mills formulae to the case when the Lie algebra is constructed from an associative one. For example, the Yang-Mills curvature is $dA+A\star A$ instead of $dA+\frac12[A,A]$ and the commutator is implicit because one-forms anti-commute. Then $\delta A=d\epsilon+A\star \epsilon-\epsilon\star A$, etc.}.
\besubeqs
\begin{align}
&\DD\omega=d\omega+\Omega\star\omega+\omega\star\Omega=
d\omega+\frac12\Omega^{AB}[T_{AB},\omega]_\star=d\omega+\{\Omega,\omega\}_\star\,,\label{UnfldsSpinorialDCA}\\
&\widetilde{\DD}C=dC+\Omega\star C-C\star \pi(\Omega)\label{UnfldsSpinorialDCB}\,,
\end{align}
where $\pi$ flips the sign of translations,
\begin{align}\label{twistedpi}
&\pi(P_{\ga\gad})=-P_{\ga\gad}\,, && \pi(L_{\ga\gb})=L_{\ga\gb}\,,
&& \pi(\bar{L}_{\gad\gbd})=\bar{L}_{\gad\gbd}\,,
\end{align}
\esubeqs
eventually turning the commutator into anti-commutator inside $\widetilde{\DD}$ in the sector of $P_{\ga\gad}$.

Now the free equations
\eqref{UnfldsSpinorial}, which to be
viewed as the linearization of a yet unknown theory, acquire plain
algebraic meaning in accordance with the general statements made about
the unfolded equations, see Section \ref{sec:Unfolding}.

There are two master-fields, the gauge field $\omega$ that takes
values in the higher-spin algebra is a usual Yang-Mills connection
of that algebra with somewhat different usage. $\DD\omega$ is a
linearization of $dW+W\star W$ with $W=\Omega+g\omega$, where
$\Omega$ is already flat, which guarantees $\DD \DD=0$. Another
master field is $C$ which also takes values in the higher-spin
algebra, but now the action of the algebra on itself is twisted by
$\pi$.

\paragraph{Twisted-adjoint representation.} Let us comment more on the twisted action. Suppose we have an
algebra, say $\mathfrak{g}$, and it acts on some linear space, say
$V$, by operators $\rho:\mathfrak{g}\rightarrow End(V)$, i.e. $V$
is a representation of $\mathfrak{g}$. Given any automorphism
$\pi$ of $\mathfrak{g}$, i.e.
$\pi:\mathfrak{g}\rightarrow\mathfrak{g}$ and
$\pi([a,b])=[\pi(a),\pi(b)]$, we can define another action on the
same space $V$, called the twisted action,
$\rho_\pi(a)=\rho(\pi(a))$. One can check that it is a
representation of $\mathfrak{g}$, i.e.
$\rho_\pi([a,b])=[\rho_\pi(a),\rho_\pi(b)]$.

Now, $\pi$ as defined in \eqref{twistedpi} is an automorphism of the anti-de Sitter algebra, $so(3,2)$. It acts on $P$ only,
so the nontrivial relations to check are (we drop the indices)
\begin{align}\notag
[\pi(P),\pi(P)]&=[-P,-P]=[P,P]=L=\pi(L)\,,\\
[\pi(L),\pi(P)]&=[L,-P]=-[L,P]=-P=\pi(P)\,.\notag
\end{align}
Formally, $\pi$ can be realized either as $y_\ga\rightarrow
-y_\ga$ or $\bar y_\gad\rightarrow -\bar y_\gad$, since $L\sim y
y$, $\bar{L}\sim \bry\bry$ and only $P\sim y\bry$ is affected by
such an action,
\begin{align}\label{auto}
&\pi (y, \bar y)=(-y, \bar y)\,,&& \pi\circ\pi=1\,,
 && \mbox{or}     &&\bar{\pi} (y, \bar y)=(y, -\bar y)\,,&& \bar{\pi}\circ\bar{\pi}=1\,.
\end{align}
With this definition
$\pi$ can be extended to the full higher-spin algebra, $\mathfrak{g}$, as an associative algebra, i.e.
$\pi(f\star g)=\pi(f)\star \pi(g)$. Indeed, $\pi$ just checks if the function is even or odd in $y$ or $\bry$ and even$\star$even=even,
odd$\star$odd=even, even$\star$odd=odd. Hence, it extends to $\mathfrak{g}$ as a Lie algebra under the commutator.

Finally, the linearized equations for massless fields of all spins
combined into a single multiplet read as in
\eqref{UnfldsSpinorial} and $\DD$,
$\widetilde{\DD}$ have the meaning of adjoint and twisted-adjoint
covariant derivatives for the master fields $\omega$, $C$ taking
values in the higher-spin algebra\footnote{That gauge fields $\omega$ and zero-forms $C$
take values in the same space is a kind of operator-state correspondence. $\pi$ is related to the Chevalley involution, on the CFT side, it is realized as an inversion.}. The nontrivial gluing term on
the r.h.s. of \eqref{UnfldsSpinorialA} is the Chevalley-Eilenberg
cocycle, see around \eqref{UnfldCocycleA} and extra Section
\ref{app:ChevalleyEilenberg} for more detail. Hence, all terms
in the equations have clear representation theory meaning.

\paragraph{'Pure gauge' solutions.} Equation \eqref{flat} is the zero-curvature condition.
Hence, any solution of \eqref{UnfldsSpinorialDCB} admits pure gauge form
\be\label{Wgauge}
\Omega=g^{-1}(Y|x)\star dg(Y|x)\,.
\ee
Equation \eqref{UnfldsSpinorialDCB} is the
covariant constancy condition in the twisted-adjoint representation of the HS
algebra. The general solution of \eqref{UnfldsSpinorialDCB} is
\be\label{Cgauge}
C(Y|x)=g^{-1}\star C_0(Y)\star \pi(g)\,,
\ee
where $C_0(Y)$ is an arbitrary $x$-independent function. Pure
gauge form of \eqref{Wgauge} and \eqref{Cgauge} may look
misleading for it seemingly suggests that one can gauge away any
solution for dynamical fields. This is not the case. There are two
restrictions that constrain gauge functions $g(Y|x)$ in
\eqref{Wgauge}. First, $g(Y|x)$ should be such that corresponding
connection $\Omega$ be of the form \eqref{con} i.e. bilinear in
$y$'s and second, it should not provide one with degenerate
vierbein $h_{\ga\gad}$. In practical calculations the gauge
function that reproduces vacuum frame fields is some exponent of
bilinears in $y$'s. The fact that higher-spin equations for zero-from $C(Y|x)$ acquire pure gauge representation is a remarkable
property of the on-shell integrability of this system uncovered
with the aid of the unfolding approach.

Let us note, that twisted-adjoint equation \eqref{UnfldsSpinorialDCB} is fixed by
representation theory. It means, in
particular, that scalar mass term \eqref{m} is completely fixed by the
HS algebra as well.

Unfolded form of dynamical equations
makes their symmetries manifest. For example, $AdS_4$ global
symmetries are governed by the symmetry parameter $\xi(Y|x)$, collective Killing, that leaves the vacuum, $\Omega$,
invariant
\be\label{glob}
0=\gd \Omega =D_\Omega\xi\,,\qquad\qquad\qquad D_\Omega=d+[\Omega,
\bullet]_\star\,,
\ee
which can be explicitly found once $g(Y|x)$ is known
\be\label{globsol}
\xi(Y|x)=g^{-1}\star \xi_{0}(Y)\star g\,.
\ee
This is a covariant generalization of the Poincare lemma, which shows
that $\DO \xi=0$ and $d\xi=0$ have isomorphic solution spaces.
It is also clear why the matter and HS curvatures module $C(Y|x)$
is described by differential zero-form, rather than some $p$-form.
It makes it gauge invariant, otherwise there would be gauge
invariance $\gd C^{(p)}=\widetilde{\DD} \xi^{(p-1)}$, where
$\widetilde{\DD}$ is the twisted-adjoint covariant derivative
given in \eqref{UnfldsSpinorialDBB}.

One may ask if it is possible to solve for $\omega$ in a 'pure
gauge' form. It cannot be just $g^{-1}\star\omega_0(Y)\star g$
since the latter is a solution to the homogeneous equation $\DD
\omega=0$. On the other hand, all gauge-invariant information is
in the zero-forms, which contain Weyl tensors and derivatives
thereof, so it must be possible to reconstruct $\omega$ from $C$
up to pure gauge solutions $\DD \xi$. The formula does exist and
can be found in \cite{Bolotin:1999fa}.

\paragraph{Klein operator.} Let us have a look at the automorphism $\pi$,
which is one of the key ingredients of the higher-spin theory. Formally, $\pi$ can be realized as
\begin{align}
\kappa&=(-)^{N_y}\,, &&\kappa y \kappa=-y\,, &&\kappa \bry \kappa=\bry\,,
\end{align}
where $N_y$ is the number operator $y^\gc\pl_\gc$ that counts $y$'s.

Let us find out whether automorphism \eqref{auto} is the inner one
or not. In other words we would like to see if it can be realized
in terms of star-product \eqref{StarProductA}-\eqref{StarProductB}. Assuming that such element
$\gk$ does exist in the star-product algebra such that \be\gk*f(y,
\bar y)*\gk=f(-y, \bar  y)\label{twistbystar}\ee and $\gk\star \gk=1$ since $\pi\pi=1$ one has
\be\label{kl}
\gk*y_{\ga}=-y_{\ga}*\gk\,,\qquad
\gk*\bar{y}_{\gad}=\bar{y}_{\gad}*\gk\,.
\ee
Using \eqref{triv} the second condition in \eqref{kl} tells us
that $\gk$ is $\bar y$-independent, while the first is equivalent
to
\be
\gk y_{\ga}=0\quad\Longrightarrow\quad \gk\sim\gd^2(y)\,.
\ee
From \eqref{kl} it follows also that $[\gk*\gk, y_{\ga}]_*=0$ and
therefore $\gk*\gk\sim{\rm{const}}$. The constant can be chosen to
be 1 which leads to
\be\label{delta}
\gk=2\pi\gd^{2}(y)\,.
\ee
Indeed,
\be\label{kltest}
\gk\star \gk=(2\pi)^2\int du\, dv\,
\gd(y+u)\gd(y+v)e^{iu_{\ga}v^{\ga}}=1\,.
\ee
Note, that in checking \eqref{kltest} one really has to use the
integral form of the star-product, for $\star$-product of
$\delta$-functions is out of reach for differential star-product.
The operator $\gk$ which satisfies the condition
\be\label{Klein}
\{\gk, y_{\ga}\}_\star=0\,,\qquad \gk\star\gk=1\,,
\ee
is called holomorphic Klein operator. Analogous operator can be
defined in the anti-holomorphic sector. As one can see, the Klein
operator, being a delta function, strictly speaking does not belong
to the $\star$-product algebra and hence the automorphism \eqref{auto} is
not the inner one. Nevertheless, representation \eqref{delta} is
very useful in practice. For example the action of Klein operator
on a function is just the Fourier transform\footnote{The appearance of Fourier transform $F$ should not be surprising
since $(F\circ F) [f(y)]=f(-y)$.} with respect to
holomorphic oscillator $y$
\be
\gk\star f(y, \bar y)=\int dv f(v, \bar y)e^{iv_{\ga}y^{\ga}}\,.
\ee
Automorphism \eqref{auto} in terms of Klein operator action is
given simply by
\be\label{pi}
\pi(f(y, \bar y))=\gk\star f\star\gk\,.
\ee
Using \eqref{pi} one can derive the Penrose transform that maps
solution of field equations \eqref{UnfldsSpinorialDCB} to $AdS_4$ HS global
symmetries \eqref{glob}. Indeed, having any HS global symmetry parameter
$\xi(y, \bar y|x)$ that satisfies \eqref{glob} one can
generate a solution to twisted-adjoint equation \eqref{UnfldsSpinorialDCB}
\be\label{Penrose}
C(y, \bar y|x)=\xi(y, \bar y|x)\star\gk=\int \xi(u,\bar
y|x)e^{-iu_{\ga}y^{\ga}}\,.
\ee
Let us note, that \eqref{Penrose} can be applied to $AdS_4$ global
symmetry parameter which being bilinear in $Y$'s will generate
some ill-defined solution via Fourier transform of quadratic
polynomial. Generic HS global symmetry parameter $\xi_0(Y)$ in
\eqref{globsol}, however, can be an arbitrary star-product
function which may lead to a well defined Fourier transform.

Let us stress that with the help of $\gk$ one can map twisted-adjoint fields in
adjoint ones and vice verse. For example,
\begin{align}\notag
d C+\Omega \star C-C\star \underbrace{\gk\star \Omega \star\gk}_{\displaystyle\pi(\Omega)}&=0 &&\Longleftrightarrow&&
d(C\star \gk)+\Omega\star (C\star \gk)-(C\star \gk)\star\Omega=0
\end{align}
where the second equation is the first one $\star$-multiplied by $\gk$ since $d\gk=0$ and $\gk\star\gk=1$.

The automorphism  $\pi$ may look like an isomorphism between the
adjoint and twisted-adjoint representations. However one has to be
careful with this point of view as the field-theoretical
interpretation is totally different. Particularly, the physical
domains of twisted-adjoint and adjoint representations do not
overlap much since polynomials are mapped by $\star \gk$ into
derivatives of $\delta$ functions and vice versa. More detail
on how to work with $\star$-products can be found in extra Section \ref{extra:Cayley}.

\paragraph{Summary.} The linearized analysis performed in the previous section gave us
a set of one-form connections. We found that there exists an infinite-dimensional algebra
that these fields are connections of. The zero-forms turns out to leave in the higher-spin algebra
too, but the representation is twisted by an automorphism of the anti-de Sitter algebra that flips translations.
The higher-spin algebra is an associative algebra that can be identified with the algebra of two pairs of usual quantum mechanical operators $q,p$. The product in the algebra can be effectively realized as the Moyal $\star$-product.



\section{Vasiliev equations}\label{sec:VasilievEquations}\setcounter{equation}{0}
In this section we consider nonlinear interactions for bosonic HS
fields governed to all orders by the Vasiliev equations. Firstly,
we are going to discuss what one should expect on the way of
constructing the interactions within the unfolded approach.
Secondly, starting from certain mild assumptions we attempt to
derive the Vasiliev equations. It is not that easy to find
reasonings that lead to these assumptions, but apart from these
gaps the derivation is quite solid. Thirdly, we look at the
$AdS_4$ vacuum solution to Vasiliev equations and show that the
linear equations derived in Section \ref{sec:AdSLinear} do emerge
in the first order perturbative expansion. Then, we discuss
certain nontrivial properties of the equations like the local
Lorentz symmetry being part of the physical requirements that
should fix the form of the equations. Next, we make some
comments on higher orders and analyze the Vasiliev system
explicitly to the second order. Lastly, we elaborate on the special property of
Vasiliev equations --- integration flow --- that allows one to reconstruct solutions
of the nonlinear system starting from a peculiar solution of the free one.

\mysubsection{Generalities}\label{subsec:Generalities}
Up to this point we have shown that the specific multiplet of free HS fields,
where a field of every spin appears in one copy, can be described in terms of two master-fields
$\omega(Y|x)$ and $C(Y|x)$, both taking values in the higher-spin algebra, with the
empty anti-de Sitter space given by the flat $so(3,2)\sim sp(4,\mathbb{R})$-connection
$\Omega$, \eqref{Cartan}, with equations being
\besubeqs
\begin{align}
&d\Omega=-\Omega\star\Omega\,,\\
&d\omega=-[\Omega,\omega]+\mathcal{V}(\Omega,\Omega,C)\label{xpaceseqEA}\,,\\
&dC=-\Omega\star C+C\star \pi(\Omega)\label{xpaceseqEB}\,,
\end{align}
\esubeqs
where there are two crucial ingredients: automorphism $\pi$,
\eqref{twistedpi}, \eqref{twistbystar}, and cocyle $\mathcal{V}(\Omega,\Omega,C)$, \eqref{UnfldsSpinorialA}. If
we believe in the existence of the consistent nonlinear theory then
these equations must be the linearization of
\besubeqs
\label{xpaceseq}
\begin{align}
&d\omega=\mathcal{V}(\omega,\omega)+\mathcal{V}(\omega,\omega,C)+
\mathcal{V}(\omega,\omega,C,C)+...=F^{\omega}(\omega,C)\label{xpaceseqA}\,,\\
&dC=\mathcal{V}(\omega,C)+\mathcal{V}(\omega,C,C)+\mathcal{V}(\omega,C,C,C)+...=F^C(\omega, C)\,,\label{xpaceseqB}
\end{align}
\esubeqs
which are a particular case of unfolded equations \eqref{UnfldEquationsA}. The vertices $\mathcal{V}(\bullet,...,\bullet)$ are linear in each
slot and correspond to certain couplings. The structure functions $F^{\omega}(\omega,C)$ and $F^C(\omega, C)$ need to obey the integrability condition \eqref{UnfldJacobi}. The first nontrivial
couplings $\mathcal{V}(\omega,\omega)$ and $\mathcal{V}(\omega,
C)$ are governed by the higher-spin algebra with the help of
$\pi$,
\begin{align*}
-\mathcal{V}(\omega,\omega)&=\omega\star\omega\,,
&-\mathcal{V}(\omega, C)&=\omega\star C-C\star \pi(\omega)\,.
\end{align*}
Linearization is carried out by replacing
$\omega\rightarrow\Omega+g \omega$ and picking up the terms of the
zeroth and first order in $g$.
The cocycle $\mathcal{V}(\omega,\omega,C)$ is not yet known to us in full, only its
value in the $AdS$ vacuum $\mathcal{V}(\Omega,\Omega,C)$ is available. We can clearly
see it not to have a simple form of $\Omega\star \Omega \star C$ that treats vielbein and spin-connection
on equal footing while $\mathcal{V}(\Omega,\Omega,C)$ does not contain spin-connection at all.
Going higher in orders we can claim that none of the vertices, apart from the bilinear one, is
related to the higher-spin algebra and its realization via the $\star$-product in any obvious sense.
This is the general property of nonlinear theories that the structure constants of Lie algebra governing lowest orders get deformed into structure functions which depend on the fields themselves. Within the unfolded approach higher interaction vertices still have an interpretation of Chevalley-Eilenberg cocycles of the higher-spin algebra.

The self-consistency of \eqref{xpaceseq} through $d^2=0$ implies that there is a gauge symmetry
\besubeqs
\label{xpaceseqGS}
\begin{align}
&\delta\omega=d\xi+\xi\frac{\pl}{\pl \omega}F^{\omega}(\omega,C)=d\xi+[\omega,\xi]_\star+O(C)\label{xpaceseqGSA}\,,\\
&\delta C=\xi\frac{\pl}{\pl\omega}F^C(\omega, C)=C\star \pi(\xi)-\xi\star C+O(C^2)\,,\label{xpaceseqGSB}
\end{align}
\esubeqs
The first terms in the gauge transformations are governed by the higher-spin algebra, but there are
corrections due to $C$ so that the higher-spin algebra as a Lie algebra does not exist at the interaction level. The best we can say is that the amount of gauge symmetry remains the same. It is a meaningful question to ask if there is a subalgebra of the higher-spin algebra that survives as a Lie algebra in interactions. Indeed, there is one and it contains Lorentz subalgebra, which we shall discuss later in Section \ref{subsec:Lorentz}.

To proceed to nonlinear level one can in principle search order by
order in perturbation theory for such a deformation of
\eqref{xpaceseqEA} and \eqref{xpaceseqEB} within the set of fields
$w(y, \bar y|x)$ and $C(y, \bar y|x)$ that is consistent with the
integrability condition $d^2=0$ (hence possessing gauge symmetry)
and reproduces correct free field dynamics. The perturbative
analysis \eqref{xpaceseq} is a challenging
technical problem and gets already cumbersome at the second order,
\cite{Vasiliev:1988sa,Vasiliev:1988xc}. This analysis was carried
out up to $\mathcal{V}(\omega,\omega,C,C)$,
\cite{Vasiliev:1989yr}, followed by the effective method of
summing all orders,
\cite{Vasiliev:1990cm,Vasiliev:1990en,Vasiliev:1990bu,Vasiliev:1992av}.

By looking at the pure gravity case we have to accept the fact
that the interaction series does not truncate to a finite number
of terms within the unfolded approach, i.e. all powers of $C$
appear. Unfortunately, no closed form for $F^{\omega}$ and $F^C$
is known. Functions $F^{\omega}$ and $F^C$ are analogous to the
expansion of the Riemann tensor in terms of the metric field.
In gravity the series goes to infinity, being though ideologically
simple as the source of the infinite tail is the expansion of inverse
metric $g_{\mu\nu}^{-1}$.

The Vasiliev equations are the generating equations written in an
extended space that sum up the infinite series into the equations
that are no more than quadratic in fields. Namely, the equations
consist of zero-curvature equation plus certain constraints that
determine the embedding of interaction vertices into the fields
obeying the zero-curvature condition. The interaction vertices
$\mathcal{V}(\bullet,...,\bullet)$ can be obtained by solving the
equations order by order. Vasiliev equations do not give
\eqref{xpaceseq} immediately! They are
formulated in certain extended space with more coordinates and
\eqref{xpaceseq} appear as solutions in the
perturbative expansion around the $AdS_4$ vacuum. In some sense
Vasiliev equations are analogous to simple differential equations
encoding complicated special functions, but in the Vasiliev case
the equations encode complicated equations as solutions of simpler equations with
respect to additional variables. The last remark is that Vasiliev equations
share many properties with integrable equations. Integrability
usually means that a theory can be put into the form of some
covariant constancy or zero-curvature equation, i.e. equations
that are no more than quadratic in fields.

As an example we will derive second order corrections in the formal
coupling constant $g$, $\omega\rightarrow\Omega+g
\omega_1+g^2\omega_2$ from Vasiliev system
\besubeqs\label{xpaceseqBQ}
\begin{align}
&\DD \omega_2=\mathcal{V}(\omega_1,\omega_1)+
\mathcal{V}(\Omega,\Omega,C_2)+\mathcal{V}(\Omega,\omega_1,C_1)+
\mathcal{V}(\Omega,\Omega,C_1,C_1)\label{xpaceseqBA}\\
&\widetilde{\DD} C_2=\mathcal{V}(\omega_1,C_1)+\mathcal{V}(\Omega,C_1,C_1)\label{xpaceseqBB}
\end{align}
\esubeqs
and we will show some of these vertices explicitly. And at the
order $N$ one can write
\besubeqs\label{xpaceseqC}
\begin{align}
&\DD \omega_N=\sum'_{n+m=N}\mathcal{V}(\omega_n,\omega_m)+\sum'_{n+m+k=N}
\mathcal{V}(\omega_n,\omega_m,C_k)+\mathcal{V}(\Omega,\Omega,C_N)\label{xpaceseqCA}\\
&\widetilde{\DD} C_N=\sum'_{n+m=N}\mathcal{V}(\omega_n,C_m)\label{xpaceseqCB}
\end{align}
\esubeqs
where the primed sum $\sum'$ designates that the background
covariant derivative has already been extracted. For future use
let us note that at order $N$ the fields of order $N$ appear the
same way as $\omega_1$ and $C_1$ appear in the linearized
equations. So the equations for $\omega_N$ and $C_N$ look like the
linear equations plus sources built of the lower order fields.

It, of course, makes sense to look for exact solutions directly in the extended space
and try to give it certain physical meaning.

\mysubsection{Quasi derivation of Vasiliev equations}
It is a sort of conventional wisdom that Vasiliev equations cannot
be derived rather one is welcome to check that they do satisfy all
the required assumptions. Below we attempt to put things in a
perspective of 'derivation' starting from the main assumption that
they should have 'almost' zero-curvature/covariant constancy form
as in integrable theories\footnote{For example, the famous
Korteweg-de Vries equation, which contains higher-derivative and
is nonlinear
--- oversimplified model of higher-spin theory, can be formulated
as a zero-curvature equation in certain extended space. Upon
solving certain first-order equations coming as the components of the zero-curvature condition one gets
the KdV equation. Vasiliev equations function analogously.}, but the field
content and the space where the equations are formulated can
possibly be extended.

Let $\WW$ be the full master-field on possibly extended space, of
which $\omega(Y|x)$ is a subspace. There must be $\WW\star
\WW$-term on the r.h.s. of \eqref{xpaceseqA} where $\star$ as well
acts on possibly extended space and reduces to the usual HS
algebra $\star$ for $\omega\in \WW$. Denoting all other
corrections on the r.h.s. of \eqref{xpaceseqA} as $\Phi$, which
must be a two-form, and using $d^2=0$ one finds
\besubeqs
\label{nhs}
\begin{align}\label{nhsA}
&d\WW+\WW\star\WW=\Phi\,,\\
&d\Phi+[\WW,\Phi]_\star=0\label{nhsB}\,,
\end{align}
\esubeqs
where the second equation comes from consistency requirement.
Unfolding tells us that this system has local gauge invariance\footnote{Mind that $[\bullet,\bullet]$ sometimes
changes to $\{\bullet,\bullet\}$ due to differential form degree.}
\be\label{locg}
\gd \WW=d\gep+[\WW,\gep]_\star+\xi\fm{1}\,,\qquad \gd \Phi=d\xi\fm{1}+\{\WW,\xi\fm{1}\}_\star-[\gep, \Phi]_\star\,.
\ee
Suppose $\Phi$ is a fundamental field, i.e. it is not expressed in
terms of some other degree-zero or degree-one fields. Then its has
its own one-form gauge parameter $\xi\fm{1}$, which is able to kill
$\WW$ by means of $\gd \WW=\xi\fm{1}$. Therefore, $\Phi$ cannot be fundamental.

From free level analysis we have found an appropriate set of
fields for the description of gauge field dynamics. These are the
master fields --- a one-form  $\omega(Y|x)$ and a zero-form $C(Y|x)$.
Equation \eqref{nhsB} hints to identify somehow $\Phi$ with
$C$ (for a moment we forget about issues with the twisted-adjoint
representation). The problem is that $\Phi$ is a two-form and hence
one needs some extra two-form to identify $\Phi\sim C$. The only
field independent two-form that one has is $dx^a\wedge dx^b$.
This, however, carrying indices should be contracted with some
derivatives in $Y$'s of $C$, thus bringing us to annoying
perturbative analysis.

A way out proposed by Vasiliev is to
introduce some auxiliary direction to space-time that will allow
one to determine the two-form carrying no indices. This can be
done one way or another, but this additional direction can be
anticipated to be auxiliary for $Y$ as well. So, in spinorial
formalism we are working with, let us choose some auxiliary
$Z_{A}=(z_\ga, \bar{z}_{\gad})$ and corresponding differential
$dZ_A$, which anticommutes with $dx_{m}$ and provides us with the
differential two-forms. We used the
same notation, $\ga$, for the auxiliary indices to anticipate that
$z$ are in close interrelation with $y$. If the auxiliary space is two-dimensional the two-form is a top-form and is unique, in
particular there are no free indices it can carry, $dz^{\ga}\wedge
dz_{\ga}$ and $d\bar{z}^{\gad}\wedge d\bar{z}_{\gad}$. That we need two auxiliary two-forms is due to
the two terms on the \rhs of \eqref{UnfldsSpinorialA}, i.e. due to the fact that the spin-$s$ Weyl tensor $C^{a(s),b(s)}$ splits into two complex-conjugate $C^{\ga(2s)}$, $C^{\gad(2s)}$ that yield two sources for
one-forms $\omega(Y|x)$. This allows
us to write down system \eqref{nhs} in the extended space by
introducing the extended connection
\be\label{defA}
\WW=W_{m}dx^{m}+\aA_{\ga}dz^{\ga}+\bar{\aA}_{\gad}d\bar{z}^{\gad}\,,
\ee
where the fields now depend on a new variable $Z_A$ as well
\beno
\omega(Y|x)\to W(Y, Z|x)\,,\qquad C(Y|x)\to B(Y,Z|x)\,,\qquad
\aA_{A}=\aA_{A}(Y, Z|x)\,.
\eeno
The exterior differential gets enhanced too and we label it with
the hat
\beno
d\to \hat{d}=d_x\oplus d_{z}\oplus d_{\bar z}\,.
\eeno
As a result, one may propose the following system
\besubeqs\label{tobehs}
\begin{align}
&\hat{d}\WW+\WW\star\WW=\Phi+\bar{\Phi}\,,\label{tobehs1}\\
&\hat{d}\Phi+[\WW,\Phi]_\star=0\,,\label{tobehs2}\\
&\hat{d}\bar{\Phi}+[\WW,\bar{\Phi}]_\star=0\,,\label{tobehs3}
\end{align}
\esubeqs
where we have implied the following identification
\be\label{PhiB}
\Phi=\ff i4dz_{\ga}\wedge dz^{\ga}B\,,\quad \bar{\Phi}=\ff
i4d\bar{z}_{\gad}\wedge d\bar{z}^{\gad}\bar{B}\,,
\ee
The system is formally consistent
and admits local gauge invariance
\be\label{tobegauge}
\gd \WW=\hat{D} \gep\equiv\hat{d}\gep+[\WW,
\gep]_{\star}\,,\qquad \gd \Phi=[\Phi, \gep]_\star\,.
\ee
Let us stress once again that eqs. \eqref{tobehs2} and
\eqref{tobehs3} arise from \eqref{tobehs1} as  consistency
conditions $d^2=0$ and are not independent.

\paragraph{Extended star-product, Klein operators.}
So far nothing is said
about extra spinor variables $z_\ga$ and $\bar{z}_{\gad}$. Let us
choose these to be dual to $Y_A$, i.e.
\be\label{comm}
[Z_A, Z_B]_\star=-2i\gep_{AB}\,,\qquad [Z_A, Y_B]_\star=0\,.
\ee
It requires the corresponding extension of the star-product, which
we will define a bit later. Recall now, that HS master field
zero-form $C(Y|x)$ is subject to twisted-adjoint flatness
condition rather than the adjoint one at free level. Therefore, at
full level we need to redefine $B(Y,Z|x)$ using the appropriate
Klein operator. Analogously to the free theory consideration, the
explicit form of the Klein operator $\gk$ depends on the form of
the automorphism through its star-product realization. At
nonlinear level, the twisted-adjoint automorphism is defined as
\besubeqs\label{auto12}
\begin{align}
\pi(y_{\ga}, \bar{y}_{\gad}, z_{\ga}, \bar{z}_{\gad})=(-y_{\ga},
\bar{y}_{\gad}, -z_{\ga}, \bar{z}_{\gad})\,,\label{auto1}\\
\bar{\pi}(y_{\ga}, \bar{y}_{\gad}, z_{\ga},
\bar{z}_{\gad})=(y_{\ga}, -\bar{y}_{\gad}, z_{\ga},
-\bar{z}_{\gad})\,.\label{auto2}
\end{align}
\esubeqs
The corresponding {to be found} Klein operators $\gk$ and
$\bar{\gk}$ are determined by the conditions
\be\label{Kleindef}
\gk\star F(Y,Z)=F(\pi(Y,Z))\star\gk\,,\qquad
\bar{\gk}\star F(Y,Z)=F(\bar{\pi}(Y,Z))\star \bar{\gk}\,,
\ee
where $F(Y,Z)$ is an arbitrary function. Just as in the free case,
the twisted-adjoint HS-curvature zero-form $B(Y,Z|x)$ is reproduced
from the adjoint one as
\be\label{subst}
B\to B\star \gk\,,\qquad \bar{B}\to B\star \bar{\gk}\,.
\ee
Let us note that the fields $B$ and $\bar B$ are not independent
as might seem from \eqref{PhiB} which is in accordance with the
linearized description where we had a single Weyl module $C(y,
\bar y|x)$. This fact imposes severe restriction on the form of
higher-spin interactions. Indeed if $B$ and $\bar B$ were
independent, equations \eqref{tobehs1} would be just a definition
of the curvature two-form in $dz_{\alpha}\wedge dz^{\alpha}$ and
$d\bar{z}_{\dot{\alpha}}\wedge d\bar{z}^{\dot{\alpha}}$ sectors on
its right hand side. The explicit form of Klein operators can be
derived not until the extended $(Y,Z)$ star-product is defined, so
let us proceed with its definition.

Commutation relations \eqref{comm} can be reached via the
following star-product
\be\label{staryz}
f(Y,Z)\star g(Y,Z)=\int dU dV f(Y+U, Z+U)g(Y+V, Z-V)e^{iU_{A}V^{A}}\,.
\ee
Mind the minus sign in the second argument of $g$-function which
guarantees that $[Y,Z]_\star =0$. Star-product can be shown to
be associative. It reduces exactly to the Moyal star-product
\eqref{StarProductAA} once functions $f$ and $g$ are
$Z$-independent. On the space of $Y$-independent functions
we find a formula similar to \eqref{StarProductAA} with an extra minus
sign in the exponent that ensures (\ref{comm}.a). Commutation
relations \eqref{comm} then can be easily reproduced from
definition \eqref{staryz}. The following simple formulas will be
useful for star-product calculations
\begin{align*}
e^{p^AY_A}\star f(Y,Z)=e^{p^AY_A}f(Y+ip, Z-ip)\,,\quad
f(Y,Z)\star e^{p^AY_A}=e^{p^AY_A}f(Y-ip, Z-ip)\,,\\
e^{p^A Z_A}\star f(Y,Z)=e^{p^AZ_A}f(Y+ip, Z-ip)\,,\quad f(Y,Z)\star e^{p^A
Z_A}=e^{p^AZ_A}f(Y+ip, Z+ip)\,.
\end{align*}
Particularly, from these relations it follows, c.f. \eqref{Yleftright},
\besubeqs\label{yz}
\begin{align}
Y_A\star f&=\Big(Y_A+i\ff{\p}{\p Y^A}-i\ff{\p}{\p Z^A}\Big)f\,,&
f\star Y_{A}&=\Big(Y_A-i\ff{\p}{\p Y^A}-i\ff{\p}{\p Z^A}\Big)f\,,\label{yz1}\\
Z_A\star f&=\Big(Z_A+i\ff{\p}{\p Y^A}-i\ff{\p}{\p Z^A}\Big)f\,,&
f\star Z_A&=\Big(Z_A+i\ff{\p}{\p Y^A}+i\ff{\p}{\p
Z^A}\Big)f\,.\label{yz2}
\end{align}
\esubeqs
It is easy to check that thus defined star-product is associative
and provides normal ordering for $a^{\pm}_{A}=Y_A\pm Z_A$. Indeed,
as follows from \eqref{yz}
\be
f\star a^{+}=fa^{+}\,,\quad a^{-}\star f=a^-f\,.
\ee

\noindent The following simple formulas we find useful below, c.f. \eqref{triv},
\besubeqs\label{YZBasic}
\begin{align}
[Y_A, Y_B]_\star&=2i\gep_{AB} \,, & [Z_A, Z_B]_\star&=-2i\gep_{AB}\,,\label{YZBasicA}\\
[Y_A, f]_\star&=2i\ff{\p}{\p Y^A}f\,,& [Z_A, f]_\star&=-2i\ff{\p}{\p Z^A}f\,,\\
\{Y_A, f\}_\star&=2\Big(Y_A-i\ff{\p}{\p Z^A}\Big)f\,,&
\{Z_A, f\}_\star&=2\Big(Z_A+i\ff{\p}{\p Y^A}\Big)f\,,
\end{align}
\esubeqs
Of crucial importance is that one can still use the same definition
\begin{align*}
T_{AB}=-\frac{i}2 Y_AY_B
\end{align*}
for the generators of $sp(4)\sim so(3,2)$. Indeed, with the above formula one finds
\besubeqs\label{YZcommutators}
\begin{align}
\left[T_{AB}, f\right]_\star&=\Big(Y_B-i\ff{\p}{\p Z^B}\Big)\ff{\p}{\p Y^A}f+
\Big(Y_A-i\ff{\p}{\p Z^A}\Big)\ff{\p}{\p Y^B}f\,,\\
\left\{T_{AB}, f\right\}_\star&=-i\Big(Y_A-i\ff{\p}{\p Z^A}\Big)\Big(Y_B-i\ff{\p}{\p Z^B}\Big)f+i\ff{\p}{\p Y^A}\ff{\p}{\p Y^B}f\,.
\end{align}
\esubeqs
Despite the fact that there are additional, second order, terms in $[T_{AB},\bullet]$ as compared to \eqref{triv}, they disappear from $[T_{AB},T_{CD}]$ and we still have $sp(4)$ realized by $T_{AB}$.

By joining $Y_A$ and $Z_A$ into a single variable $X_{\mathbf{A}}$, $\mathbf{A}=1...8$ we can see that $\star$-product \eqref{staryz} delivers a particular realization of $sp(8)$ on bilinears $X_{\mathbf{A}}X_{\mathbf{B}}$ because of $[X_{\mathbf{A}},X_{\mathbf{B}}]=2iC_{\mathbf{A}\mathbf{B}}$ for certain $C_{\mathbf{A}\mathbf{B}}$. There is $sp(4)\oplus sp(4)$ subalgebra, which is embedded as $-\frac{i}2Y_AY_B$ and $\frac{i}2Z_A Z_B$ (the sign differs because of \eqref{YZBasicA}).

Klein operators can now be easily derived
from their definition \eqref{Kleindef}. We already know from
\eqref{Klein} that in case of $Z$-independence the Klein operator
is a delta-function of (anti)holomorphic oscillator. Similar
consideration for $Y$-independent functions leads to
delta-functions $\gk_z=2\pi\gd^2(z)$ and
$\bar{\gk}_{\bar{z}}=2\pi\gd^{2}(\bar z)$. As a result, the
complete Klein operators that satisfy \eqref{Kleindef} are given
by
\besubeqs
\begin{align}
\gk=\gk_y\star \gk_z=(2\pi)^2\gd^{2}(y)\star \gd^{2}(z)=e^{iz_{\ga}y^{\ga}}\,,\label{Klein1}\\
\bar\gk=\bar\gk_y\star \bar\gk_z=(2\pi)^2\gd^{2}(\bar y)\star \gd^{2}(\bar
z)=e^{i\bar{z}_{\gad}\bar{y}^{\gad}}\,.\label{Klein2}
\end{align}
\esubeqs
Note, that the Klein operators turned out to be regular functions
in the extended star-product algebra and therefore automorphism
\eqref{auto12} is inner. One can check the
following straightforward properties
\be\begin{array}{ccc}
\gk\star \gk=1\,,&\qquad\qquad & \gk\star f(y,z)=f(z,y)e^{iz_{\ga}y^{\ga}}\,,\\
\gk\star f(y,z)=f(-y, -z)\star \gk\,, &\qquad\qquad & f(y,z)\star \gk=f(-z,-y)e^{iz_{\ga}y^{\ga}}\,.
\end{array}\label{kl1}
\ee
analogously for antiholomorphic Klein $\bar\gk$. Note the
interchange of variables $y$ and $z$ within the arguments of
$f(y,z)$.

\paragraph{Back to HS equations.}
The equations that account properly for the twist
automorphism can be now obtained via substitution \eqref{subst}
and \eqref{Klein1}, \eqref{Klein2} into
\eqref{tobehs}
\besubeqs\label{hs}
\begin{align}
&\hat{d}\WW+\WW \star \WW=\Phi\star \gk+\bar\Phi\star \bar{\gk}\,,\label{hs1}\\
&\hat{d}\Phi+\WW\star \Phi-\Phi\star \pi(\WW)=0\,,&& \Phi=\ff i4dz_{\ga}\wedge dz^{\ga}B\,, \label{hs2}\\
&\hat{d}\bar{\Phi}+\WW\star \bar{\Phi}-\bar\Phi\star \bar{\pi}(\WW)=0\,,&& \bar{\Phi}=\ff
i4d\bar{z}_{\gad}\wedge d\bar{z}^{\gad}B\,.\label{hs3}
\end{align}
\esubeqs
The first equation tells us that curvature two-from \eqref{hs1} is
only allowed to be nonzero in auxiliary $dZ\wedge dZ$ sector being
pure gauge in space-time. This is a generic feature of the
unfolded equations that tend to get rid of space-time dependence
and reformulate the dynamics in the auxiliary ``twistor'' space.
Using definition \eqref{defA}, let us rewrite the above equations
in the component form.

Before doing this we make one comment. In obtaining
\eqref{hs} we carried out field redefinition $\Phi\to
\Phi\star \gk$, $\bar\Phi\to\bar\Phi\star \bar{\gk}$ to meet the
twisted-adjoint requirement. While it is fine to do so in
\eqref{tobehs1}, the substitution into \eqref{tobehs2} and
\eqref{tobehs3} seemingly produces terms of the form
$\p_{z}(B\star \gk)$ and $\p_{\bar z} (B\star \bar{\gk})$ which do not allow
one to drag the Kleinians through the $Z$--derivative because of
their $Z$--dependence. It does not happen though as these terms
simply do not show up. Indeed the three-forms appearing on l.h.s.
of \eqref{hs2} and \eqref{hs3} are identically zero for $dz\wedge
dz\wedge dz\equiv 0$ and ${d\bar z}\wedge{d\bar z}\wedge{d\bar
z}\equiv 0$. Another type of potentially dangerous terms $\p_{\bar
z}(B\star \gk)$ and $\p_{z} (B\star \bar{\gk})$ are harmless since $\p_{\bar
z}(\gk)=0$ and $\p_{z} (\bar{\gk})=0$.

The fact that there are no integrability conditions in $zzz$ and
$\bar z\bar{z}\bar{z}$ sectors for dimensional reason (spinorial
indices take two values) suggests that it might be possible to
impose some extra constraints consistent with equations
\eqref{hs}. It turns out this is what one actually
should do to describe irreducible nonlinear equations for HS
bosonic fields for $d=3$ \cite{Prokushkin:1998bq} and arbitrary
$d$ \cite{Vasiliev:2003ev} systems. In other words, systems
\eqref{tobehs} typically have some spurious
solutions due to lack of constraint in extra twistor space. Indeed
the way it is written in \eqref{tobehs} the
system just tells us in which sector the HS curvature is allowed
to be non-zero, \eqref{tobehs1}, without specifying the curvature
itself --- since the rest conditions \eqref{tobehs2} and
\eqref{tobehs3} are simply the integrability consequences. The
four dimensional case is to some extent peculiar for it was
already pointed out that fields $\Phi$ and $\bar{\Phi}$ are not
independent rather related to each other through the Klein
operators. This fact stipulates some restriction on the possible
form of higher-spin curvature that enters r.h.s. of
\eqref{tobehs1} and takes place only in the case of four
dimensional spinorial system where the two types of spinor fields
--- holomorphic and antiholomorphic are available. Eventually, the
$4d$ extra constraint will turn out to be equivalent to kinematic
condition for the system \eqref{hs} to be bosonic.
The required kinematic condition is not yet fully there, but it is
not going to be a problem to identify it. Impressive is the fact
that the desired $4d$ kinematic condition will provide us with
some nontrivial algebraic constraint which is typical for all
available nonlinear HS systems.

That the two-form $\Phi$ is expressed in terms of
zero-form $B$ makes $\Phi$ a composite field, which solves the
problem of the extra gauge symmetry, $\xi\fm{1}$, \eqref{locg} that
would be associated with $\Phi$ if it were a fundamental field.
But for now let us proceed with
component form of \eqref{hs} which reads
\begingroup\allowdisplaybreaks\besubeqs\label{eq}
\begin{align}
&dW+W\star W=0\,,\label{eq1}\\
&dB+W\star B-B\star \pi(W)=0\,, && dB+W\star B-B\star \bar{\pi}(W)=0\,,\label{eq2}\\
&\ff{\p \bar{\aA}_{\gad}}{\p z^{\ga}}-\ff{\p
\aA_{\ga}}{\p\bar{z}^{\gad}}+[\aA_{\ga}, \bar{\aA}_{\gad}]_\star =0\,,\label{eq3}\\
&d\aA_{\ga}+[W,\aA_{\ga}]_\star -\ff{\pl W}{\pl z^\ga}=0\,,
&&d\bar{\aA}_{\gad}+[W,\bar{\aA}_{\gad}]_\star -\ff{\pl W}{\pl\bar{z}^\gad}=0\,,\label{eq4}\\
&\ff{\p \aA^{\ga}}{\p
z^{\ga}}+\aA_{\ga}\star \aA^{\ga}=\ff{i}{2}B\star \gk\,,
&&\ff{\p\bar{A}^{\gad}}{\p \bar
z^{\gad}}+\bar{\aA}_{\gad}\star \bar{\aA}^{\gad}=\ff
i2B\star \bar{\gk}\,,\label{eq5}\\
&\ff{\p B}{\p
z^{\ga}}+\aA_{\ga}\star B-B\star \bar{\pi}(\aA_\ga)=0\,,
&&\ff{\p B}{\p
\bar{z}^{\gad}}+\bar{\aA}_{\gad}\star B-B\star \pi(\bar{\aA}_{\gad})=0\,,\label{eq6}
\end{align}
\esubeqs\endgroup
which results from \eqref{eq1}-\eqref{eq3} as coefficients of $dx\wedge dx$, \eqref{eq1};
$dx \wedge dz\wedge dz$, (\ref{eq2}.1),
$dx \wedge d\bar{z}\wedge d\bar{z}$, (\ref{eq2}.2); $dz\wedge d\bar{z}$, \eqref{eq3};
$dx\wedge dz$, (\ref{eq4}.1), $dx\wedge d\bar{z}$, (\ref{eq4}.2);
$dz\wedge dz$, (\ref{eq5}.1), $d\bar{z}\wedge d\bar{z}$, (\ref{eq5}.2); $dz\wedge d\bar{z}\wedge \bar{z}$, (\ref{eq6}.1),
$d\bar{z}\wedge d{z}\wedge {z}$, (\ref{eq6}.2). While system \eqref{nhs}
is manifestly integrable, it is not so obvious for the component form \eqref{eq}.

One may argue that the form of equations contradicts the
perturbative scheme laid out in \eqref{xpaceseq} as \eqref{eq1}
seemingly contains no higher-spin corrections that appear on the
r.h.s. of \eqref{xpaceseqA} already at free level, \eqref{xpaceseqEA}. Recall, however, that
master fields $W$ and $B$ having got an extra dependence on extra
$Z$-variable would provide the desired HS corrections for
dynamical fields $w(Y|x)$ and $C(Y|x)$ through star-product
\eqref{staryz}. Let us now enlist some obvious properties of the
obtained equations
\begin{itemize}
\item The equations are to be bosonic. Note, that there is
$\pi$-automorphism that enters \eqref{eq2}. There is however
similar equation arising from \eqref{hs3} with $\pi$ replaced by
$\bar{\pi}$. Therefore the system makes sense only for
$\pi(W)=\bar{\pi}(W)$, or equivalently
\begin{align}\label{Weven}
W(-Y, -Z|x)&=W(Y,Z|x)&&\Longleftrightarrow&&
\gk\star \bar{\gk}\star W=W\star \gk\star \bar{\gk}\,.
\end{align}
The parity property for master field $W(Y,Z|x)$ is a manifestation
of the bosonic nature of the system\footnote{It is possible
to include fermions into nonlinear system by introducing some
extra Klein operators \cite{Vasiliev:1999ba}. The resulting equations are
supersymmetric and contain two copies of each spin.}. It would imply $\omega(Y|x)=\omega(-Y|x)$ for the physical field.

Correspondingly, the gauge symmetry parameter $\gep$ must be an
even function too $\gep(Y, Z|x)=\gep(-Y, -Z|x)$. It has an immediate consequence
\begin{align}
\aA_{A}(Y,Z|x)&=-\aA_{A}(-Y,
-Z|x)&&\Longleftrightarrow&&
\gk\star \bar{\gk}\star \aA_{A}=-\aA_{A}\star \gk\star \bar{\gk}\,,\label{SoddA}
\end{align}
since $W$ and $\aA_A=(\aA_\ga, \bar{\aA}_\gad)$ are parts of the same
connection. Fields and gauge parameters take values in the same
space, now it is the space of even functions $\epsilon(Y,Z)$. That
the gauge transformation for $\aA$, $\delta \aA_A=\frac{\pl}{\pl
Z^A}\epsilon+...$, contains a derivative along $Z$ direction
changes the parity of $\aA_A$ as compared to $W$ for which the gauge
transformation $\delta W=d\epsilon+...$ does not affect the parity
in $Y,Z$ space.

The bosonic projection \eqref{SoddA} immediately implies from
\eqref{eq4} the corresponding projection for the
zero-forms\footnote{When no bosonic constraints imposed one has a
strange theory containing bosons described at the free level by
gauge fields and fermions described by gauge-invariant generalized
Weyl tensors. }
\begin{align}
B(Y, Z|x)&=B(-Y,
-Z|x)&&\Longleftrightarrow
&& \gk\star \bar{\gk}\star B=B\star \gk\star \bar{\gk}\,.\label{BevenA}
\end{align}
Let us stress that unlike \eqref{Weven}, condition \eqref{SoddA}
and its consequence \eqref{BevenA} do {\it not} follow from the
equations. It is this missing kinematic condition \eqref{SoddA}
that will be equivalent to some extra algebraic constraint
consistent with \eqref{eq}.

\item Gauge
symmetry. Component form of gauge transformations results from
\eqref{tobegauge} upon redefinition \eqref{subst}
\begin{align*}
&\gd W=d\gep+[W, \gep]\,,\\
&\gd \aA_{A}=\ff{\p \gep}{\p Z^{A}}+[\aA_A, \gep]\,,\\
&\gd B=B\star \pi(\gep)-\gep\star B
\end{align*}

\item Purely
gauge space-time dependence. From \eqref{eq1}, \eqref{eq2} and
\eqref{eq4} one can always determine space-time dependence of the
master fields in a pure gauge fashion
\begin{align*}
&W=g^{-1}\star dg\,,\\
&B=g^{-1}\star B_{0}(Y, Z)\star \pi(g)\,,\\
&\aA_{A}=g^{-1}\star \ff{\p g}{\p Z^A}+g^{-1}\star \aA_{A}^0(Y, Z)\star g\,,
\end{align*}
where $g=g(Y,Z|x)$ is an arbitrary even function. This is a
general statement that any covariant constancy/zero curvature
equations can be at least formally solved in the pure gauge form.
Note that the curvature for $\aA_\ga, \bar{\aA}_\gad$ is not entirely
zero, which explains the extra term in the last line.

These formulas suggest, for example, that one can gauge away
$W$--field that is supposed to encode HS gauge potentials. While
formally it looks as really the case, do not forget that such
``gauging'' when applied washes away $AdS$ space-time itself
making its frame field and Lorentz connection equal to zero, see discussion after \eqref{Wgauge}. It raises an important question
of admissible gauge transformations, which draw a line between
small gauge transformations, which are true gauge
transformations, and large gauge transformations that relate
physically distinguishable solutions. Space-time independent
functions $B_0$ and $\aA_0$ play a role of initial data imposed at a
given space-time point $x_0$ where $g(Y, Z|x_0)=1$.

\item It is now obvious that $F^\omega(\omega,C)$ and $F^C(\omega,C)$ in \eqref{xpaceseq} are more constrained than just by $d^2=0$, see discussion around \eqref{UnfldGenCA}-\eqref{UnfldGenCA1}. Indeed $dW+W\star W=0$ yields $dB+W\star B-B\star \pi(W)=0$ upon applying $B\star \gk\frac{\delta}{\delta W}$ to the former. This proves \eqref{UnfldGenCA}-\eqref{UnfldGenCA1} to all orders.

\end{itemize}
As it was mentioned already, the system we are dealing with
requires the bosonic kinematic constraint \eqref{SoddA} which is
not a consequence of the equations. Our goal is to rewrite
\eqref{SoddA} in some algebraic way and add it to the system so as
to make the bosonic nature of the equations intrinsic and
manifest. To proceed in this direction let us first perform some
harmless field redefinition with $A$-field
\be\label{Ashift}
\aA_{\ga}=\ff{i}{2}(S_{\ga}-z_{\ga})\,,\qquad \bar{\aA}_{\gad}=\ff
i2(\bar{S}_{\gad}-\bar{z}_{\gad})\,,
\ee
where $S$ is some new field. The shift of vacuum value of
$A$-field is designed to eliminate partial derivatives with
respect to $z$-variable in \eqref{eq3}-\eqref{eq6}, the
coefficient $i/2$ is chosen appropriately to account for interplay
between $[z, f]_\star =-2i\p_z f$ and $\p_z f$ terms. Shift \eqref{Ashift} yields
\besubeqs\label{osc123}
\begin{align}
&[S_{\ga}, \bar{S}_{\gad}]_\star =0\,,\label{osc1}\\
&[S_{\ga},
S_{\gb}]_{\star }=-2i\gep_{\ga\gb}(1+B\star \gk)\,,
&&[\bar{S}_{\gad},
\bar{S}_{\gbd}]_\star =-2i\gep_{\gad\gbd}(1+B\star \bar{\gk})\,,\label{osc2}\\
&[S_{\ga}, B\star \bar{\gk}]_\star =0\,,
&&[\bar{S}_{\gad}, B\star \gk]_\star =0\label{osc3}\,,
\end{align}
\esubeqs
Note, that \eqref{osc3} is a consequence of \eqref{osc1} and
\eqref{osc2}. Clearly, should one had $B\star \gk=B\star \bar{\gk}=0$, the above
commutation relations would simply correspond to two copies of
Weyl algebra generated by $S_{\ga}$ and $\bar{S}_{\gad}$. If
$B$ is not equal to zero we see, that $B\star \gk$ is a central element
for $\bar{S}_{\gad}$ and $B\star \bar{\gk}$ --- for $S_{\ga}$. From \eqref{osc3} and
\eqref{SoddA} it immediately follows
\be\label{right}
\{S_{\ga}, B\star \gk\}_\star =0\,,\qquad \{\bar{S}_{\gad}, B\star \bar{\gk}\}_\star =0\,.
\ee
Algebraic condition \eqref{right} eventually brings us to
explicitly bosonic and complete nonlinear HS equations.
That \eqref{right} respects the Jacobi identities deserves special
attention.

\paragraph{Deformed oscillators.} Set of equations \eqref{osc123} supplemented
with \eqref{right} represents two copies of what is known as
deformed oscillators. Consider one copy that can be defined as
follows. Let the generating elements $\hat{y}_{\ga}$ and $K$
satisfy the relations
\be\label{deform}
[\hat y_{\ga}, \hat{y}_{\gb}]=-2i\gep_{\ga\gb}(1+K)\,,\qquad\qquad
\{\hat y_{\ga}, K\}=0\,.
\ee
The deformed oscillators \eqref{deform} were originally discovered
by Wigner in \cite{WignerDef} and happened to be related to Calogero model \cite{Brink:1993sz}. It
is interesting that if one looks at the Jacobi $[[\hat{y}_{\ga},
\hat{y}_{\gb}], \hat{y}_{\gc}]+cycle=0$ it is not going to hold in
general unless indices of $\hat y_{\ga}$ take two values so that
antisymmetrization of any three gives identically zero. This is
the case for two copies of deformed oscillators generated by HS
master fields \eqref{osc123} and \eqref{right}. Another
very important property is that the deformation of Heisenberg
algebra \eqref{deform} respects $sp(2)$ symmetry. Indeed it is
easy to see that generators $T_{\ga\gb}=\ff i4\{\hat{y}_{\ga},
\hat{y}_{\gb}\}$ form $sp(2)$ algebra
\be\label{sp2}
[T_{\ga\gb},
T_{\gc\gd}]=\gep_{\ga\gc}T_{\gb\gd}+(\ga\leftrightarrow\gb)+
(\gc\leftrightarrow\gd)
\ee
which yet rotates $\hat{y}_{\ga}$ as a vector
\be\label{sp2vec}
[T_{\ga\gb},
\hat{y}_{\gc}]=\gep_{\ga\gc}\hat{y}_{\gb}+\gep_{\gb\gc}\hat{y}_{\ga}\,.
\ee
Deformed oscillator properties \eqref{sp2} and \eqref{sp2vec} will
be of utter importance in identifying local Lorentz symmetry for
HS system which guarantees tensor interpretation of the dynamical
fields.

It is worth mentioning that the associative algebra generated by $\hat{y}_{\ga}$ and $K$ is
the universal enveloping algebra of $osp(1|2)$, \cite{Bergshoeff:1991dz},
\begin{align}\label{Uosp}
U(\hat{y}_{\ga},K)=U(osp(1|2))
\end{align}
where $\hat{y}_{\ga}$ are to be identified with the odd generators. Then $T_{\ga\gb}=\ff i4\{\hat{y}_{\ga}, \hat{y}_{\gb}\}$ is the expected relation, which in fact is a definition. Together with \eqref{sp2} and \eqref{sp2vec} we have all defining relations of $osp(1|2)$. The only
Casimir operator of $osp(1|2)$ is related to $K$
\begin{align}
C_2=-\frac12 T_{\ga\gb} T^{\ga\gb}+\frac{i}4 \hat{y}_\ga \hat{y}^\gb = \frac14 (K^2-1)
\end{align}
Since $K$ is an independent generator and $K^2$ is not a $c$-number the Casimir is free. The Casimir operator of the $sp(2)$ subalgebra is
\begin{align}
C_2=-\frac12 T_{\ga\gb} T^{\ga\gb} = \frac14 (K+1)(K-3)
\end{align}

Taking into account relation \eqref {Uosp} between deformed oscillators and $U(osp(1|2))$ and comparing them with \eqref{osc123} and \eqref{right} we conclude that the subset of Vasiliev equations is equivalent to defining relations for two copies of $osp(1|2)$ with $B\star \gk$ and $B\star \bar{\gk}$ playing the role of $K$, while $S_\ga$ and $S_\gad$ play the role of $\hat{y}_\ga$. This
has lead to interpretation of higher-spin theory as a fuzzy sphere, \cite{Vasiliev:1989re, Vasiliev:1989qh}.

\paragraph{Vasiliev equations.} Remarkably, it is the deformed oscillator constraint \eqref{right}
that one can additionally impose to \eqref{hs} leads
to the nonlinear system for bosonic massless fields which we can
finally write down in the form
\besubeqs\label{VasAll}
\begin{empheq}[box=\fbox]{align}
\,\,\rule{0pt}{14pt}&dW+W\star W=0\,,\label{1}\\
&dB+W\star B-B\star \pi(W)=0\,,\label{2}\\
&dS_{\ga}+[W,S_{\ga}]_\star =0\,,
&&d\bar{S}_{\gad}+[W, \bar{S}_{\gad}]_\star =0\,,\label{3}\\
&[S_{\ga},
S_{\gb}]_\star =-2i\gep_{\ga\gb}(1+B\star \gk)\,,
&&[\bar{S}_{\gad},
\bar{S}_{\gbd}]_\star =-2i\gep_{\gad\gbd}(1+B\star \bar{\gk})\,,\label{4}\\
&\{S_{\ga}, B\star \gk\}_{\star }=0\,,
&&\{\bar{S}_{\gad}, B\star \bar{\gk}\}_{\star }=0\,,\label{5}\\
&[S_{\ga}, \bar{S}_{\gad}]_\star =0\,.\label{6}
\end{empheq}
\esubeqs
The system of equations \eqref{VasAll} is known as Vasiliev
nonlinear equations for HS bosonic fields in four dimensions. It
has a form of \eqref{eq} upon field redefinition
\eqref{Ashift} with an extra constraint \eqref{5} that makes this
system explicitly bosonic. Written this way it contains some
flatness conditions in space-time \eqref{1}-\eqref{3} and a set of
algebraic constraints \eqref{4}-\eqref{6} which are nothing but a
direct sum of two deformed oscillators given by $S_{\ga}$ and
$\bar{S}_{\gad}$. The system correspondingly inherits local gauge
invariance
\besubeqs\label{gaugeTransAll}
\begin{align}
&\gd W = d\xi+[W,\xi]_\star \,,\\
&\gd B=B\star \pi(\xi)-\xi\star B&&\Longleftrightarrow&&
\gd(B\star \gk)=[B\star \gk, \xi]_\star \,,\label{twistedadjB}\\
&\gd S_{\ga}=[S_{\ga}, \xi]_\star \,,\\
&\gd\bar{S}_{\gad}=[\bar{S}_{\gad}, \xi]_\star \,.
\end{align}
\esubeqs
As it was already mentioned, introducing the deformed oscillator
anticommutator condition \eqref{5} one imposes extra kinematic
constraints \eqref{SoddA}, \eqref{BevenA} and vice versa. Eq.
\eqref{SoddA} can be easily obtained using e.g., $[S_{\ga},
B\star \bar\gk]_\star =0$ and $\{S_{\ga}, B\star \gk\}_\star =0$. Once it is proved,
eq. \eqref{BevenA} follows immediately from the fact that $S\star S$ is
an even function. Together \eqref{Weven} and \eqref{SoddA},
\eqref{BevenA} consistently imply that the system under
consideration is bosonic.

An important difference between the free and interacting equations
is the doubling of oscillators $Y\to (Y, Z)$ and appearance of
extra $S$-field. A question is whether this procedure preserves
physical degrees of freedom which as we have seen are encoded in a
single function $C(Y|x)$ and whether the linearized approximation
results in \eqref{UnfldsSpinorial}. We will see, that while
from \eqref{4} one expresses $B$ in terms of $S\star S$, perturbatively
it is the $S$-field that appears to be totally auxiliary and is
expressed on-shell in terms of $B$-field modulo gauge ambiguity.
As for an extra $Z$-dependence, it turns out to be fixed again up
to a gauge ambiguity by the extra condition \eqref{5}.

\paragraph{Reality conditions.}
One thing that was missed so far is the reality conditions for
master fields. These are dictated by the star-product properties
and free level analysis. Without going into details we give the
final result
\begin{align*}
&y_{\ga}^{\dagger}=\bar{y}_{\gad}\,,\qquad
z_{\ga}^{\dagger}=-\bar{z}_{\gad}\,,\\
&W^{\dagger}=-W\,,\\
&S_{\ga}^{\dagger}=-\bar{S}_{\gad}\,,\\
&B^{\dagger}=\pi(B)\,.
\end{align*}
Once the hermitian properties of $Y_A$ and $Z_A$ are fixed, the last three conditions simply tell
us what are the properties of the Taylor coefficients. For example, at the free level we find that the vielbein component $e^{\ga\gad}$ is hermitian and spin-connections $\omega^{\ga\gb}$, $\omega^{\gad\gbd}$ are conjugate to each other.

\paragraph{Generalizations and reductions.}

We can ask ourselves to what extent the form of the equations
\eqref{VasAll} is unique and what possible generalizations
or ambiguities in higher-spin interactions are. One assumption
that was used in writing down system \eqref{hs} is that
there are no mixing terms $dz_{\ga}\wedge d\bar{z}_{\gad}$ in the
auxiliary space curvature sector. These terms when present violate
local Lorentz symmetry since to convert spinor indices one has to
introduce some field $H_{\ga\gad}(B)$ that breaks down the
symmetry manifestly. So, this possibility is forbidden by the
equivalence principle. Another possible modification which does
not ruin formal integrability of the Vasiliev equations is to
change $B\star\gk$ and $B\star\bar{\gk}$ on the r.h.s. of
\eqref{4} to any odd $\star$-product functions $f_{\star}(B\star\gk)$ and
$\bar{f}_{\star}(B\star\bar{\gk})$, correspondingly. While it is
possible to make such a modification one can partially eliminate
its effect by a field redefinition $B\to F_\star(B)$ which leaves
nontrivial only the phase in complex function $f$. In other words
the ambiguity that cannot be eliminated by field redefinitions is
given by an arbitrary real function $\phi(B)$ that enter \eqref{4} as
$\exp_{\star}(i\phi_\star(B\star\gk))\star B\star\gk$ in
holomorphic part and $\exp_{\star}(-i\phi_\star(B\star\gkb))\star
B\star\gkb$ in the antiholomorphic. The cases of $\phi=0, \pi/2$
corresponds to $A$ and $B$ models correspondingly. Let us note that
there are reasonable doubts concerning redefinitions $B\to F_\star(B)$ where
$F_\star(B)$ is a $\star$-product Taylor series since $\star$-products
of master-fields are nonlocal in the space-time. That the correlation functions
can be represented in the form of $\star$-products, \cite{Didenko:2012tv}, is
a serious argument that forbids such redefinitions.

Finally, there is a way to introduce fermions in the system by
doubling the set of fields. Another option is to further extend the spectrum by adding some nondynamical moduli that play the role of different parameters of interaction, see
\cite{Vasiliev:1999ba}. It is also possible to truncate the bosonic system to the fields with even spins only,
$s=0,2,4,6,...$.

\mysubsection{Perturbation theory}

In perturbation theory one starts with an appropriate exact vacuum
solution. For a field theory, a classical vacuum is some
background having no fields propagating on it. In the case of HS gauge
theory, the proper vacuum is the $AdS$ space-time as we already
know, therefore the vacuum $W_0$ one-form should be taken from
\eqref{con} such that \eqref{flat} is satisfied and as long as no
dynamical fields are around we take $B=0$ and $A_A=0$. One has
then the following exact solution of \eqref{VasAll}
\besubeqs\label{VacuumSol}
\begin{align}
&W_0=\Omega=-\ff
i4(\go_{\ga\gb}y^{\ga}y^{\gb}+\bar{\go}_{\gad\gbd}
\bar{y}^{\gad}\bar{y}^{\gbd}+2
h_{\ga\gad}y^{\ga}\bar{y}^{\gad})=\frac12 T^{AB}\Omega_{AB}\,,\label{w0}\\
&B_0=0\,,\label{B0}\\
&S_{0\ga}=z_{\ga}\,,\qquad\qquad \bar S_{0\gad}=\bar
z_{\gad}\,,\label{S0}
\end{align}
\esubeqs
where $\Omega$ is a good old $AdS_4$ flat connection,
\eqref{AllFlat}. The vacuum for $S_A$ is
designed to undo the shift \eqref{Ashift} and to get back to
\eqref{eq}, for which it is obvious that any flat
$\Omega$, $B=0$ and $A_\ga=\bar{A}_\gad=0$ is an exact solution.

Having fixed the vacuum we can look at perturbative expansion
\begin{align*}
W&=W_0+W_1+\dots\,\,=\Omega+W_1+\dots\,,\\
B&=B_0+B_1+\dots\,\,\,\,=0+B_1+\dots\,,\\
S_{A}&=S_{0A}+S_{1A}+\dots=Z_A+S_{1A}+\dots\,.
\end{align*}
A general scheme for looking at perturbative series for Vasiliev
equations is to first solve for the algebraic constraints
\eqref{4}-\eqref{6} and then substitute the solution into
space-time part of equations \eqref{VasAll}. At first order
from \eqref{3} we have
\beno
z_{\ga}\star B_1\star \gk+B_1\star \gk\star z_{\ga}=0\,,\qquad \bar
z_{\gad}\star B_1\star \bar{\gk}+B_1\star \bar\gk\star \bar z_{\gad}=0\,,
\eeno
that using \eqref{Kleindef} gives
\beno
[z_{\ga}, B_1]_\star =[\bar z_{\gad},
B_1]_\star =0\qquad\Longleftrightarrow\qquad \ff{\p B_1}{\p Z^A}=0\,.
\eeno
Its generic solution is
\beno
B_1=C(y, \bar y |x)\,,
\eeno
where $C$ is an arbitrary $z$-independent function. We see that at
first order $B$-field appeared to be $z$-independent by \eqref{5}
--- being a manifestation of generic property stating that
$z$-dependence of $B$-field is always perturbatively fixed by
\eqref{5}. Now the space-time evolution of $C(Y|x)$ is governed by
\eqref{2} with $W=\Omega$ resulting in twisted-adjoint flatness
condition \eqref{UnfldsSpinorialC}. This way we find that at free level master
field $B$ indeed describes HS curvatures via \eqref{UnfldsSpinorialC}, i.e. $\DOt C=0$. The next
step is to reproduce the on-mass-shell theorem \eqref{UnfldsSpinorialA} from
the linearized equations. In order to do so, we first evaluate
$S_{1A}$ from \eqref{4} which in our case gives
\beno
[S_{0\ga}, S_{1}^{\ga}]_\star =-2iC\star \gk
\eeno
and similarly for $\bar{S}_{\gad}$. Substituting $S_{0\ga}$ from
\eqref{S0} and using \eqref{kl1}, we get
\be\label{S1eq}
\ff{\p S_{1}^{\ga}}{\p z^{\ga}}=C\star \gk=C(-z, \bar
y)e^{iz_{\gb}y^{\gb}}\,.
\ee
Before we proceed let us note that there are two types of
equations that steadily appear in perturbation analysis, see Appendix \ref{sec:HomotopyIntegrals}. These are
\besubeqs\label{homeq}
\begin{align}
&\ff{\p f^{\ga}}{\p z^{\ga}}=g(z)\,,\label{homeq1}\\
&\ff{\p f}{\p z^{\ga}}=g_\ga\,.\label{homeq2}
\end{align}
\esubeqs
There is a consistency constraint for the left hand side of
\eqref{homeq2} which requires $\ff{\p g^{\ga}}{\p z^{\ga}}=0$.
Generic solutions to \eqref{homeq1} and \eqref{homeq2} can be
written down in the form of homotopy integrals
\besubeqs
\begin{align}
&f_{\ga}=\ff{\p\eta}{\p z^{\ga}}+z_{\ga}\int_{0}^{1}dt\, t\, g(tz)\,,\label{homsol1}\\
&f=c+z^{\ga}\int_{0}^{1} dt\, g_{\ga}(tz)\,,\label{homsol2}
\end{align}
\esubeqs
where $c$ is $z$-independent and $\eta$ is an arbitrary function, which gives
\be\label{S1sol}\begin{cases}
&S_{1\ga}=-2i\frac{\p}{\p z^{\ga}}\xi_1+z_{\ga}\int_{0}^{1}dt\, t\, C(-tz, \bar
y)e^{itz_{\gb}y^{\gb}}\,,\\
&\bar{S}_{1\gad}=-2i\frac{\p}{\p z^{\gad}}{\xi}_{1}+\bar{z}_{\gad}\int_{0}^{1}dt\,
t\, C(y, -t\bar z)e^{it\bar z_{\gbd}\bar y^{\gbd}}\,,\end{cases}
\ee
where $\xi_1(Y, Z|x)$ is arbitrary function that plays a role of
gauge ambiguity and can be set to zero for convenience. We prefer,
however, to keep it nonzero so as to make sure that it will not
affect on-mass-shell theorem \eqref{UnfldsSpinorialA}. One can see
at this level, that $S$ indeed is auxiliary being a functional of
$B$. The next step is to substitute \eqref{S1sol} into \eqref{3}
that gives the following equation
\be\label{W1eq}
\DO S_{1A}+[W_1, S_{0A}]=0\,,
\ee
where $\DO =d+[\Omega, \bullet]_\star $, which is nilpotent $\DO ^2=0$ on
account of \eqref{w0} and its explicit action can be easily found
from \eqref{YZcommutators} to be
\be\label{D0}
\DO  f=df+\Omega^{AB}\Big(Y_A-i\ff{\p}{\p Z^{A}}\Big)\ff{\p}{\p
Y^{B}}f\,.
\ee
Substituting \eqref{S0} into \eqref{W1eq} we have
\begin{align}
\frac{\pl}{\p Z^A}W_1&=\ff i2 \DO  S_{1A}  && \Longleftrightarrow&&
\begin{cases} \p_{\ga}W_1=\ff i2 \DO  S_{1\ga}\,,\\ \p_{\gad}W_1=\ff i2
\DO \bar S_{1\gad}\,.\end{cases}\label{diffWSeq}
\end{align}
These are the equations of  type \eqref{homeq2} and we can solve
them as
\be\label{W1sol}
W_{1}=\DO \xi_1+\ff i2z^\ga\,\int_{0}^{1}dt\,\left.\DO\hat
S_{1\ga}\right|_{z\to tz}+c.c.+\omega(y, \bar y|x)\,,
\ee
where $\omega(y, \bar y|x)$ plays a role of a constant with respect to
$z$ in \eqref{homsol2}, while $\hat S_1$ means that we explicitly
extracted pure gauge dependence from \eqref{S1sol} which is given
by the first term in \eqref{W1sol}. Indeed, $\DO\frac{\p}{\p Z^A}\xi_1= \frac{\p}{\p Z^A}\DO\xi_1$ and the $\xi_1$-part in \eqref{diffWSeq} can be solved for without homotopy integrals.  The abbreviation $c.c.$ means that all undotted
quantities need to be replaced with dotted ones and it does not conjugate complex numbers.

Note that $\hat S_{1\ga}$ is proportional
to $z_\ga$ and, since $z^\ga z_\ga\equiv0$, the only terms that survive are
those for which $\DO$ hits $z_\ga$ inside $\hat S_{1\ga}$. There are several such terms,
\be \label{secondder}\DO\ni  -i\omega^{\ga\gb}\frac{\pl^2}{\pl y^\ga\pl z^\gb}-ih^{\ga\gad}\frac{\pl^2}{\pl \bar{y}^\gad\pl z^\ga}-i\bar{\omega}^{\gad\gbd}\frac{\pl^2}{\pl \bar{y}^\gad\pl \bar{z}^\gbd}
-ih^{\ga\gad}\frac{\pl^2}{\pl \bar{z}^\gad\pl y^\ga}\,,\ee
The first two contribute to $z^\ga \DO \hat S_{1\ga}$ and the last two to its conjugate. Simple computation, which applies \eqref{RepeatedHomotopy} to reduce repeated homotopy integrals to a single one, yields
\begin{align} \begin{split}W_1&= \DO \xi_1+\omega(y, \bar y|x)+\\
&-\ff 12\,\int_{0}^{1}(1-t)\, dt\, z_\ga \left(\omega^{\ga\gb}z_\gb t+
h^{\ga\gad}\frac{\pl}{\pl \bar{y}^{\gad}}\right)C(-zt,\bar{y})e^{itz_{\gb}y^{\gb}}+c.c.\end{split}\label{Wfirst}\end{align}
In accordance with our expectation that physical fields are the initial conditions at $Z=0$ we find that
$W_1(Y,0)$ gives gauge connections $\omega(Y)$ up to a gauge transformation
\begin{align}\left.W_1\rule{0pt}{12pt}\right|_{Z=0}&=\left. \rule{0pt}{12pt}\DO \xi_1\right|_{Z=0}+\omega(y, \bar y|x)\,.\end{align}

Final step is to substitute
\eqref{W1sol}, or more explicitly \eqref{Wfirst}, into \eqref{1} to determine space-time evolution of
$\omega(y, \bar y|x)$. At the first order this substitution gives for $\DO W_1=0$
\be\label{wyeq}
\DO  \omega(y, \bar y|x)=-\DO \Big(\ff i2z^\ga\int_{0}^{1}dt\, \left.\DO\hat
S_{1\ga}\right|_{z\to tz}+c.c.\Big)\,.
\ee
Note, that gauge $\xi$-terms do not contribute to \eqref{wyeq} as
they enter the right hand side via $(\DO)^2\xi\equiv 0$. So, indeed,
we see that arbitrary function that arises in $S$ field as an
integration constant corresponds to a gauge freedom. Now, the left
hand side of \eqref{wyeq} is explicitly $z$-independent and so
should be its right hand side. In other words, the r.h.s.
$z$-dependence is fictitious as becomes obvious after integration by
parts. Equivalently,
\be \label{trivialZdep}\frac{\pl}{\pl z^\ga} \DO W_1=
\DO\frac{\pl}{\pl z^\ga}  W_1=\DO \left(\frac{i}2 \DO S_{1\ga}\right)=\frac{i}{2} \DO\DO S_{1\ga}=0\,,\ee
i.e. $\DO W_1$ is $z$-independent. It makes it convenient to calculate \eqref{wyeq} at $z=0$
and be sure that the result is correct. For that reason we only
need 2nd derivative terms in \eqref{D0}, i.e. \eqref{secondder},  because otherwise there
will be the terms on the r.h.s. of \eqref{wyeq} proportional to
$z$ or $zz$ prior to homotopy integration that do not contribute
at $z=0$ anyway. Up to irrelevant after substitution
in \eqref{wyeq} terms we find
\be\label{W1re}
W_{1}=\omega(Y)-\ff{1}{2}h^{\ga\gad}z_{\ga}\ff{\p}{\p\bar
y^{\gad}}\int_0^1 dt\, (1-t)\, C(-tz, \bar
y)h^{itz_{\gb}y^{\gb}}+c.c.+O(z^2)\,.
\ee
Now we need to apply $\DO $ to \eqref{W1re} and set $z=0$. Again,
the nontrivial part is reached only for the second derivative term
in \eqref{D0}
\begin{align*}
\eqref{wyeq}=\DO  \omega(y, \bar y|x)=&-h^{\gb\gbd}\wedge
h^{\ga\gad}\ff{i}{2}\ff{\p}{\p z^{\gb}}z_{\ga}
\ff{\p}{\p\bar{y}^{\gbd}\p
\bar{y}^{\gad}}\int_0^1dt\, (1-t)\, C(0,
\bar{y})+c.c.=\\=&-\ff{i}{4}h_{\ga}{}^{\gad}\wedge
h^{\ga\gbd}\ff{\p^2C(0,\bar
y)}{\p\bar{y}^{\gad}\p\bar{y}^{\gbd}}+c.c.\,.\end{align*}
So one arrives at the central on-mass-shell theorem \eqref{UnfldsSpinorialA}
\be\label{coms}
\DO \omega(y, \bar{y}|x)=-\ff{i}{4}h_{\ga}{}^{\gad}\wedge
h^{\ga\gbd}\ff{\p^2}{\p\bar{y}^{\gad}\p\bar{y}^{\gbd}} C(0,\bar
y)-\ff{i}{4}h^{\ga}{}_{\gad}\wedge
h^{\gb\gad}\ff{\p^2}{\p{y}^{\ga}\p{y}^{\gb}} C(y,0)\,.
\ee
This completes the free level analysis. It has shown that at
linearized approximation the equations do describe bosonic
Fronsdal fields along with spin-zero free scalar being a part of
HS multiplet in accordance with Sections \ref{sec:AdSLinear}, \ref{sec:HSAlgebra}. Moreover, from
perturbation theory it is clear that all degrees of freedom are
encoded in a single function $C(Y|x)$. Those are as many as for the free
fields governed by the Weyl module. It guarantees that at
nonlinear level one has perturbatively the same amount of degrees
of freedom, yet the unfolded form of equations \eqref{VasAll}
prevents any field redefinitions that could possibly reduce
nonlinear equations to the linear ones.

The important question is
whether the components of master fields $W$, $B$ and $S$ can be
eventually treated as space-time tensors in accordance with the
equivalence principle or not. Recall that within the unfolded
approach the dictionary between fiber fields and world tensors is
achieved by the local Lorentz symmetry. This raises the question
whether the Lorentz symmetry acts on equations
\eqref{VasAll}.

At free level the equation that we got
\eqref{coms}, contained background Lorentz connections
$\omega_{\alpha\beta}$ and $\bar\omega_{\dot{\alpha}\dot{\beta}}$
via the Lorentz covariant derivative $\DL$ as a part of the $AdS_4$ covariant derivative $\DO $ and therefore in a
covariant fashion. That makes it possible to convert components of
$\omega(y,\bar y|x)$ into space time tensors with the aid of frame
field. The fact that equation \eqref{coms} would be Lorentz
covariant was not at all obvious from the point of view of
initial equations \eqref{VasAll} in the first place. Indeed,
when fields depend on both of star-product variables $Y$ and $Z$, covariant derivative $\DO$,
\eqref{D0}, no longer acts covariantly. As an example, take
$f=f_{\ga\gb}(x)\,y^{\ga}z^{\gb}$, then $\DO
f=\DL f-i\omega^{\ga\gb}f_{\ga\gb}$ which contains Lorentz connection in a
noncovariant way. The same problem would be with \eqref{Wfirst} unless noncovariant
terms disappeared from \eqref{coms}.

The fact that at free level noncovariant terms
for one-forms in $W$ dropped away turned out to be totally
coincidental and is not going to take place any longer already at
second order. This brings us to the problem of local Lorentz
symmetry and laborious search for the field redefinition that
would respect it. Luckily, the Lorentz symmetry happens to be
intrinsically resided in the Vasiliev system and this fact is
crucially related to the property of deformed oscillators
\eqref{sp2}. Moreover, the very existence of Lorentz symmetry to
much extent fixes the equations of motion at the end of the day
and could have been a guideline for their derivation. This important property
we shall discuss next.

\mysubsection{Manifest Lorentz symmetry}\label{subsec:Lorentz}
\paragraph{Generalities.}
Let us recall the notion of the Lorentz connection. One-form
$\varpi^{\ga\gb}$ is said to be a spin-connection if it transforms
under the local Lorentz transformations with parameter
$\Lambda^{\ga\gb}$ as
\begin{align*}
\delta \varpi^{\ga\gb}&=d \Lambda^{\ga\gb}+\varpi\fud{\ga}{\gc}
\Lambda^{\gc\gb}+\varpi\fud{\gb}{\gc} \Lambda^{\ga\gc}
\end{align*}
Having some generators $t_{\ga\gb}$ of $sp(2)$ we can introduce
$\varpi=\frac12\varpi^{\ga\gb}t_{\ga\gb}$,
$\Lambda=\frac12\Lambda^{\ga\gb}t_{\ga\gb}$ and rewrite it as
\begin{align*}
\delta \varpi = d\Lambda +[\varpi,\Lambda]
\end{align*}
An object $w^{\ga(k)}$ is called a rank-$k$ spin-tensor if it
transforms as
\begin{align}\label{goodtensor}
\delta w^{\ga(k)}&= \Lambda\fud{\ga}{\gc}  w^{\gc\ga(k-1)}\,.
\end{align}
The derivative of $w^{\ga(k)}$ must be always accompanied by a
term with the spin-connection to give Lorentz-covariant derivative
\begin{align}\label{goodtensord}
\DL w^{\ga(k)}=dw^{\ga(k)}+\varpi\fud{\ga}{\gc}  w^{\gc\ga(k-1)}\,.
\end{align}
We are being somewhat clingy in defining Lorentz connection and
local spin-tensors on purpose as we will face a problem that not
every object with indices can be called a spin-tensor. Sometimes
it can transform in a wrong way under local Lorentz
transformations. For example, instead of \eqref{goodtensord} one
may have something like, c.f. \eqref{Wfirst},
\begin{align}\label{badtensor}
&\varpi_{\gb\gc}  w^{\gb\gc\ga(k-2)} && \mbox{or} &&
\varpi^{\ga\ga}  w^{\ga(k)}
\end{align}
as a contribution to one of the nonlinear equations. These terms
can in principle appear and they do appear. Having such terms
would lead to problems in interpretation as they violate the
equivalence principle discussed at the end of the previous
section. The crucial statement about the Vasiliev equations is
that one can find a field redefinition that removes
\eqref{badtensor}-like terms and the spin-connection can then be
found to appear in the form of Lorentz-covariant derivative only.

Let us find some explanation for why the problem of Lorentz symmetry not being manifest can arise.
Suppose we found some consistent unfolded system of the form \eqref{xpaceseq} that has manifest Lorentz symmetry. Then we can split
connection $\omega$ into spin-connection $\varpi$ and the rest $\omega^t$, which transforms as a
bunch of tensors. With $D=d+\varpi$ being the Lorentz covariant derivative, the system necessarily has the following form
\besubeqs
\label{xpaceseqLor}
\begin{align}
\DL\omega^t&=F^{\omega^t}(\omega^t,\omega^t,C)\label{xpaceseqLorA}\,,\\
d\varpi+\varpi\varpi=\DL^2&=F^{\varpi}(\omega^t,\omega^t,C)\label{xpaceseqLorB}\,,\\
\DL C&=F^C(\omega^t, C)\,,\label{xpaceseqLorC}
\end{align}
\esubeqs
where we split $F^\omega$ of \eqref{xpaceseqA} into $F^{\omega^t}$ and $F^{\varpi}$. We also emphasized that $F^\omega(\omega,\omega,C)$ is bilinear in $\omega$ and $F^C(\omega,C)$
is only linear in $\omega$. As usual $D^2$ is a Riemann two-form, which by virtue of equations of motion should be equal to the Weyl tensor and have certain nontrivial contribution from stress-tensors of higher-spin fields and scalar. The restrictions on $F^\omega(\omega,\omega,C)$ and $F^C(\omega,C)$ that imply that the theory is manifestly Lorentz-covariant are
\begin{align*}
F^{\omega}(\varpi,\omega^t,C)&=-[\varpi,\omega^t]\,,\\
F^{\omega}(\varpi,\varpi,C)&=-\varpi\varpi\,,\\
F^{C}(\varpi,C)&=-[\varpi,C]\,,\end{align*}
and there are no $C$-corrections to the expressions above. Having such a consistent system we can perform a harmless field-redefinition, which is a freedom allowed in nonlinear theories,
\beno \omega\rightarrow \omega+\mathcal{L}(\omega,C)\eeno
Function $\mathcal{L}(\omega,C)$ may involve spin-connection in any possibly non-covariant form. The result of such redefinition will be the equations that may not have manifest Lorentz symmetry. This is what happens to Vasiliev equations, where, apart from solving them perturbatively, one has to undo the field redefinition in order to make equations Lorentz-covariant. Problems like this perhaps can never appear in theories with a finite number of fields, where such redefinitions are easy to control.

We would like to stress that the problem is technical and not a conceptual one --- it just happens that the Vasiliev equations are not given in the Lorentz frame, where the Lorentz-covariance is manifest, but in some other frame, where the interactions have a simpler form. There is no general proof but it seems likely that the Lorentz symmetry cannot be broken at all within the unfolded approach unless it is done explicitly by means of some external tensorial objects (at least this is obvious at the linearized level where all noncovariant terms can be absorbed by field redefinition). This is the case with the $so(3,2)\sim sp(4)$ symmetry that has been explicitly broken by having to split $sp(4)$ indices $A,B,...$ into $(\ga,\gad)$. Such splitting requires an additional $so(3,2)$-vector, compensator, represented by an $sp(4)$ anti-symmetric tensor $V_{AB}=diag(\epsilon_{\ga\gb},-\epsilon_{\gad\gbd})$, which can be combined with the $sp(4)$ metric tensor $C_{AB}=diag(\epsilon_{\ga\gb},\epsilon_{\gad\gbd})$ in order to project onto (un)dotted indices.

\paragraph{Implementation.}
To proceed, let us search for the Lorentz generators that rotate
master fields properly. At free level we know that these are given
by
\begin{align}
\label{freeLor}
L^y_{\alpha\beta}=-\frac{i}{2}y_{\alpha}y_{\beta}\,,\qquad
\bar{L}^y_{\dot{\alpha}\dot{\beta}}=-\frac{i}{2}\bar{y}_{\dot{\alpha}}\bar{y}_{\dot{\beta}}\,.
\end{align}
Their action on the free master fields is indeed that of Lorentz
transformation
\begin{align*}
\Lambda=\frac12 \Lambda^{\ga\gb}L^y_{\ga\gb}+c.c
\end{align*}
\begin{align*}
\gd_{\Lambda} C(y, \bar y|x)=[\Lambda,
C]_{\star}=\left(\Lambda^{\ga\gb}y_{\ga}\frac{\partial}{\partial
y^{\gb}}+
\bar{\Lambda}^{\dot{\ga}\dot{\gb}}\bar{y}_{\dot\ga}\frac{\partial}{\partial
\bar{y}^{\dot\gb}}\right)C(y, \bar y|x)\,,
\end{align*}
which when rewritten in component form gives precisely
\eqref{goodtensor}. The free HS connections encoded in $\omega(y, \bar
y|x)$ transform analogously. Indeed, from
\beno
\gd_{\Lambda}\omega(y, \bar y|x)=d\Lambda+[\omega,
\Lambda]
\eeno
it follows that if one decomposes
\beno
\omega={\omega}^{tensor}+\varpi\,,\qquad\qquad \varpi=\frac12\varpi^{\ga\gb}L^y_{\ga\gb}+c.c.
\eeno
in other words if one separates the bilinear in oscillators part
of $\omega$ from the rest, then one arrives at the Lorentz connection
and Lorentz tensor fields transformation
\beno
\gd_{\Lambda}\varpi=d\Lambda+[\varpi,
\Lambda]\,,\qquad\qquad \gd_{\Lambda}{\omega}^{tensor}=[{\omega}^{tensor},
\Lambda]\,.
\eeno
Note also that, while $C$ transforms in the twisted-adjoint under
HS transformations \eqref{twistedadjB}, the Lorentz subalgebra
action reduces to the adjoint one, since $\pi$ does not affect
Lorentz generators. Transformations \eqref{freeLor} clearly do not
extend to nonlinear level as they do not act on $Z$-variable and
hence fields that carry indices contracted with $Z^{A}$ will not
be affected by \eqref{freeLor}. This problem can be easily fixed
by appending $L^y$ with similar generators $L^z$ that rotate $z$
\begin{align}\label{Lor0}
L^y_{\ga\gb}\to
L^{0}_{\ga\gb}&=L^y_{\ga\gb}+L^z_{\ga\gb}=-\ff{i}{2}(y_{\alpha}y_{\beta}-z_{\ga}z_{\gb})\,,\\
[L^0_{\ga\ga}, f]_\star&=\left(y_{\ga}\frac{\partial}{\partial y^{\ga}}+
z_{\ga}\frac{\partial}{\partial z^{\ga}}\right)f\,,
\end{align}
analogously for $\bar{L}$. Note, that the sign in \eqref{Lor0}, which is due to \eqref{YZBasicA}, is
important as it guarantees cancelation of second derivative
terms. Thus defined generators form Lorentz algebra which properly
rotates spinor oscillators
\begin{align*}
&[L^{0}_{\ga\gb},
L^{0}_{\gc\gd}]_{\star}=\gep_{\gb\gc}L^{0}_{\ga\gd}+\dots\,,\\
&[L^{0}_{\ga\gb},
y_{\gc}]_{\star}=\gep_{\gb\gc}y_{\ga}+\gep_{\ga\gc}y_{\gb}\,,\\
&[L^{0}_{\ga\gb},
z_{\gc}]_{\star}=\gep_{\gb\gc}z_{\ga}+\gep_{\ga\gc}z_{\gb}
\end{align*}
and implies that field $B(Y,Z|x)\star\gk$ does transform in
Lorentz covariant fashion ($[L^0_{\ga\gb},\gk]_\star=0$). The nonlinear system
\eqref{VasAll} contains field $S$ apart from master fields
$B$ and $W$ that on-shell is perturbatively expressed via $B$-field.
We know that in solving for $Z$-dependence $S_\ga$ is
reconstructed in terms of $B\star\gk$ up to a gauge freedom, which at the first order
is represented by $\xi_1$ in \eqref{S1sol}.
The subtle point is that the gauge chosen in perturbative
expansion of $S_\ga$ matters a lot for the discussion of Lorentz symmetry.

Hereafter, we would like to discuss the simplest class of the gauges that do not introduce any
external objects, e.g. spinors, vectors, etc.. This is what makes field $S_\ga=S_\ga[B]$ purely
auxiliary. In other words, at each stage of perturbative expansion
with $S_{\ga}$ defined up to a $\p_{\ga}\gep^N(Y, Z)$-term at order-$N$ the gauge
parameter $\gep$ is supposed to contain no extra fields in its
$z$-dependence. There is a gauge invariance for physical
fields which is given by $Y$-dependent functions only. That
invariance is of course unconstrained. The restriction in question
concerns the extra gauge freedom in auxiliary $Z$-sector. This
is obviously true if we impose $\pl^z_\ga \epsilon^N=0$ at each
stage\footnote{Another option is to express $\epsilon$ as some functional of $B$.}, so that the exact forms $\pl^z_\ga \epsilon^N(Y,Z)$ do not
contribute to $S^N_A$ at any order $N$. If that is the case
$S$-field is purely auxiliary and should be properly rotated under
Lorentz transformation
\be\label{Stransformation}
\gd_{\Lambda} S_{\ga}=\frac{\gd S_{\ga}}{\gd B}\gd_{\Lambda}B\,.
\ee
We will soon see that it is this property that guarantees the existence of proper
field-redefinition and that the proper Lorentz transformations belong to the residual symmetries, which preserve the gauge. However, this is not the case with \eqref{Lor0}. On the contrary one has
\begin{align}\label{Srot}
[L^0_{\ga\gb},S_\gc]=S_\ga\epsilon_{\gb\gc}+S_\gb\epsilon_{\ga\gc}+
\frac{\gd S_{\gc}}{\gd B}[L^0_{\ga\gb}, B]_{\star}\,,
\end{align}
where the first two terms on the r.h.s. of \eqref{Srot} arise from
acting with $L^0$ on the spinor index of $S_{\ga}$, which can be
carried by oscillators and their derivatives. In particular, \eqref{Srot} is obvious for $S_{1\ga}$, \eqref{S1sol}, provided that $\pl_z \xi_1=0$. This is the point
where the requirement of no extra fields gets crucial. Therefore,
even though \eqref{Lor0} properly rotates master field $B$, it
still does not provide one with Lorentz generators for it acts on
$S_{\ga}$ inconsistently with the requirement
$S_{\ga}=S_{\ga}[B]$. As a result we face the problem of proper
deformation $L^0\to L^{int}$ in such a way as to compensate the
extra terms in \eqref{Srot} and yet preserve canonical Lorentz
action on physical fields. A priori it is not guaranteed, that
such a deformation exists and in that case one loses physical
interpretation based on equivalence principle. Luckily, for HS
system Lorentz symmetry does exist thanks to the property of the
deformed oscillators \eqref{sp2} and \eqref{sp2vec} constructed
from $S$-field. Indeed to compensate unwanted terms we subtract
\beno
L^{S}_{\ga\gb}=\ff{i}{4}\{S_{\ga}, S_{\gb}\}_\star
\eeno
from $L^{0}_{\ga\gb}$ and define
\be\label{Lor}
L_{\ga\gb}=L^0_{\ga\gb}-L^{S}_{\ga\gb}
\ee
Note that in the vacuum $S_\ga=z_\ga$ we have $L^S=L^z$ and hence we are back to $L=L^y$.
Using \eqref{4}-\eqref{6} one can check the following commutation
relations
\besubeqs
\begin{align}
[L^S_{\ga\gb},S_\gc]&=S_\ga\epsilon_{\gb\gc}+S_\gb\epsilon_{\ga\gc}\,, \\
[L^S_{\ga\gb},L^S_{\gc\gd}]&=L^S_{\ga\gd}\epsilon_{\gb\gc}+... \\
[L^S_{\ga\gb}, B\star\gk]&=0\label{Bkapcommute}
\end{align}
\esubeqs
that give the desired
\beno
[L_{\ga\gb}, S_{\gc}]_\star=\ff{\gd S_{\gc}}{\gd
B}[L^{0}_{\ga\gb}, B]_\star\,.
\eeno
It is easy to check that $L$ acts on $B\star\gk$ in the right way
too due to \eqref{Bkapcommute}
\be
\gd_{\Lambda} (B\star\gk)\equiv \left[\frac12\Lambda^{\ga\gb}L_{\ga\gb},
B\star\gk\right]_\star=\left[\frac12\Lambda^{\ga\gb}L^0_{\ga\gb},
B\star\gk\right]_\star\,.\label{LorentzBtrans}
\ee
Let us make an important comment. The generators \eqref{Lor}
themselves formally do not form Lorentz algebra in a sense, that
$[L,L]_\star\neq L$. Indeed, it is straightforward to check
\beno
[L_{\ga\gb},
L_{\gc\gd}]_\star=(\gep_{\gb\gc}L_{\ga\gd}+\dots)-\ff{\gd
L^{S}_{\gc\gd}}{\gd B}[L^0_{\ga\gb}, B]_\star+\ff{\gd
L^{S}_{\ga\gb}}{\gd B}[L^0_{\gc\gd}, B]_\star\,,
\eeno
where the two last terms break down Lorentz symmetry. This
apparent contradiction is in fact false. The reason is that generators
\eqref{Lor} are field dependent and therefore when successively
applied get differ by the corresponding change of the $B$-field.
In other words the action of Lorentz generators \eqref{Lor} must
account for the change of generators themselves\footnote{We are
very grateful to M.A. Vasiliev for explaining this point to us.}.
Let us show that being properly treated the successive application
of \eqref{Lor} on $B$-field is equivalent to successive Lorentz
transformations. Apply \eqref{Lor} to $B\star\gk$ one time
\beno
\gd_{\Lambda_1}(B\star\gk)=[L_{\Lambda_1}[B\star\gk],
B\star\gk]_\star=[L^0_{\Lambda_1}, B\star\gk]_\star\,,
\eeno
which as was shown is a proper Lorentz rotation. Now, applying \eqref{Lor} for the
second time
\beno
\gd_{\Lambda_2}\gd_{\Lambda_1}(B\star\gk)=[L_{\Lambda_2}[B\star\gk],
\gd_{\Lambda_1}(B\star\gk)]_\star+[\gd_{\Lambda_1}L^{S}_{\Lambda_2}[B\star\gk],
B\star\gk]_\star\,,
\eeno
where the second term has been added to account for the change of
field dependent generator. Now, using that $[L^{S},
B\star\gk]_\star=0$ and therefore $[\gd L^{S},
B\star\gk]_\star+[L^S, \gd(B\star\gk)]_\star=0$ one finds
\beno
(\gd_{\Lambda_2}\gd_{\Lambda_1}-\gd_{\Lambda_1}\gd_{\Lambda_2})(B\star\gk)=\gd^{0}_{[\Lambda_2,
\Lambda_1]}(B\star\gk)\,.
\eeno
So, one concludes that local Lorentz symmetry is restored. The
situation with one-forms $W(Y,Z|x)$ is a bit trickier. With $\Lambda=\frac12\Lambda^{\ga\gb}L_{\ga\gb}+c.c.$ looking at
\be\label{LorW}
\gd_{\Lambda}W\equiv d\Lambda+[W,
\Lambda]_{\star}+c.c.=\frac12d\Lambda^{\ga\gb}L_{\ga\gb}+\left[W,
\frac12\Lambda^{\ga\gb}L^0_{\ga\gb}\right]_\star+c.c.\,,
\ee
where we made use of \eqref{3}, we see that to extract Lorentz
tensors and Lorentz connection from $W$ we need to decompose it as
follows
\be\label{LorTen}
W=W^{tensor}+\varpi\,,\qquad\qquad
\varpi=\frac12\varpi^{\,\ga\gb}L_{\ga\gb}+\frac12\varpi^{\gad\gbd}\bar{L}_{\gad\gbd}\,,
\ee
such that transformation \eqref{LorW} reduces to
\begin{align}
\gd_{\Lambda}\varpi&=d\Lambda+[\varpi,\Lambda]_\star,
+c.c.\,,\label{Lorc}\\
\gd_{\Lambda}W^{tensor}&=\left[W^{tensor},
\frac12\Lambda^{\ga\gb}L^{0}_{\ga\gb}\right]_\star+c.c.\,,
\end{align}
where again we used that in making field dependent Lorentz
transformation one has to compensate the change of generators
which effectively cancels the otherwise appearing extra term in \eqref{Lorc}
\beno
\Lambda^{\ga\gb}\varpi^{\gc\gd}\ff{\gd L^{S}_{\gc\gd}}{\gd
(B\star\gk)}[L^0_{\ga\gb}, B\star\gk]_\star\,.
\eeno
Decomposition \eqref{LorTen} may look like some arbitrary
separation of $W$ into two terms which particularly does not
constraint the form of connection fields $\varpi_{\ga\gb}$ and
$\varpi_{\gad\gbd}$. These are nonetheless get totally
fixed by the requirement that the spin-two contribution is absent
in $W^{tensor}(Y,Z|x)$ master field, i.e. there are no $yy$ and $\bar{y}\bar{y}$
contributions in $W^{tensor}(Y,Z|x)$.

In analyzing the problem of Lorentz symmetry we had to impose a
requirement of no external fields that carry indices to appear in
reconstruction of $Z$-dependence. Does it mean that otherwise the
system possesses no local Lorentz symmetry? No it does not,
because these extra fields that break down explicit Lorentz
symmetry appear in a pure gauge fashion. We have proven the
existence of local Lorentz symmetry within certain setup which
guarantees \eqref{Srot} and allows one to explicitly identify the Lorentz
generators. In case of extra fields there is a gauge transformation
that brings solution to the form that possesses explicit Lorentz
symmetry. It is just that a concrete form of that transformation
depends on the details of particular solution. Or other way
around, the form of Lorentz generators in that case contains
these extra fields in question. One can also have a look at \cite{Prokushkin:1998bq,Vasiliev:1999ba,Sezgin:2002ru,Sezgin:2011hq} where
the issue of Lorentz symmetry is discussed.

At the end, let us mention that there are similarities between $S_A$-field and compensator field $V^\aAs$ of MacDowell-Mansouri-Stelle-West approach, which is reviewed in Section \ref{extra:MMSW}.
In both cases the physical world is identified with the stability algebra of the compensator field.
The similarities/differences of $V^\aAs$ with $S_A$ include: (i) $S_A$ is a connection-like field while $V^\aAs$ is matter-like field; (ii) $S_A$ is expressed in terms of the other field, $B$, that is why the condition that defines the physical world is not $\delta S_A=0$ but \eqref{Stransformation}; (iii) in both cases we need to employ additional transformations, diffeomorphisms for $V^\aAs$ and certain field-dependent transformations for $S_A$; (iv) the physical world is orthogonal to $V^\aAs$  for MMSW-gravity and $Z$-independent for Vasiliev HS theory; (v) in both cases there is a nontrivial expression for the spin-connection that depends on compensators $V^\aAs$ or $S_A$; (vi) in both cases the symmetry of the physical world is the residual symmetry that preserves the gauge chosen for $V^\aAs$ or $S_A$.

\mysubsection{Higher orders and gauge fixing} Let us briefly consider how to operate with Vasiliev equations at
higher orders. Before going into the computations let us rewrite the equations in a more compact form by taking advantage of $sp(4)$-indices and shifting all the fields by the vacuum values thereof
\besubeqs\label{VasilievEqB}
\begin{align}
\DO W+W\wedge \star W&=0\,,  \label{VasilievEqBA}\\
\DOt B+W\star B-B\star\tilde{W}&=0\,,\label{VasilievEqBB}\\
\pl_A W&= \DO \aA_{A}+[W,\aA_{A}]\label{VasilievEqBC}\\
\pl_A B&=R_A\label{VasilievEqBD}\\
\pl_A \aA_{B}-\pl_B \aA_{A}&=R_{AB}\label{VasilievEqBE}
\end{align}
\esubeqs
where $\p_{A}=\ff{\p}{\p Z^{A}}$ and the \rhs are defined as
\begin{align*}
R_{AB}=&\,\,\frac{i}2\begin{vmatrix}
         \epsilon_{\ga\gb}B\star \kappa & 0 \\
         0 & \epsilon_{\gad\gbd} B\star \gkb \\
       \end{vmatrix}
-[\aA_{A},\aA_{B}]\\
=&\,\,\phantom{\frac{i}2}\begin{vmatrix}
\epsilon_{\ga\gb}(\frac{i}2B\star \kappa- \aA_{\nu}\star \aA^{\nu}) &  - [\aA_{\ga},\aA_{\gbd}]\\
- [\aA_{\gad},\aA_{\gb}] & \epsilon_{\gad\gbd}(\frac{i}2B\star \gkb- \aA_{\gnd}\star \aA^{\gnd}) \\
\end{vmatrix}\\
R_A=& -\begin{vmatrix}
                 \aA_{\ga}\star B+ B\star \gk \star \aA_{\ga}\star \gk  \\
                 \aA_{\gad}\star B+ B\star \gkb \star \aA_{\gad}\star \gkb  \\
               \end{vmatrix}
\end{align*}
We can solve for the evolution along $Z$-direction as
\besubeqs\label{VasilievEqC}
\begin{align}
W&= \omega(Y)+Z^A\int_0^1 dt\,\left.\rule{0pt}{14pt}\left(\DO \aA_{A}+[W,\aA_{A}]\right)\right|_{Z\to tZ}\label{VasilievEqCC}\\
B&=C(Y)+Z^A\int_0^1 dt\,\left.\rule{0pt}{14pt}R_A{}\right|_{Z\to tZ}\label{VasilievEqCD}\\
\mathcal{A}_A&=Z^B\int_0^1 t\,dt\,\left.\rule{0pt}{14pt}R_{BA}\right|_{Z\to tZ}+\pl_A \xi(Y,Z)\label{VasilievEqCE}
\end{align}
\esubeqs
which is to be understood in the realm of perturbative expansion. The iterative process of solving equations starts with the initial condition $B^0=0$, $\aA_{A}^0=0$ and $W^0=0$. Suppose we have already solved for all orders up to $N-1$. At the order $N$ we proceed as follows. Firstly, one gets $B^N$ from \eqref{VasilievEqCD}, which turns out to be expressed in terms of lower order fields. Secondly, one obtains certain $\aA_{A}^N$ from \eqref{VasilievEqCE}. Next, we solve for $W^N$ from \eqref{VasilievEqCC}. Finally, we plug everything into the two \eqref{VasilievEqBA}-\eqref{VasilievEqBB} to get equations in the physical $x$-space. These
are to be of the form \eqref{xpaceseqC}.

At every order of perturbative expansion one gathers $\omega^N$ and $C^N(Y)$ as integration constants in accordance with the formal expansion \eqref{xpaceseqC} of \eqref{xpaceseq}. The proliferation of degrees of freedom is because of somewhat formal expansion --- we expand nonlinear equations in orders of interactions. In practice, we either collect all integration constants together with the true fields being $\omega(Y)=\sum_k\omega^k(Y)$ and $C(Y)=\sum_kC^k(Y)$ or impose some boundary conditions to get rid of extra degrees of freedom. From explicit solutions \eqref{VasilievEqC} we see that the following identification is possible
\begin{align*}
&\omega(Y)=\sum_k\omega^k(Y)=\left.W(Y,Z)\right|_{Z=0} &&
C(Y)=\sum_kC^k(Y)=\left.B(Y,Z)\right|_{Z=0}
\end{align*}
which is because the extra pieces in \eqref{VasilievEqC} are proportional to $Z$. The same idea can be rewritten as
\begin{align}\label{WBfromOC}
&W(Y,Z)=\omega(Y)+O(Z,\omega, C) && B(Y,Z)=C(Y)+O(Z,C)
\end{align}
By plugging these expressions into \eqref{VasilievEqBA}-\eqref{VasilievEqBB} we find something like
\be\label{reconstr}\begin{split}
d\omega+\omega\star\omega-\mathcal{V}(\omega,\omega,C)-...+O(Z,\omega,\omega,C)&=0\\
dC+\omega\star C-C\star \pi(\omega)-\mathcal{V}(\omega,C,C)-...+O(Z,\omega,C)&=0\end{split}
\ee
where we wrote the terms that are obviously there at first. By $\mathcal{V}(...)$ we denoted possible interaction vertices that are $Z$-independent. Let us note that the appearance of $Z$-independent
terms can happen because for any two functions, say $f(Y,Z)$ and $g(Y,Z)$, that vanish at $Z=0$, $f(Y,0)=g(Y,0)=0$ we in general find $(f\star g)(Y,0)\neq0$, so that $O(Z)$ terms in \eqref{WBfromOC} can generate nontrivial $\mathcal{V}(...)$. This is in fact the whole idea of the Vasiliev equations.  Possible $Z$-dependent corrections to \eqref{reconstr} are written at the end. The crucial fact is that there are no $Z$-dependent corrections in \eqref{reconstr}. This is because we have already solved all equations along $Z$-direction. Therefore, applying $\pl_A$ to \eqref{reconstr}, i.e. to \eqref{VasilievEqBA}-\eqref{VasilievEqBB}, we get \eqref{VasilievEqBC}-\eqref{VasilievEqBD}, respectively, but these equations hold true identically. The same reasoning was used around \eqref{trivialZdep} to show that $DW_1$ is $Z$-independent.

Finally, we see that we get some consistent equations in $x$-space (they result from partial solving of consistent equations for $Z$-dependence). These equations may contain nontrivial interaction vertices and are $Z$-independent. Therefore, we get something like \eqref{xpaceseq}, which is considered as a solution of the higher-spin problem at the level of equations of motion.

Let us comment on the pure gauge freedom present in solution \eqref{VasilievEqCE} for $\aA_{A}$. We have already seen that the simplest field redefinition that maps the equations to the manifestly Lorentz covariant ones requires the absence of $\xi$. As we solve for the $Z$-dependence we may fix the gauge in some way. The simplest solution is to impose a gauge such that $\pl_A \xi(Y,Z)=0$. First, let us note that $Z^A \aA_{A}=Z^A \pl_A \xi$ because $R_{AB}$ in \eqref{VasilievEqCE} is antisymmetric and has one index already contracted with $Z$. We need to fix one function, so one gauge condition should be enough. One may choose\footnote{Thinking of $\aA_{A}$ as a Yang-Mills type connection, gauge $Z^A\aA_{A}=0$ or rewritten using standard letters $x^m A_m=0$ is known under many names in electrodynamics and Yang-Mills theory, viz. Schwinger-(Fock) gauge, coordinate gauge, multipolar gauge, Poincare gauge. It is particulary useful, see e.g. \cite{Shuryak:1981pi} for applications, thanks to the simple expression for the gauge potential in terms of (non-Abelian) field strength $F_{mn}$,
\beno A_m=x^n\int_0^1 dt\,t\, F_{nm}(xt)\,,\eeno which is not surprising now as the whole perturbation
theory is built of homotopy integrals.}
\begin{align}
Z^A \aA_{A}=0\label{bestgauge}
\end{align}
This gauge is admissible and complete since $Z^A \pl_A \xi=0$ admits constant solutions only in the class
of analytic functions. Constants in $Z$ are in fact arbitrary functions $\xi(Y|X)$, which is the required amount of residual gauge symmetry. In perturbation theory we have to solve for residual gauge symmetries order by order from
\begin{align}\label{residualsymmetry}
0&=Z^A\delta \aA_{A}=Z^A\pl_A \epsilon(Y,Z)+ Z^A[\aA_{A},\epsilon]_\star
\end{align}
whose general solution is
\begin{align*}
\epsilon(Y,Z)=- Z^A\int_0^1 dt\,\left. \rule{0pt}{14pt}[\aA_{A},\epsilon]_\star\right|_{zt}+\epsilon(Y)
\end{align*}
For example, at the first order we have
\begin{align*}
\delta \aA_{A}^1=\pl_A \epsilon_1(Y,Z)
\end{align*}
which exactly compensates the freedom in $\xi_1$. The residual gauge symmetry $Z^A\pl_A \epsilon_1=0$
yields $Z$-independent $\epsilon_1(Y)$. At the second order we find
\begin{align*}
\delta \aA^2_{A}=  \pl_A\epsilon_2(Y,Z)+[\aA_{A}^1,\epsilon_1]
\end{align*}
The residual gauge symmetry is
\begin{align*}
\epsilon_2(Y,Z)=- Z^A\int_0^1 dt\,\left. \rule{0pt}{14pt}[\aA_{A}^1,\epsilon_1]_\star\right|_{zt}+\epsilon_2(Y)
\end{align*}
At every order of perturbation theory we find $\epsilon_N(Y)$, which is the gauge parameter for $\omega_N(Y)$. Of course the residual symmetry we are solving for order by order is nothing but the genuine gauge symmetry of unfolded equations \eqref{xpaceseq}. The latter symmetry can be also obtained by restricting the transformations of master-fields to $Z=0$ and further restricting them to the residual symmetry
\begin{align*}
&\delta \omega=\left. \rule{0pt}{14pt}\delta W\right|_{Z=0,\, \eqref{residualsymmetry}} && \delta C=\left. \rule{0pt}{14pt}\delta B\right|_{Z=0,\, \eqref{residualsymmetry}}
\end{align*}

One of the ways to understand the issue of Lorentz covariance is that there is a problem with the naive Lorentz transformations of compensator field $S_A$. Indeed, with \eqref{Srot} we find
\begin{align*}
z^\gc\left[\frac12\Lambda^{\ga\ga}L^0_{\ga\ga}, S_\gc\right]=-\Lambda^{\ga\ga} z_\ga S_\ga +z^\gc\frac{\delta S_\gc}{\delta B}\left[\frac12 \Lambda^{\ga\ga}L^0_{\ga\ga},B\right]_\star
\end{align*}
so that gauge condition \eqref{bestgauge} is not preserved by the tentative transformations with $L^0$, which must not affect the compensator field,
\begin{align*}
Z^C\left[\frac12\Lambda^{\ga\ga}L^0_{\ga\ga}+c.c., S_C\right]&=-\Lambda^{\ga\ga} z_\ga S_\ga +c.c.+Z^C\frac{\delta S_C}{\delta B}\left[\frac12 \Lambda^{\ga\ga}L^0_{\ga\ga}+c.c.,B\right]_\star\\
&=-\Lambda^{\ga\ga} z_\ga S_\ga +c.c.\neq0
\end{align*}
The corrected Lorentz generators defined in \eqref{Lor} do preserve the gauge since the problematic term above is eaten up by the $L^S$-part in $L=L^0-L^S$. Therefore, the Lorentz transformations with $L$ do belong to the set of residual symmetries.

Let us finally note, that other gauge conditions are also possible, but this may nontrivially affect the field-redefinition needed to restore Lorentz covariance. The simplest gauge, which generalizes \eqref{bestgauge} while still covariant, is $\{Z^A, S_A\}_\star=0$.

\paragraph{Second-order example.}
It is trivial to expand to the second order. We denote the second order fields as $B^2$, $W^2$ and $S^2_A$. Using \eqref{VasilievEqCD} we can first solve for $Z$-dependence of $B^2$ field
\begin{align}\label{B2}
2iB_2=z^\ga\int_{0}^{1} dt \left.(S^1_\ga\star
C(Y|x)+C(Y|x)\star\tilde{S}^1_\ga)\right|_{z\to tz}+c.c.+C^2(Y|x)\,,
\end{align}where tilde means the twisting,
e.g., $\tilde{S}_{\ga}=\pi(S_{\ga})$,
$\tilde{\bar{S}}_{\dot{\ga}}=\bar{\pi}({\bar{S}}_{\dot{\ga}})$,
and $S^{1}_{\ga}$ is given by \eqref{S1sol}. Substituting
\eqref{B2} and \eqref{W1sol} into \eqref{VasilievEqBB} and after
some algebra and integration by parts we get the equation for the second order perturbation $C_2$
\begin{align*}
&D^{L}C_2-ih^{\ga\gad}\Big(y_{\ga}\bar{y}_{\gad}-\ff{\p^2}{\p
y^{\ga}\p\bar{y}^{\gad}}\Big)C_2
+\omega\star C-C\star\tilde{\omega}+\notag\\
&-\ff{1}{2i}h^{\ga\gad}\Big\{ \bar{y}_{\gad}\int_{0}^{1}dt\,
t\,\Big(z_{\ga}C(-tz, \bar y)e^{itz_{\gb}y^{\gb}}\star C-C\star
z_{\ga}C(tz, \bar
y)e^{itz_{\gb}y^{\gb}}\Big)+c.c. \Big\}_{Z=0}+\notag\\
&+\ff12h^{\ga\gad}\int_{0}^{1}dt\, (1-t)\Big(
z_{\ga}\ff{\p}{\p\bar{y}^{\gad}}C(-tz,
\bar{y})e^{itz_{\gb}y^{\gb}}\star C+C\star
z_{\ga}\ff{\p}{\p\bar{y}^{\gad}}C(tz,
\bar{y})e^{itz_{\gb}y^{\gb}}+c.c. \Big)_{Z=0}=0\,,
\end{align*}
where $C$ and $\omega$ are the first order fields and we also made use of the following formula
\beno
\int\limits_{[0,1]\times [0,1]}dt\, dt'\, t\, f(tt')=
\int\limits_{[0,1]\times [0,1]} f(s)\theta(\tau-s)d\tau
ds=\int_{0}^{1} (1-s) f(s) ds\,,
\eeno
which a particular case of \eqref{RepeatedHomotopy}. Expanding $\star$-products one finds
\begin{align*}
&\DOt C_2 +\omega\star C-C\star \tilde{\omega}+\frac{i}2h^{\ga\gad}\int_0^1 dt \int dU dV\,K_{\ga\gad}(U,V;Y;t) C(-ut,\bar{y}+\bar{u})C(Y+V) =0\\
&K_{\ga\gad}(U,V;Y;t)=u_\ga  \left\{(t \bar{y}_\gad  -(1-t) v_\gad )e^{ity^\gb u_\gb +iV^B U_B}+  (t \bar{y}_\gad  +(1-t) v_\gad)e^{-ity^\gb u_\gb-iV^AU_A}\right\}
\end{align*}
which is to be compared with \eqref{xpaceseqBB}, allowing us to identify two interaction vertices
\begin{align*}
\mathcal{V}(\omega_1,C_1)&=\omega\star C-C\star \tilde{\omega}\\
\mathcal{V}(\Omega,C_1,C_1)&=\frac{i}2h^{\ga\gad}\int_0^1 dt \int dU dV\,K_{\ga\gad}(U,V;Y;t) C(-ut,\bar{y}+\bar{u})C(Y+V)
\end{align*}

Note, that at the given order noncovariant terms with respect to
Lorentz connection $\omega^{\ga\gb}$ that in principle could have
appeared cancel, which relies on some nontrivial $\star$-product identities. This
is to be expected since the corrections to the Lorentz generators disappear
from $\delta B$, \eqref{LorentzBtrans}, but they are relevant for $S_A$ and $W$.

To proceed to $W$-sector one solves
\eqref{VasilievEqBE} for $S^{2}_{A}$,
which gives
\begin{align*}
S^2_\gb=z^\ga\int_{0}^{1}dt\, t\,\left.(\epsilon_{\ga\gb}B^2\star\gk-\frac{i}{4}
[S^1_\ga, S^1_\gb])\right|_{z\to tz}
-\frac{i}{4}\bar{z}^\gad\int_{0}^{1} dt\, t\, [\bar{S}^1_\gad,
S^1_\gb]_{\bar z\to t\bar z}
\end{align*}
and allows one to determine $Z$-dependence of $W$ field from
\eqref{VasilievEqBC}
\begin{align*}
-2iW_2=z^\ga\int_{0}^{1}dt\,\left.(\DO S^2_\ga+[W^1,S^1_\ga])\right|_{z\to
tz}+c.c.+\omega_2(Y|X)\,.
\end{align*}
The latter when substituted into \eqref{VasilievEqBA} results in
second order space-time equation for $\omega^2(Y|x)$ that we omit
for brevity.

\mysubsection{Topological fields and integrating flow} Remarkable
property of the Vasiliev equations is the existence of the
integrating flow that maps interacting system to the linearized
one around certain vacuum with topological fields activated. The
existence of such a flow was discovered in \cite{Prokushkin:1998vn} for
three-dimensional case, but the very construction is essentially
the same for $d=4$ and arbitrary $d$ equations. The flow exists
for the extended by auxiliary fields HS equations. We have briefly
mentioned such possibility when discussed possible generalizations
of Vasiliev equations. Let us now proceed in some detail. First,
let us note that there is a way to rewrite eqs \eqref{VasAll}
in the form that does not explicitly contain $\pi$-automorphism.
To do it we introduce some extra Klein operators $k$ and $\bar k$
such that
\begin{align*}
&\{k, dz\}=\{\bar k, d\bar z\}=0\,,\qquad\qquad \{k, z\}=\{k, y\}=0\,,\\
&\{\bar k, \bar z\}=\{\bar k, \bar y\}=0\,,\qquad\qquad\qquad k^2=\bar k^2=1
\end{align*}
in addition $k$ commutes with all antiholomorphic variables and
$\bar k$ with the holomorphic ones. Let us note, that $k$ and
$\bar k$ are extra variables and do not belong to the star-product
algebra. This makes it possible to rewrite \eqref{VasAll} as
follows
\besubeqs\label{HSgen}
\begin{align}
&d\WW+\WW\star \WW=0\,,\label{HSgen1}\\
&d\Scal+[\WW,\Scal]_\star =0\,,\label{HSgen2}\\
&d\BB+[\WW,\BB]_\star =0\,,\label{HSgen3}\\
&\Scal\star \Scal=-idz_{\ga}\wedge dz^{\ga}(1+\BB\star k\gk)+c.c.\,,\label{HSgen4}\\
&[\Scal,\BB]_\star =0\,,\label{HSgen5}
\end{align}
\esubeqs
where
\beno
\BB=B p\,,\qquad\qquad \Scal=S_{\ga}dz^{\ga} P+c.c.\,,\qquad\qquad \WW=W P
\eeno
and we have introduced the operators
\be\label{kproj}
P=\ff12(1+k\bar k)\,,\qquad p=\ff12(k+ \bar k)\,,\qquad
P^2=P\,,\qquad p^2=P\,.
\ee
It is a simple exercise to check that
\eqref{HSgen} indeed reduces to
\eqref{VasAll}. The only thing those extra Kleinians do is
account properly for the twisting automorphism. System
\eqref{HSgen} is however consistent for any
dependence of the fields $\WW$, $\BB$, $\Scal$ on $k$ and $\bar k$
\footnote{Any such dependence can be at most bilinear in $k$ and
$\bar k$.} and this is the way to introduce fermions, for example,
by doubling the spectrum of fields with the aid of extra Klein
operators\footnote{Operators \eqref{kproj} are designed to project
HS system to the bosonic sector where no doubling takes place.}.
Another option within the bosonic case we are dealing with is to
consider
\beno
\BB=B(Y,Z) p+ b(Y,Z) P\,,\qquad \WW=W(Y,Z) P+w(Y,Z) p\,.
\eeno
At linearized level the effect of new fields $b$ and $w$ does not
influence free field dynamics of HS fields $W$ and $B$ and it
does not introduce new degrees of freedom. Without going into much
of detail let us write down the linearized equations for
$b_1=c(Y|x)$ and $w_1=w(Y|x)$ fields that arise from
\eqref{HSgen}
\besubeqs\begin{align}
&D_{\Omega}c(y, \bar y|x)=0\,,\label{top1}\\
&\tilde{D}_{\Omega}w=-\ff{i}{4}h^{\ga}{}_{\gdd}\wedge
h^{\gb\gbd}y_{\ga}y_{\gb}\,c(y, 0|x)+c.c.\,.\label{top2}
\end{align}\esubeqs
From \eqref{top1} we see that $c(y, \bar y|x)$ satisfies the
adjoint $AdS_4$ covariant constancy condition rather than the
twisted-adjoint as in the case of dynamical field $C(Y|x)$. This
means that $c(Y|x)$ carries no physical degrees of freedom being
just a set of global symmetry parameters. Analogously $w(Y|x)$
turns out to be nonpropagating too.
Perturbatively, these new fields are not dynamical, \cite{Vasiliev:1987hv}, and for that
reason can be called {\it topological fields}. Nevertheless the
dynamics beyond free level essentially differs in the presence of
topological fields. Therefore, they provide in general infinitely
many (tensorial) parameters that correspond to inequivalent HS
theories.

There is an important gauge invariant vacuum solution in the
topological sector of the theory
\be\label{vac}
b_0=\nu=const\,.
\ee
It is gauge invariant since the gauge transformation for
topological zero-forms has no twisting
\beno
b\to g^{-1}\star b\star g\,.
\eeno
The exact vacuum solution for $\WW$--field that corresponds to
\eqref{vac} is known only in $3d$, \cite{Prokushkin:1998vn}, thanks to the
properties of the deformed oscillators. Consider now perturbative
expansion around the vacuum \eqref{vac} in $d=4$
\beno
\BB=\BB_0+\mu \BB'(\mu, \nu)\,,\qquad\qquad \WW=\WW(\mu, \nu)\,,\qquad\qquad
\Scal=\Scal(\mu, \nu)\,,
\eeno
where $\mu$ is a formal  expansion parameter in perturbation
theory and the dependence on $\nu$ is due to chosen vacuum
\eqref{vac}. The linearized perturbations for $\BB$-field reside
in $\CC(\nu)=\BB'(0, \nu)$. Remarkable observation of
\cite{Prokushkin:1998vn} states that the following integrating flow
\besubeqs
\begin{align}
&\ff{\p\WW}{\p\mu}=(1-\eta)\BB'\star \ff{\p\WW}{\p\nu}+\eta\ff{\p\WW}{\p\nu}\star \BB'\,,\\
&\ff{\p \BB'}{\p\mu}=(1-\eta)\BB'\star \ff{\p
\BB'}{\p\nu}+\eta\ff{\p
\BB'}{\p\nu}\star \BB'\,,\label{flowB}\\
&\ff{\p \Scal}{\p\mu}=(1-\eta)\BB'\star \ff{\p
\Scal}{\p\nu}+\eta\ff{\p \Scal}{\p\nu}\star \BB'\,,
\end{align}
\esubeqs
where $\eta$ is an arbitrary constant, is consistent with the
equations of motion \eqref{HSgen}. The power of
the integrating flow equations is that the whole complicated
dynamics of nonlinear HS system can be step by step recovered from
linearized solution around topological vacuum \eqref{vac}. For
example, for HS system without topological fields the second order
value of the $B$-field can be found from \eqref{flowB} as follows
\beno
B^{ph}_{(2)}=\ff12\ff{\p}{\p\nu}\left.\Big(B^{ph}_{(1)}(\nu)\star
\pi(b^{top}_{(1)}(\nu))+ b^{top}_{(1)}(\nu)\star
B^{ph}_{(1)}(\nu)\Big)\right|_{\nu=0}\,,
\eeno
here we set $\eta=1/2$ for convenience.

Integrating flow machinery
potentially provides very powerful tool for analysis of nonlinear
interactions since it reduces the perturbation theory to the free
level data that reconstructs nonlinearities algebraically. The
problem is that the explicit expression for the vacuum $\WW_0$
field is not known in four dimensions. And even if one will manage
to find the vacuum, the free level analysis would be highly
nontrivial due to the presence of infinitely many HS fields within
$\WW_0$. Still, we believe the integrating flow approach is very
promising for applications.

Lastly, let us mention that the integration flow is similar to the Seiberg-Witten map
that relates fields of Yang-Mills theory to those of non-commutative Yang-Mills. The integration flow
relates a simple but still noncommutative model to a more complicated non-commutative one --- the full Vasiliev
equations.



\mysectionstar{Aknowledgements}
\label{sec:Aknowledgements}
We would like to thank Kostya Alkalaev, Nicolas Boulanger, Stefan Fredenhagen,  Gustavo Lucena G\'{o}mez, Carlo Iazeolla, Pan Kessel,
Jianwei Mei, Rongxin Miao, Teake Nutma, Per Sundell, Massimo Taronna, Stefan Theisen, Mirian Tsulaia, all the participants of the Higher Spins, Strings and Duality school that took place at Galileo Galilei Institute, Florence. We are especially grateful to Mikhail Vasiliev for enlightening discussions, encouragement and comments. We also would like to thank Nikita Misuna, Dario Francia and especially Alexander Chekmenev, Egor Korovins, Ali Rezaie, Tran Tung, Massimo Taronna, M.Sivakumar and Samuel Monier for reading the draft, pointing to us many misprints and making valuable suggestions. We are indebted to Gustavo Lucena G\'{o}mez and Pan Kessel
for hours of discussions that have had considerable impact on the presentation in Section \ref{sec:VasilievEquations}.
We thank Galileo Galilei Institute and organizers of the school for the invitation to give lectures and for creating a wonderful atmosphere. We thank Kavli Institute for the theoretical Physics, Santa-Barbara, for hospitality during the final stage of this work. This research was supported in part by the National Science Foundation under Grant No. NSF PHY11-25915. The work of E.S. was supported by the Alexander von Humboldt Foundation. The work of
V.D. was supported in part by the grant of the Dynasty Foundation.
The work of E.S. and V.D. was supported in part by RFBR grants
No. 11-02-00814, 12-02-31837.



\mysection{Extras}
\mysubsection{Fronsdal operator on Riemannian manifolds}\setcounter{equation}{0} \label{extra:FronsdalonRiemann}
Let us take the kinetic part of the Fronsdal operator and put it on a general Riemannian manifold
\begin{align*}
F[\phi]^{\ua(s)}&=\square \phi^{\ua(s)}-\nabla^\ua \nabla_\mm\phi^{\mm \ua(s-1)}+\frac12\nabla^\ua \nabla^\ua \phi\fud{\ua(s-2)\mm}{\mm}\,,
\end{align*}
where $\nabla_\mm$ is a covariant derivative with respect to some metric $g_{\mm\nn}$. First, we can check that the Fronsdal operator still satisfies the double trace constraint,
\be F[\phi]^{\ua(s-4)\mm\mm\nn\nn}g_{\mm\mm}g_{\nn\nn}=0\,,\ee
which might not have been the case. At least we have the same number of equations as the number of fields. The Fronsdal operator is not gauge invariant in general, which is a sign of a serious problem. Once the gauge symmetry is lost, or weakened, we gained new degrees of freedom. New degrees of freedom come usually in the form of ghosts. Let us also note that the Fronsdal operator is not what one gets from the covariantization of the Fronsdal action since one has to commute the derivatives in taking the variation. The source of non-invariance is
\be \nabla^\ua [\nabla^\ua,\nabla_\mm]\xi^{\mm \ua(s-2)}-[\nabla^\ua,\square]\xi^{\ua(s-1)}\label{noninvarterm}\,.\ee
When the Riemann tensor has only scalar curvature nonvanishing, i.e. we are in (anti)-de Sitter, which is displayed by \eqref{covderadsrule}, the non-invariance is of special form, the same terms originate from mass-like terms, as we have already seen. Let us now see what happens if the only non-zero part of the Riemann tensor is the Weyl tensor (which is introduced systematically in Section \ref{sec:UnfldGravity}). Then \eqref{noninvarterm} reduces to
\be2 C\fud{\ua\ua\ua,}{\mm\mm}\xi^{\ua(s-3)\mm\mm}+
 C\fud{\ua\ua,}{\mm\mm}\nabla^\ua\xi^{\ua(s-3)\mm\mm}-
  2C\fud{\ua\ua,}{\mm\mm}\nabla^\mm\xi^{\ua(s-2)\mm}\label{nonivariance}\,,\ee
where $C^{\ua\ua,\ub\ub}$ is the Weyl tensor and $C^{\ua\ua\ua,\ub\ub}$ is its first derivative, which is constrained by the differential Bianchi identity to have \YoungmCB-symmetry. We may try to cancel this terms by adding to the Fronsdal operator
\be \label{goodtry} C\fud{\ua\ua,}{\mm\mm}\phi^{\ua(s-2)\mm\mm}+g^{\ua\ua}C\fud{\ua\ua,}{\mm\mm}\phi\fud{\ua(s-4)\mm\mm\nn}{\nn}\ee
but find this impossible for $s>2$, because of the first term in \eqref{nonivariance}, which does not have derivative on the gauge parameter. The last two terms of \eqref{nonivariance} are problematic too because the relative coefficient does not allow one to cancel them by \eqref{goodtry}. So we see that it is not a piece of cake to make higher-spin fields live on a manifold, which is different from Minkowski or (anti)-de Sitter. In particular generic Ricci flat or Einstein backgrounds are not accessible by the Fronsdal theory. The solution, which is a part of the Vasiliev theory, is non-minimal in the sense that it cannot be achieved via simple modifications of the Fronsdal operator.

\mysubsection{Other gravity-like actions}\label{extra:Lovelock}

In some applications the gravity Lagrangians are allowed to have a more general form, suggested first by Lovelock, \cite{Lovelock:1971yv}. One adds scalar polynomials in the Riemann tensor such that the equations of motion are still second-order. The Lovelock terms have the following simple form within the frame-like approach
\be S_{L,n}= \int \underbrace{F^{a,b} \wedge...\wedge F^{c,d}}_{n}\wedge e^f\wedge e^g\wedge...\wedge e^u\, \epsilon_{ab...cdfg...u}\,.\label{lovelockn}\ee
In particular it is easy to see that there are $[d/2]$ Lovelock terms in dimension $d$ with $S_{L,1}$ corresponding to the pure Einstein-Hilbert. The equations of motion involve at most two time derivatives. Indeed, each $F$, $F=d\omega+...$, is of the second order with respect to $e$, $F_{0\ii}=\dot{\omega}_\ii+...$, $\ii=1,...,d-1$. Since the $\wedge$-product anti-symmetrizes over the indices, $F_{0\ii}$ can appear only once in $F\wedge...\wedge F$.

In even dimension $d=2n$ the top Lovelock term is topological
\be S_{L,n}= \int \underbrace{F^{a,b} \wedge...\wedge F^{c,d}}_{n} \epsilon_{ab...cd}\,,\ee
its variation vanishes up to boundary term because of $\delta
F=D\delta \omega$ and $DF\equiv0$.

Let us take the most general action composed of Lovelock-type terms, \eqref{lovelockn}, which reads
\be S=\sum_n \frac{c_n}{d-2n} S_{L,n}\,.\ee
It is interesting that it can have (anti)-de Sitter spaces with different cosmological constants as solutions. Indeed, assuming that the torsion is zero and taking the variation with respect to $e^a$
we get
\be \delta S=\sum c_n \int F^{a,b}\wedge...\wedge F^{c,d}\wedge \delta e^f\wedge e^g\wedge...\wedge e^u\epsilon_{abcdfg...u}\,.\ee
In looking for the (anti)-de Sitter type solutions we replace $F^{a,b}$ with $\Lambda e^a\wedge e^b$ to find
\be 0=\sum c_n \Lambda^n=(\Lambda-\Lambda_1)...(\Lambda-\Lambda_n)\,.\ee
Other possibilities include Yang-Mills-type action
\be S_{L,n}= \int F^{a,b}_{\mm\nn}  F^{c,d}_{\rr\kk} \eta_{ac}\eta_{bd} g^{\mm\rr} g^{\nn\kk}=
\int tr(F_{\mm\nn}  F_{\rr\kk}) g^{\mm\rr} g^{\nn\kk}\,,\label{YMGravity}\ee
which turns out to be higher-derivative (it is not topological or of Lovelock type) and does not lead to the Einstein-Hilbert action.

\mysubsection{MacDowell-Mansouri-Stelle-West}\label{extra:MMSW}
That (anti)-de Sitter algebra ($so(d-1,2)$) $so(d,1)$ is semi-simple as compared to the Poincare one allows us to improve on the interpretation of gravity as a gauge theory even further. Recall that the curvature corresponding to $L_{ab}$ generators of the (anti)-de Sitter algebra, \eqref{dsadspoincare}, is $R^{a,b}$ and $R^{a,b}=F^{a,b}-\Lambda e^a\wedge e^b$, where $F^{a,b}$ is the $L_{ab}$-part of the curvature for the Poincare algebra, it is related to the Riemann tensor. It was the observation by MacDowell and Mansouri, \cite{MacDowell:1977jt} that the following $4d$-action
\begin{align}
S_{MM}&=-\frac1{2\Lambda}\int R^{a,b}\wedge R^{c,d} \epsilon_{abcd}
\end{align}
is equivalent to the Einstein-Hilbert one with the cosmological term. Indeed, expanding $R=F-\Lambda e e$ we find
\begin{align}
S_{MM}&=\int \left(-\frac1{2\Lambda}F^{a,b}\wedge F^{c,d}+  F^{a,b}\wedge e^c\wedge e^d-\frac{\Lambda}{2} e^a\wedge e^b\wedge e^c\wedge e^d\right)\epsilon_{abcd}\,.
\end{align}
The first term is of Lovelock type, \eqref{lovelockn}, and in fact topological in $4d$
and hence does not contribute to the equations of motion. The last
two terms sum up to \eqref{FrameLikeActionC} with a slightly
different cosmological constant, $\Lambda\rightarrow\Lambda/2$. Note that $\Lambda\rightarrow0$
limit is singular in the action but not in the equations of
motion since the singularity multiplies the topological term.

The MacDowell-Mansouri action looks similar to the Yang-Mills one, now it is quadratic in the field-strength, but it is not exactly of the Yang-Mills form, \eqref{YMGravity}. It can be formally rewritten as
\begin{align}
S_{MM}&=\int R^{a,b}\wedge R^{c,d} \epsilon_{abcd5}\,,\label{MMB}
\end{align}
where we introduced the $5d$-epsilon symbol and
$\epsilon_{abcd}=\epsilon_{abcd5}$. One may wish to do that in
order to keep the symmetries of the most symmetric solution, ($so(3,2)$) $so(4,1)$, which is
the (anti)-de Sitter space, so it has something to do with rotations in
$5d$. It does not look satisfactory yet as one would like to make
all symmetries manifest and the presence of $5$ in
$\epsilon_{abcd5}$ breaks down the symmetry. This can be fixed
with a little more work.

First of all, since (anti)-de Sitter algebra ($so(d-1,2)$) $so(d,1)$ is the algebra of infinitesimal rotations, it is convenient not to split generators into $L_{ab}$ and $P_a$, which already breaks the anti-de Sitter symmetry down to the Lorentz one. To accomplish this, analogously to the Lorentz generators $L_{ab}$ themselves, we define $T_{\aAs\aBs}=-T_{\aBs\aAs}$, where capital Latin indices $\aAs,\aBs,...$ run over $d+1$ values, $\aAs=\{a,5\}$, where $5$ refers to the extra $(d+1)$-th direction as compared to the Lorentz algebra. The generators $T_{\aAs\aBs}$ obey
\begin{align}
[T_{\aAs\aBs},T_{\aCs\aDs}]=T_{\aAs \aDs}\eta_{\aBs\aCs}-T_{\aBs\aDs}\eta_{\aAs \aCs}-T_{\aAs \aCs}\eta_{\aBs\aDs}+T_{\aBs\aCs}\eta_{\aAs \aDs}\,,
\end{align}
where $\eta_{\aAs\aBs}$ is the ($so(d-1,2)$) $so(d,1)$ invariant metric.
Lorentz-covariant formulas \eqref{dsadspoincare} can be recovered
upon identifying $L_{ab}$ and $P_a$ with $T_{ab}$ and $T_{a5}$. Not surprisingly,
if we define the Yang-Mills connection $\Omega=\frac12
\Omega^{\aAs,\aBs}T_{\aAs\aBs}$ of the (anti)-de Sitter algebra then the
curvature
\begin{align}
R^{\aAs,\aBs}=d\Omega^{\aAs,\aBs}+\Omega\fud{\aAs,}{\aCs}\wedge \Omega^{\aCs,\aBs}
\end{align}
reduces to \eqref{adsLorentzCurvatures} for $R^{a,b}$ and $R^{a,5}$ as well as $\delta \Omega^{\aAs,\aBs}=D\epsilon^{\aAs,\aBs}$ reduces to \eqref{adsLorentzGauge}.
In particular the zero-curvature equation
\begin{align}\label{zc}
d\Omega^{\aAs,\aBs}+\Omega\fud{\aAs,}{\aCs}\wedge \Omega^{\aCs,\aBs}&=0
\end{align}
describes (anti)-de Sitter space analogously to
\eqref{adsMetricLike}-\eqref{adsFrameLike} provided that we
specified the way of splitting $\Omega^{\aAs,\aBs}$ into vielbein $e^a$
and spin-connection $\omega^{a,b}$, e.g. as $\Omega^{a,5}$ and
$\Omega^{a,b}$, and required the vielbein to be nondegenerate. Let
us note that \eqref{zc} being a certain flatness condition
generates its solution space locally in a pure gauge form. Indeed,
one can easily check that any function $g=g(T_{\aAs\aBs}|x)$ with values in the (anti)-de Sitter group gives rise to
a solution\footnote{The expression is somewhat formal. For example, $g(T|x)$ can be taken to be the usual $\exp$-map from a Lie algebra to certain neighborhood of the unit element of the group. With the help of $g^{-1}g=1$ one checks that $d\Omega+\Omega\Omega=0$, where the product $\Omega\Omega$ is the usual group product since $\Omega$ are given in terms of group element $g$.}
\be
\Omega=g^{-1}(T|x)dg(T|x)\,.
\ee
This fact might suggest an erroneous interpretation, namely that
all such solutions are gauge equivalent to $\Omega=0$ and,
particularly, one can ``gauge away'' the $AdS$ space-time itself.
While formally it seems to be the case, this argument suffers from
a flaw that makes the vielbein vanish. Recall that only
nondegenerate frame fields admit physical interpretation. This is
another illustration that gauge formalism applied to pure gravity
should be used with great care.

We would like to extend \eqref{MMB} to any $d$ and rewrite it in manifestly (anti)-de Sitter invariant form. A natural extension seems to be of the form, \cite{Vasiliev:2001wa},
\begin{align}
S&=\int R^{a,b}\wedge R^{c,d} \wedge e^f\wedge ...e^u \epsilon_{abcdf...u 5}\label{MMBa}
\end{align}
where $5$ again means $d+1$. However, the Lovelock term $\int
FFe...e$ is not topological in $d>4$, which brings in nonlinear
corrections to the Einstein equations.  It is hard to tell if
these corrections, however beautiful, are phenomenologically acceptable since we live,
at least effectively, in $4d$.

We rewrite \eqref{MMBa}  by using as many uppercase indices as we can
\begin{align}
S&=\int R^{\aAs,\aBs}\wedge R^{\aCs,\aDs} \wedge e^f\wedge ...e^u \epsilon_{\aAs \aBs\aCs\aDs f...u 5}\,.
\end{align}
Since the value $5$ is already occupied $\aAs,\aBs,\aCs,\aDs$ are constrained
to the Lorentz index range simply because two $5$ indices cannot
appear simultaneously in the $\epsilon$-symbol. Formally we can
extend $e^a$ to the extra direction too, e.g. to define $E^\aAs$ such
that $E^a=e^a$ and $E^5=0$ since $E^5$ does not contribute to the
action anyway. Then we get
\begin{align}
S&=\int R^{\aAs,\aBs}\wedge R^{\aCs,\aDs} \wedge E^\aFs\wedge ...E^\aUs \epsilon_{\aAs \aBs\aCs\aDs \aFs...\aUs 5}\,.
\end{align}
The anti-de Sitter symmetry is still explicitly broken by $5$ in
the $\epsilon$-symbol and by embedding of $e^a$ into $E^\aAs$. A
natural way out is to think of $5$ as of the vacuum expectation of
some compensator vector field, \cite{Stelle:1979aj}, say $V^\aAs$,
and we have $V^\aAs=\delta^\aAs_5$. Now it is better,
\begin{align}
S_{MMSW}&=\int R^{\aAs,\aBs}\wedge R^{\aCs,\aDs} \wedge E^\aFs\wedge ...E^\aUs V^\aWs \epsilon_{\aAs \aBs\aCs\aDs\aFs...\aUs \aWs}\label{MMSW}\,.
\end{align}
We can make all definitions $5$-independent by using $V^\aAs$
that has to have some vacuum value, e.g. $\delta^\aAs_5$. We can
force $V^\aAs$ to be non-zero by imposing the following constraint \be
V^\aAs V^\aBs\eta_{\aAs\aBs}=\sigma\,,\ee where $\sigma=1$ for de Sitter and
$\sigma=-1$ for anti-de Sitter. We refer to the choice
$V^\aAs=\delta^\aAs_5$ as the {\it standard gauge}.

Introducing a new object $V^\aAs$ may seem to be too naive --- for
some reason we have to restrict ourselves to the range $a$ rather
than the full range $\aAs$, this can be always achieved by
introducing a (number of) vectors $V^\aAs$ whose purpose is to span
the extra directions of $\aAs$, so we  just have a split
$\mathbb{R}^{n+m}$ as $\mathbb{R}^{n}\oplus\mathbb{R}^{m}$ where
(a number of) $V^\aAs$'s span the basis of $\mathbb{R}^{m}$ (in our
case $n=d$, $m=1$). Such restoration of 'broken symmetry' can always
be achieved. In particular instead of extending $so(d-1,1)$ to
(anti)-de Sitter algebra we could choose $so(d+m)$ (the signature
is immaterial now) with $m>1$, which require $m$ linearly
independent $V^\aAs$'s to be introduced. The reason not to go beyond
the (anti)-de Sitter algebra is our desire to study field theory
over (anti)-de Sitter space and we have no evidence so far that a
larger symmetry that contains $so(d-1,2)$ or $so(d,1)$ is present
in the theory. From this perspective it would be great to have
(anti)-de Sitter symmetry manifest. Actually, we find in Section \ref{sec:HSAlgebra}
that a larger symmetry does exists and
acts on the infinite multiplet of fields of all spins, but not a spin-two alone. But there
are no signs of higher symmetry when dealing with pure
gravity\footnote{In the vicinity of cosmological singularity the
Einstein equations were shown to exhibit a higher symmetry, which
is a certain infinite-dimensional affine algebra of hyperbolic
type, see, e.g, \cite{Damour:2002et}. The question of whether this
symmetry acts in the gravity in higher orders and far from the
singularity remains open.}. Indeed, (anti)-de Sitter and Minkowski
are known to be maximally-symmetric backgrounds. Therefore, just single $V^\aAs$ should be enough.

We need some $E^\aAs=E^\aAs_\mm dx^\mm$ such that $E^5=0$, $E^a=e^a$ and
$e^a=\Omega^{a,5}$ in the standard gauge. We can state
this as $E^\aAs=\Omega^{\aAs,\aBs}V_\aBs$ but it is better to use
\be E^\aAs=\DO V^\aAs=dV^\aAs+\Omega\fud{\aAs,}{\aBs}V^\aBs\,,\ee
because it transforms covariantly.
Now all constituents of \eqref{MMSW} are well-defined. Action
\eqref{MMSW} is manifestly invariant under local (anti)-de Sitter
transformations, i.e. Lorentz plus translations, and diffeomorphisms if
we assume that the compensator transforms as a fiber vector
\begin{align}\delta \Omega^{\aAs,\aBs}&=\DO \epsilon^{\aAs,\aBs}\,,& \delta V^\aAs&=-\epsilon\fud{\aAs,}{\aBs}V^\aBs\,, & \delta E^\aAs&=-\epsilon\fud{\aAs,}{\aBs}E^\aBs\,.\end{align}
The local (anti)-de Sitter invariance follows from the fact that
all the (anti)-de Sitter indices in \eqref{MMSW} are contracted
using one $\eta_{\aAs\aBs}$ or another invariant
tensor $\epsilon_{\aAs B...U}$ of the (anti)-de Sitter algebra and all the constituents transform covariantly. The diffeomorphism
invariance is explicit thanks to differential forms.

Our previous experience shows that having local (anti)-de Sitter
transformations and diffeomorphisms simultaneously is too much.
Recall that \eqref{FrameLikeActionB} was not invariant under local
translations unless torsion is zero and at vanishing torsion the
local translations are identical to diffeomorphisms when acting on
the frame field $e^a$. Hence we expect $d+d(d-1)/2=d(d+1)/2$ local gauge
parameters in total.

The extension of the local symmetry algebra arises due to the
presence of the compensator field. As we already mentioned, given
an action/equations of motion that are invariant under local
transformations belonging to some algebra $\mathfrak{h}$ and are
not invariant under a bigger algebra
$\mathfrak{g}\supset\mathfrak{h}$, it is always possible to
restore $\mathfrak{g}$-symmetry by introducing new fields that
transform in such a way as to compensate for
$\mathfrak{g}/\mathfrak{h}$-noninvariance. Of course, this does
not mean that the theory is invariant under $\mathfrak{g}$. A
simple example was given above where we could try to extend the
local Lorentz symmetry to any $so(d+m)$ with $m>1$ by introducing
$m$ compensator fields.

The genuine symmetry is the stability
algebra of the compensator field. The condition for the
compensator to remain invariant reads
\begin{align}
(\mathcal{L}_\xi+\delta_\epsilon)V^\aAs=0\label{stabV}\,,
\end{align}
where we employed both the diffeomorphisms and the local (anti)-de Sitter symmetry. In components it reads
\begin{align}
\xi^\mm \pl_\mm V^\aAs-\epsilon\fud{\aAs,}{\aBs}V^\aBs=0\,,
\end{align}
equivalently it can be rewritten as
\begin{align}
i_\xi E^\aAs-(i_\xi\Omega\fud{\aAs,}{\aBs}+ \epsilon\fud{\aAs,}{\aBs})V^\aBs=0\,,
\end{align}
It is instructive to look at this condition in the {\it standard gauge} $V^\aAs=\delta^\aAs_5$,
\begin{align}
\epsilon\fud{\aAs,}{\aBs}V^\aBs=\epsilon\fud{\aAs,}{5}=0\,.
\end{align}
The stability condition for $V^\aAs$ kills $\epsilon^{a,5}$
components, reducing the local symmetry group to diffeomorphisms
and Lorentz rotations with $\epsilon^{a,b}$. We are left with the
expected amount of symmetry, i.e. diffeomorphisms and local
Lorentz transformations. Hence, in the {\it standard gauge} we
roll back to \eqref{MMBa}.

One can choose a nonconstant compensator as well. In the latter
case \eqref{stabV} starts to mix $\xi^\mm$ and local translations
$\epsilon\fud{\aAs,}{\aBs}V^\aBs$. Our physical world (in the tangent space)
is to be identified with the subspace orthogonal to the
compensator. In particular we have to ensure that all the
quantities, e.g. the vielbein and spin-connection, do not feel the
presence of a compensator. Firstly, note that the generalized vielbein is still effectively a $d\times d$ matrix thanks to
\be E^\aAs V_\aAs\equiv0\,,\ee
where we used $V^\aAs d V_\aAs=0$ and the fact that $\Omega^{\aAs,\aBs}$ is
antisymmetric. What is the generalized spin-connection? It can be
defined as follows, \cite{Vasiliev:2001wa},
\begin{align} \Omega_L{}^{\aAs,\aBs}&=\Omega^{\aAs,\aBs}-\frac1{V^2}(E^\aAs V^\aBs-E^\aBs V^\aAs)\,, &\DL&=d+\Omega_L\,.\end{align}
It has some nice properties, which fix its form completely,
\begin{align}
D V^\aAs&\equiv0\,, &D E^\aAs&\equiv0\,,
\end{align}
i.e. the compensator $V^\aAs$ is insensible with respect to the
physical covariant derivative, which is $\DL$. The generalized
vielbein is covariantly constant, which is an analog of $De^a=0$.
In the {\it standard gauge} one recovers
$\Omega_L^{a,b}=\Omega^{a,b}$.

The meaning of the compensator field can be understood from the
following picture. The (anti)-de Sitter space is replaced with the
sphere for simplicity reason. Thinking of AdS or sphere as $d$-dimensional hypersurface
embedded into $(d+1)$-dimensional (ambient) space, there is another natural gauge for
the compensator, which is to make it lie along the radial direction. The stability algebra of the
compensator is then the algebra of rotations in the plane orthogonal to
$V^\aAs$ and tangent to the sphere --- it is an equivalent of the Lorentz algebra.
\putfigureifpdf{\AmbientSpace}

\mysubsection{Interplay between diffeomorphisms and gauge symmetries}
\label{extra:semidirect}
A generic feature of gauge formulation is that diffeomorphisms
are entangled with gauge transformations and hence we cannot treat
them separately. When considering gauge fields on manifolds we
have to deal with the group of
diffeomorphisms\footnote{Diffeomorphisms deliver a very difficult
infinite-dimensional group to work with.} and group of local gauge transformations. It
is easier to talk about the corresponding Lie algebras. The
diffeomorphisms $diff(M)$ as a Lie algebra are given by vector
fields $\xi^\mm\in diff(M)$ with the Lie commutator defined to be
the Lie bracket of vector fields
$[\xi,\eta]^\mm=\xi^\nn\pl_\nn\eta^\mm-\eta^\nn\pl_\nn\xi^\mm$.
Gauge parameters, say $\epsilon^\aI(x)$, are the maps from the
given manifold to some Lie algebra. Call this linear space of maps $\mathfrak{h}$.
The bracket is given by the Lie algebra commutator, i.e.
$[\epsilon, \theta]^\aI(x)= f^\aI_{\aJ\aK}\epsilon^\aJ(x)
\theta^\aK(x)$.

Gauge parameters are not left intact by
$diff(M)$, rather they transform as a number of scalars under
diffeomorphisms, which affect $x$ but not the Lie algebra
index $\aI$. Therefore, the elements of the gauge algebra are
affected by diffeomorphisms.  Indeed, the Lie derivative obeys the Leibnitz law, i.e. $\mathcal{L}_\xi (f(x)g(x))=(\mathcal{L}_\xi f(x)) g(x)+f(x)\mathcal{L}_\xi g(x)$, where $f(x)$, $g(x)$ are two functions and $\mathcal{L}_\xi f(x)=\xi^\mm\pl_\mm f(x)$. Then, it is obvious that \be\mathcal{L}_\xi([\epsilon, \theta]^\aI(x))=[\mathcal{L}_\xi(\epsilon), \theta]^\aI(x)+[\epsilon, \mathcal{L}_\xi(\theta)]^\aI(x)\,,\ee
which means that $\mathcal{L}_\xi$ acts as a derivation on $\mathfrak{h}$. The action of vector fields on gauge parameters, i.e. the
infinitesimal change of coordinates on $M$, can be understood as
the homomorphism $diff(M)\rightarrow Der(\mathfrak{h})$. Hence pairs $(\xi^\mm,
\epsilon^\aI)$ belong to the semidirect product $diff(M)\ltimes
\mathfrak{h}$ and the full algebra of symmetries is the semidirect product of diffeomorphisms and gauge transformations\footnote{There are simpler finite-dimensional
examples, the affine group and the Poincare group. The affine group
$Aff(d)$ consists of pairs $(A,a)$ with $A$ belonging to $GL(d)$
and $a$ being a $d$-dimensional vector. The group law
$(A,a)\circ(B,b)=(AB, Ab+a)$ can be read off the action on vectors
$(A,a)(\vec{x})=A\vec{x}+a$. It resembles the semidirect origin of
the affine group $Aff(d)=GL(d)\ltimes T_d$, where $T_d$ is the
group of translations in $d$ dimensions, $a\circ b=a+b$, $a,b\in
\mathbb{R}^d$.

Going to the Lie algebras of affine or Poincare group we find $[(A,a), (B,b)]=([A,B],Ab-Ba)$. Note that $A,B$ are now any $d\times d$ matrices, not necessary invertible. The translations play a role analogous to that of 'gauge symmetries' in the sense that they are affected by coordinate transformations associated with $A$. Clearly the vector of translation has to be transformed in accordance with the change of the basis induced by $A$.}.

Given the semidirect structure of the full gauge symmetry algebra it is easy to see that the two types of symmetries do not commute with the interesting part of the commutator residing in the gauge algebra sector,
\be [(\xi_1, \epsilon_1),(\xi_2, \epsilon_2)]=([\xi_1,\xi_2], [\epsilon_1,\epsilon_2]+\mathcal{L}_{\xi_1}\epsilon_2-\mathcal{L}_{\xi_2}\epsilon_1)\,.\ee

As we already know, when the Yang-Mills curvature vanishes, every diffeomorphism can be represented as a gauge transformation with the help of
\be \mathcal{L}_\xi A=D (i_\xi A) +i_\xi F(A)\,. \ee
The gauge algebra at a point (just the algebra we gauge itself) can be larger or smaller than $d$, which is the dimension of a vector field, $\xi^\mm(x)$.
When the gauge algebra is smaller we cannot identify diffeomorphisms with gauge transformations without having to trivialize the dynamics of gauge fields due to the necessity of $F(A)=0$. For example, this is true for $u(1)$ gauge field, $A_\mm$. When the gauge algebra is larger one may set only part of the curvature to zero. For example, vanishing of the torsion allows us to treat diffeomorphisms as the subalgebra of local gauge transformations. Note however, that this is true only for the part of the gauge field for which the curvature vanishes, i.e. the vielbein, while the action of a diffeomorphism on the spin-connection is a sum of a gauge transformation and a curvature piece.

Let us consider an example\footnote{We are grateful to T.Nutma and M.Taronna for discussions.}. The standard gauge transformation law of a spin-one gauge field $A\equiv A_\mm dx^\mm$ under gauge symmetries with $\epsilon(x)$ and
diffeomorphisms with $\xi^\mm(x)$ reads
\be\delta_{\xi,\epsilon}A=d\epsilon+ \mathcal{L}_\xi A\ee
The commutator of two transformations is a gauge transformation of the same type, which leads to the following algebra,
\be\label{comma}[\delta_{\xi',\epsilon'},\delta_{\xi,\epsilon}]=\delta_{\mathcal{L}_{[\xi',\xi]}, \mathcal{L}_{\xi'}\epsilon-\mathcal{L}_{\xi}\epsilon'}\ee
This form of commutator is typical of the semidirect product. Within field theory we allow for redefinitions of fields and gauge parameters,
\besubeqs\begin{align}
\xi&\rightarrow\xi+g(\xi,\pl\xi,...;\phi,\pl\phi,...)\,,\\
\phi&\rightarrow\phi+f(\phi,\pl\phi,...)\,,
\end{align}\esubeqs
where the redefinition of the gauge parameters can be field-dependent and involve derivatives of the gauge parameters and fields. Such redefinitions can destroy the nice Lie algebra structure found above, but they are allowed in the realm of field theory, \cite{Berends:1984rq}. There is a redefinition that makes the commutator look almost like a direct product. Indeed, inserting $\epsilon-i_\xi A$ instead of $\epsilon$ one finds
\be\hat{\delta}_{\xi,\epsilon}A=d\epsilon+ i_\xi F(A)\,,\ee
where $F(A)=dA$. The commutator now reads
\be[\hat{\delta}_{\xi',\epsilon'},\hat{\delta}_{\xi,\epsilon}]=\hat{\delta}_{{[\xi',\xi]}, i_\xi i_{\xi'} F(A)}\,.\ee
The diffeomorphism algebra can be seen in the first argument through the Lie commutator. Miraculously, the gauge parameters $\epsilon'$, $\epsilon$ disappeared on the right-hand side. So the commutator of diffeomorphisms and gauge transformations vanishes now. Unfortunately, we are not able to apply the Lie algebra terminology anymore. Firstly, the 'structure constants' became field-dependent via $i_\xi i_{\xi'} F(A)$. Secondly, diffeomorphisms do still contribute to the sector of gauge transformations via the same $i_\xi i_{\xi'} F(A)$.

Therefore, even ignoring the field-dependence of the structure constants we cannot say that the algebra is a direct product. Though we can say that the deformation of the gauge transformations induced by coupling of $A$ to gravity is abelian in a sense that there are no $\epsilon-\xi$-mixing terms in the commutator (but we cannot say that it it a direct product in any sense), which is not however the case in \eqref{comma}.

Consequently, within the Lie theory the coupling is non-abelian while within the more general framework the coupling is abelian. It is a general situation that the 'structure constants' may not be constant and can be field-dependent, so there is no way to interpret it in the language of Lie algebras directly. It is the case for the higher-spin theory. It turns out that with the help of the unfolded approach, one can assign representation theory meaning to such cases as well, see Sections \ref{sec:Unfolding} and \ref{app:ChevalleyEilenberg}. The unfolded approach disentangles nontrivial algebraic structure of field theory and diffeomorphisms. A good example of the theory where 'structure constants' are field dependent is provided by an attempt to rewrite $3d$ higher-spin theory in terms of Fronsdal fields, \cite{Campoleoni:2012hp}.

\mysubsection{Chevalley-Eilenberg cohomology and interactions}\label{app:ChevalleyEilenberg}
\paragraph{Note on cohomologies of Lie algebra.} Before going deep into the structure constants we would like to remind the basic definitions from the theory of Lie algebras. Given a Lie algebra $\mathfrak{g}$ together with some its representation $V$ we can construct a cochain complex as follows. $k$-chains are elements of $V\otimes \mathfrak{g}^*\wedge...\wedge \mathfrak{g}^*$, where $\mathfrak{g}^*$ is a dual of $\mathfrak{g}$ as a vector space and $\wedge$ denotes the exterior power of $\mathfrak{g}^*$ as a vector space. So $k$-chain $C^k$ is a skew-symmetric functional of $k$ elements from $\mathfrak{g}$ with values in $V$
\be \notag C^k: \mathfrak{g}\wedge...\wedge \mathfrak{g}\longrightarrow V\,,\ee
\be c\in C^k, \qquad\qquad c(a_1,...,a_k)\in V, \qquad\qquad a_i\in \mathfrak{g}\,,\notag\ee
and $c(a_1,...,a_k)$ is antisymmetric in its arguments. The differential $d_k$ takes $C^k$ to $C^{k+1}$
\begin{align*}
(d_k c)(a_1,...,a_{k+1})&=\sum_i (-)^i a_i c(a_1,...,\hat{a}_i,...,a_{k+1})+\\ &+\sum_{i<j} (-)^{i+j+1}c([a_i,a_j],a_1,...,\hat{a}_i,...,\hat{a}_j,...,a_{k+1})\,,
\end{align*}
where $a_i c(...)$ means that $a_i$ acts on the value of $c(...)$
since it belongs to $V$, which is a $\mathfrak{g}$-module.
Therefore, given a skew-symmetric functional in $k$ variables from
$\mathfrak{g}$ we can construct a skew-symmetric functional in
$k+1$ variables. One can check that $d_q\circ d_{q-1}=0$ so
cohomology groups are defined in a standard way as
\be H^q=Ker\, d_q / Im\, d_{q-1}\,.\ee
To compute $H^q$ we have to find a general solution to $d_q c(a_1,...,a_q)=0$ for $c\in C^q$ and identify two solutions if they differ by $(d_{q-1} b)(a_1,...,a_q)$ for some $b\in C^{q-1}$. Classes $H^q$ are called $\mathfrak{g}$-cohomology with values in $V$.


\paragraph{Cocyles and couplings.}\label{sec:ChevalleyEilenberg}
Suppose we have an unfolded system with underlying Lie algebra
$\mathfrak{g}$, which is defined via equations
\eqref{UnfldFlatness} for $\Omega^I$. We filter the rest of the
fields by their form degree, i.e. we have sectors of $p$ forms,
$q$ forms etc., and without loss of generality we assume that
$p$-forms $W^\aAs\fm{p}$ take values in some,
perhaps trivial, $\mathfrak{g}$-module defined by the structure
constants of one-forms via \eqref{UnfldCovariantConstancy}, {\it
idem} for $q$-forms etc. Consider the most general unfolded
equations that are still linear in $W^\aAs\fm{p}$, but can be nonlinear in $\Omega^I$, i.e. we think of $\Omega^I$ as the vacuum and consider linear perturbation over it.
So the unfolded equations for $W^\aAs\fm{p}$ have the form of a
covariant constancy condition \eqref{UnfldCovariantConstancy}
whose \rhs can be sourced by other fields.

Suppose $W^\aA\fm{p}$ take values in $\mathfrak{g}$-module $\module_1$. The simplest option, which we never find in practice, is to write ($p+1=k$)
\begin{align} \label{UnfldCovconst}
D_\Omega W^\aA\fm{p} &\equiv d W^\aA\fm{p}+{f_I}^\aA_{\phantom{\aA}\aB}\,\Omega^I\wedge W^\aB\fm{p}={f_{I_1...I_k}}^\aA\, \Omega^{I_1}\wedge...\wedge\Omega^{I_k}\,.
\end{align}
The \rhs can be thought of as a skew-symmetric functional $f^\aA(\Omega,...,\Omega)$ with values in $\module_1$. The integrability \eqref{UnfldJacobi} implies that
\be {f_I}^\aA_{\phantom{\aA}\aB}\,\Omega^I f^\aA(\Omega,...,\Omega)=k f^\aA(f^I_{JK}\Omega^J\wedge\Omega^K,...,\Omega)\,,\ee
i.e. $f^\aA(\Omega,...,\Omega)$ is the cocycle of $\mathfrak{g}$ with values in $\module_1$. It is a cocycle but not necessary nontrivial, i.e. not of the form $d b_{k-1}$ in the notation as above. The trivial cocyles can be removed by field redefinitions, \cite{Vasiliev:2007yc}. Therefore, we see that nontrivial \rhs are in one-to-one correspondence with degree-$k$ cohomology of $\mathfrak{g}$ with values in $\module_1$, i.e. are given by $H^k$.

More realistically, suppose we have two sectors $W^\aA\fm{p}$ and $W^\aAs\fm{q}$ of $p$- and $q$-forms, each taking values in some $\mathfrak{g}$-modules, say $\module_1$ and $\module_2$. Assuming $p+1=q+k$ the most general equations that are linear in $W^\aA\fm{p}$ and $W^\aAs\fm{q}$ read
\besubeqs\begin{align} \label{UnfldCocycle}
D_\Omega W^\aA\fm{p} &\equiv d W^\aA\fm{p}+{f_I}^\aA_{\phantom{\aA}\aB}\,\Omega^I\wedge W^\aB\fm{p}=
{f_{I_1...I_k}}^\aA_{\phantom{\aA}\aBs}\, \Omega^{I_1}\wedge...\Omega^{I_k}\wedge W^\aBs\fm{q}\,,\\
D_\Omega W^\aAs\fm{q} &\equiv d W^\aAs\fm{q}+{f_I}^\aAs_{\phantom{\aAs}\aBs}\,\Omega^I\wedge W^\aBs\fm{q}=...\,,
\end{align}\esubeqs
and we ignore the possible \rhs in the second equation that will have analogous interpretation. The integrability \eqref{UnfldJacobi} implies that ${f_I}^\aA_{\phantom{\aA}\aB}(\Omega,...,\Omega)$ is a $\mathfrak{g}$-cocycle with values in $\module_1\otimes \module_2^*$, where $\module_2^*$ is the dual of $\module_2$.

More generally, nonlinear deformations can look like
\begin{align*}
D_\Omega W^\aA\fm{p} &\equiv d W^\aA\fm{p}+{f_I}^\aA_{\phantom{\aA}\aB}\,\Omega^I\wedge W^\aB\fm{p}=
{f_{I_1...I_k}}^\aA_{\phantom{\aA}\aBs_1...\aBs_m}\, \Omega^{I_1}\wedge...\Omega^{I_k}\wedge W^{\aBs_1}\fm{q_1}\wedge ...\wedge W^{\aBs_m}\fm{q_m}\,,
\end{align*}
where $p$-form $W^\aA\fm{p}$ with values in some $\module$ is sourced by a number of other forms $W^{\aBs_1}\fm{q_1}$ with values in $\module_1$,..., $\module_m$. The integrability condition implies that ${f_{I_1...I_k}}^\aA_{\phantom{\aA}\aBs_1...\aBs_m}$ is a cocycle with values in $\module\otimes\module_1^*\otimes...\otimes\module_m^*$.

\mysubsection{Universal enveloping realization of HS algebra}\label{sec:UEA}
Since everything below applies equally well to $sp(2n)$ for any $n>1$, while we need $sp(4)$, we shall consider $sp(2n)$ with invariant metric $\epsilon_{AB}$. The starting point is the $sp(2n)$ commutation relations ($\star$ will denote the product in $U(sp(2n))$\,)
\be[T_{AB},T_{CD}]_\star=T_{AD}\epsilon_{BC}+T_{BD}\epsilon_{AC}+T_{AC}\epsilon_{BD}+
T_{BC}\epsilon_{AD}\,.\label{spfour}\ee
Consider the universal enveloping algebra $U(sp(2n))$ of $sp(2n)$, i.e. the algebra of all polynomials $P(T_{AB})$ in the generators $T_{AB}$ modulo the commutation relations \eqref{spfour} of $sp(2n)$. It is easy to work out first several levels by decomposing the Taylor coefficients
\be P(T_{AB})=P_0 +P_1^{AB}T_{AB}+P_2^{AB|CD}T_{AB}\star T_{CD}+...\ee
into $sp(2n)$ irreducible tensors. One finds
\begin{align}
U(sp(2n))=\underbrace{\bullet}_0\oplus\underbrace{\left(\parbox{20pt}{\YoungB}\,\right)}_1\oplus\underbrace{\left(\parbox{20pt}{\YoungBB}\oplus
\parbox{40pt}{\YoungD}\oplus\parbox{10pt}{\YoungAA}\oplus\bullet\right)\,}_2\oplus...
\end{align}
The lowest component is the unit $\mathbf{1}$ of $U(sp(2n))$, which is an $sp(2n)$ singlet. At the first level we find just $T_{AB}$. At the second level there is a number of components: the singlet $\bullet$ is the Casimir operator $C_2=-\frac12 T_{AB}\star T^{AB}$, it corresponds to $P_2^{AB|CD}=\epsilon^{AC}\epsilon^{BD}+\epsilon^{BC}\epsilon^{AD}$; $\parbox{40pt}{\YoungD}$ is a totally symmetric $T_{AA}\star T_{AA}$ and automatically traceless since $\epsilon^{AB}$ is antisymmetric; the antisymmetric component is
\be\parbox{10pt}{\YoungAA}=\frac12 \{T_{AB},T\fdu{C}{B}\}_\star+\frac1{n} \epsilon_{AC} C_2\,.\ee
There is also a window component, $\smallpic{\YoungpBB}$.

In the case of the HS theory in $AdS_4$ we need various totally symmetric $sp(4,\mathbb{R})$-tensor of even ranks, which are equivalent to $so(3,2)$-tensors with the symmetry of two-row rectangular Young diagram,
\be so(5): \parbox{40pt}{\RectBRow{4}{4}{$s-1$}{}} \qquad \Longleftrightarrow\qquad sp(4): \parbox{40pt}{\RectARow{5}{$2s$}}\ee

In the HS algebra we are looking for every symmetric $sp(4)$-tensor of even rank (generators for bosonic fields) must appear once.
Already at the second level there are unwanted tensor types. The window component and the rank-two antisymmetric simply do not have the type we need. The singlet component will lead to proliferation of fields. Indeed, given some generator $T$ and a singlet $C$, for any $k$ the monomial $T \star C^k$ is
a generator again.

To get rid of the unwanted diagrams one can define a two sided ideal as follows
\begin{align}
I_\lambda=U(sp(2n))\otimes \left(\parbox{10pt}{\YoungAA}\oplus\parbox{20pt}{\YoungBB}\oplus(C_2-\lambda \mathbf{1})\right)\otimes U(sp(2n))
\end{align}
and then try to define the HS algebra $\mathfrak{g}$ as the quotient
\be\mathfrak{g}=U(sp(2n))/I_\lambda\,,\ee
with the hope that $\mathfrak{g}$ is nontrivial and free from unwanted diagrams.
Note that while we have to get rid of certain nonsinglet components the singlet generator, which is the Casimir, does not necessary have to be put to zero, rather it can be made equivalent to the unit element. As we see soon, $\lambda$ turns out to be fixed by the self-consistency of the procedure.

In practice it means that we try to quotient out or to put to zero certain generators within the universal enveloping algebra.
However, $U(sp(2n))$ is not a free algebra, so at some point we can come to a contradiction that requiring certain generators to be zero entail via the commutation relations that $T_{AB}\sim0$ and hence the quotient algebra is empty.

Indeed, let us look at the relations among the low lying  generators. First, we combine the window component and
$\parbox{10pt}{\YoungAA}$ into a single $I_{UU,VV}$
\begin{align}
I_{UU,VV}=T_{UV}\star T_{UV}+\gamma C_2\left(\epsilon_{UU}\epsilon_{VV}+\frac14\epsilon_{UV}\epsilon_{UV}\right)\,,
\end{align}
where the anti-symmetrization over $UU$ and $VV$ is implied wherever necessary. $I_{UU,VV}$ has the symmetry of the window Young diagram but it is not completely traceless, containing $\parbox{10pt}{\YoungAA}$ as a trace. The coefficient $\gamma=4/(n(2n+1))$ is fixed by requiring
$I_{UU,VV}$ not to contain $C_2$, i.e. $\epsilon^{UU}\epsilon^{VV}I_{UU,VV}=0$. Using the commutation relations we find the following identities
\besubeqs\begin{align}
[T_{AB},T\fdu{C}{B}]_\star&=2(n+1)T_{AC}\,,\\
T_{AB}\star T\fdu{C}{B}&=\alpha\, T_{AC}+\beta\, \epsilon_{AC} C_2+\parbox{10pt}{\YoungAA}\,, &&\alpha=n+1\,, &&\beta=-\frac1n \,,\\
[T_{AB},T_{CD}]_\star\star T^{AB}&=4(n+1)T_{CD}\,.
\end{align}\esubeqs
Now we would like to study the self-consistency of the procedure. Since $I_{UU,VV}$ belongs to the ideal,
any element of $U(sp(2n))\otimes I_{UU,VV}$ must belong too. In particular,
\begin{align}
0\approx T^{UU} \star I_{UU,VV}=T_{UV} \star\left(C_2\left(\frac32\gamma -2\beta-4\right)-2\alpha(\alpha-2)\right)\,.
\end{align}
Either we have to fix the Casimir to be $C_2-\lambda\mathbf{1}\sim 0$ with $\lambda=-\frac14n(2n+1)$ or $T_{AB}$ itself has to be quotient out. The first option leads to a nontrivial solution and fixes $\lambda$. If we ignore
that $\lambda=-\frac14n(2n+1)$ then the resulting two-sided ideal takes away everything, so the quotient is trivial.

It is far from obvious that we will not meet any inconsistencies by considering other identities in $U(sp(2n))/I_\lambda$, but it starts looking as there exists an algebra with exactly the spectrum we need
\begin{align}
\mathfrak{g}=U(sp(2n))/I_\lambda=\bullet\oplus\parbox{20pt}{\YoungB}\oplus\parbox{40pt}{\YoungD}\oplus...
\end{align}
More complicated Young shapes have been taken away by \YoungpBB, \YoungpAA{} and their descendants obtained by sandwiching them with $U(sp(2n))$.

We discussed the invariant definition of the HS algebra on the language of the universal enveloping algebra. In
particular it is now obvious that HS algebra is an associative one since it was obtained as a quotient of the associative algebra. However, this realization is difficult to deal with in practice. There are two sources of  complexity. First of all, a universal enveloping algebra is a complicated object by itself, taking into account it is not free. Secondly, we quotiented it by a two-sided ideal, which entails many more relations between the generators. Other works along the same lines include \cite{Vasiliev:2004cm,Iazeolla:2008ix,Boulanger:2011se,Fernando:2009fq,Govil:2013uta}.

The advantage of the $\star$-product realization discussed in Section \ref{sec:HSAlgebra} is that the ideal is automatically resolved. We already noticed that only symmetric tensors appear as Taylor
coefficients in $Y_A$, which implies more complicated Young shapes, including the window and the rank-two antisymmetric, be projected out.

Let us also note that the case of $sp(2)$ is special in that there are no unwanted diagrams at all, so we do not have to define an ideal to remove them. In order to avoid proliferation of generators due to the Casimir operator we can quotient by $(C_2-\lambda \mathbf{1})$ resulting in Feigin's $gl_\lambda$,  \cite{Feigin}. Algebra $gl_\lambda$ has the same spectrum as above, i.e. all even rank tensors of $sp(2)$ with the multiplicity one. In addition it has a free parameter. For those values $\lambda$ that correspond to finite-dimensional representations of $sp(2)$, say to $V_n$ of dimension $n$, one finds that $gl_\lambda$ develops a two-sided ideal. If we quotient this ideal out the resulting algebra is finite-dimensional and is in fact $gl(n)$ where $n$ is the dimension $V_n$. By contrast in the case of $sp(2n)$, $n>1$, there is no free parameter and no finite-dimensional truncations.

\mysubsection{Advanced \texorpdfstring{$\star$}{star}-products: Cayley transform}\label{extra:Cayley}
There are useful tricks that sometimes make star-product
calculations quite simple, \cite{Didenko:2003aa, Didenko:2012tv}. Particularly, the repetitive gaussian
integration drastically simplifies with the aid of the so called
Cayley transform. Suppose one wants to compute a bunch of gaussian
integrals of the form
\be\label{phi}
\Phi=\Phi(f_1, \xi_1, q_1)\star\dots \star\Phi(f_n, \xi_n, q_n)\,,
\ee
where
\be
\Phi(f,\xi, q) = \exp i{\Big(
\ff{1}{2}f_{AB}Y^{A}Y^{B}+\xi^{A}Y_{A}+q\Big)}\,,
\ee
Star-product of two such $\Phi$'s is already complicated and
given explicitly by
\begin{align}\label{StarProductGeneric}
\Phi(f_{1},\,\xi_{1},\,0)\star \Phi(f_{2},\,\xi_{2},\,0)
=\ff{1}{\sqrt{\det{|1+f_1f_2|}}}\Phi(f_{1,2},\,\xi_{1,2},\,q_{1,2})\,,
\end{align}
where
\begin{align}
&f_{1,2}=\ff{1}{1+f_2f_1}(f_2+1)+\ff{1}{1+f_1f_2}(f_1-1)\,,\label{f12}\\
&\xi_{1,2}^{A}=\xi_{1}^{B}\Big(\ff{1}{1+f_2f_1}{(f_2+1)}
\Big)_{B}{}^{A}+\xi_{2}^{B}\Big(\ff{1}{1+f_1f_2}{(1-f_1)}
\Big)_{B}{}^{A}\,,\label{xi12}\\
&q_{1,2}=\ff{1}{2}\Big(\ff{1}{1+f_2f_1}f_2\Big)_{AB}\xi_{1}^{A}\xi_{1}^{B}+
\ff{1}{2}\Big(\ff{1}{1+f_1f_2}f_1\Big)_{AB}\xi_{2}^{A}\xi_{2}^{B}-
\Big(\ff{1}{1+f_2f_1}\Big)_{AB}\xi_{1}^{A}\xi_{2}^{B}\label{q12}\,.
\end{align}
Further multiplication with some other $\Phi$ gives cumbersome
result. A systematic way to proceed is to use the map of elements
\eqref{phi} into $SpH(2n)$ group which is the semidirect product of
$Sp(2n)$ and Heisenberg group, \cite{Gelfond:2008td, Gelfond:2010xs}. Its elements consist of triplets
$\mathcal{G}=(U_{A}{}^{B}, x_{A}, c)$, where $U_{A}{}^{B}\in
Sp(2n)$ with the following product
\be
\mathcal{G}_1\circ\mathcal{G}_2=\Big((U_1U_2)_{A}{}^{B},
x_{1A}+U_{1A}{}^{B}x_{2B},
c_1+c_2+x_{1}^{A}U_{1A}{}^{B}x_{2B}\Big)\,.\label{SpH}
\ee
The embedding into the star-product algebra which respects the
$SpH(2n)$ group law
\begin{align*}
& r(\mathcal{G}_1)\,\Phi\Big(
f(\mathcal{G}_1),\,\xi(\mathcal{G}_1),\,q(\mathcal{G}_1)\Big)\star
r(\mathcal{G}_2)\,\Phi\Big(
f(\mathcal{G}_2),\,\xi(\mathcal{G}_2),\,q(\mathcal{G}_2)\Big)=\\
&\qquad\quad =r(\mathcal{G}_1\mathcal{G}_2)\,\Phi\Big(
f(\mathcal{G}_1\mathcal{G}_2),\,\xi
(\mathcal{G}_1\mathcal{G}_2),\,q(\mathcal{G}_1\mathcal{G}_2)\Big)
\end{align*}
can be shown to be of the following simple form
\besubeqs\begin{align}
&f_{AB}(\mathcal{G})=\Big(\ff{U-1}{U+1}\Big)_{AB}\,,\label{f(U)}\\
&r(\mathcal{G})=\ff{2^{n/2}}{\sqrt{\det{|1+U|}}}\,,\\
&\xi_{A}(\mathcal{G})=\pm 2\Big(\ff{1}{1+U}\Big)_{A}{}^{B}x_B\,,\\
&q(\mathcal{G})=c+\ff{1}{2}\Big(\ff{U-1}{U+1}\Big)_{AB}x^Ax^B\,.
\end{align}\esubeqs
This Cayley transform\footnote{There are two well-known equivariant maps from a Lie algebra to the group, the $\exp$-map and the Cayley one. The $\exp$-map is in general neither injective nor surjective. In many cases the Cayley map is more useful since it is a rational map.} allows one to express \eqref{phi} as
\be
\Phi=\frac{r(G_1...G_n)}{r(G_1)...r(G_n)}\Phi\Big(f(G_1...G_n),
\xi(G_1...G_n), q(G_1...G_n)\Big)\,.
\ee
Let us note, that Cayley transform is not always well defined.
Particularly, it is not defined if for some $f_{AB}$,
$f_{A}{}^{C}f_{C}{}^{B}=\gd_{A}{}^{B}$. In that case the
corresponding $Sp(2n)$ group element $U$ does not exist (formally it is at the infinity).
Nevertheless, the star-product with such elements is perfectly
well defined as can be seen from \eqref{f12}-\eqref{q12}. Such
elements with $f^2=1$ turns out to be star-product projectors
\be
2^n e^{\frac{i}{2}f_{AB}Y^A Y^B}*2^n e^{\frac{i}{2}f_{AB}Y^A Y^B}=
2^n e^{\frac{i}{2}f_{AB}Y^A Y^B}\,,\qquad f^2=1\,.
\ee
These play an important role in construction of boundary-to-bulk
propagators for HS fields and in HS black hole solutions, \cite{Didenko:2009td,Iazeolla:2011cb,Didenko:2012tv}.

\mysubsection{Poincare Lemma, Homotopy integrals} \label{sec:HomotopyIntegrals}
There is a standard problem how to solve equations of the form
\begin{align}
df\fm{k}&= g\fm{k+1} & dg\fm{k+1}&=0
\end{align}
where $f\fm{k}\equiv f_{a[k]}(x)\, dx^a\wedge ...\wedge dx^a$ is a degree-$k$ differential form and $g\fm{k+1}$ has degree $(k+1)$. This is the simplest system of unfolded type, a contractible pair unless $g\fm{k+1}\equiv0$. The second equation is the integrability condition for the first one, \eqref{UnfldCocycle}. If it is not true the system is inconsistent.

The general solution reads
\begin{align}
f_{a[k]}=\int_0^1 t^{k}\, dt\, x^c g_{ca[k]}(tx)+
\begin{cases} c,& k=0,\\
dc\fm{k-1}, & k>0,\end{cases}
\end{align}
In checking that this is indeed a solution we used that the number operator $x^c\frac{\pl}{\pl x^c}$ gives the same result as $t\frac{\pl}{\pl t}$ on functions $g(x t)$, then one gets a total derivative in $t$. Also, the integrability condition needs to be used to transform $\pl_a g_{ca[k]}$ (anti-symmetrization over all $a$'s is implied) into $\pl_c g_{a[k]}$.

Viewing the system of equations as an unfolded one there is a gauge symmetry, $\delta f =\xi\fm{k}$, $\delta g\fm{k+1}=d\xi\fm{k}$ and we see that all solutions are pure gauge from the unfolded perspective\footnote{It is a choice as to whether think of the equations as having gauge symmetries or not. For example, one can
reconstruct Maxwell gauge potential from the field strength by solving $dA=F$. Despite the fact that $F$ is
treated as a two-form, it does not have its own one-form gauge parameter that is capable of gauging $A$ away completely. So this system is not of unfolded type, see Section \ref{subsec:spinone} for the unfolding of a spin-one field. Unfolded equations always assume the richest gauge symmetry possible.}.

\paragraph{Two-dimensional case.} In Section \ref{sec:VasilievEquations} we will face a particular case of the above equations, which
are the equations along the auxiliary $z$-direction. That $z^\ga$ is two-dimensional leaves us with two types of equations, for $k=0$ and $k=1$ ($k=2$ does not have any $g\fm{k+1}$ on the {\it r.h.s.}, so the solution is pure gauge $d\xi$). The first one, $k=0$ reads as
\begin{align}
\pl_\ga f(z,y)&=g_\ga(z,y)\,,
\end{align}
where $\pl_\ga=\frac{\pl}{\pl z^\ga}$ and $y$ denotes collectively all other variables functions can depend on. It is implied that $g_\ga(z,y)$ must obey the integrability condition $\pl^\ga g_\ga(z,y)=0$. The general solution can be represented as
\begin{align}
f(z,y)=z^\ga\int_0^1 dt\, {g_\ga}(zt,y)+c(y)
\end{align}
The second one we meet, $k=1$ reads as
\begin{align}
\pl_\ga f_\gb(z,y) \epsilon^{\ga\gb}&=g(z,y)\,,
\end{align}
where $g(z,y)$ is a two form, $\frac12dz_\ga\wedge dz^\ga\, g(z,y)$ and the integrability imposes no restrictions on $g(z,y)$. The general solution reads
\begin{align}
f_\ga(z,y)=z_\ga\int_0^1 t\, dt\, {g}(zt,y)+\pl_\ga c(z,y)
\end{align}

\paragraph{Repeated homotopy integrals.} In the practical computations with Vasiliev equations
one finds nested homotopy integrals. These can be reduced to single integrals. Let us define operator $\Gamma_k$ as
\begin{align}
\Gamma_k\left[f\right](z)=\int_0^1 t^k\,dt\, f(zt)
\end{align}
then one can prove (for example, by looking at monomials) that there is the following associative and commutative composition law
\begin{align}\label{RepeatedHomotopy}
(\Gamma_n\circ\Gamma_m)[f]=(\Gamma_m\circ\Gamma_n)[f]=\begin{cases}
                          \frac1{m-n}\left(\Gamma_n-\Gamma_{m}\right)[f], & n\neq m , \\
                          \int_0^1 t^n \log{\frac1{t}}\, f(zt), & n=m.
                        \end{cases}
\end{align}



\begin{appendix}
\renewcommand{\thesection}{\Alph{section}}
\renewcommand{\theequation}{\Alph{section}.\arabic{equation}}
\setcounter{equation}{0}\setcounter{section}{0}
\mysection{Indices}
\begin{tabular}{|x{4cm}|w{10cm}|}
  \hline
  indices & \centering affiliation \tabularnewline\hline
  $\mm,\nn,\rr$ & world indices of the base manifold, are mostly implicit thanks to the differential form language \tabularnewline
  $a,b,c,...$ & fiber vector indices of $so(d-1,1)$\tabularnewline
  $\ga,\gb,...$, $\gad,\gbd,...$ & two-component spinor indices \tabularnewline
  $A,B,...$ & four-component indices of $sp(4)$-vectors or more generally $2n$-component indices of $sp(2n)$ \tabularnewline
  $\aAs,\aBs,...$ & (anti)-de Sitter algebra ($so(d-1,2)$)  $so(d,1)$ indices, range
  over $d+1$ values\tabularnewline
  \hline
\end{tabular}

\mysection{Multi-indices and symmetrization}\label{app:multiindex}
Before going to higher-spin fields we need to introduce certain condensed notation for
indices. The higher the spin the more indices is needed, so it can become a waste of letters. In many cases tensor expressions are symmetric in all the indices, e.g. $T^{abc...u}=T^{bac...u}=T^{acb...u}=...$, so it is natural to write $T^{a_1a_2...a_s}$ instead of $T^{abc...}$
assuming that the tensor is totally symmetric in all indices '$a$'. Moreover, since all indices are denoted by the same letter now, there is no need to keep them all, to indicate the number thereof is sufficient. So let $a(s)$ denote a group of $s$ indices $a_1a_2...a_s$ such that an object (tensor) is totally symmetric with respect to all permutations of $a_1a_2...a_s$.

We still need to improve notation a little bit. Sometimes a tensor
we find is not totally symmetric, e.g. $\pl^m \xi^{a(s-1)}$ is
only partly symmetric, and we may need to make it symmetric by summing
over all permutations. Ideologically the right way to achieve this
is to apply the symmetrizator
$\boldsymbol{P}=\displaystyle\frac{1}{s!} \sum_{\mbox{\tiny all
permutations}}$, which has a nice property of being a projector
$\boldsymbol{P} \boldsymbol{P}=\boldsymbol{P}$. However, in
practice it is more useful to adopt a convention where the sum is
taken over {\it all necessary} permutations without dividing by
$s!$. To simplify notation, all the indices of the group of
indices to be symmetrized are denoted by the same letter, so a string of indices $a(k)...a..a$
means that either the tensor is already symmetric in all of them or needs to be symmetrized. For
example, (the hatted indices are omitted)
\besubeqs\begin{align}
\pl^{a}\xi^{a(s-1)}&=\sum_{i=1}^{i=s} \pl^{a_i} \xi^{a_1...\hat{a}_i...a_s}\,,\\
\pl^a \pl^a \phi\fud{a(s-2)m}{m}&=\sum_{i<j} \pl^{a_i}\pl^{a_j} \phi\fud{a_1...\hat{a}_i...\hat{a}_j...a_sm}{m}\,.
\end{align}\esubeqs
One should be careful with using this convention, still it leads
to simpler formulas and most of the normalization coefficients
simply do not appear.

For example, let us check the gauge invariance of the Fronsdal operator, \eqref{FronOper},
\besubeqs\begin{align}
\delta\square \phi^{a(s)}&= \square\pl^{a}  \xi^{a(s-1)}\,,\\
\pl_m\delta\phi^{ma(s-1)}&=\square \xi^{a(s-1)}+\pl^a\pl_m\xi^{a(s-2)m}\,,\\
\delta\phi\fud{a(s-2)m}{m}&=2 \pl_m\xi^{a(s-1)m}\,,\\
\pl^a(\pl^a\pl_m\xi^{a(s-2)m})&=2\pl^a\pl^a\pl_m\xi^{a(s-2)m}\,.
\end{align}\esubeqs
Combining all terms together we find they cancel each other. Notice the bracket removal in the last line brings a factor of
two, this is an example of a nested symmetrization. The subtle
point is that in $\pl^a(\pl^a\pl_m\xi^{a(s-2)m})$ one should think
of the inner expression as a generic rank-$(s-1)$ symmetric
tensor, (despite the fact that it was obtained by summing over
$s-1$ permutations). It came from $\pl^a(\pl_m\phi^{a(s-1)m})$ for
which $s$ permutations are needed. Hence, to symmetrize it with
$\pl^a$ one needs $s$ permutations. In total it gives $s(s-1)$
permutations. However, on the \rhs we symmetrize $\pl^a\pl^a$,
which is already symmetric, with $\pl_m\xi^{a(s-2)m}$, so
$s(s-1)/2$ permutations are needed. This brings a factor of two
for the measure of all permutations that have been idle on the
\lhs because we were not aware of the internal structure of
$(\pl^a\pl_m\xi^{a(s-2)m})$ and the fact that $\pl^a\pl^a$ is
already symmetric. Nested symmetrizations are the sources of some
simple factors that are important for the final cancelation of
terms.

For antisymmetric or to be anti-symmetrized indices we adopt the same rules but with indices
enclosed in square brackets, e.g. $u[q]$.

\mysection{Solving for spin-connection} \label{app:spinconnection}
Christoffel symbols $\Gamma^\rr_{\mm\nn}$ can be solved for as usual. Let us solve for $\omega^{a,b}$.
Contracting \eqref{vpostulateA} with $e^{\mm b} e^{\nn c}$ and
defining $\Upsilon^{a|bc}=(\pl_\nn e^a_\mm-\pl_\mm e^a_\nn)e^{\mm
b} e^{\nn c}$ and $\omega^{a,b|c}=\omega^{a,b}_\mm\, e^{\mm c}$ we
find
\be \Upsilon^{a|bc}+\omega^{a,c|b}-\omega^{a,b|c}=0\,.\ee
As in the case of $\Gamma^{\rr}_{\nn\mm}$ we add two more equations that differ by cyclic permutation of $abc$.
Then it is easy to see that
\be\omega^{a,b|c}=\frac12 \left(\Upsilon^{a|bc}+\Upsilon^{b|ca}-\Upsilon^{c|ab}\right)\,,\ee
and finally
\be\omega^{a,b}_\mm=\frac12 \left(\Upsilon^{a|bc}+\Upsilon^{b|ca}-\Upsilon^{c|ab}\right)e_{c\mm}\,.\ee
Let us note that there is a simple relation between Christoffel symbols and spin-connection
\be \Gamma^{\rr}_{\nn\mm}=-e_a^\rr\left(\pl^{\vphantom{a}}_\nn e^a_\mm
+\omega^{\vphantom{a}}_\nn{}\fud{a,}{b} e^b_\mm\right)\,,\ee
which can be obtained from the vielbein postulate \eqref{De}.

\mysection{Differential forms} \label{app:diffforms}
Among
all tensors there is a subclass of covariant
totally-antisymmetric tensors that are special in many respects.
Given such a tensor, say $T_{\mm_1...\mm_q}$,
$T_{\mm_1...\mm_i\mm_{i+1}...\mm_q}=-T_{\mm_1...\mm_{i+1}\mm_i...\mm_q}$
for all $i$, it is called a differential form and its rank, $q$,
is referred to as the form degree. Since the index structure is
fixed by anti-symmetry it is useful just to indicate the form
degree as a subscript, e.g. $T\fm{q}$, without having to write down
the indices.

A useful way to ensure that the tensor is antisymmetric is to use
the Grassmann algebra. This is an associative algebra with a unit
generated by $\theta^\mm$ obeying
$\theta^{\mm}\theta^\nn=-\theta^{\nn}\theta^\mm$. Consider
polynomials in the Grassmann algebra
\be P(\theta)=\phi+A_\mm \theta^\mm+ \frac12 F_{\mm\nn}\theta^\mm\theta^\nn+...+\frac1{d!}\omega_{\mm_1...\mm_d}\theta^{\mm_1}...\theta^{\mm_d}\,.\ee
The expansion coefficients are forced to be antisymmetric tensors and hence the expansion stops at the form of the highest degree possible, which is $d$. There are three important operations on the class of differential forms.

(i) exterior product, which is denoted usually by $\wedge$. This is just the product in the Grassmann
algebra. In terms of Taylor coefficients it corresponds to first
taking the tensor product and then anti-symmetrizing all the
indices. It takes degree-$q$ form $T\fm{q}$ and degree-$p$ form
$R\fm{p}$ to a degree-$(p+q)$ form $(T\fm{q}\wedge R\fm{p})$ (note that
the order matters) $\frac1{(p+q)!}\sum_{\sigma\in
S_n}(-)^{|\sigma|}\sigma\circ T_{\mm_1...\mm_q}
R_{\mm_{q+1}...\mm_{p+q}}$. In the main text we sometimes omit $\wedge$ symbol
when one of the factors is a zero-form since zero-forms do not carry any differential
form indices and the anti-symmetrization is trivial. Zero-forms serve as purely numerical factors
that can be commuted without producing a sign. In general we have $T\fm{q}\wedge R\fm{p}=(-)^{pq}R\fm{p}\wedge T\fm{q}$.

(ii) exterior derivative, $d$. Operator $d$ is defined as
$d=\theta^\mm\pl_\mm$. It is nilpotent $d d\equiv0$ and applying
$\theta^\mm\nabla_\mm$ produces the same result as
$\theta^\mm\pl_\mm$, i.e. Christoffel symbols drop out from final
expressions because of anti-symmetrization and are irrelevant in
definition of differentiation for differential forms.

(iii) inner derivative. Given a vector field $\xi^\mm$ one defines
the inner derivative $i_\xi$ as
$\xi^\mm\frac{\pl}{\pl\theta^\mm}$. The Lie derivative
$\mathcal{L}_\xi$ is then $\mathcal{L}_\xi=d i_\xi+i_\xi d$.

In the literature instead of Grassmann algebra, symbols $dx^\mm$
are mostly used which are required to anti-commute $dx^\mm\wedge
dx^\nn=-dx^\nn\wedge dx^\mm$ with respect to exterior product
$\wedge$. The inner derivative is defined as
$\xi^\mu\frac{\pl}{\pl dx^\mu}$, correspondingly. We will also use
$dx^\mm$ having in mind the Grassmann algebra interpretation.

\paragraph{Integration and Einstein-Hilbert action.}
The naive way to rewrite $\int \sqrt{\det{g}}\, R$ is to note that\footnote{To be honest, $\det g=\det^2{e} \,\det \eta$.} $\sqrt{\det{g}}=\det{e}$, and $R=F_{\mm\nn}{}^{a,b} e^\mm_a
e^\nn_b$. There is a fancier way by taking the advantage of the
language of differential forms. Let us recall how to integrate
over manifolds. The measure $d^dx$ is not invariant under a change of coordinates, $d^dx'=J d^dx$, where $J=\det{|\pl x'/\pl x|}$. One has
to cure this non-invariance by multiplying it by something that
compensates $J$. One can take any covariant rank-two tensor, not
necessarily metric and not necessarily symmetric, because its
determinant transforms as $J^{-2}$, so $\sqrt{\det{}}$ does the
job. Another way is to take a rank-$d$ totally-antisymmetric
covariant tensor, say $\omega_{\mm_1...\mm_d}$, i.e. a
differential form of top degree. As a tensor it has only one
independent component and can be represented as
$\omega_{\mm_1...\mm_d}=C(x)\epsilon_{\mu_1...\mu_d}$, where the
totally antisymmetric tensor $\epsilon_{\mu_1...\mu_d}$ is
normalized as $\epsilon_{12...d}=1$. Under change of coordinates it
transforms as $C'=\det{|\pl x/\pl x'|}C=J^{-1}C$. Therefore, given
a top form, called the volume form sometimes, we can integrate
over the manifold. To be precise one may use $ \omega_{12...d}\,
d^dx $ as an invariant measure.

\mysection{Young diagrams and tensors}\label{app:Young}
We enter the world of Young diagrams through the $GL(d)$-tensors' door and then discuss what needs to be added to
cover the case of $SO(d)$ tensors. We do not aim at comprehensive review and present mostly the facts useful for this particular course.

\mysubsection{Generalities}\label{app:YoungGeneral}
\paragraph{\texorpdfstring{$\boldsymbol{GL(d)}$}{GL(d)}.} As is well-known,
a rank-two tensor $T^{a|b}$ (for simplicity we deal only with either contravariant or covariant tensors) can be decomposed into its symmetric and antisymmetric components
\begin{align}
T^{a|b}&=T_S^{ab}+T_A^{a,b}\,,& \\ T_S^{ab}=T_S^{ba}&=\frac12(T^{a|b}+T^{b|a}) \,, &T_A^{a,b}=-T_A^{b,a}&=\frac12(T^{a|b}-T^{b|a})\,.\label{ranktwodec}
\end{align}
If we deal with $GL(d)$, one can easily see that under any $GL(d)$ transformation
\be T^{ab...c}\longrightarrow S\fud{a}{a'} S\fud{b}{b'}...S\fud{c}{c'} T^{a'b'...c'}\,,\qquad\qquad S\fud{a}{m}\in GL(d)\,,\ee
(i) the two parts do not mix and (ii) they preserve their symmetry type, i.e. remain (anti)symmetric. It is also obvious that there are no more $GL(d)$-invariant conditions one can impose to decompose $T_S^{ab}$, $T_A^{a,b}$ even further.

It is convenient to employ Young diagrams as a tool to encode
graphically the symmetry type of tensors. A Young diagram is a
picture made of boxes, one box per each tensor index. The rank-two
(anti)-symmetric tensors are pictured by the following Young
diagrams
\begin{align} &T_S^{ab}\sim
\parbox{20pt}{\bep(20,10)\put(0,0){\YoungB}\put(2,2){a}\put(12,2){b}\eep}\,, &&
T_A^{a,b}\sim
\parbox{10pt}{\bep(10,20)\put(0,0){\YoungAA}\put(2,12){a}\put(2,2){b}\eep}\,.\end{align}
A vector and scalar are denoted by $\YoungpA$ and an empty diagram $\bullet$, respectively.

As for rank-three tensors one finds something new in addition to totally (anti)-symmetric parts. Since the decomposition of rank-two tensors into irreducibles is known, we can take a rank-three tensor which is already irreducible in some two indices, say $T^{ab|c}=T^{ba|c}$. Then one finds,
\begin{align}
T^{ab|c}&=T_S^{abc}+H_S^{ab,c}\,,\notag\\
T_S^{abc}&=\frac13(T^{ab|c}+T^{bc|a}+T^{ca|b})\,, &T_S^{abc}&=T_S^{bac}=T_S^{acb}\,,\label{hookcondition}\\
H_S^{ab,c}&=\frac13(2T^{ab|c}-T^{bc|a}-T^{ca|b})\,,& H_S^{ab,c}&=H_S^{ba,c}\,, & &H_S^{ab,c}+H_S^{bc,a}+H_S^{ca,b}=0\,.\notag
\end{align}
The appearance of the totally symmetric component, $T_S^{abc}$, was expected. There is one more, $H_S^{ab,c}$, that is neither totally symmetric nor antisymmetric. It is symmetric in the first two indices, $ab$ and the condition for it not to contain the totally symmetric component, which is already covered by $T_S^{abc}$, implies that the symmetrization over all three indices vanishes. With our convention on symmetrization we can rewrite it as
\begin{align} H_S^{aa,b}:& &H_S^{aa,a}=0\,.\end{align}
Again one can check that under any $GL(d)$ transformations both
$T_S^{abc}$ and $H_S^{ab,c}$ preserve the symmetry conditions they
obey and do not mix. In the language of Young diagrams these are
pictured as follows
\begin{align} &T_S^{abc}\sim
\parbox{30pt}{\bep(30,10)\put(0,0){\YoungC}\put(2,2){a}\put(12,2){b}\put(22,2){c}\eep} &&
H_S^{ab,c}\sim
\parbox{20pt}{\bep(20,20)\put(0,0){\YoungBA}\put(2,2){c}\put(2,12){a}\put(12,12){b}\eep}\label{youngBAsym}\end{align}
Because of the shape, $H_S^{ab,c}$ is sometimes called 'hook'-diagram. We see tensors tend to be symmetric in the indices associated with rows of Young diagrams. Not surprisingly, a rank-$s$ totally symmetric tensor is denoted by
\begin{align} &T^{a(s)}\sim
\underbrace{\,\parbox{60pt}{\bep(60,10)\put(0,0){\YoungB}\put(2,2){a}\put(12,2){a}
\put(24,3){$\ldots$}\put(40,0){\YoungB}\put(42,2){a}\put(52,2){a}\eep}\,}_{\displaystyle s\mbox{ boxes}}\end{align}

If instead we take a rank-three tensor that is already antisymmetric in the first two indices,
$T^{ab|c}=-T^{ba|c}$, we will find a similar decomposition into $GL(d)$ irreducible components
\begin{align}
T^{ab|c}&=T_A^{abc}+H_A^{ab,c}\,,\notag\\
T_A^{abc}&=\frac13(T^{ab|c}+T^{bc|a}+T^{ca|b})\,, &T_A^{abc}&=-T_A^{bac}=-T_A^{acb}\,,\label{hookconditionA}\\
H_A^{ab,c}&=\frac13(2T^{ab|c}-T^{bc|a}-T^{ca|b})\,,& H_A^{ab,c}&=-H_A^{ba,c}\,, & &H_A^{ab,c}+H_A^{bc,a}+H_A^{ca,b}=0\,.\notag
\end{align}
In addition to the totally antisymmetric component $T_A^{abc}$
one finds $H_A^{ab,c}$ that is antisymmetric in $ab$ and does not
contain the totally antisymmetric component, which is the last
condition\footnote{Coincidentally, for rank-three tensors
\eqref{hookcondition} and \eqref{hookconditionA} look identical,
the symmetry conditions they obey are different though. This is
due to the fact that the cyclic permutation taking place in the
definition of $H_S^{ab,c}$ and $H_A^{ab,c}$ (anti)-symmetrizes over
the three indices depending on whether the original tensor is
symmetric or antisymmetric in the first two indices. For tensors
of higher rank one finds more difference between symmetric and
antisymmetric presentations.}. In the language of Young diagrams
we have
\begin{align} &T_A^{abc}\sim
\parbox{10pt}{\bep(10,30)\put(0,0){\YoungAAA}\put(2,2){c}\put(2,12){b}\put(2,22){a}\eep} &&
H_A^{ab,c}\sim
\parbox{20pt}{\bep(20,20)\put(0,0){\YoungBA}\put(2,2){b}\put(2,12){a}\put(12,12){c}\eep}\label{youngBAsymB}\end{align}
Tensors tend to be antisymmetric in the indices associated with the columns. A rank-$q$ totally antisymmetric tensor is denoted by
\begin{align} &T^{u[q]}\sim
\left.\,\parbox{10pt}{\bep(10,50)\put(0,0){\YoungAA}\put(2,2){u}\put(2,12){u}
\put(3,24){$\vdots$}\put(0,40){\YoungA}\put(2,42){u}\eep}\,\right\}{\displaystyle q\mbox{ boxes}}\end{align}
Tensors that are neither symmetric nor antisymmetric are referred to as having mixed-symmetry. $H_S^{ab,c}$ and $H_A^{ab,c}$ have mixed-symmetry. There is a strange thing one might have noticed that the second diagram in \eqref{youngBAsymB} is identical to that of \eqref{youngBAsym}. How come that $H_S^{ab,c}$ and $H_A^{ab,c}$ have the same symmetry type but obey different symmetry conditions?

This is an essential feature of genuine mixed-symmetry tensors, i.e. the tensors with the symmetry of a Young diagram that is different from one-row or one-column diagram. The symmetries of (one-column) one-row Young diagrams are always presented by (anti)-symmetric tensors. There are in general many ways to present a tensor with the symmetry of more complicated Young diagrams. The simplest mixed-symmetry tensor has the symmetry of \YoungmBA. In components we have either some $H_S^{aa,b}$, which is symmetric in the first two indices and obeys the Young condition $H_S^{aa,a}=0$, or some $H_A^{ab,c}$, which is antisymmetric in the first two indices and obeys $H_A^{[ab,c]}=0$.

Given $H_S^{ab,c}$ one can always construct an object of type $H_A^{ab,c}$ as
\be H_A^{ab,c}=H_S^{ac,b}-H_S^{bc,a}\,.\ee
The map is invertible. Indeed, one can go back by
\be H_S^{ab,c}=\alpha(H_A^{ac,b}+H_A^{bc,a})\,,\ee
where $\alpha$ turns out to be $1/3$. Since the map is an
isomorphism, it just relates two bases. We can say that there are
several ways to implement symmetries of some Young diagram into a
tensor. In practice it is sometimes useful to switch from one base
to another one to simplify computations.  It is not possible to
have the symmetry properties both of $H_S^{ab,c}$ and of
$H_A^{ab,c}$ realized simultaneously.

There are two bases for mixed-symmetry tensors that are most natural, symmetric and antisymmetric.
In the (anti)-symmetric base tensors are explicitly (anti)-symmetric in the indices corresponding to the (columns) rows of Young diagrams, with some additional relations imposed. For example, $H_S^{ab,c}$ was given in the symmetric base, while $H_A^{ab,c}$ in the antisymmetric one.

Bearing in mind the option of having several ways to present a
tensor, we usually do not fill the boxes in Young diagrams with
indices. For example, there are five different symmetry types
possible for rank-four tensors
\begin{align} &\YoungpD &&\YoungpCA && \YoungpBB && \YoungpBAA && \YoungpAAAA\label{rankfourtensors}\end{align}
The rows in a Young diagram are left aligned. The length of the rows in a proper Young diagram cannot increase downwards (the upper rows are not shorter than the lower ones).

The first and the last diagrams correspond to totally
(anti)-symmetric tensors, for which there is no ambiguity in the
choice of a base. There are two different presentations for the
three diagrams in the middle. For example, the Riemann and Weyl
tensors have the symmetry of the diagram in the exact center.
Usually the Riemann tensor is defined in the antisymmetric base,
\begin{align} R^A_{ab,cd}&=-R^A_{ba,cd}=-R^A_{ab,dc}=R^A_{cd,ab}\,, & R^A_{[ab,c]d}&=0\,.\end{align}
Seldom used, but not in this review, is the symmetric base
\begin{align} \label{RiemannSym}R^S_{ab,cd}&=R^S_{ba,cd}=R^S_{ab,dc}=R^S_{cd,ab}\,, & R^S_{(ab,c)d}&=0\,.\end{align}
Note that $R^{A(S)}_{ab,cd}=R^{A(S)}_{cd,ab}$ is not an independent relation and follows from the rest. That the two ways are equivalent is shown by writing the linear transformation explicitly
\begin{align} R^S_{ac,bd}&=R^A_{ab,cd}+R^A_{cb,ad}\,, & R^A_{ac,bd}&=\beta(R^S_{ab,cd}-R^S_{cb,ad})\,,\end{align}
where the first formula is treated as a definition of $R^S$, then the coefficient in the second formula is found to be $\beta=1/3$. Note that in this particular case it is sufficient to (anti)-symmetrize over two indices, the (anti)-symmetry in the other two indices then follows from the properties of the original tensors.

For a tensor having the symmetry of the second diagram from \eqref{rankfourtensors} in the symmetric base one has $T^{aaa,b}$ that obeys $T^{aaa,a}=0$.

Let us consider one more example of a tensor having $\BYoungp{4}{1}{k}{}$ symmetry type. We begin with $T^{a(k-1)|b}$, which is symmetric in $a(k)$ and there are no symmetry conditions among $b$ and $a(k)$. On subtracting the totally symmetric component one finds a remnant,
\begin{align*}
T^{a(k)|b}&=T^{a(k)b}+T^{a(k),b}\,,\\
T^{a(k)b}&=\frac1{k+1}\left(T^{a(k)|b}+T^{a(k-1)b|a}\right)\,,\\
T^{a(k),b}&=\frac1{k+1}\left(kT^{a(k)|b}-T^{a(k-1)b|a}\right)\,, &&T^{a(k),a}=0\,.
\end{align*}

It is not that difficult to get the complete classification of symmetry types. For the purpose of totally symmetric higher spin fields it is sufficient to restrict ourselves to the class of two-row Young diagrams with the tensors presented in the symmetric base,
\begin{align} &T^{a(k),b(m)}\sim
\BYoungp{6}{4}{k}{m}=
\overbrace{\underbrace{\,\parbox{80pt}{\bep(80,20)\put(0,0){\YoungCB}\put(2,12){a}\put(12,12){a}\put(22,12){a}
\put(12,2){b}\put(2,2){b}\put(34,13){$\ldots$}\put(72,12){a}\put(52,2){b}
\put(24,3){$\ldots$}\put(50,0){\YoungCA}\put(62,12){a}\put(52,12){a}\eep}\,}_{\displaystyle m}}^{\displaystyle k}\end{align}
The tensor $T^{a(k),b(m)}$ has two groups of indices $a(k)$ and $b(m)$; it is symmetric in $k$ indices $a$ and $m$ indices $b$; it is irreducible under $GL(d)$ iff the symmetrization of all indices $a$ with at least one index from the second group vanishes, i.e.
\begin{align} &T^{a(k),ab(m-1)}\equiv0\,.\label{tworowdefiningrel}\end{align}
Note that one cannot impose more symmetry conditions in general as it would be equivalent to requiring a tensor to be symmetric and antisymmetric in some indices at the same time, which implies the tensor is identically zero. If $k<m$ the tensor vanishes identically, which explains the condition for the length of rows not to increase downwards.

For $k=2$, $m=0$ we have $T^{a_1 a_2}=T^{a_2 a_1}$, i.e. symmetric. For $k=m=1$ we find $T^{a,b}+T^{b,a}=0$ as Young condition \eqref{tworowdefiningrel}, i.e. $T^{a,b}$ is antisymmetric. For $k=2$, $m=1$ we find \eqref{hookcondition}. For $k=2$, $m=2$ we find the symmetry conditions for the Weyl/Riemann tensor, \eqref{RiemannSym}. There is a slightly degenerate case of rectangular Young diagrams. In the latter case the two groups of indices in the tensor, $T^{a(k),b(k)}$, are equivalent. In particular, the following relation is true
\be T^{a(k),b(k)}=(-)^kT^{b(k),a(k)}\,,\label{exchangeidentity}\ee
which is obvious for $k=1$ and for $k=2$, the Riemann tensor, the condition is also known to be true.
The proof is easy in the antisymmetric presentation of tensors, where one has $k$ pairs of antisymmetric indices and swapping the indices inside all pairs yields $(-)^k$.

At the end of the $GL(d)$ section let us give several identities that are frequently used to rearrange indices. Suppose we are given $T^{a(k),b(m)}$ and there is a symmetrization imposed over $k$ indices with one of them taken from the second group of indices, i.e.
\be T^{a(k-1)c,ab(m-1)}= \sum_i T^{a_1...\hat{a}_i...a_kc, a_i b_2...b_m}\label{symAA}\ee
This is what is effectively imposed when $T^{a(k),b(m)}$ is contracted with another tensor $V_{a(k)}$ that is symmetric
\be T^{a(k-1)c,ub(m-1)} V_{a(k-1)u}\label{symAB}\,.\ee
The defining relation \eqref{tworowdefiningrel} then tells that
\begin{align} T^{a(k-1)c,ab(m-1)}= -T^{a(k),b(m-1)c}\label{niceident}\end{align}
but
\begin{align} T^{a(k-1)c,ub(m-1)} V_{a(k-1)u}= -\frac1{k}T^{a(k),b(m-1)c}V_{a(k)}\,,\end{align}
the difference in $\frac1{k}$ is because \eqref{symAA} contains $k$ terms explicitly, while \eqref{symAB} does not,
\begin{align} T^{a(k-1)c,ub(m-1)} V_{a(k-1)u}=\frac1{k} T^{a(k-1)c,ab(m-1)} V_{a(k)}\end{align}
All $k$ terms in the last expression are identical thanks to the symmetry of $V_{a(k)}$, hence they cancel $\frac1{k}$. A more general relation holds true
\be T^{a(k-n)c(n),u(n)b(m-n)}V_{a(k-n)u(n)}= \frac{(-)^n n!(k-n)!}{k!}T^{a(k),c(n)b(m-n)}V_{a(k)}\ee
and allows one to always put all symmetrized indices into the first group of indices. Note that one can roll symmetrized indices to the group of indices corresponding to a longer row of Young diagram, but not in the opposite direction.  Also note that \eqref{exchangeidentity} is a particular case of the identity above for $k=n=m$.

The property of being a Young diagram tells us that it is the first row/column that is the longest one.
Obviously, the height of the columns in a Young diagram cannot exceed $d$, for we can choose the antisymmetric base to present a tensor with such a symmetry and it then will carry more than $d$ antisymmetric indices, i.e. it has to vanish.

To draw a line, $GL(d)$ irreducibility requirements impose certain symmetry conditions on the indices carried by a tensor. Irreducibility conditions are nicely encoded by Young diagrams. There are in general several ways to present an irreducible $GL(d)$ tensor, one can transfer between different bases by (anti)-symmetrizing indices.

\paragraph{\texorpdfstring{$\boldsymbol{SO(d)}$}{SO(d)}.} In case $T^{a|b}$ is an $so(d)$ tensor (signature is irrelevant) we have an invariant tensor, which is the metric $\eta_{ab}$. With the help of the metric one can do more and extract the trace
\begin{align}
T^{a|b}&=T_S^{ab}+T_A^{a,b}+\frac1d\eta^{ab} T\,, \label{ranktwodecso}\\
T_S^{ab}&=\frac12(T^{a|b}+T^{b|a}-\frac2d \eta^{ab} T^{c|d}\eta_{cd})\,, & T_S^{ab}&=T_S^{ba}\,, & T_S^{ab}\eta_{ab}&=0\,, \notag\\
T_A^{a,b}&=\frac12(T^{a|b}-T^{b|a})\,,  &  T_A^{a,b}&=-T_A^{b,a}\,,\notag\\
T&= T^{c|d}\eta_{cd}\,.\notag\end{align}
Again, one can check that the decomposition is stable under any $SO(d)$ transformations.

It is obvious that $SO(d)$-irreducibility requires $GL(d)$-irreducibility, i.e. Young symmetry, and one needs to supplement Young symmetry conditions with the trace constraints. The full set of constraints for tensors with the symmetry of two-row Young diagrams includes
\begin{align} T^{a(k),b(m)}&\sim
\BYoungp{6}{4}{k}{m}\,, & T^{a(k),ab(m-1)}&=0\,, \\
T\fudu{a(k-2)c}{c}{,b(m)}&=0\,, & T\fudu{a(k-1)c,}{c}{b(m-1)}&=0\,, & T\fud{a(k),b(m-2)c}{c}&=0\,.
\end{align}
There are three types of traces one can take, depending on how the two contracted indices are distributed over the two groups of indices. Not all of these traces are independent. Indeed, assuming that $T\fudu{a(k-2)c}{c}{,b(m)}=0$ we can symmetrize all $a(k-2)$ with one of the $b$'s to see, applying \eqref{niceident},
\begin{align}
T\fudu{a(k-2)c}{c}{,ab(m-1)}=-T\fudu{a(k-1)c,}{c}{b(m-1)}\,.
\end{align}
Symmetrizing now $a(k-1)$ with one of the $b$'s once again we find
\begin{align}
T\fudu{a(k-1)c,}{c}{ab(m-2)}=-T\fudu{a(k),c}{c}{b(m-2)}\,.
\end{align}
We see that the first trace condition implies the second and the second then implies the third, but not in the opposite direction --- it is possible to have the third trace conditions satisfied without enforcing the first and the second.

Young symmetry plus trace constraints furnish the complete set of irreducibility conditions in most cases. However, when a tensor has $d/2$ antisymmetric indices (it is better to refer to the number of rows in the Young diagram), in particular we are in even dimension, one can impose (anti)-selfduality conditions,
\begin{align}
T^{u[q]}&=\pm(i) \epsilon\fud{u[q]}{v[q]}T^{v[q]}\,, &d&=2q\,,
\end{align}
where $\epsilon_{u[d]}$ is the totally antisymmetric tensor, which is also an invariant tensor of $so(d)$. Whether one can impose the (anti)-self duality condition with $\pm1$ or $\pm i$ depends on the dimension, $d$, modulo $4$. In particular one can impose $(\pm i)$ (anti)-selfduality for tensors with the symmetry of two-row rectangular Young diagrams in the case of $so(3,1)$. This explains why a single higher-spin connection $\omega^{a(s-1),b(k)}$ splits into two
complex conjugate connections $\omega^{\ga(s+k-1),\gad(s-k-1)}$, $\omega^{\ga(s-k-1),\gad(s+k-1)}$ --- two-row Young diagrams do not correspond to irreducible tensors in $4d$ once we allow for a pair of complex conjugated tensors.

To deal with $so(d)$ is even more restrictive. It is useful to prove the following result (we do not use this anywhere in the lectures): a traceless tensor with the symmetry of a Young diagram for which the sum of heights of the first two columns exceeds $d$ must vanish identically. Note that nothing prevents the first column to exceed $[d/2]$ at the price of all other columns being shorter than $[d/2]$, for example totally antisymmetric tensors of any rank $q=0,...,d$ do exist.

The existence of $\epsilon_{u[d]}$ imposes more restrictions on $so(d)$-Young diagrams. Any tensor having a symmetry of a Young diagram whose first column, say of height $q$, exceeds $[d/2]$ is equivalent to a tensor whose first column does not exceeds $[d/2]$, its height is $d-q$. For example, given a totally antisymmetric tensor $T^{u[q]}$, i.e. its symmetry is given by a Young diagram made of a single column, we can dualize it to a rank-$(d-q)$ tensor $T'^{u[d-q]}$
\be T'^{u[d-q]}=\epsilon\fud{u[d-q]}{v[q]}T^{v[q]}\,.\ee

This implies that any Young diagram of $so(d)$ should have no more than $[d/2]$ rows. As was just mentioned, this does not mean that all other Young diagrams correspond to identically vanishing tensors, but those that correspond to nontrivial tensors are equivalent via $\epsilon_{u[d]}$-dualization to Young diagrams with no more than $[d/2]$ rows.

\paragraph{Why Young diagrams?} The rationale behind Young diagrams lies in a close connection of representation theory of $GL(d)$ and the symmetric group. The symmetric group in $n$ letters, $S_n$, acts naturally on
the $n$-th tensor power, $T^n V$ of a vector space, $V$, by
permuting the factors $v_1\otimes...\otimes v_n$. Hence $T^n V$ is
a representation of $S_n$, a reducible one. One can project onto
various $S_n$-irreducible subspaces, which simultaneously projects onto $GL(d)$
irreducible subspaces. As is well-known the irreducible
representations of $S_n$ are in one-to-one correspondence with
conjugacy classes. They can be enumerated by partitions of $n$
into nonnegative integers, say $n=s_1+...+s_n$, which can be
ordered $s_1\geq s_2\geq...\geq s_n\geq0$. Each partition can be
encoded by a Young diagram that has rows of lengths $s_1,...,s_n$.
If we go on further we face the representation theory of Lie
algebras and find out that Verma modules are parameterized by a
number of constants, the weights, with the number of weights equal
to the rank of a given Lie algebra. In particular the rank of
$so(d)$ is $[d/2]$, so the number of parameters is in accordance
with the number of rows in $so(d)$ Young diagrams. The general
theory is discussed in many textbooks, see e.g. \cite{Barut} for the summary. The Young
diagrams are not just nice pictures and appear naturally in many
different topics, \cite{Fulton}.

\mysubsection{Tensor products}\label{app:YoungTensorProd}
We more or less understand now what are the conditions for a
tensor to be irreducible. Let us discuss the inverse problem of
how to decompose a reducible tensor into its irreducible
components. This amounts to computing tensor products. The main
properties of the tensor product are associativity, commutativity\footnote{By saying that it is commutative we mean the that the decomposition of the tensor product $V\otimes W$ of two representations $V$ and $W$ into irreducible representations does not depend on the order, i.e. it is the same for $W\otimes V$, while it does not make much sense to compare $v\otimes w$ and $w\otimes v$ where $v\in V$, $w\in W$. } and
distributivity. Actually, this is what we have already done for
the tensors of simplest types. In the case of $GL(d)$ we manually
found that
\besubeqs\begin{align}
\label{simplestTA}\eqref{ranktwodec}:& &&T^{a|b}=T_S^{ab}+T_A^{a,b} & \YoungpA\otimes\YoungpA&=\YoungpB\oplus\YoungpAA\,,\\
\label{simplestTB}\eqref{hookcondition}:& &&T^{(ab)|c}=T_S^{abc}+H_S^{ab,c} & \YoungpB\otimes\YoungpA&=\YoungpC\oplus\YoungpBA\,,\\
\label{simplestTC}\eqref{hookconditionA}: & &&T^{[ab]|c}=T_A^{abc}+H_A^{ab,c} & \YoungpAA\otimes\YoungpA&=\YoungpAAA\oplus\YoungpBA \,.
\end{align}\esubeqs
Had we started with the most general rank-three tensor $T^{a|b|c}$ without any symmetry conditions imposed we would have to compute, $V\otimes V\otimes V$,
\begin{align}
\YoungpA\otimes\YoungpA\otimes\YoungpA
\end{align}
We first find that it decomposes according to \eqref{simplestTA} in any pair of indices, say in $a$ and $b$ (tensor product is an associative, commutative operation, so we can insert brackets wherever we like as well as to permute the factors),
\begin{align}
\left(\YoungpA\otimes\YoungpA\right)\otimes\YoungpA=\left(\YoungpB\oplus\YoungpAA\right)\otimes\YoungpA
\end{align}
Then, with the help of distributivity and \eqref{simplestTB}, \eqref{simplestTC} we get
\begin{align}\label{rankthreemost}
\left(\YoungpB\otimes\YoungpA\right)\oplus\left(\YoungpAA\otimes\YoungpA\right)=
\YoungpC\oplus2\,\YoungpBA\oplus\YoungpAAA
\end{align}
Using \eqref{ranktwodec}, \eqref{hookcondition},
\eqref{hookconditionA} we could obtain a more detailed structure
\footnote{There is an analogy with the canonical QM textbook
exercise on the $su(2)$ representation theory, namely the
multiplication of quantum angular momentum. The decomposition in terms of
Young diagrams contain the same information as the statement
$$j_1\otimes
j_2=|j_1-j_2|\oplus...\oplus |j_1+j_2|\,.\qquad\qquad\qquad(1)\label{sutwo}$$
But representation denoted loosely by $j_1$ is a
$(2j_1+1)$-dimensional vector space, analogously for $j_2$ and any
of the spins on the r.h.s of (1). In principle we can write the
decomposition in more detail, showing exactly how each of the base
vectors belonging to one of $|j_1-j_2+2i|$ on the \rhs decomposes into a sum
of $|m_1,j_1\rangle\otimes|m_2,j_2\rangle$. This is analogous to
what we did in \eqref{ranktwodec}, \eqref{hookcondition},
\eqref{hookconditionA}. This more detailed information is not
captured by (1). Luckily, in many cases knowing (1) or the
decomposition in terms of Young diagrams is sufficient.}. Note
that \eqref{rankthreemost} does not contain any information about
the particular choice of indices that we made and just states that
there is a totally symmetric and a totally antisymmetric
component and two independent components with the symmetry of
\YoungmBA. An exercise analogous to \eqref{hookcondition} but for $so(d)$ reads
\besubeqs\begin{align}
T^{aa|c}&=T_S^{aac}+H_S^{aa,c}-\frac2{(d+2)(d-1)}\eta^{aa}T^b+\frac{d}{(d+2)(d-1)}\eta^{ab}T^a\,,\notag\\
T_S^{aaa}&=\frac13\left(T^{aa|a}-\frac2{d+2}\eta^{aa}T^a\right)\,, & \notag\\
T_S^{abc}&=T_S^{bac}=T_S^{acb}\,, \qquad\qquad\qquad T_S\fud{ab}{b}=0\,,\label{hookconditionso}\\
H_S^{aa,c}&=\frac13\left(2T^{aa|c}-T^{ac|a}+\frac1{d-1}(2\eta^{aa}T^c-\eta^{ac}T^a)\right)\,,&\notag\\ H_S^{ab,c}&=H_S^{ba,c}\,, \qquad \qquad H_S^{ab,c}+H_S^{bc,a}+H_S^{ca,b}=0\,, \qquad\qquad  H_S\fudu{c}{c,}{a}=H_S\fud{ac,}{c}=0\,.\notag
\end{align}\esubeqs
Note the appearance of an additional component, the trace $T^a=T\fud{am|}{m}$. We can summarize \eqref{ranktwodecso}, \eqref{hookconditionso}, and the undone exercise for $T^{[ab]|c}$ with two antisymmetric indices as follows
\begin{align}
\eqref{ranktwodecso}:& &&T^{a|b}& \YoungpA\otimes\YoungpA&=\YoungpB\oplus\YoungpAA\oplus\bullet\,,\\
\eqref{hookconditionso}:& &&T^{(ab)|c} & \YoungpB\otimes\YoungpA&=\YoungpC\oplus\YoungpBA\oplus\YoungpA\,,\\
\mbox{left as an exercise}: & && T^{[ab]|c}& \YoungpAA\otimes\YoungpA&=\YoungpAAA\oplus\YoungpBA\oplus\YoungpA \,.
\end{align}
The manipulations with Young diagrams are simpler than writing
down the decomposition into irreducibles in the language of
tensors explicitly. For example, for the most general rank-three
tensor $T^{a|b|c}$ we find for $V\otimes V\otimes V$
\begin{align}
\YoungpA\otimes\YoungpA\otimes\YoungpA&=
\left(\YoungpB\oplus\YoungpAA\oplus\bullet\right)\otimes\YoungpA=
\YoungpC\oplus2\,\YoungpBA\oplus3\,\YoungpA\oplus \YoungpAAA
\end{align}

As the rank of tensors grows the computation of tensor products become more and more involved. We will need\footnote{The tensor product rules do not depend on the signature of the metric, $\eta_{ab}$.}
\begin{align}
gl(d):& &\AYoungp{6}{k} \otimes \YoungpA&=\BYoungp{6}{1}{k}{}\oplus \AYoungp{7}{k+1} \\
so(d):& &\AYoungp{6}{k} \otimes \YoungpA&=\BYoungp{6}{1}{k}{}\oplus \AYoungp{7}{k+1}\oplus \AYoungp{5}{k-1}
\end{align}
and just for fun
\begin{align}
gl(d):& &\BYoungp{6}{4}{k}{m}  \otimes \YoungpA=&\CYoungp{6}{4}{1}{k}{m}{}\oplus \BYoungp{6}{4}{k+1}{m}\oplus \BYoungp{6}{5}{k}{m+1}\\
so(d):& &\BYoungp{6}{4}{k}{m}  \otimes \YoungpA=&\CYoungp{6}{4}{1}{k}{m}{}\oplus \BYoungp{6}{4}{k+1}{m}\oplus \BYoungp{6}{5}{k}{m+1}\\ &&&\oplus \BYoungp{5}{4}{k-1}{m}\oplus \BYoungp{6}{4}{k}{m-1}\notag
\end{align}
The general rule for multiplying by $\YoungpA$ is quite simple.
For the $gl(d)$ case one tries to add one cell to the diagram
in all possible ways such that the result is again a Young
diagram. The $so(d)$-rule is a combination of the $gl(d)$-rule with
an additional cycle where we try to remove (take the trace) one
cell in all admissible ways. One has to remember that if the
height of some of the Young diagrams on the r.h.s. exceeds $[d/2]$
for $so(d)$ (or $d$ for $gl(d)$) then one has to use the properties
mentioned at the end of $gl(d)$ and $so(d)$ sections. In particular
some diagrams may just vanish or should be dualized to fit in
$[d/2]$ height restrictions in the case of $so(d)$. The general rules, when both factors are generic Young diagrams, are quite complicated and can be found, for example, in \cite{QuantGrCrystals}.

\mysubsection{Generating functions}\label{app:YoungGenfunc}
Since in the higher-spin theory we have to work with infinite
collections of tensors we find it convenient, if not necessary, to
contract all tensor/spinor indices with some auxiliary variables.
We would like to discuss how
various Young/trace constraints can be implemented on appropriate
functional space.

For example, suppose we need all symmetric tensors, say $C^{a(k)}$, with tensor of each rank appearing once, i.e. the space of tensors is multiplicity free. Then all these tensors can be collected into just one function of auxiliary variables $y^a$
\begin{align}
C^{a(k)}\,,& \qquad k=0,1,... &&\Longleftrightarrow&& C(y)=\sum_k \frac1{k!}C^{a(k)} y_a...y_a\,.
\end{align}
The Taylor coefficients are our original tensors. If our tensors are all traceless then the appropriate functional space is the space of harmonic functions in $y^a$
\begin{align}
C\fud{a(k-2)m}{m}=0 &&\Longleftrightarrow && \square C(y)=0\,.
\end{align}
Indeed, on computing $\square$ termwise and equating each Taylor coefficient to zero we get the desired
\begin{align}
\square C(y)&=\frac{\pl^2}{\pl y_m \pl y^m} \sum_k \frac1{k!}C^{a(k)} y_a...y_a = \sum_k \frac1{(k-2)!} C\fud{a(k-2)m}{m} y_a...y_a=0\,.
\end{align}

Suppose we need a space of tensors with the symmetry of all two-row Young diagrams, again each symmetry type appearing once, i.e.
\begin{align}
C^{a(k),b(m)}& &C^{a(k),ab(m-1)}=0& &k&=0,1,... &m&=0,...,k
\end{align}
Two auxiliary vector-like variables are required now, say $y^a$ and $p^a$, with the help of which we can build a generating function
\begin{align}
C(y,p)=\sum_{k,m}\frac{1}{k!m!} C^{a(k),b(m)} y_a...y_a\, p_b...p_b\,.
\end{align}
The Taylor coefficients of a generic function of $y$ and $p$ do not obey any Young symmetry conditions, these are tensors $C^{a(k)|b(m)}$ symmetric in $a(k)$ and $b(m)$ with no conditions that entangle $a$'s and $b$'s. With a little thought the right additional restrictions on the functional space are found to be
\begin{align}
y^c\frac{\pl}{\pl p^c} C(y,p)=0\,.
\end{align}
Indeed,
\begin{align}
y^c\frac{\pl}{\pl p^c}\sum_{k, m}\frac{1}{k!m!} C^{a(k),b(m)} y_a...y_a\, p_b...p_b=
\sum_{k, m}\frac{1}{k!(m-1)!} C^{a(k),cb(m-1)} y_a...y_ay_c\, p_b...p_b
\end{align}
all indices contracted with the same commuting variable $y^a$ appear automatically symmetrized, i.e.
\begin{align}
C^{a(k),cb(m-1)} y_a...y_ay_c = \frac1{k+1} C^{(a(k),c)b(m-1)} y_a...y_ay_c\,.
\end{align}
The original sum was over all values of $k$ and $m$ but the Young symmetry condition $C^{a(k),ab(m-1)}=0$ defines an identically zero tensor for $k<m$ in accordance with the restriction on Young diagrams not to have shorter rows on top of longer ones.

If tensors need to be traceless we can impose trace constraints as harmonicity with respect to
\begin{align}
\frac{\pl^2}{\pl y_c \pl y^c}\,,& & \frac{\pl^2}{\pl y_c \pl p^c}\,,& & \frac{\pl^2}{\pl p_c \pl p^c}\,.&
\end{align}

\mysection{Symplectic differential calculus}\label{app:diffcalc}
Mastering symplectic calculus requires some time and a handful of examples to compare with. There are
formulas that work for any dimension, i.e. with the only assumptions about the symplectic metric being
\begin{align}
\epsilon_{\ga\gb}&=-\epsilon_{\gb\ga}\,, & \epsilon^{\ga\gb}&=-\epsilon^{\gb\ga}\,,& \epsilon_{\ga\gb}\epsilon^{\gc\gb}&=\delta_\ga^\gc\,,
\end{align}
where $\delta_\ga^\gc$ us the usual identity matrix. There are also formulas that work in $2d$ only,
i.e. when $\ga,\gb,...$ runs over two values, we shall stress this below. The main rules are on how to raise and lower indices
\begin{align}
y^\ga&=\epsilon^{\ga\gb}y_\gb\,, & y_\gc&=y^{\ga}\epsilon_{\ga\gc}\,,
\end{align}
with this one checks $y^\ga=\epsilon^{\ga\gb}y_\gb=\epsilon^{\ga\gb} (y^\gc\, \epsilon_{\gc\gb})=y^\ga$,
where we used $\epsilon^{\ga\gb}\epsilon_{\gc\gb}=\delta^\ga_\gc$.

If we apply raising/lowering rules to $\epsilon_{\ga\gb}$ itself we find first of all that $\epsilon^{\ga\gb}$ is identical to $\epsilon_{\ga\gb}$ with all indices raised. Moreover,
\begin{align}
\epsilon\fdu{\ga}{\gb}&=\delta_\ga^\gb\,, & \epsilon\fud{\ga}{\gb}&=-\epsilon\fdu{\ga}{\gb}=-\delta_\ga^\gb\,.
\end{align}
If we remember these rules we can forget about $\delta^\ga_\gb$ as an independent object. The minus sign in the last expression manifests the antisymmetry of scalar products, e.g. for two vectors $v_\ga$, $u_\gb$, we have
\begin{align}
v_\ga u_\gb \epsilon^{\ga\gb}=v_\ga u^\ga= -v_\ga u_\gb \epsilon^{\gb\ga}=-v^\gb u_\gb\,.
\end{align}
In particular, for any vector $v_\ga v^\ga\equiv0$. Partial derivative $\pl_\ga=\frac{\pl}{\pl y^\ga}$ is defined by the following rules
\besubeqs\begin{align}
\pl_\ga y_\gb&=\epsilon_{\ga\gb}\,, & \pl_\ga y^\gb&=\epsilon\fdu{\ga}{\gb}\equiv \delta^\ga_\gb\,, & \\
\pl^\ga y^\gb&=\epsilon^{\ga\gb}\,, & \pl^\ga y_\gb&=\epsilon\fud{\ga}{\gb}=-\delta^\ga_\gb\,,
\end{align}\esubeqs
so we can think of anyone as the definition, the rest being consequences of raising/lowering rules. The second one, $\pl_\ga y^\gb=\epsilon\fdu{\ga}{\gb}$, is the most natural.
We would like to warn everybody against using blindly the chain rule to relate $\pl_\ga$ to $\pl^\ga$.
The feature of symplectic calculus is that $\pl^\ga$ is defined as $\pl_\ga$ with the index raised according
to the rules above and it does not coincide with $\frac{\pl}{\pl y_\ga}$! This is because raising index of $\pl_\ga$ follows a different rule than lowering index of $y^\ga$ inside $\pl/\pl y^\ga$. It is much more convenient to adopt the same raising/lowering rules for all objects than mess up when the rules for variables and derivatives are different. In particular, $\pl_\ga$ is required to behave in the same way as any other vector. Another way to overcome the difficulty is to always use $\pl_\ga$ and never try to raise the index. In practice the chain rule is not needed, the formulae above are sufficient.

The Euler or number operator is useful sometimes
\begin{align}
N&=y^\gc \pl_\gc\,, & [N,y_\ga]&=y_\ga\,, & [N,\pl_\ga]=-\pl_\ga&\,,
\end{align}
notice the position of indices since $N=-y_\gc \pl^\gc$.

In particular, in $2d$ we have $y^1=y_2$, $y^2=-y_1$ with the canonical choice $\epsilon_{12}=1$. The main property of $2d$ symplectic world is that any tensor that is antisymmetric in two indices is proportional to $\epsilon_{\ga\gb}$
\begin{align}
T_{\ga\gb}&=-T_{\gb\ga} && \Longrightarrow&& T_{\ga\gb}=\frac12\epsilon_{\ga\gb} T_{\gc\gd}\epsilon^{\gc\gd}=
\frac12\epsilon_{\ga\gb} T\fdu{\gc}{\gc}
\end{align}
Therefore, every two indices can be decomposed as follows
\begin{align}
F_{\ga\gb}&=\frac12(F_{\ga\gb}+F_{\gb\ga})+\frac12(F_{\ga\gb}-F_{\gb\ga})=S_{\ga\gb}+\frac12\epsilon_{\ga\gb} F\fdu{\gc}{\gc}\,,\\
S_{\ga\gb}&=\frac12(F_{\ga\gb}+F_{\gb\ga})\,.
\end{align}
The big consequence is that in $2d$ all nontrivial tensors are symmetric. Whatever antisymmetric part we find can be expressed in terms of a number of $\epsilon$ factors and a totally symmetric tensor, which is the symplectic trace of the original one.

The last thing is that in higher-spin theory one finds dotted and undotted symplectic indices running
over two values, e.g. $y_\ga$, $y_\gad$. These are considered as totally independent objects/indices and they
will never mix together (there is no such object as $\epsilon_{\ga\gad}$) unless one considers particular solutions to the theory, where the choice of coordinates can break Lorentz symmetry, e.g. \eqref{flatP}. In fact $\ga$ and $\gad$ are different components of the $sp(4)$ index $A$, $A=\{\ga,\gad\}$. Sometimes we use $x^{\ga\gad}$ as a spinorial avatar for $x^m$, they are related by $\sigma_m^{\ga\gad}$. The rules for $\ga,\gad,...$ are the same, in particular
\begin{align}
\pl_{\ga\gad} x^{\gb\gbd}&= \epsilon\fdu{\ga}{\gb}\epsilon\fdu{\gad}{\gbd}
\end{align}
hence, $\pl_{\ga\gad} x^{\ga\gad}=4$ in accordance with $\pl_m x^m=4$ in $4d$ where it applies.

\mysection{More on \texorpdfstring{$\boldsymbol{so(3,2)}$}{so(3,2)}}\label{app:moreonso}
Since the anti-de Sitter algebra $so(3,2)$ is at the core of the $4d$ higher-spin theory,
in particular the higher-spin algebra can be described in terms of the universal enveloping algebra $U(so(3,2))$,
and there is a special isomorphism $so(3,2)\sim sp(4,\mathbb{R})$ that simplifies things a lot and make the $4d$ theory special, we would like to give more info on $so(3,2)$. This section could be too much as at the end we will have representation theory of $so(3,1)\sim sl(2,\mathbb{C})$ and $so(3,2)\sim sp(4,\mathbb{R})$  expressed in two different forms each.

\mysubsection{Restriction of \texorpdfstring{${so(3,2)}$ to ${so(3,1)}$}{so(3,2) to so(3,1)}}

First of all, general arguments from the unfolded approach, see \eqref{UnfldqFormsEquations} and after, tells us
that whenever we find a closed subset of fields of the same form degree, it must form certain representation
of the background space symmetry algebra.

For example this should apply to the set of one-forms $\omega^{a(s-1),b(k)}$, $k=0,...,s-1$. We considered the $d$-dimensional Minkowski space, but the set does not change when we switch on the cosmological constant, \cite{Lopatin:1987hz}. So, this set of fields must belong to certain finite-dimensional representations of the anti-de Sitter algebra $so(d-1,2)$, which are known to be tensors or spin-tensors\footnote{Recall that Poincare algebra is not semi-simple and talking about Poincare tensors sounds unnatural.}.

The simplest example of this kind is the pure gravity, where
MacDowell-Mansouri-Stelle-West results, see Section \ref{extra:MMSW}, show
that vielbein $e^a$ and spin-connection $\omega^{a,b}$ can be
viewed as different components of a single $so(d-1,2)$-connection
$\Omega^{\aAs,\aBs}$. In this case, the statement is that a $(d+1)\times
(d+1)$ antisymmetric matrix can be viewed as $d\times d$ one, $\Omega^{a,b}$, and a
$d$-dimensional vector\footnote{Again, $5$ denotes the
extra value of $\aAs$ index that is complementary to the $d$-dimensional $a$ index, $\aAs=\{a,5\}$.}, $\Omega^{a,5}$. In the language of group theory we say
that
\begin{align}
\left. \YoungpAA \right |_{so(d+1)\downarrow so(d)} \sim \YoungpAA \oplus \YoungpA
\end{align}
which is called the branching rules or restriction rules. It is also easy to see that
\begin{align}
\left. \YoungpB \right |_{so(d+1)\downarrow so(d)} \sim \YoungpB \oplus \YoungpA \oplus \bullet
\end{align}
Indeed, given a symmetric traceless $so(d+1)$ tensor $T^{\aAs\aBs}$ we can decompose it as the traceless symmetric tensor $T^{ab}-\frac1{d} \eta^{ab} T\fud{m}{m}$, vector $T^{a 5}$ and scalar $T\fud{m}{m}$. Note that $so(d+1)$ tracelessness implies $0\equiv T\fud{\aAs}{\aAs}=T\fud{a}{a}+\eta_{55}T^{55}$, i.e. $T\fud{a}{a}$ and $T^{55}$ are the same up to a sign.
A less trivial example, which is related to the spin-three case, is
\begin{align}
\left. \YoungpBB \right |_{so(d+1)\downarrow so(d)} \sim \YoungpBB \oplus \YoungpBA \oplus \YoungpB
\end{align}
On the r.h.s we see exactly the Young symmetry types that are needed for the frame-like formulation of a spin-three field.

The full dictionary up to two-row Young diagrams is as follows\\
\phantom{a}\\
\noindent\begin{tabular}{|x{0.1cm}x{6cm}|x{8.6cm}|}
\hline
&$so(d+1)$ tensor & $so(d)$ tensor content\tabularnewline\hline
&$\bullet$ & $\bullet$ \tabularnewline
&$T^\aAs\sim\YoungpA$ & $\YoungpA\oplus\bullet$ \tabularnewline
\rule{0pt}{26pt}& $T^{\aAs\aAs}\sim\YoungB$&  $\YoungpB\oplus\YoungpA\oplus\bullet$ \tabularnewline
\rule{0pt}{26pt}& $T^{\aAs(k)}\sim\AYoungp{4}{k}$ & $\AYoungp{4}{k}\oplus\AYoungp{4}{k-1}\oplus...\oplus\YoungpA\oplus\bullet$ \tabularnewline
\rule{0pt}{26pt}& $T^{\aAs,\aBs}\sim\YoungpAA$ & $\YoungpAA\oplus\YoungpA$\tabularnewline
\rule{0pt}{26pt}& $T^{\aAs(k),\aBs(k)}\sim\BYoungp{4}{4}{k}{}$ &
$\BYoungp{4}{4}{k}{}\oplus\BYoungp{4}{3}{k}{}\oplus...\oplus\BYoungp{4}{1}{k}{}\oplus\AYoungp{4}{k}$  \tabularnewline
\rule{0pt}{26pt}& $T^{\aAs(m+k),\aBs(m)}\sim\BYoungp{6}{4}{m+k}{m}$ &
$\displaystyle\bigoplus_{i=0}^{i=k}\bigoplus_{j=0}^{j=m}\BYoungp{6}{4}{m+i}{j}$  \tabularnewline
\hline
\end{tabular}
\vspace{0.3cm}

The most important line is the last but one, which shows that all higher-spin connections $\omega^{a(s-1),b(k)}$
needed for a spin-$s$ field can be packed into just one connection of $so(d-1,2)$
\be W^{\aAs(s-1),\aBs(s-1)}\ee
which has the symmetry of the two-row rectangular Young diagram of length-$(s-1)$, \cite{Vasiliev:2001wa}.

Again let us note that the branching rules expressed in terms of
Young diagrams are much simpler than the equivalent statements in
the language of tensors.

\mysubsection{\texorpdfstring{$so(3,2)\sim sp(4,\mathbb{R})$}{so(3,2) is sp(4,R)}}\label{app:sofivespfour}
One more miraculous isomorphism is between the anti-de Sitter algebra $so(3,2)$ and $sp(4,\mathbb{R})$. The signature is irrelevant in the section, so one can think of complex Lie algebras, but we do not change the notation. Let us first note that
according to the general relation between structure constants of unfolded equations and representation theory
the set of one-forms needed for a spin-$s$ field, i.e.
\begin{align}
&\omega^{\ga(m),\gad(n)} && m+n=2(s-1)\,,
\end{align}
must belong to some finite-dimensional representation of $so(3,2)$, see previous section. The same time, as we noticed around \eqref{spfourpacking}, the same field content can be packed as
\begin{align}
&\omega^{\Omega(2s-2)}\,Y_\Omega...Y_\Omega\,,
\end{align}
where $\Omega$ runs\footnote{In the main text we use $A,B,...$ as $sp(4,\mathbb{R})$ indices and $\aAs,\aBs,...$ as $so(3,2)$ indices, but this could cause a confusion when $so(3,2)$ and $sp(4,\mathbb{R})$ are confronted. Therefore, $A,B,...$ are changed to $\Lambda,\Omega,...$.} over four values, which cover $\{\ga,\gad\}$. The reason is that $so(3,2)\sim sp(4,\mathbb{R})$ and $Y_\Omega$ is a vector of $sp(4,\mathbb{R})$, so both Lorentz algebra $so(3,1)$ and anti-de Sitter algebra $so(3,2)$ are special.

To prove the isomorphism and build the dictionary one introduces $so(3,2)$ Dirac $\gamma$-matrices $\gamma_\aAs\equiv \gamma_\aAs{}\fud{\Lambda}{\Omega}$, $\Lambda,\Omega,...=1,...,4$, $\aAs,\aBs,...=0,...,4$.
\be(\gamma_\aAs\gamma_\aBs{})\fud{\Lambda}{\Omega}+(\gamma_\aBs\gamma_\aAs){}\fud{\Lambda}{\Omega}
=2\delta{}\fud{\Lambda}{\Omega}\eta_{\aAs \aBs}\ee
The generators of $so(3,2)$ in the spinorial representation
\be T_{\aAs\aBs}=-T_{\aBs\aAs}=\frac14 [\gamma_\aAs,\gamma_\aBs]\ee
can be observed to have a special structure,
\begin{align}
T_{\aAs\aBs}{}^{\Lambda\Omega}&=T_{\aAs\aBs}{}^{\Omega\Lambda}\,,
\end{align}
where we raise and lower $\Lambda,\Omega,...$-indices with the charge-conjugation matrix $C_{\Lambda \Omega}=-C_{\Omega\Lambda}$ using the standard symplectic rules. The charge-conjugation matrix is going to be the invariant tensor of $sp(4,\mathbb{R})$. Usually the above relations are written as $(\gamma_{\aAs\aBs}C)^T=(\gamma_{\aAs\aBs}C)$.

In addition $\gamma$-matrices can be shown to be all antisymmetric and $C_{\Lambda \Omega}$-traceless
\begin{align}
\gamma_{\aAs}{}^{\Lambda \Omega}&=-\gamma_{\aAs}{}^{\Omega\Lambda}\,, & \gamma_{\aAs}{}^{\Lambda \Omega}C_{\Lambda \Omega}&=0\,,
\end{align}
which is usually written as $(\gamma_{\aAs}C)^T=-(\gamma_{\aAs}C)$.
The latter property implies that one can use $\gamma$-matrices to map an $so(3,2)$-vector, say $V_\aAs$,
into antisymmetric rank-two tensor $V^{\Lambda \Omega}=-V^{\Omega\Lambda}$, $V^{\Lambda \Omega}=\gamma_{\aAs}{}^{\Lambda \Omega} V^\aAs$. This is an isomorphism, which can be proven by observing that $V^{\Lambda \Omega}$ has the same number of components, $4\cdot 3/2-1=5$, as $V^\aAs$ and one can find a backward transform using the properties of $\gamma$-matrices.

Analogously, a rank-two antisymmetric tensor of $so(3,2)$, say $B^{\aAs,\aBs}=-B^{\aBs,\aAs}$ can be mapped into
rank-two symmetric tensor $B^{\Lambda \Omega}=B^{\Omega\Lambda}$, $B^{\Lambda \Omega}=T^{\Lambda \Omega}{}_{\aAs\aBs}B^{\aAs,\aBs}$. This is an isomorphism again.

Continuing along the same lines, one can derive the following $so(3,2)-sp(4,\mathbb{R})$ dictionary \\
\phantom{a}\\
\begin{tabular}{|x{0.1cm}x{6cm}|x{6cm}|x{2cm}|}
\hline
&$so(3,2)$ tensor & $sp(4,\mathbb{R})$ tensor & dim\tabularnewline\hline
&$\bullet$ & $\bullet$ & 1 \tabularnewline
&Dirac spinor & $T^\Lambda\sim\YoungpA$ & 4 \tabularnewline
\rule{0pt}{26pt}& $T^{\aAs}\sim\YoungpA$ & $T^{\Lambda,\Omega}\sim\YoungpAA$ & 5\tabularnewline
\rule{0pt}{26pt}& $T^{\aAs,\aBs}\sim\YoungpAA$ & $T^{\Lambda\Lambda}\sim\YoungpB$ & 10\tabularnewline
\rule{0pt}{26pt}& $T^{\aAs\aAs}\sim\YoungB$&  $T^{\Lambda\Lambda,\Omega\Omega}\sim\YoungBB$ & 15\tabularnewline
\rule{0pt}{26pt}& $T^{\aAs(k)}\sim\AYoungp{4}{k}$ & $T^{\Lambda(k),\Omega(k)}\sim\BYoungp{4}{4}{k}{}$ &  \tabularnewline
\rule{0pt}{26pt}& $T^{\aAs(k),\aBs(k)}\sim\BYoungp{4}{4}{k}{}$ &
$T^{\Lambda(2k)}\sim\AYoungp{6}{2k}$ &  \tabularnewline
\rule{0pt}{26pt}& $T^{\aAs(k),\aBs(m)}\sim\BYoungp{6}{4}{k}{m}$ &
$T^{\Lambda(k+m),\Omega(k-m)}\sim\BYoungp{6}{4}{k+m}{k-m}$ &  \tabularnewline
\hline
\end{tabular}
\vspace{0.3cm}

Irreducible $so(3,2)$ tensors are traceless with respect to symmetric $\eta_{AB}$ and irreducible $sp(4,\mathbb{R})$ tensors are traceless with respect to antisymmetric $C_{\Lambda\Omega}$.

To summarize, we have the following equivalent ways to describe the space of higher-spin connections of
a spin-$s$ field in $4d$
\besubeqs\begin{align}
so(3,1):& &\bigoplus_{k=0}^{s-1}&\,\omega^{a(s-1),b(k)} \label{LorGen}\\
sl(2,\mathbb{C}):&  &\bigoplus_{i+j=2(s-1)}&\omega^{\ga(i),\gad(j)}\\
so(3,2):&  &&\omega^{A(s-1),B(s-1)}\label{AdSGen}\\
sp(4,\mathbb{R}):&  &&\omega^{\Lambda(2s-2)}
\end{align}\esubeqs
\eqref{LorGen} and \eqref{AdSGen} are valid in any dimension $d\geq4$ provided we replace $so(3,1)$ and $so(3,2)$ with $so(d-1,1)$ and $so(d-1,2)$, respectively.

\end{appendix}


    \bibliographystyle{JHEP-2}%
    \bibliography{megabib}%
\end{document}

%% file: YoungDiagrams.tex
\newcommand{\bep}{\begin{picture}}
\newcommand{\eep}{\end{picture}}

\newcommand{\smallpic}[1]{{\unitlength=0.2mm#1}}

\newcounter{YoungHeight}\newcounter{YoungWidth}

\newcounter{Mul1}\newcounter{Mul2}\newcounter{Mul3}\newcounter{Mul4}
\newcounter{A0}\newcounter{A1}\newcounter{A2}
\newcounter{B3}
\newcounter{C3}\newcounter{C4}
\newcounter{D1}\newcounter{D2}\newcounter{D3}
\newcounter{T0}\newcounter{T1}

\newlength{\txtHShift}

\newlength{\txtWidth}

\newcommand{\HalfLength}[2]{\setcounter{Mul1}{#1}\setcounter{Mul2}{#1}\addtocounter{Mul1}{\value{Mul2}}\addtocounter{Mul1}{\value{Mul2}}%
\addtocounter{Mul1}{\value{Mul2}}\addtocounter{Mul1}{\value{Mul2}}\setcounter{#2}{\value{Mul1}}}

\newcommand{\Add}[3]{\setcounter{#1}{#2}\addtocounter{#1}{#3}}

\newcommand{\Length}[1]{#10}
\newcommand{\YoungScale}{}

\newcommand{\shiftedText}[2]{{\hspace{#1}#2}}
\newcommand{\calcHShift}[1]{\settowidth{\txtWidth}{#1}\setlength{\txtHShift}{-0.5\txtWidth}}

\newcommand{\TextTop}[3]{{\calcHShift{#1}\HalfLength{#2}{T0}\Add{T1}{\Length{#3}}{-9}\put(\value{T0},\value{T1}){\shiftedText{\txtHShift}{#1}}}}

\newcommand{\BlockA}[2]{{\YoungScale\bep(\Length{#1},\Length{#2}){\Add{A1}{#1}{1}\Add{A2}{#2}{1}}%
\multiput(0,0)(10,0){\value{A1}}{\line(0,1){\Length{#2}}}\multiput(0,0)(0,10){\value{A2}}{\line(1,0){\Length{#1}}}%
\setcounter{YoungHeight}{\Length{#2}}\setcounter{YoungWidth}{\Length{#1}}\eep}}

\newcommand{\BlockB}[4]{{\YoungScale\Add{B3}{\Length{#2}}{\Length{#4}}%
\bep(\Length{#1},\value{B3})\put(0,\Length{#4}){\BlockA{#1}{#2}}%
\put(0,0){\BlockA{#3}{#4}}\setcounter{YoungHeight}{\value{B3}}\setcounter{YoungWidth}{\Length{#1}}\eep}}

\newcommand{\RectT}[3]{\bep(\Length{#1},\Length{#2})\put(0,0){\line(1,0){\Length{#1}}}\put(0,0){\line(0,1){\Length{#2}}}%
\put(\Length{#1},\Length{#2}){\line(-1,0){\Length{#1}}}\put(\Length{#1},\Length{#2}){\line(0,-1){\Length{#2}}}#3{#1}{#2}\eep}

\newcommand{\RectARow}[2]{{\bep(\Length{#1},10)\put(0,0){\RectT{#1}{1}{\TextTop{#2}}}\eep}}

\newcommand{\RectBRow}[4]{{\bep(\Length{#1},20)\put(0,0){\RectT{#2}{1}{\TextTop{#4}}}%
\put(0,10){\RectT{#1}{1}{\TextTop{#3}}}\eep}}

\newcommand{\RectCRow}[6]{{\bep(\Length{#1},30)\put(0,0){\RectT{#3}{1}{\TextTop{#6}}}%
\put(0,10){\RectT{#2}{1}{\TextTop{#5}}}\put(0,20){\RectT{#1}{1}{\TextTop{#4}}}\eep}}

\newcommand{\RectAYoung}[3]{{\bep(0,0)#3\eep\Add{A0}{\value{YoungWidth}}{\Length{#1}}%
\Add{A1}{\value{YoungHeight}}{-10}\bep(\value{A0},\value{YoungHeight})%
\put(\value{YoungWidth},\value{A1}){\RectT{#1}{1}{\TextTop{#2}}}\eep}}

\newcommand{\RectBYoung}[3]{{\bep(0,0)\put(0,0){#3}\eep\Add{A0}{\value{YoungHeight}}{10}%
\bep(\Length{#1},\value{A0})%
\put(0,\value{YoungHeight}){\RectT{#1}{1}{\TextTop{#2}}}\eep}}

\newcommand{\YoungA}{\BlockA{1}{1}}
\newcommand{\YoungB}{\BlockA{2}{1}}
\newcommand{\YoungC}{\BlockA{3}{1}}
\newcommand{\YoungD}{\BlockA{4}{1}}

\newcommand{\YoungAA}{\BlockA{1}{2}}
\newcommand{\YoungBA}{\BlockB{2}{1}{1}{1}}
\newcommand{\YoungBB}{\BlockA{2}{2}}
\newcommand{\YoungCA}{\BlockB{3}{1}{1}{1}}
\newcommand{\YoungCB}{\BlockB{3}{1}{2}{1}}

\newcommand{\YoungAAA}{\BlockA{1}{3}}

\newcommand{\BlockApar}[2]{\parbox{\Length{#1}pt}{\YoungScale\bep(\Length{#1},\Length{#2}){\Add{A1}{#1}{1}\Add{A2}{#2}{1}}%
\multiput(0,0)(10,0){\value{A1}}{\line(0,1){\Length{#2}}}\multiput(0,0)(0,10){\value{A2}}{\line(1,0){\Length{#1}}}%
\setcounter{YoungHeight}{\Length{#2}}\setcounter{YoungWidth}{\Length{#1}}\eep}}

\newcommand{\BlockBpar}[4]{\parbox{\Length{#1}pt}{\YoungScale\Add{B3}{\Length{#2}}{\Length{#4}}%
\bep(\Length{#1},\value{B3})\put(0,\Length{#4}){\BlockA{#1}{#2}}%
\put(0,0){\BlockA{#3}{#4}}\setcounter{YoungHeight}{\value{B3}}\setcounter{YoungWidth}{\Length{#1}}\eep}}

\newcommand{\YoungpA}{\BlockApar{1}{1}}
\newcommand{\YoungpB}{\BlockApar{2}{1}}
\newcommand{\YoungpC}{\BlockApar{3}{1}}
\newcommand{\YoungpD}{\BlockApar{4}{1}}

\newcommand{\YoungpAA}{\BlockApar{1}{2}}
\newcommand{\YoungpBA}{\BlockBpar{2}{1}{1}{1}}
\newcommand{\YoungpBB}{\BlockApar{2}{2}}
\newcommand{\YoungpCA}{\BlockBpar{3}{1}{1}{1}}
\newcommand{\YoungpCB}{\BlockBpar{3}{1}{2}{1}}

\newcommand{\YoungpDB}{\BlockBpar{4}{1}{2}{1}}

\newcommand{\YoungpAAA}{\BlockApar{1}{3}}
\newcommand{\YoungpBAA}{\BlockBpar{2}{1}{1}{2}}
\newcommand{\YoungpBBA}{\BlockBpar{2}{2}{1}{1}}

\newcommand{\YoungpAAAA}{\BlockApar{1}{4}}

\newcommand{\BlockAm}[2]{\mbox{\smallpic{\YoungScale\bep(\Length{#1},\Length{#2}){\Add{A1}{#1}{1}\Add{A2}{#2}{1}}%
\multiput(0,0)(10,0){\value{A1}}{\line(0,1){\Length{#2}}}\multiput(0,0)(0,10){\value{A2}}{\line(1,0){\Length{#1}}}%
\setcounter{YoungHeight}{\Length{#2}}\setcounter{YoungWidth}{\Length{#1}}\eep}}}

\newcommand{\BlockBm}[4]{\mbox{\smallpic{\YoungScale\Add{B3}{\Length{#2}}{\Length{#4}}%
\bep(\Length{#1},\value{B3})\put(0,\Length{#4}){\BlockA{#1}{#2}}%
\put(0,0){\BlockA{#3}{#4}}\setcounter{YoungHeight}{\value{B3}}\setcounter{YoungWidth}{\Length{#1}}\eep}}}

\newcommand{\YoungmA}{\BlockAm{1}{1}}
\newcommand{\YoungmB}{\BlockAm{2}{1}}

\newcommand{\YoungmAA}{\raisebox{-2pt}{\BlockAm{1}{2}}}
\newcommand{\YoungmBA}{\raisebox{-2pt}{\BlockBm{2}{1}{1}{1}}}
\newcommand{\YoungmBB}{\raisebox{-2pt}{\BlockAm{2}{2}}}

\newcommand{\YoungmCB}{\raisebox{-2pt}{\BlockBm{3}{1}{2}{1}}}

%% file: notation.tex
\newcolumntype{x}[1]{%
>{\centering\hspace{0pt}}m{#1}}%
\newcolumntype{w}[1]{%
>{\raggedright\hspace{0pt}}m{#1}}%
\newcolumntype{z}[1]{%
>{\raggedleft\hspace{0pt}}m{#1}}%

\newcommand{\putfigureifpdf}[1]{%
\ifthenelse{\isundefined{\OneChapterOnly}}{#1}{
\ifpdf%
{#1}%
\else%
\fi}%
}%

\newcommand{\CYoungp}[6]{\parbox{\Length{#1}pt}{{\RectCRow{#1}{#2}{#3}{${#4}$}{${#5}$}{${#6}$}}}}
\newcommand{\BYoungp}[4]{\parbox{\Length{#1}pt}{{\RectBRow{#1}{#2}{${ #3}$}{${ #4}$}}}}
\newcommand{\AYoungp}[2]{\parbox{\Length{#1}pt}{{\RectARow{#1}{${#2}$}}}}

\newcommand{\BYoung}[4]{{{\smallpic{\RectBRow{#1}{#2}{${\scriptstyle #3}$}{${\scriptstyle #4}$}}}}}

\newcommand{\module}{{\ensuremath{\mathcal{R}}}}

\newcommand{\DL}{{D}}
\newcommand{\DD}{{{\mathcal{D}}}}
\newcommand{\DO}{{D_{\Omega}}}

\newcommand{\WW}{{\ensuremath{\mathcal{W}}}}

\newcommand{\Sigm}{{\ensuremath{{\boldsymbol \sigma}_-}}}

\newcommand{\pl}{\partial}

\newcommand{\fm}[1]{_{\boldsymbol{{#1}}}}

\newcommand{\be}{\begin{equation}}
\newcommand{\ee}{\end{equation}}
\newcommand{\beno}{\begin{equation*}}
\newcommand{\eeno}{\end{equation*}}

\newcommand{\besubeqs}{\begin{subequations}}
\newcommand{\esubeqs}{\end{subequations}}

\newcommand{\Aa}{{\ensuremath{\mathsf{a}}}}
\newcommand{\Ab}{{\ensuremath{\mathsf{b}}}}

\newcommand{\ua}{{\ensuremath{\underline{a}}}}
\newcommand{\ub}{{\ensuremath{\underline{b}}}}

\newcommand{\mm}{{\ensuremath{\underline{m}}}}
\newcommand{\nn}{{\ensuremath{\underline{n}}}}
\newcommand{\rr}{{\ensuremath{\underline{r}}}}
\newcommand{\kk}{{\ensuremath{\underline{k}}}}
\newcommand{\uu}{{\ensuremath{\underline{u}}}}
\newcommand{\ii}{{\ensuremath{\underline{i}}}}

\newcommand{\aI}{{\ensuremath{\mathcal{I}}}}
\newcommand{\aJ}{{\ensuremath{\mathcal{J}}}}
\newcommand{\aK}{{\ensuremath{\mathcal{K}}}}

\newcommand{\aAt}{{\ensuremath{\mathtt{A}}}}

\newcommand{\aAs}{{\ensuremath{\mathsf{A}}}}
\newcommand{\aBs}{{\ensuremath{\mathsf{B}}}}
\newcommand{\aCs}{{\ensuremath{\mathsf{C}}}}
\newcommand{\aDs}{{\ensuremath{\mathsf{D}}}}
\newcommand{\aFs}{{\ensuremath{\mathsf{F}}}}

\newcommand{\aUs}{{\ensuremath{\mathsf{U}}}}

\newcommand{\aWs}{{\ensuremath{\mathsf{W}}}}

\newcommand{\aA}{{\ensuremath{\mathcal{A}}}}
\newcommand{\aB}{{\ensuremath{\mathcal{B}}}}
\newcommand{\aC}{{\ensuremath{\mathcal{C}}}}
\newcommand{\aD}{{\ensuremath{\mathcal{D}}}}
\newcommand{\aE}{{\ensuremath{\mathcal{E}}}}

\newcommand{\aAb}{{\ensuremath{\boldsymbol{\mathcal{A}}}}}
\newcommand{\aBb}{{\ensuremath{\boldsymbol{\mathcal{B}}}}}
\newcommand{\aCb}{{\ensuremath{\boldsymbol{\mathcal{C}}}}}
\newcommand{\aDb}{{\ensuremath{\boldsymbol{\mathcal{D}}}}}

\newcommand{\gad}{{\dot{\alpha}}}
\newcommand{\gbd}{{\dot{\beta}}}
\newcommand{\gdd}{{\dot{\gamma}}}
\newcommand{\gdl}{{\dot{\delta}}}

\newcommand{\gnd}{{\dot{\nu}}}

\newcommand{\ga}{\alpha}
\newcommand{\gb}{\beta}
\newcommand{\gc}{\gamma}
\newcommand{\gd}{\delta}
\newcommand{\gl}{\lambda}

\newcommand{\fud}[2]{{}^{#1}{}_{#2}\,}
\newcommand{\fdu}[2]{{}_{#1}{}^{#2}\,}
\newcommand{\fudu}[3]{{}^{#1}{}_{#2}{}^{#3}\,}

\newcommand{\DOt}{{\ensuremath{\tilde{D}_\Omega}}{}}

\newcommand{\rhs}{{\it r.h.s. }{}}
\newcommand{\lhs}{{\it l.h.s. }{}}

\newcommand{\bry}{{{\bar{y}}}}

\newcommand{\gk}{\varkappa}
\newcommand{\gkb}{{\bar{\varkappa}}}
\newcommand{\gep}{\epsilon}

\newcommand{\gs}{\sigma}
\newcommand{\go}{\omega}

\newcommand{\ff}{\frac}
\newcommand{\p}{\partial}

\newcommand{\Scal}{{\cal S}}
\newcommand{\BB}{{\cal B}}
\newcommand{\CC}{{\cal C}}

%% file: figures.tex
\newcommand\pgfmathsinandcos[3]{%
  \pgfmathsetmacro#1{sin(#3)}%
  \pgfmathsetmacro#2{cos(#3)}%
}
\newcommand\LongitudePlane[3][current plane]{%
  \pgfmathsinandcos\sinEl\cosEl{#2} 
  \pgfmathsinandcos\sint\cost{#3} 
  \tikzset{#1/.estyle={cm={\cost,\sint*\sinEl,0,\cosEl,(0,0)}}}
}
\newcommand\LatitudePlane[3][current plane]{%
  \pgfmathsinandcos\sinEl\cosEl{#2} 
  \pgfmathsinandcos\sint\cost{#3} 
  \pgfmathsetmacro\yshift{\cosEl*\sint}
  \tikzset{#1/.estyle={cm={\cost,0,0,\cost*\sinEl,(0,\yshift)}}} %
}
\newcommand\DrawLongitudeCircle[2][1]{
  \LongitudePlane{\angEl}{#2}
  \tikzset{current plane/.prefix style={scale=#1}}
  \pgfmathsetmacro\angVis{atan(sin(#2)*cos(\angEl)/sin(\angEl))} %
  \draw[current plane] (\angVis:1) arc (\angVis:\angVis+180:1);
  \draw[current plane,dashed] (\angVis-180:1) arc (\angVis-180:\angVis:1);
}
\newcommand\DrawLatitudeCircle[2][1]{
  \LatitudePlane{\angEl}{#2}
  \tikzset{current plane/.prefix style={scale=#1}}
  \pgfmathsetmacro\sinVis{sin(#2)/cos(#2)*sin(\angEl)/cos(\angEl)}
  \pgfmathsetmacro\angVis{asin(min(1,max(\sinVis,-1)))}
  \draw[current plane] (\angVis:1) arc (\angVis:-\angVis-180:1);
  \draw[current plane,dashed] (180-\angVis:1) arc (180-\angVis:\angVis:1);
}

\newcommand{\TikzRect}[2]{\filldraw[color=black,fill=red]  (#1-\R,#2-\R) rectangle (#1+\R,#2+\R);}

\newcommand{\AmbientSpace}{
\begin{figure}[h!]
  \caption{\it Sphere in ambient space and the compensator}
  \centering
\begin{tikzpicture}

\def\R{2.5} 
\def\angEl{35} 
\def\angAz{-105} 
\def\angPhi{-40} 
\def\angBeta{19} 
\pgfmathsetmacro\H{\R*cos(\angEl)}
\tikzset{xyplane/.estyle={cm={cos(\angAz),sin(\angAz)*sin(\angEl),-sin(\angAz),
                              cos(\angAz)*sin(\angEl),(0,\H)}}}

\filldraw[ball color=white] (0,0) circle (\R);
\foreach \t in {-80,-60,...,80} { \DrawLatitudeCircle[\R]{\t} }
\foreach \t in {-5,-35,...,-175} { \DrawLongitudeCircle[\R]{\t} }
\draw[-] (0,0) -- (0,1.6*\R) node[above] {};

\draw[xyplane] (-1.5*\R,-1.5*\R) rectangle (1.5*\R,1.5*\R);

\draw[xyplane][fill=lightgray,draw=red,opacity=.5,very thin,line join=round]
 (-1.5*\R,-1.5*\R,0) --
 (-1.5*\R,1.5*\R,0) --
 (1.5*\R,1.5*\R,0) --
 (1.5*\R,-1.5*\R,0) --cycle ;
\draw[->] (0,0.8*\R) -- (0,2.5*\R) node[above] {$V^A$};
\draw[xyplane][->] (0,0) -- (0,1.3*\R) node[above] {$x$};
\draw[xyplane][->] (0,0) -- (1.3*\R,0) node[left] {$y$};

\draw[-latex,thick](4,5)node[above,scale=1.3]{Physical 'Lorentz' subspace}
        to[out=-90,in=45] (2,3.0);
\end{tikzpicture}
\end{figure}}

\newcommand{\UnfoldedIdea}{\begin{figure}[h!]
  \caption{\it The idea of the unfolded approach. }
  \centering
\begin{tikzpicture}
 	\draw[semithick] (0,0) ellipse (1 and 0.5);
	\draw[semithick] (0,0) ellipse (2 and 1.0);
	\draw[semithick] (0,-2) ellipse (0.32 and 0.16);
	\draw[semithick] (0,-2) ellipse (1 and 0.5);
	\draw[semithick] (1,0) -- (0,-3);
    \draw[semithick] (-1,0) -- (0,-3);
	\draw[semithick] (1.96,-0.2) -- (0,-4);
    \draw[semithick] (-1.96,-0.2) -- (0,-4);

    \draw[-latex,thick](2,-3)node[above,scale=1.3]{$\phi$}
        to[out=-90,in=0] (0,-4.0);
     \draw[-latex,thick](3,-0.5)node[above,scale=1.3]{$\partial^k\phi$}
        to[out=-90,in=0] (1.0,-2.0);
   \draw[-latex,thick](0,4)node[above,scale=1.3]{$Eq(\phi)=0$, $\delta\phi=\partial\xi$}
        to[out=-90,in=80] (1.5, 0);
    \draw[-latex,thick](-2.5,2)node[above,text width=10em]{degrees of freedom, i.e. zero-forms}
        to[out=-90,in=100] (0, 0);
        \draw[color=black,thick,->] (4,-4,0) -- (4,1,0) node[anchor=south]{derivatives};
\end{tikzpicture}
\end{figure}}

\newcommand{\UnfoldedEqFieldMap}{\begin{figure}[h!]
\caption{\it  A picture showing the position of various fields and their mixing via unfolded equations. The coordinates are the number of undotted/dotted indices carried by a field. The fields connected by links talk
to each other via the unfolded equations \eqref{UnfldsSpinorial}.}

\begin{tikzpicture}[scale=0.8]
\tikzset{%
  >=latex, 
  inner sep=0pt,%
  outer sep=2pt,%
  mark coordinate/.style={inner sep=0pt,outer sep=0pt,minimum size=3pt,
    fill=black,circle}%
}
\def\R{0.15}
\def\mar{0.15}

\draw[->,black,thick] (0,0) -- (0,8) node[left]{dotted};
\draw[->,black,thick] (0,0) -- (8,0) node[right]{undotted} ;

\filldraw[color=black,fill=green]  (9,6) circle (\R);
\node[right] at (9.3,6) {gauge module, one-forms};

\filldraw[color=black,fill=red]  (9-\R,7-\R) rectangle (9+\R,7+\R);
\node[right] at (9.3,7) {Weyl module, zero-forms};

\filldraw[color=black,fill=green]  (4,0) circle (\R);
\filldraw[color=black,fill=green]  (3,1) circle (\R);
\filldraw[color=black,fill=green]  (2,2) circle (\R);
\filldraw[color=black,fill=green]  (1,3) circle (\R);
\filldraw[color=black,fill=green]  (0,4) circle (\R);
\draw[green,thick] (4,0) -- (0,4);

    \draw[-latex,thick] (3.5,3.5) node[above,scale=1.0]{$e^{\alpha(s-1),\dot{\alpha}(s-1)}$}
        to[out=-90,in=45] (2+\mar, 2+\mar);

    \draw[-latex,thick] (4,6) node[right,scale=1.0]{$C^{\dot{\alpha}(2s)}$}
        to[out=180,in=0] (\mar, 5);

    \draw[-latex,thick] (-2,6) node[above,scale=1.0]{$\omega^{\dot{\alpha}(2s-2)}$}
        to[out=-90,in=180] (-\mar, 4);

    \draw[-latex,thick] (6,4) node[above,scale=1.0]{$C^{\alpha(2s)}$}
        to[out=-90,in=90] (5, \mar);

    \draw[-latex,thick] (6,-2) node[right,scale=1.0]{$\omega^{{\alpha}(2s-2)}$}
        to[out=180,in=-90] (4, -\mar);

\TikzRect{0}{5}
\TikzRect{1}{6}
\TikzRect{2}{7}
\TikzRect{3}{8}
\draw[red,thick] (5,0) -- (9,4);

\TikzRect{5}{0}
\TikzRect{6}{1}
\TikzRect{7}{2}
\TikzRect{8}{3}
\draw[red,thick] (0,5) -- (4,9);

    \draw[-latex,thick] (-1.5,-1.5)  to[out=20,in=-90] (4.5, -\R);
    \draw[-latex,thick] (-1.5,-1.5)  to[out=70,in=180] (-\R, 4.5);

\filldraw[color=black,fill=black]  (-\R,4+2*\R) rectangle (\R,5-2*\R);
\filldraw[color=black,fill=black]  (4+2*\R,-\R) rectangle (5-2*\R,\R);

\node[below] at (-1.5,-1.5) {\large Chevalley-Eilenberg cocycle};

\end{tikzpicture}
\end{figure}}

\newcommand{\figureTangetBase}{
\begin{figure}[h!]
\caption{\it The basis vectors $\vec{e}_i$in the tangent plane $TM_O$ at some point $O$ are by definition the vectors that are tangent to the coordinate lines. Let $p(x_1,...,x_n)$ be the point on the manifold parameterized by Cartesian coordinates $(x_1,...,x_n)$ in some chart, we can think of it as a point in a bigger Euclidian space where the given manifold is embedded. Then, $\vec{e}_i=\frac{d}{dt}p(x_1,...,x_i+t,...,x_n)|_{t=0}$, which are shown below. }

\begin{tikzpicture}
\tikzset{%
  >=latex, 
  inner sep=0pt,%
  outer sep=2pt,%
  mark coordinate/.style={inner sep=0pt,outer sep=0pt,minimum size=3pt,
    fill=black,circle}%
}
 \tikzstyle{ann} = [fill=white,font=\footnotesize,inner sep=1pt]
\def\R{2.5} 
\def\angEl{35} 
\def\angAz{-105} 
\def\angPhi{-40} 
\def\angBeta{19} 
\pgfmathsetmacro\H{\R*cos(\angEl)}
\tikzset{xyplane/.estyle={cm={cos(\angAz),sin(\angAz)*sin(\angEl),-sin(\angAz),
                              cos(\angAz)*sin(\angEl),(0,\H)}}}

\filldraw[ball color=white] (0,0) circle (\R);
\foreach \t in {-80,-60,...,80} { \DrawLatitudeCircle[\R]{\t} }
\DrawLongitudeCircle[\R]{-30}
\DrawLongitudeCircle[\R]{60}

\draw[xyplane] (-1.5*\R,-1.5*\R) rectangle (1.5*\R,1.5*\R);

\draw[xyplane][fill=lightgray,draw=red,opacity=.3,very thin,line join=round]
 (-1.5*\R,-1.5*\R,0) --
 (-1.5*\R,1.5*\R,0) --
 (1.5*\R,1.5*\R,0) --
 (1.5*\R,-1.5*\R,0) --cycle ;
\draw[xyplane][->] (0,0) -- (0.6,1.3*\R) node[above] {$\vec{e}_2$};
\draw[xyplane][->] (0,0) -- (1.3*\R,-0.6) node[left] {$\vec{e}_1$};

\draw[xyplane][-latex,thick] (-2,-2) node[above,scale=1.0]{$O$}
        to[out=90,in=180] (0, 0);

\draw[xyplane] node at (-1,-3) {$TM_O$};

\node[below] at (1.085,1.3) {$B$};
\node[draw,circle,minimum size=1mm,inner sep=0pt,outer sep=0pt,fill=black] at (1.085,1.4) {};
\node[below] at (-0.55,1.1) {$A$};
\node[draw,circle,minimum size=1mm,inner sep=0pt,outer sep=0pt,fill=black] at (-0.63,1.15) {};

\begin{scope}[xshift=8cm,yshift=0cm]
\draw[->,black,thick] (-2,1) -- (-4,1.5) ;
\node[above] at (-2.5,1.5) {$p(x_1,...,x_n)$};

\draw[->,black,thick] (-3,0) -- (3,0) node[right]{$x_2$};
\draw[->,black,thick] (0,3) -- (0,-3) node[right]{$x_1$} ;
\draw[color=black]  (0,0) circle (1);
\draw[color=black]  (0,0) circle (2);
\node[right] at (0,0.2) {$O$};
\node[above right] at (2,0) {$B$};
\node[draw,circle,minimum size=1mm,inner sep=0pt,outer sep=0pt,fill=black] at (2,0) {};
\node[below right] at (0,-2) {$A$};
\node[draw,circle,minimum size=1mm,inner sep=0pt,outer sep=0pt,fill=black] at (0,-2) {};
\end{scope}

\end{tikzpicture}
\end{figure}}